\documentclass{article}
\usepackage{geometry}
\usepackage{graphicx}
\usepackage{hyperref}
\usepackage{subcaption}
\usepackage{multirow}
\usepackage{makecell}
\usepackage[numbers,comma,square,sort&compress]{natbib}
\graphicspath{{cll_cg_images/}}
\begin{document}

\begin{center}
\LARGE Analytical coarse grained potential parameterization by\\
Reinforcement Learning for anisotropic cellulose\\
\end{center}

\begin{center}
\renewcommand{\thefootnote}{\fnsymbol{footnote}}
Xu Dong$^{1,}$\footnote{Corresponding author, Email: donx@zuaa.zju.edu.cn}\\
$^{1}$\textit{Department of Engineering Mechanics, Zhejiang University, Hangzhou 310027, China}
\end{center}

\begin{center}
Keywords: Reinforcement Learning, Boltzmann Inversion, coarse grained, cellulose, anisotropy, hydrogen bonds
\end{center}

\section*{Abstract}
\addcontentsline{toc}{section}{\protect\numberline{}Abstract}\indent

Cellulose nanocrystals (CNCs) are a type of cellulose with excellent mechanical performance and other advantageous attributes.
According to previous reports, hydrogen bonds play a pivotal role in the anisotropic structure of the CNC.
Understanding the structure and mechanical behavior of CNC on a mesoscopic scale is critical for the development and manufacture of cellulose materials.
However, experimental observations and atomistic simulations are not appropriate on the mesoscopic scale.
In this study, we introduce an analytical coarse-grained (CG) potential following an extended bottom-up approach that is directly parameterized using Reinforcement Learning (RL).
RL is a powerful tool for industrial and academic applications in various fields.
Nevertheless, the potential of RL has not yet been fully exploited in the field of molecular dynamics.
The RL and Boltzmann Inversion methods were employed to develop a novel CG model of cellulose to represent its anisotropy and polymer stiffness.
The resulting approximate CG model is not limited to the specific properties used for training; it can reproduce the dynamic mechanical properties under various conditions without further optimization.
This model confirms that RL can construct a CG potential that is both physically interpretable and powerful.
The training code is available at $\textrm{https://github.com/EiPiFun/rl-cll-cg}$.

\section*{Introduction}
\addtocounter{section}{1}
\setcounter{subsection}{0}
\addcontentsline{toc}{section}{\protect\numberline{}Introduction}\indent

\subsection{Cellulose nanocrystals}\indent

As the most abundant natural polymer on Earth, cellulose is a widely available and environmentally friendly resource, obtained from wood, bamboo, algae, and bacteria\cite{moon2016overview,kim2015review}.
Renewability, low cost, biocompatibility, biodegradability, and good mechanical performance\cite{teo2020towards,moon2011cellulose} are the outstanding merits of cellulose, making it a promising alternative to petroleum-based materials.
As a nano-form of cellulose, cellulose nanocrystals (CNCs) can exhibit superior properties, such as high elastic modulus (approximately 200 GPa), low mass density (1.6 $\rm{g/cm^3}$), high specific surface area (several hundreds of $\rm{m^2/g}$), and low coefficient of thermal expansion\cite{moon2011cellulose}.
Owing to their excellent mechanical properties and interactions with other ions and polar molecules\cite{zhu2020stimuli}, CNCs have been widely studied and employed in biomedicine\cite{du2019cellulose}, water treatment\cite{yu2021water}, energy storage\cite{liu2021cellulose}, and humidity sensors\cite{wang2020flexible}.
Computational simulations, including ab initio simulations\cite{reis2020dft}, molecular dynamics\cite{zhu2020evaluation}, and finite element method simulations\cite{chen2016linear}, were performed to reveal the molecular details and kinetics of CNC.
Experimental\cite{diddens2008anisotropic,iwamoto2009elastic,wagner2016mechanical}, computational\cite{dri2013anisotropy,wu2014tensile,chen2016linear}, and analytical\cite{meng2018effects,meng2017multiscale} approaches have been exploited to understand the mechanical properties of CNC.
Hydroxyls and hydrogen bonds (HBonds) play pivotal roles in these behaviors\cite{shen2009stability,djahedi2016role,wohlert2022cellulose}.

\begin{figure}[htbp]
    \centering
    \begin{subfigure}[b]{0.24\textwidth}
        \centering
        \includegraphics[width=0.96\textwidth]{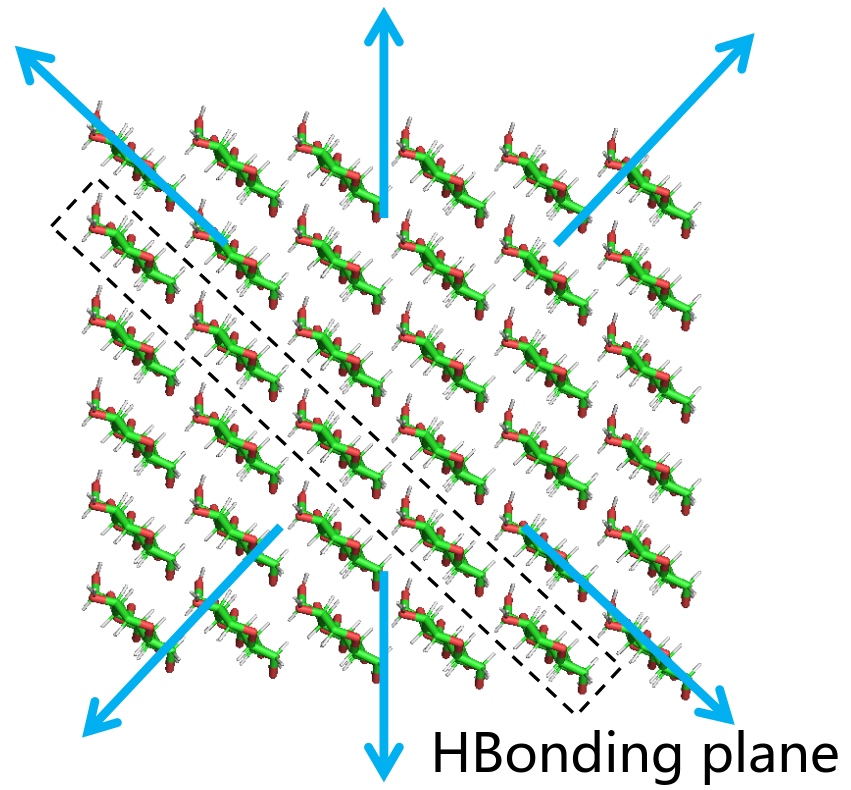}
        \subcaption{}
    \end{subfigure}
    \begin{subfigure}[b]{0.72\textwidth}
        \centering
        \scriptsize
        \begin{minipage}[b]{0.32\textwidth}
            \centering
            \includegraphics[width=0.35\textwidth]{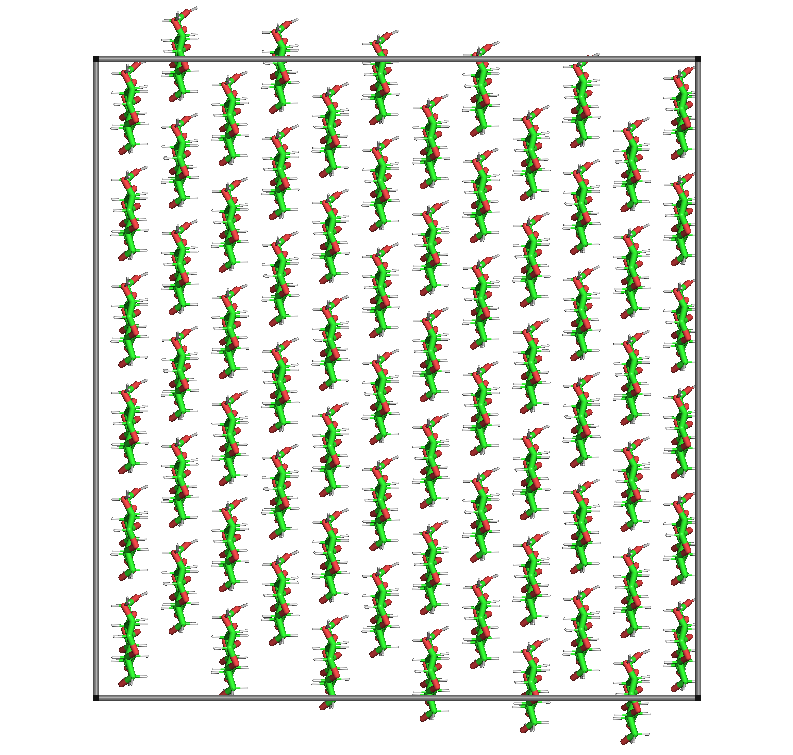}
            \includegraphics[width=0.35\textwidth]{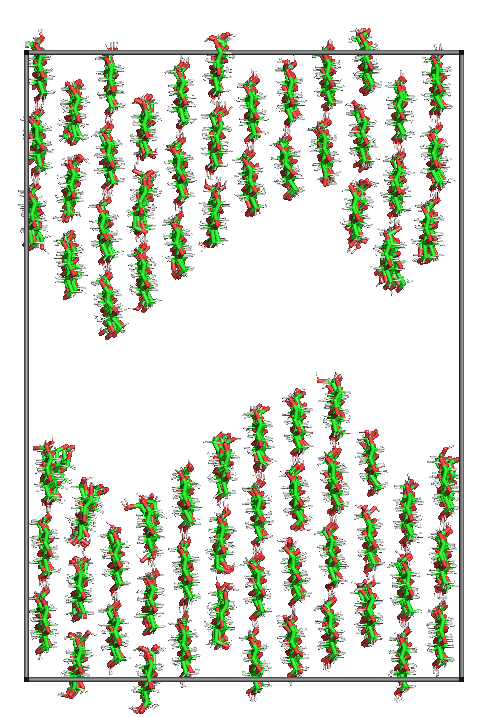}
            \\
            Vertical
        \end{minipage}
        \begin{minipage}[b]{0.32\textwidth}
            \centering
            \includegraphics[width=0.38\textwidth]{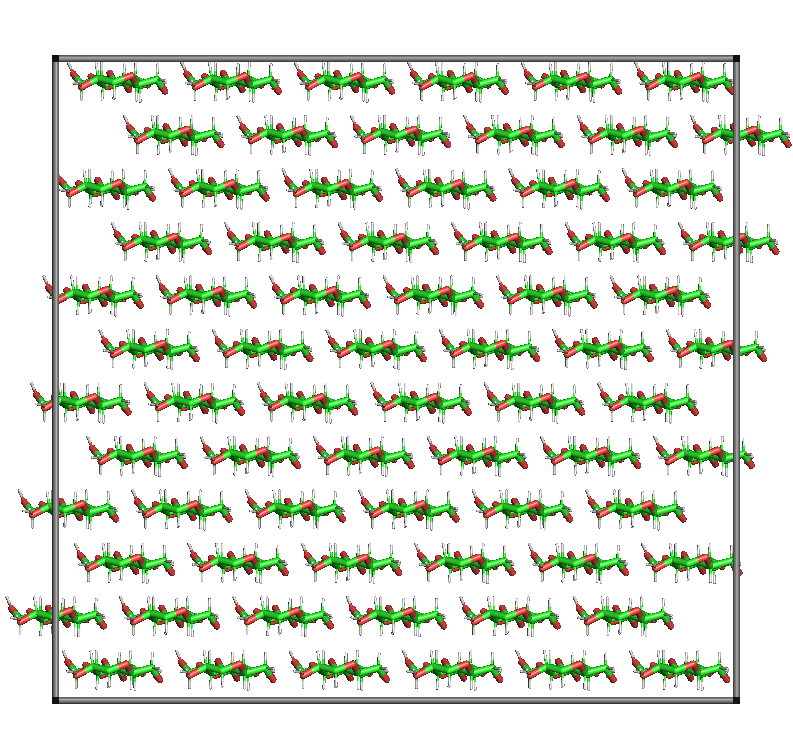}
            \includegraphics[width=0.38\textwidth]{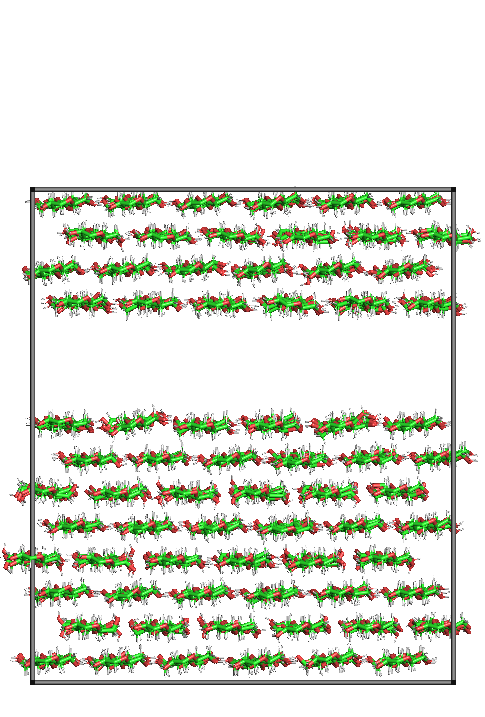}
            \\
            Horizontal
        \end{minipage}
        \begin{minipage}[b]{0.32\textwidth}
            \centering
            \includegraphics[width=0.48\textwidth]{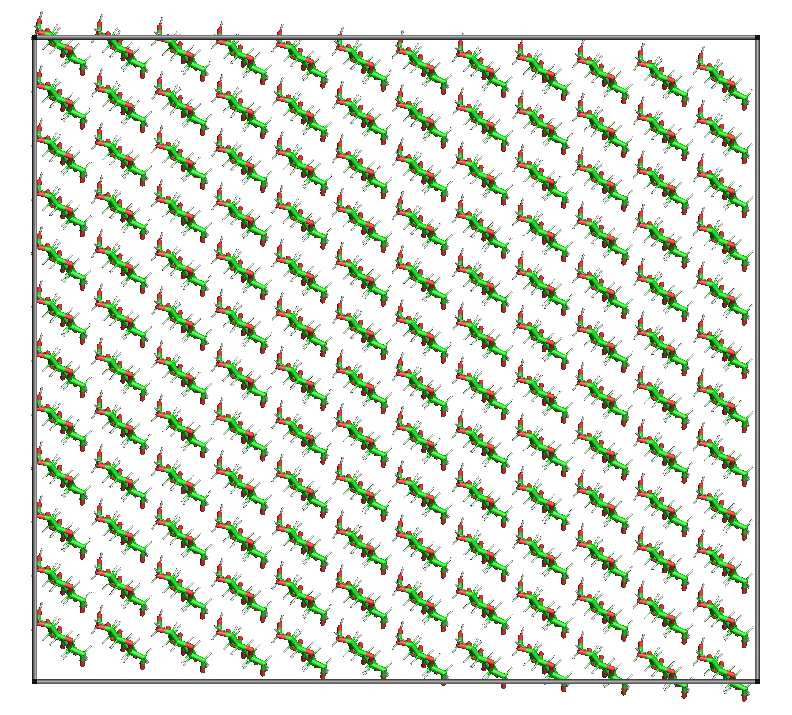}
            \includegraphics[width=0.48\textwidth]{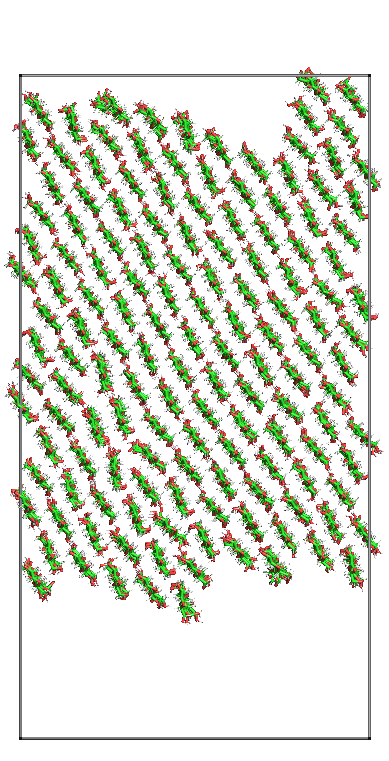}
            \\
            Slant
        \end{minipage}
        \subcaption{}
    \end{subfigure}
    \captionsetup{font=scriptsize}
    \caption{
    Orthotropic transverse section and characteristic directions.
    (a) Characteristic directions of orthotropic transverse sections dominated by HBonding planes.
    (b) Behaviors in characteristic directions.
    Flat residues and HBonds are critical for cellulose anisotropy.
    The vertical and horizontal models are dominated by HBonds and van der Waals interactions, resulting in brittle failure.
    The load of the slant model induced friction and rotation of cellulose residues.
    }
    \label{fig:orthotropic_cross_section_and_characteristic_directions}
\end{figure}

CNCs are composed of parallel cellulose chains and their axial elastic modulus is much larger than their transverse modulus.
The covalent bonds along the cellulose chains are much stronger than the van der Waals and hydrogen bonding (HBonding) interactions in the transverse section.
The axial modulus and transverse modulus of CNC are approximately 100-200 GPa and 10-50 GPa, respectively, as confirmed by experiments\cite{moon2011cellulose} and simulations\cite{dri2013anisotropy,wu2014tensile}.
In the transverse section, the CNCs are orthotropic and exhibit three characteristic load directions: along, perpendicular to, and slanted to the HBonding planes, as shown in Figure~\ref{fig:orthotropic_cross_section_and_characteristic_directions}.
Flat residues and HBonds are the key factors in the transverse anisotropy of cellulose.
Wu et al. emphasized the anisotropy of CNC, and they simulated the stretch in three orthogonal directions (along the chain, parallel, and perpendicular to the HBonding planes) at three different strain rates and found that the mechanical properties (elastic modulus and Poisson's ratio) are highly anisotropic and independent of the strain rate\cite{wu2014tensile}.
Under vertical loading, the slant CNC model exhibits ductility owing to frictional sliding, whereas the vertical and horizontal models exhibit brittle failure.
Friction is an important mechanical behavior of CNC and has been elucidated in previous studies.
The friction of the nearby cellulose layers at different velocities, normal loads, and relative angles was also simulated.
Wu et al. found that the number of HBonds and the distances between layers were important factors in this case\cite{wu2013atomistic}.
Zhang et al. systematically simulated and examined the impact of the loading direction, interfacial moisture, and misalignment on the CNC-CNC interface friction mechanical behavior.
The density of HBonds is strongly correlated with the interaction energy in their results\cite{zhang2021hydrogen}.
The frictional sliding of CNC is critical to their mechanical performance when the loading direction is slanted toward the HBonding planes.

To reflect the performance of CNC on a larger length scale, many coarse-grained (CG) models have been developed with different CG levels for different purposes such as mechanical properties and molecular interactions\cite{mehandzhiyski2021review}.
Phenomena such as self-assembly or cellulose nano networks require supra-molecular CG models, in which several cellulose residues are mapped into a single bead\cite{rolland2020new,mehandzhiyski2020novel,li2020coarse}
If the mechanical properties of the CNC are considered, one cellulose residue (one residue or two nearby residues in the same chain) is usually mapped to a bead\cite{qin2017optimizing,ray2021mechanics}.
The directionality of the transverse sections was neglected.
Some researchers have emphasized molecular details and interactions; thus, they will preserve more molecular details and map one cellulose residue to more than one bead instead of several residues into one bead\cite{lopez2015martini,bu2009energy,wohlert2011coarse,goundla2014coarse}.

However, the CG models of cellulose mentioned previously cannot reveal the HBond interactions and resultant directionality.
Several studies have presented directionality using anisotropic particles\cite{rolland2020new,nguyen2022systematic} or additional bonds that do not correspond to real molecular structures to compensate for the directionality \cite{fan2015coarse,ramezani2019mechanical,shishehbor2019effects,shishehbor2021influence}.
Shishehbor et al. developed a CNC CG model by matching the axial elastic modulus and transverse stress-strain curve to investigate the interface interaction and failure mechanism\cite{shishehbor2019effects}.
The CG mappings used by Shishehbor et al. neglected the flat structure of cellulose residue.
Anisotropic particles and bonded (BD) directionality are unsuitable when a transverse failure mechanism with frictional sliding is considered in the CG model.
Nonetheless, presenting the kinetics, energy, and mechanical properties by CNC CG model is challenging\cite{wohlert2011coarse}.

\subsection{Machine Learning parameterization and Reinforcement Learning}\indent

The data-driven Machine Learning potential is often the primary consideration for combining of Machine Learning models and force fields.
Machine learning models, particularly neural network models, can describe particle interactions without substantial physical or chemical intuitions\cite{noe2020machine,martin2024overview}.
However, significant problems remain for Machine Learning potentials, such as long-range interactions, transferability, and kinetics (particularly for CG models)\cite{noe2020machine}.
Neural network models also have problems with explainability\cite{samek2021explaining} and computational cost.
Machine learning methods can also be used to parameterize analytical force fields\cite{batra2020machine,befort2021machine}.
Optimization methods, including Bayesian Optimization\cite{sestito2020coarse}, particle swarm optimization\cite{duan2019machine}, genetic algorithm\cite{chan2019machine}, and Gaussian process regression\cite{razi2020force} have been used to parameterize different analytical CG potentials.

The use of Reinforcement Learning (RL) in CG simulations and potential parameterization is still in its early stages\cite{chandra2023reinforcement}.
Wu et al. integrated graph attention neural networks and RL to learn CG mapping\cite{wu2022rlcg}.
Chandra et al. used a continuous action Monte Carlo Tree Search\cite{manna2022learning} powered by RL to parameterize an analytical hybrid bond-order CG force field to simulate the aggregation of liquid-liquid mixtures\cite{chandra2023reinforcement}.
RL imitates the learning of creatures and is promising for control and optimization\cite{recht2019tour,liu2024predicting}.
The model-free RL algorithm can deal with strongly nonlinear problems without prior expert knowledge and automatically perform trial-and-error learning procedures.
More information on the RL scheme is presented in the methods section.

In summary, flat residues and HBond interactions are key features of CNC orthotropic transverse sections and their corresponding transverse mechanical behaviors, which have not been well described in previous CG models.
With a tight budget for computational cost and a desire for better explainability, transferability, and kinetics, a carefully designed model with explicit analytical potential, instead of a data-driven potential model, was developed to reproduce the orthotropic mechanical behavior of the CNC.
In this study, we introduce a novel CNC CG model parameterized by RL that considers flat residues, HBonds, polymer stiffness, and anisotropic mechanical performance.
HBonds were implemented as nonbonded (NB) interactions with directionality in this model.

\subsection{Coarse grained molecular dynamics}\indent

CG models focus on essential features defined by researchers, and compromise between accuracy/scale and computational efficiency compared with all atom (AA) models, such as the CG models of CNC on larger spatial and time scales\cite{noid2024rigorous}.
Together with the reduction of the degrees of freedom by the mapping process (replacing several atoms with a bead) and the often elimination of electrostatic interactions, the smoother and shorter energy surface from averaging over the atomic structures allows them to use a much longer integration timestep.
The combination of reduced degrees of freedom, a smoother energy surface, and a longer timestep significantly accelerates simulations and facilitates the sampling of collectively emergent mesoscopic behaviors or observable macroscopic properties\cite{jin2022bottom}.
Therefore, CG models demonstrate the relationship between the molecular scale and performances on larger scale.
Although ``coarse-grained'', they are ``fine-grained'' compared to continuum models and provide insights into molecular fluctuations and interactions.
Well-designed CG models are particularly helpful in understanding the essential aspects of physical phenomena.
However, the shortcomings also arise from the reductions: dynamics, non-equilibrium processes, consistency, representability, and transferability\cite{kmiecik2016coarse}.
Most CG models concentrate on equilibrium trajectories.
Pressure is an example of these problems: with a much lower number of particles and velocities, and even with the help of Brownian motions/Langevin dynamics, it is still difficult to confirm the pressure consistency considering entropy and enthalpy, particularly when transferred to a new environment.
The representability problems of CG models particularly for thermodynamics are the research fronts of many scientists\cite{dannenhoffer2019compatible,dunn2016van,jin2019understanding,johnson2007representability,wagner2016representability}.

Top-down and bottom-up approaches are the two main paradigms for CG modeling\cite{noid2013perspective}.
Roughly, top-down emphasizes more on the emergent phenomena whereas bottom-up concentrates on microscopic origins.
Top-down methods are not directly related to microscopic details, they typically utilize simple analytical functions and are difficult to provide quantitative insights.
Bottom-up approaches are usually structure-based, aiming at the properties of statistical mechanics including many body potential of mean force and mapped probability distributions.
To trace them more accurately, bottom-up methods may utilize more complex functions to provide quantitative descriptions.
The distinction between these two paradigms has become increasingly blurred.
For instance, as a famous top-down CG potential, MARTINI also incorporate elements of bottom-up strategies and reproduced the molecular structure of bilayers, hydrocarbons, and short peptides\cite{souza2021martini}.

To better illustrate the characteristics and advantages of this CG model, several conventional CG modeling methods were compared as baselines in the following sections.
In addition to the top-down MARTINI, bottom-up Iterative Boltzmann Inversion (IBI)\cite{muller2002coarse}, Relative Entropy (RE)\cite{shell2008relative}, and Force Matching (FM)\cite{izvekov2005multiscale} were also compared as baselines.
IBI is a structure-based correlation function approach that reproduces the atomic radial distribution function.
In IBI, all interactions are assumed to be independent; thus, IBI is often appropriate for ``dilute'' NB pairs and ``stiff'' BD interactions.
RE is a variational approach to iteratively minimize the Kullback-Leibler divergence of the mapped probability distribution (the configuration-dependent entropy) between AA and CG.
FM is a non-iterative variational method and can be directly solved, which was initially designed for the mean force of NB pairs and can be extended for BD interactions\cite{noid2008multiscale,das2012multiscale}.
Therefore, the FM does not guarantee the reproduction of atomic distributions.
More descriptions and comparisons of CG modeling methods can be found in reviews\cite{noid2024rigorous,jin2022bottom,kmiecik2016coarse,noid2013perspective}.

Although MARTINI and IBI have already been applied to cellulose\cite{wohlert2011coarse,lopez2015martini,grunewald2022martini,srinivas2011solvent}, prior researches have not concentrated on the anisotropic mechanical properties, which will be further presented in the following sections.
Considering the necessity of flat residues and HBond interactions, this CG model of CNC follows an extended bottom-up approach, focusing on the residue shape, directional many body interactions, and the mapped probability distributions of bead pair geometries.
The bottom-up approach in this work is extended by mixing top-down performance-oriented philosophy: it uses simple analytical potential functions and parameterizes them to reproduce the mechanical properties under external loads, in an attempt to explore the challenging dynamics on the CG scale.
Owing to the representability problem and theoretical difficulty in providing a precise description of the dynamics behavior rationality in the CG model, this is only an approximate approach to preserve performances under equilibrium simulations and stretch processes.
From the perspective of CG modeling, this approximate approach combines the idea of bottom-up and top-down, and considers the properties of CNC under both equilibrium and non-equilibrium conditions.

\section*{Methods}
\addtocounter{section}{1}
\setcounter{subsection}{0}
\addcontentsline{toc}{section}{\protect\numberline{}Methods}\indent

\subsection{Mapping and topology}\indent

\begin{figure}[htbp]
    \centering
    \begin{subfigure}[b]{0.32\textwidth}
        \centering
        \includegraphics[width=0.96\textwidth]{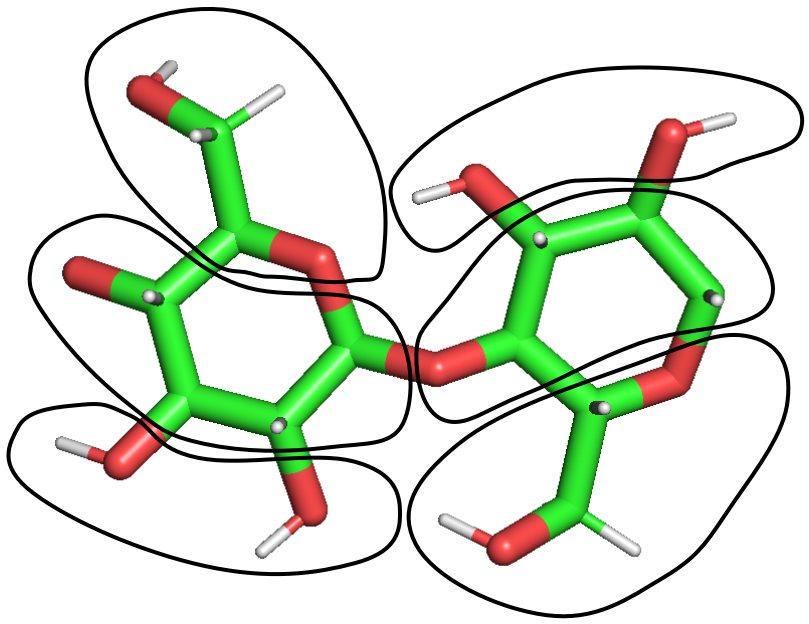}
        \subcaption{}
    \end{subfigure}
    \begin{subfigure}[b]{0.32\textwidth}
        \centering
        \includegraphics[width=0.96\textwidth]{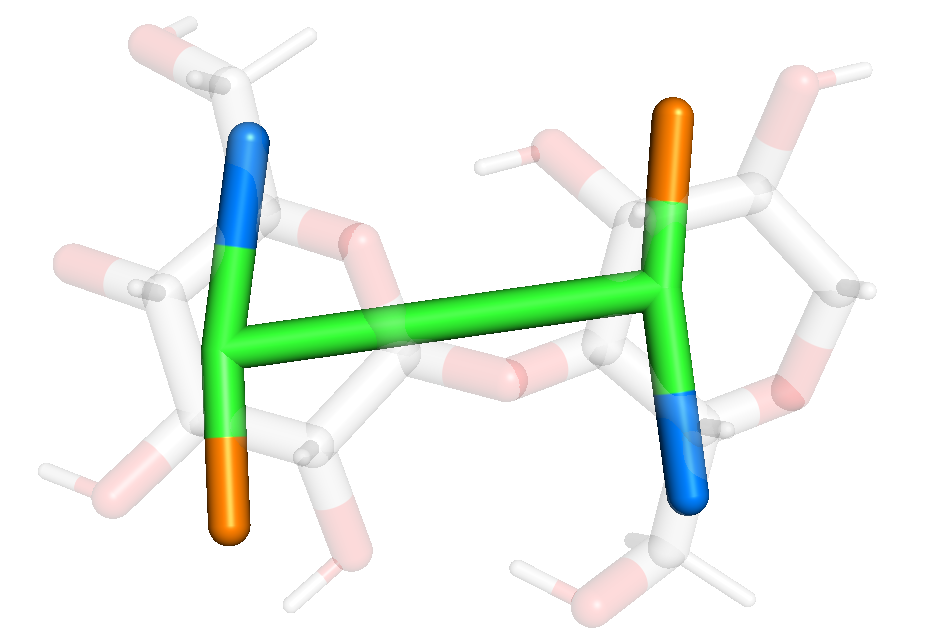}
        \subcaption{}
    \end{subfigure}
    \begin{subfigure}[b]{0.32\textwidth}
        \centering
        \includegraphics[width=0.96\textwidth]{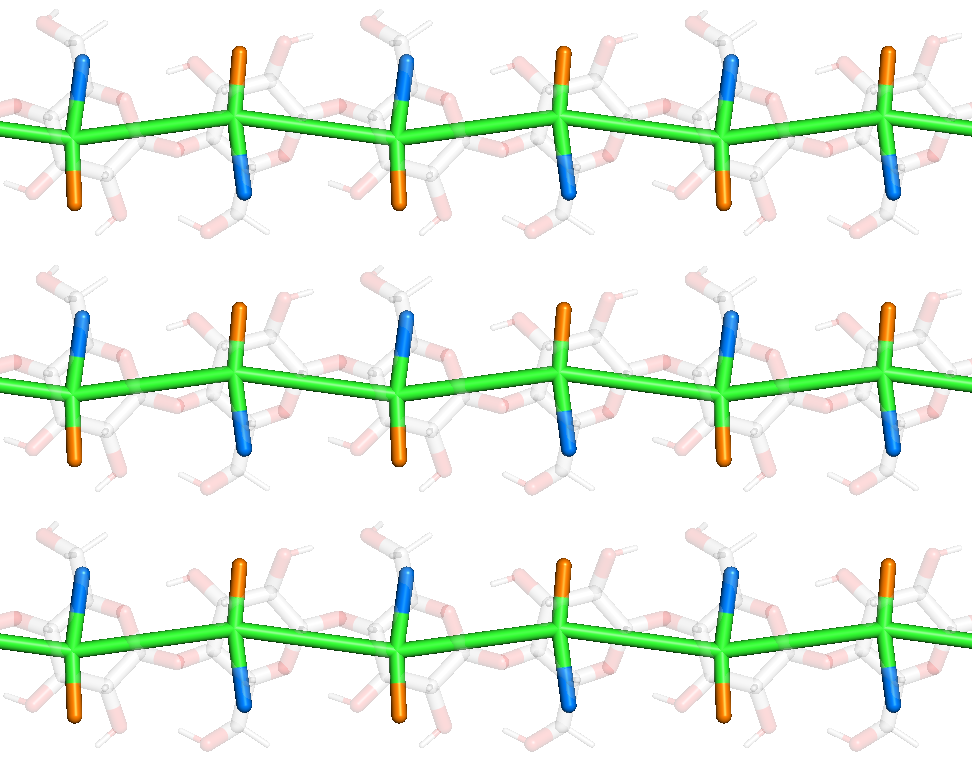}
        \subcaption{}
    \end{subfigure}
    \\
    \begin{subfigure}[b]{0.60\textwidth}
        \centering
        \scriptsize
        \begin{minipage}[b]{0.29\textwidth}\centering\includegraphics[width=0.96\textwidth]{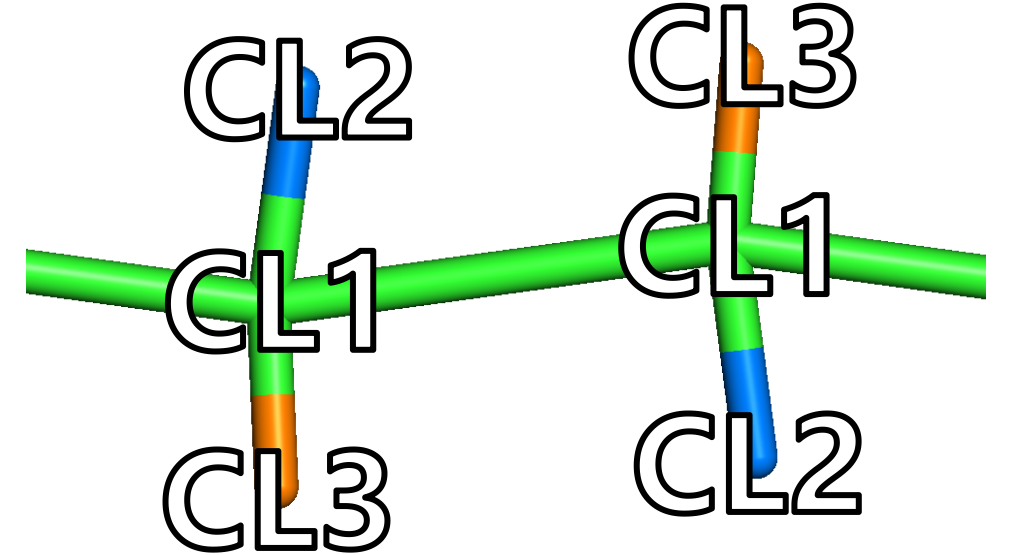}\\Bead\end{minipage}
        \begin{minipage}[b]{0.29\textwidth}\centering\includegraphics[width=0.96\textwidth]{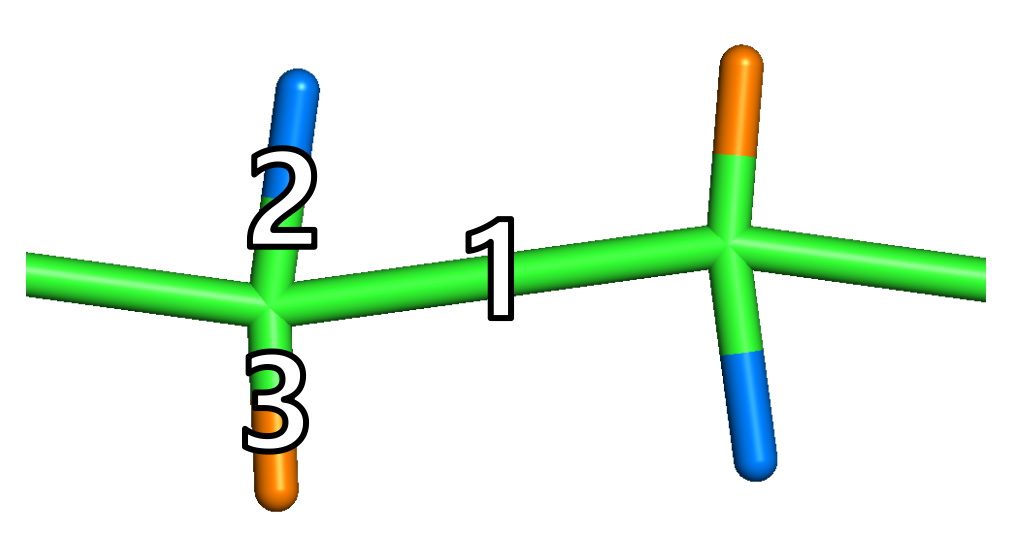}\\Bond\end{minipage}
        \begin{minipage}[b]{0.29\textwidth}\centering\includegraphics[width=0.96\textwidth]{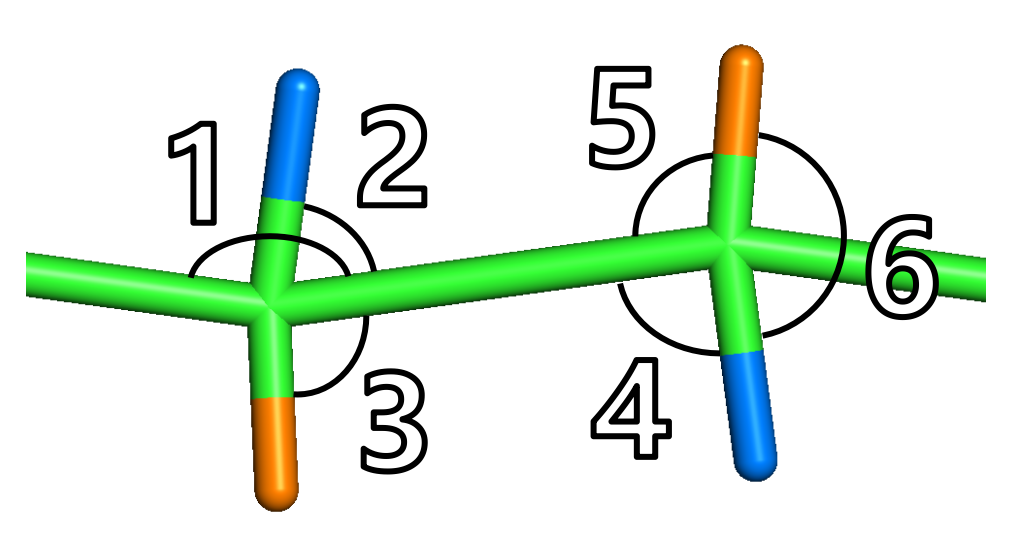}\\Angle\end{minipage}
        \\
        \begin{minipage}[b]{0.96\textwidth}\centering\includegraphics[width=0.90\textwidth]{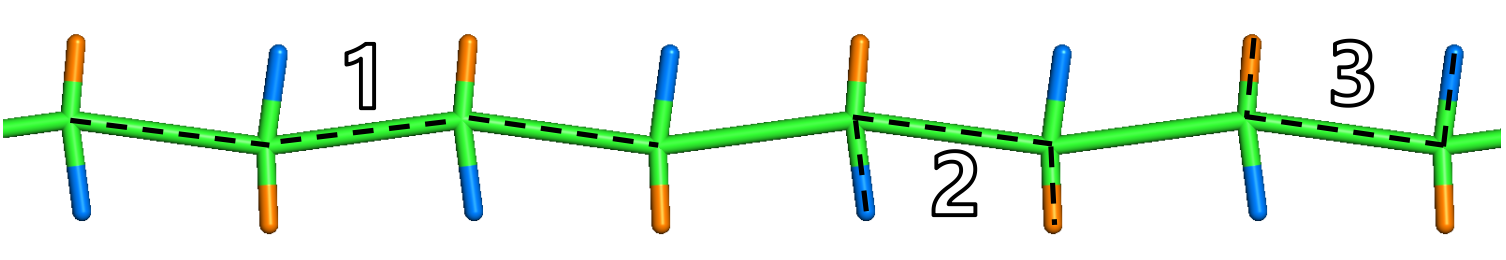}\\Improper\end{minipage}
        \subcaption{}
    \end{subfigure}
    \begin{subfigure}[b]{0.38\textwidth}
        \centering
        \includegraphics[width=0.54\textwidth]{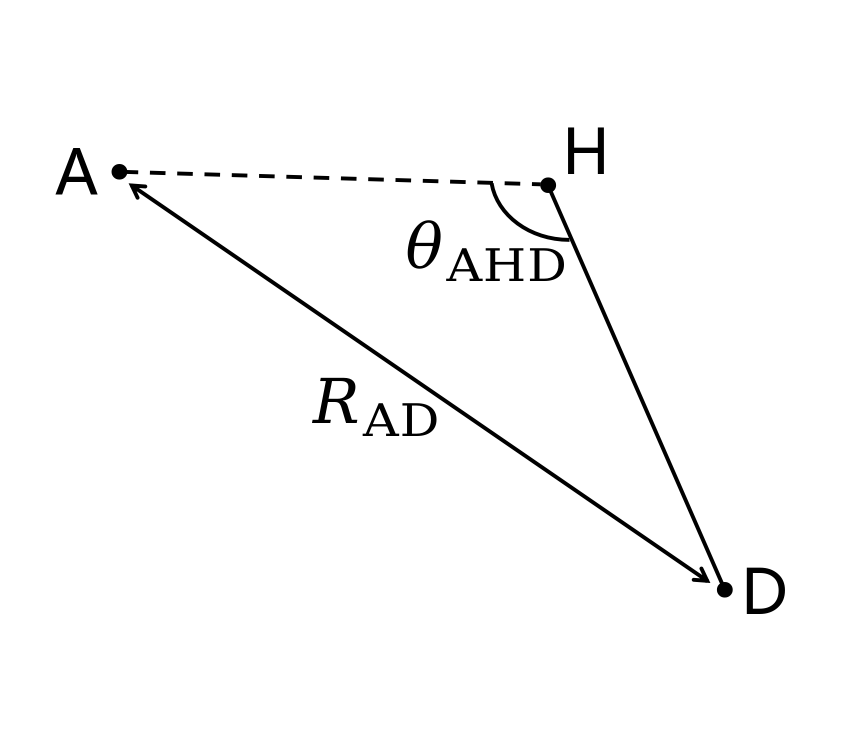}
        \includegraphics[width=0.36\textwidth]{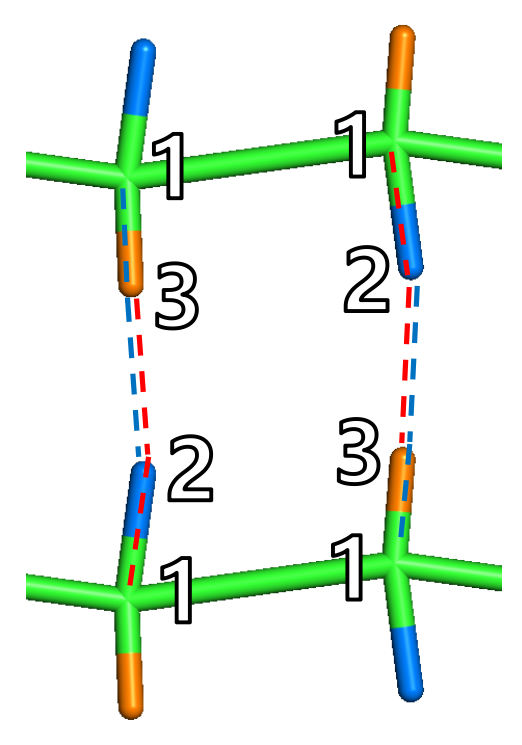}
        \vspace{8.0pt}
        \subcaption{}
    \end{subfigure}
    \captionsetup{font=scriptsize}
    \caption{
    Mapping and topology.
    (a) Mapping and (b) mapped overlays for the cellulose residue and (c) laminar HBonding plane.
    One cellulose residue was mapped to one backbone CL1 bead and two branched CL2 and CL3 beads.
    The basic idea behind this mapping is to preserve the flat structure and HBonds within the HBonding layer.
    Qualitatively, the distance between CL2 and CL3 from the same residue was sufficient and the branched beads were expected to be perpendicular to the backbones.
    Under this mapping, a layer of CNC was mapped as a network structure, where the HBond interactions were stressed.
    (d) BD topology.
    BD interactions are required to preserve molecular structure and performance under stretching, bending, and twisting loads.
    (e) NB topology.
    For HBond (HB) potentials, two symmetric A:H-D relations (CL2:CL3-CL1 and CL3:CL2-CL1) were defined to describe HBonds.
    HB potentials were defined by using both the A:D distance and A:H-D angle.
    }
    \label{fig:mapping_and_topology}
\end{figure}

The mapping emphasizes the flat residues of the cellulose chains and the HBond interactions.
Each cellulose residue was mapped to a single backbone bead (CL1) and two branched beads (CL2 and CL3, Figure~\ref{fig:mapping_and_topology}).
Two features are considered: the distance between CL2 and CL3 in a residue and the angle of CL2-CL3 relative to the backbone.
The first distance feature must be sufficiently large to provide structural rigidity to withstand the torque and present flat geometry of the residues, whereas the second angle feature should be approximately 90$^\circ$.
In the crystalline state, the angles of the nearby CL2:CL3-CL1(blue:orange-green) and CL3:CL2-CL1(orange:blue-green) (colons indicate nonbonded pairs and dashes represent bonds) are close to $180^\circ$, as shown in Figure~\ref{fig:mapping_and_topology}.

Based on this mapping, we applied three types of bonds, six types of angles, and three types of improper dihedrals as the BD interactions (Figure~\ref{fig:mapping_and_topology}).
The BD interactions were designed to describe and restrain the deformations under stretching, bending, and twisting loads.
To ensure restraint in the aperiodic chain (which was used for the polymer stiffness simulations during training), the CL2-CL1-CL1 and CL3-CL1-CL1 angles were categorized into two types, respectively.
A type of directional NB interaction was introduced to describe HBonds, which requires the Acceptor, Hydrogen and Donor beads.
Two A:H-D topologies were determined based on the importance of HBonding and symmetries: CL2:CL3-CL1 and CL3:CL2-CL1.
This directional interaction is affected by the A:H-D angle and A-D distance\cite{thompson2022lammps}.

\subsection{Potentials}\indent

In this model, the potential energy contained both BD and NB interactions:
\begin{equation}V_{System} = V_{Bond}+V_{Angle}+V_{Improper}+V_{Nonbonded}\end{equation}
. The BD interactions were harmonic:
\begin{equation}V_{Bond}=\sum_{} k_{B,i}(r-r_{0,i})^2\end{equation}
\begin{equation}V_{Angle}=\sum_{} k_{A,i}(\theta-\theta_{0,i})^2\end{equation}
\begin{equation}V_{Improper}=\sum_{} k_{I,i}(\chi-\chi_{0,i})^2\end{equation}
. The NB interaction contained Lennard-Jones (LJ) and HBond (HB) potentials\cite{mayo1990dreiding}:
\begin{equation}V_{Nonbonded}=\sum_{} V_{LJ,i}+\sum_{} V_{HB,i}\end{equation}
, in which $V_{LJ}$ and $V_{HB}$ were in the forms
\begin{equation}V_{LJ}=\cases{\epsilon_{LJ}[4{(\frac{\sigma_{LJ}}{r})}^{12}-4{(\frac{\sigma_{LJ}}{r})}^{6}]+V_{shift} &, $0<r \leq r_{in}$\cr
                                                  \int_{r_{in}}^{r}[C_3(x-r_{in})^3+C_2(x-r_{in})^2+C_1(x-r_{in})+C_0]dx+V_0 &, $r_{in}<r \leq r_{c}$\cr}\end{equation}
\begin{equation}V_{HB}=\cases{\epsilon_{HB}[5{(\frac{\sigma_{HB}}{r})}^{12}-6{(\frac{\sigma_{HB}}{r})}^{10}]{{\rm cos}^n (\theta-\theta_0+\pi)} &, $0<r \leq r_{in}\, \&\, \theta-\theta_0+\pi \geq \theta_{c}$\cr
                                                  S(i)\cdot\epsilon_{HB}[5{(\frac{\sigma_{HB}}{r})}^{12}-6{(\frac{\sigma_{HB}}{r})}^{10}]{{\rm cos}^n (\theta-\theta_0+\pi)} &, $r_{in}<r \leq r_{c}\, \&\, \theta-\theta_{0}+\pi \geq \theta_{c}$\cr}\end{equation}
, where:
\begin{equation}S(i)=\frac{{(r_{c}^2-r^2)}^2[r_{c}^2+2r^2-3r_{in}^2]}{{(r_{c}^2-r_{in}^2)}^3}\end{equation}
.

The LJ potential is a 12-6 formula of Lennard-Jones interactions, and the HB potential is a 12-10 formula with scaling by the power of the cosine of the A:H-D angle.
Technically, these are LAMMPS\cite{thompson2022lammps} pair styles.
The LJ potential is the lj/smooth with a shift, whereas the HB potential is a modified version of the hbond/dreiding/lj with a shift.
lj/smooth is designed to smooth force to zero at the cutoff radius using a polylinear expression;
this modified version of hbond/dreiding/lj ensures that potential energy reach minima at $\theta_0$ instead of $\pi$.
It should be emphasized that the smoothing part between $r_{in}$ and $r_{cut}$ ensures the continuity of both force and energy.
The energy and force curves of the HB and LJ potentials are shown in Figure~\ref{fig:nonbonded_energy_curve}.

The equilibrium A:H-D angles in the CG mapping were designed to be close to $\pi$.
Even though we generalize the HB potential by adding an equilibrium angle of A:H-D, an equilibrium angle close to $\pi$ is suggested to stabilize the conical energy surface.

The Lennard-Jones potential was employed at the CG level for simplicity and to facilitate structural stability through its single stationary point.
The 4th power of the cosine in the HB potential was the same as that in the original study which first proposed this formula\cite{mayo1990dreiding}.
For the artifacts (cut, smoothing, and directional HB potential), the sensitivity and ablation checks were performed based on trained coefficients represented in the following sections:
half/double energy coefficients (while distance coefficients were kept), 0.8/1.2 times distance coefficients (while energy coefficients were kept), different exponents of cosine, resetting $\theta_0$ to $\pi$, and removal of HB potential (Table~\ref{tab:sensitivity_and_ablation_check_equilibrium_thermodynamics}, Table~\ref{tab:sensitivity_and_ablation_check_nonbonded_properties}, and Figure~\ref{fig:sensitivity_and_ablation_check_structure_relaxation}).
Sensitivity and ablation tests confirmed the effectiveness of these potentials.

\subsection{Boltzmann Inversion for harmonic potential}\indent

Boltzmann Inversion is structure-based and is widely utilized to determine BD potentials.
Boltzmann Inversion can also be integrated with Machine Learning methods to construct potentials\cite{duan2019machine,matin2024machine} or used as a baseline\cite{khot2023delta,wang2023learning}.

For a harmonic potential following the Boltzmann distribution, where the bond length (or angle, dihedral) is a deviated $x$ around the equilibrium length $x_0$:
\begin{equation}\rho(x+x_0)\propto e^{-\frac{V}{k_BT}}=e^{-\frac{V_x-V_0}{k_BT}}=e^{-\frac{kx^2}{2k_BT}}\end{equation}
. Let $\rho(x_0)=\rho_0$:
$${\rm ln} \rho(x+x_0)=-\frac{kx^2}{2k_BT}+{\rm ln} \rho_0$$
\begin{equation}k=-2k_BT\cdot\frac{d{[{\rm ln} \rho(x+x_0)]}}{d{[x^2]}}\end{equation}
. We then let $\frac{d{[{\rm ln} \rho(x+x_0)]}}{d{[x^2]}}=\kappa$:
\begin{equation}k=-2k_BT\cdot\kappa\end{equation}
. However, the $k$ derived must be reparameterized to represent the mechanical performance\cite{tozzini2007flap,ruhle2009versatile,chandra2023reinforcement} of the CG models with fewer degrees of freedom.
The effective $k$ is $k_0$ times of $\kappa$:
\begin{equation}k_i=k_0\kappa_i\end{equation}
. Based on $\kappa$ values, three $k$ values were used for the bond, angle, and improper force constants.
Considering the axial elastic modulus and polymer stiffness, $k_{B0}$, $k_{A0}$, and $k_{I0}$ were determined via RL optimization.

\subsection{Reinforcement Learning and degenerate RL}\indent

In RL, the entire problem is called the environment and the optimization program is called the agent.
The environment provides at least three types of information: observation, action space, and reward.
In an interaction episode with the environment, the agent selects an action $a_t$ in the action space based on the obtained partial observation $o_t$ of the state of the environment.
The action of the agent impacts the environment evolution and is evaluated using the reward $r_t$ function.
By continuously interacting with the environment, the agent gradually learns to perform actions corresponding to a given observation to maximize the cumulative reward.
The cumulative reward is defined as
\begin{equation}R(\tau)=\sum_{t=0}^{\infty}\gamma^t r_t\end{equation}
with $\gamma\in(0,1)$.
The discount factor $\gamma\in(0,1)$ emphasizes the latest episode and ensures numerical convergence.

\begin{figure}[htbp]
    \centering
    \begin{subfigure}[b]{0.40\textwidth}
        \centering
        \includegraphics[width=0.96\textwidth]{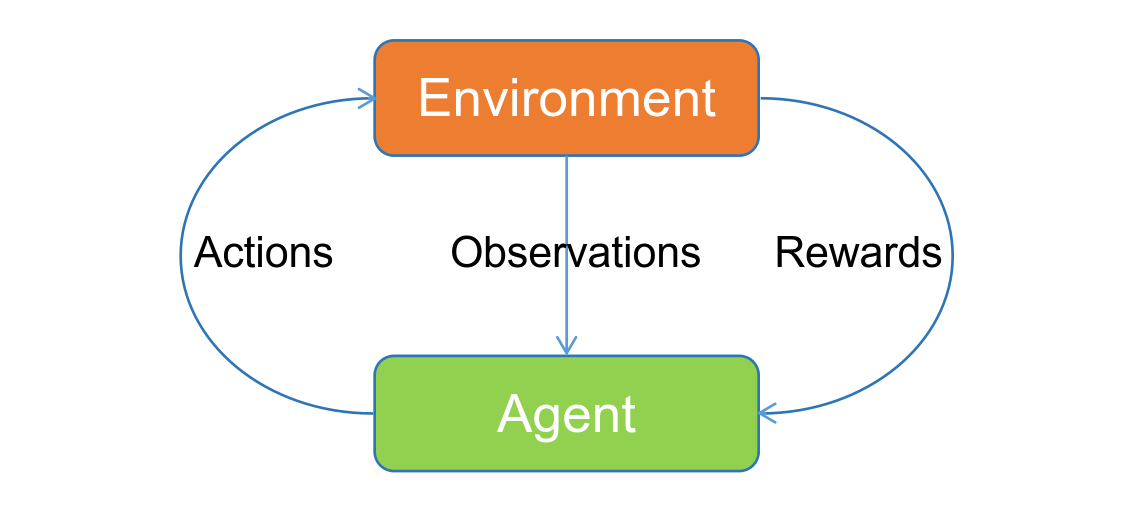}
        \subcaption{}
    \end{subfigure}
    \begin{subfigure}[b]{0.40\textwidth}
        \centering
        \includegraphics[width=0.96\textwidth]{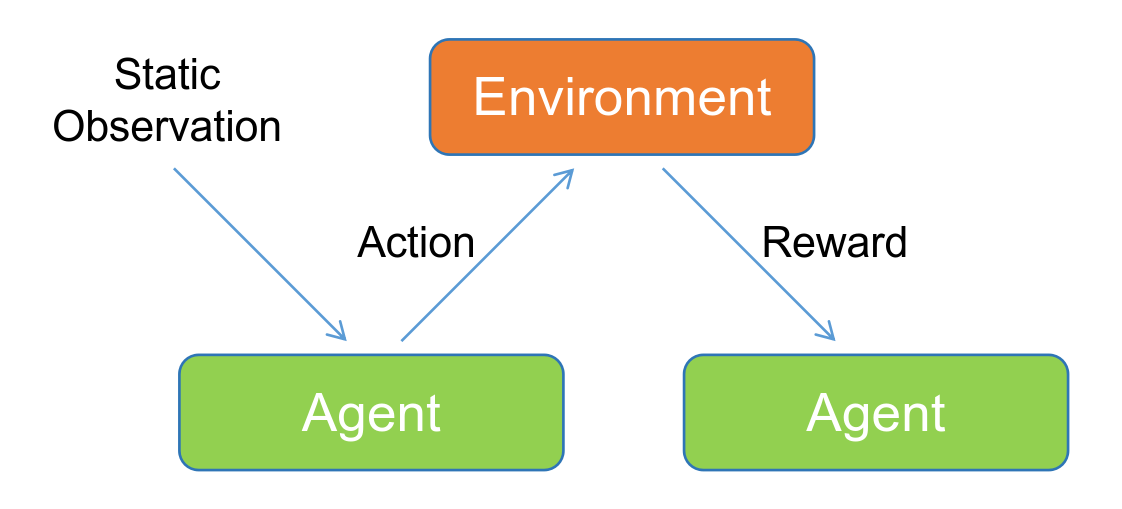}
        \subcaption{}
    \end{subfigure}
    \captionsetup{font=scriptsize}
    \caption{
    RL diagrams.
    (a) Standard RL diagram.
    In the standard RL diagram, the agent interacts with the environment continuously and performs actions corresponding to observations.
    Its optimization target is to maximize the cumulative reward $R(\tau)=\sum_{t=0}^{\infty}\gamma^t r_t$.
    (b) One-shot RL diagram.
    However, a one-shot diagram was applied, and only one single-step episode was processed (marking ``done'' signal to ``True'' for every training step).
    In this study, RL was employed as a nonlinear optimizer, guided by its reward function.
    }
    \label{fig:rl_diagrams}
\end{figure}

In this study, a degenerate version of RL was leveraged in a one-shot diagram (Figure~\ref{fig:rl_diagrams}).
Each learning episode consists of a single step with one sampled reward, because the performance of a coefficient set is independent of the others.
Therefore, the $\gamma$ is effectively set to 0 and the reward can be simplified as
\begin{equation}R(\tau)=r_0\end{equation}
. Consequently, the RL algorithm was exploited as a nonlinear optimizer, guided under indirect supervision by a reward function.
The use of degenerate RL was inspired by the work of Viquerat et al.\cite{viquerat2021direct}, which is rare, particularly for potential parameterizations.

In this study, the RL agent was implemented using the Soft Actor Critic (SAC) algorithm\cite{haarnoja2018soft}.
SAC is the state-of-the-art algorithm recommended for continuous action space by the doc of OpenAI Stable-Baselines3\cite{stable-baselines3}.
The key feature is the reconciliation between the expected reward and entropy, which implies that stochasticity is included in the SAC policy to a certain extent.
SAC is a gradient-free and model-free method that does not rely on a specific problem environment, and can address strongly nonlinear problems after sufficient training.
Detailed information on the SAC algorithm can be obtained in the doc of OpenAI Stable-Baselines3\cite{stable-baselines3}.

In practice, $\gamma$ was maintained at the default value; however, the ``done'' signal was set to ``True'' at every step to restrict the agent to perform a single-step episode.
For the RL agent, the cumulative reward value of an episode is the target to maximize.
In Stable-baselines3\cite{stable-baselines3}, when ``done'' is set to ``True'', one episode was considered to be finished and the RL agent will learn based on its cumulative reward.
While marking ``done'' as ``True'' for every step restricts the agent to a single-step episode;
setting $\gamma$ to 0 is computationally problematic as the cumulative reward is only affected by the last step and previous training steps were meaningless, and the learning is stagnated.
Therefore, stopping and restarting an episode immediately after a step is a practical and accurate method for the degenerate RL approach.
With the only exception of the ``done'' signal, hyperparameters such as the learning rate (0.0003), batch size (256), buffer size (1000000), discount factor (0.99), entropy coefficient (auto), and any other parameters of SAC were not explicitly set and maintained at their default values.

\begin{figure}[htbp]
    \centering
    \includegraphics[width=0.80\textwidth]{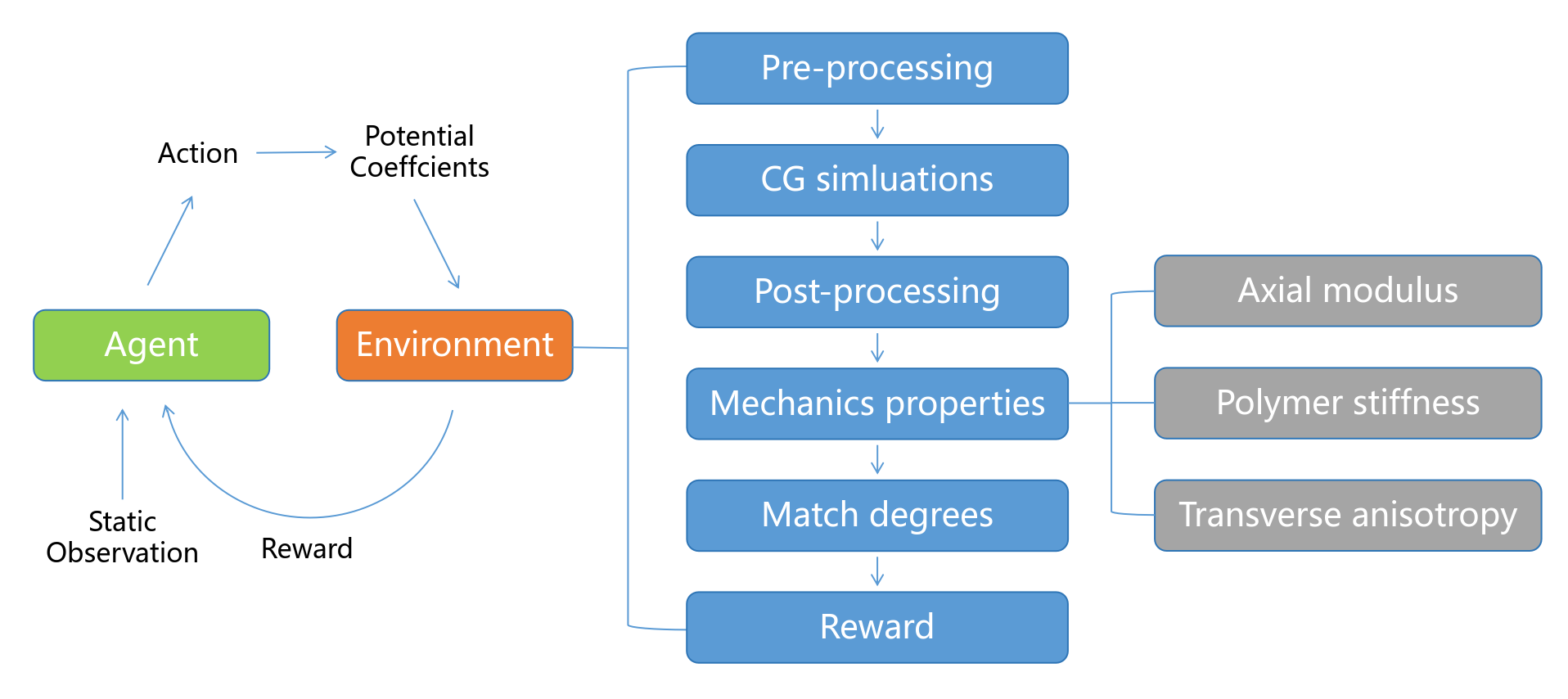}
    \captionsetup{font=scriptsize}
    \caption{
    RL degenerate interaction for this CG model.
    As emphasized in the degenerate diagram, the observation was static for a constant problem, and the crucial interaction is Action-Reward translation.
    The normalized action array was linearly transformed to potential coefficients before being passed to the environment.
    In this environment, simulations with pre- and post-processings were executed to extract the corresponding mechanical properties.
    After that, the match degrees and reward were computed upon the AA reference data.
    Once the observation and reward were received, the agent adjusted its policy and initiated another trial.
    }
    \label{fig:rl_degenerate_interaction}
\end{figure}

During training, the action consists of 17 normalized coefficients ranging from [-1,1] because most RL algorithms rely on standard normal distributions\cite{stable-baselines3}.
The actions are linearly transformed to a physical space, which corresponds to 3 BD $k$, 1 HB $\epsilon$, 1 HB $\sigma$, 6 LJ $\epsilon$ and 6 LJ $\sigma$.
The reward function is a critical factor that guides the learning of the agent, and is defined by a series of matching degrees.
The matching degree $m$ for a specific performance is defined as
\begin{equation}m=\frac{1}{1+|\frac{y_1-y_0}{y_0}|}\end{equation}
with a threshold $m_t$
\begin{equation}M=\cases{m &, $m \geq m_t$\cr
                                           0 &, $m < m_t$}\end{equation}
, where $y_0$ and $y_1$ are the reference and sampled values, respectively.
The reward function $R$ is the sum of a series of matching degrees, except for the transverse strength $M_s$ and toughness $M_t$:
\begin{equation}R=\sum_{}M_i+w\cdot{\rm min}(M_s, M_t)\end{equation}
, because transverse strength and toughness were the hardest to match during the training procedures.
In RL, the reward functions must be accurate and discriminant sufficiently\cite{razin2025makes}.
The weights of $M_s$/$M_t$ and the thresholds were designed to enhance discrimination.
The interaction process between the agent and the environment in this degenerate approach is shown in Figure~\ref{fig:rl_degenerate_interaction}.

\subsection{Simulation setup}\indent

For both the AA and CG simulations, three models in different arrangement directions under the NPT ensemble at 0.1 MPa and 300 K with a stretch load in the vertical direction were simulated.
All reference AA simulations were performed using GROMACS\cite{abraham2015gromacs,bjelkmar2010implementation} and CHARMM36\cite{huang2013charmm36,huang2017charmm36m} force fields (force field files were generated using CHARMM-GUI tools\cite{jo2008charmm,lee2016charmm}).
All the CG simulations were performed using LAMMPS\cite{thompson2022lammps} with a self-defined force field.
The AA systems were governed by Leap-frog, V-Rescale ($\tau_t$ = 0.5~ps), and Berendsen ($\tau_p$ = 1.0~ps), whereas the CG systems were governed by Velocity Verlet, Langevin (damping parameter = 4000~dt), and Nose-Hoover (pressure damping factor = 100~dt).
The molecular structures were visualized using Open-Source PyMOL.

Referring to the performance-size curves obtained by Sinko et al.\cite{sinko2014dimensions}, we used a CNC model comprising 144 cellulose chains.
Fragile vertical and horizontal models with 72 cellulose chains were used to reduce the computational cost of the simulations during training.
Vertical stretches were implemented using cell deformation loads at a constant speed of 10.0~nm/ns.
100 replica AA simulations with varying stretch speed ranging from 0.1~nm/ns to 40.0~nm/ns were performed.
The AA simulation results confirmed that the sampled strength and toughness were stable within this speed range (with deviations of less than 15\%, as shown in Figure~\ref{fig:aa_transverse_stretch_performance_with_speed}).
Figure~\ref{fig:aa_transverse_stretch_curve_collection} shows the stress-strain curves of the 100 replica simulations at 10.0~nm/ns.
The timesteps used in the CG simulations were 4~fs for the BD properties and 12~fs for the NB properties (transverse stretch), chosen as a result of bisection between 10 and 15~fs.
The validation for the CG timestep was provided by a 12~ns CNC equilibrium thermodynamics under NVE and NPT ensembles using a 12~fs timestep (Figure~\ref{fig:timestep_validation_thermodyanmics}).

\subsection{Training procedures and statistics-guided reduction}\indent

\begin{figure}[htbp]
    \centering
    \includegraphics[width=0.80\textwidth]{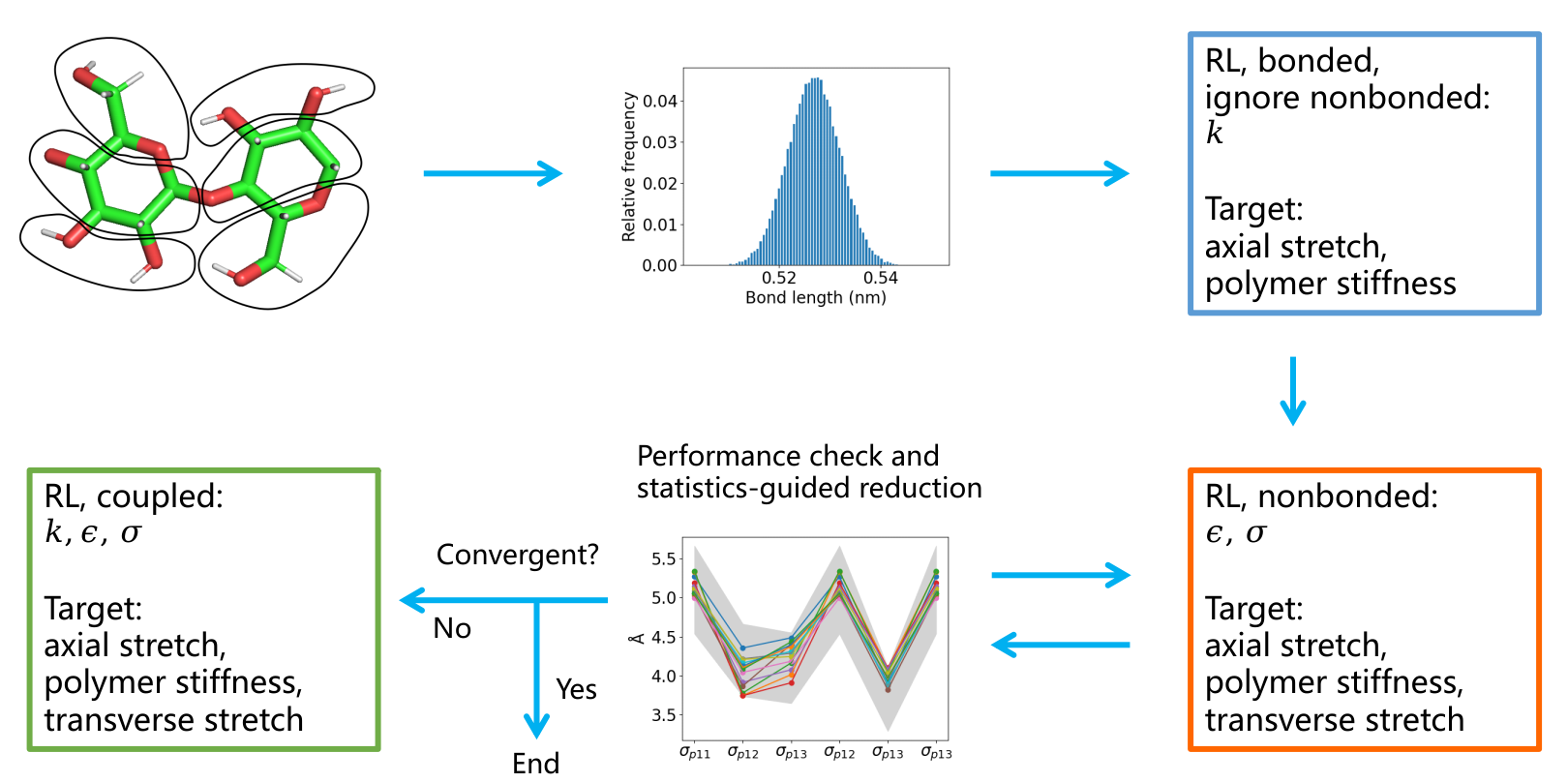}
    \captionsetup{font=scriptsize}
    \caption{
    Training procedures.
    The BD properties (axial elastic modulus and polymer stiffness) and NB properties (transverse strength and toughness in characteristic directions) were the target attributes.
    Before training, the equilibrium BD geometry parameters and force constants of the BD harmonic potentials were estimated from the mapped equilibrium AA trajectories via the Boltzmann Inversion method.
    In the first stage, the force constants were scaled by the RL agent in the BD training to reproduce the BD properties.
    In BD training, all NB interactions were ignored because we assumed that the influences of NB interactions were not significant in this case.
    Given both BD and NB properties, NB training using the previously obtained BD coefficients was performed to match both BD and NB properties.
    However, only one batch of simulations was performed for the target properties during the training.
    Thus, the transverse stretch performance data of the coefficients with the highest reward value were manually verified through replica simulations to confirm the statistical agreement.
    Multiple sets of coefficients were collected to reduce the number of independent coefficients for the computational cost and physical explainability.
    These results confirm that our assumption of the BD properties is reasonable.
    Therefore, the coupled training was unnecessary.
    }
    \label{fig:training_procedures}
\end{figure}

Before training, the equilibrium geometry parameters (bond length $l^e$, angle size $\theta^e$, and dihedral size $\chi^e$) were derived from the mapped equilibrium AA trajectories.
The estimated force constants for the BD harmonic interactions were determined using the same trajectories and Boltzmann Inversion method.
Three rescale factors were required for the bonds, angles, and improper dihedrals in this model.
Although rescaled, the relative ratios of the force constants were maintained (e.g., the force constants for improper dihedrals were rescaled with the same factor).
As previously mentioned, there are 17 coefficients, where 3 are BD coefficients and 14 are NB coefficients.
To simplify the problem and reduce computational cost, three training stages were designed (Figure~\ref{fig:training_procedures}).
The first stage was BD training for three BD $k$ only, ignoring all NB interactions and targeting at the BD properties (axial elastic modulus and polymer stiffness).
We cannot determine the influence of NB interactions in the first stage, and we assume that they are not critical for the BD properties.
The second stage was NB training for 14 NB coefficients with the BD coefficients obtained from the first stage, targeting at both the BD and NB properties.
The third stage was preserved only if the performance of the second stage was unsatisfactory, in which all the properties were targeted to parameterize the 17 coefficients at every turn.

The training results indicated that NB interactions did not play a pivotal role in the target BD properties.
Therefore, our training ended in the second stage.

However, there was also important artificial intervention during the training process.
The transverse stretch performance was sampled only once to control the cost, which was insufficient to ensure accuracy of the distribution levels between the AA and CG models.
Therefore, manual tests for the optimized coefficients with the highest reward value were required to verify the performance distribution, and the optimized coefficients were analyzed, inspired by Befort et al.\cite{befort2021machine}.
Only coefficients with significant tendencies rather than those without clear expectations were retained as independent coefficients.
These statistics-guided reduction procedures are described in the ``Results'' section.

\section*{Results}
\addtocounter{section}{1}
\setcounter{subsection}{0}
\addcontentsline{toc}{section}{\protect\numberline{}Results}\indent

\subsection{Coefficients}\indent

The distance coefficients ($\sigma$) of the NB interactions were estimated using the equilibrium distance of the mapped AA trajectories.
The equilibrium distance $d^e$ is defined as the average of the minimum distances for the NB bead pairs:
\begin{equation}d^e=<d^e_t>=<<d^e_{tm}>>=<<{\rm min}(d^e_{tmn})>>\end{equation}
. For the No. $m$ CL1 bead at time $t$, there are $n$ CL1 beads that are not connected to it by bonds, and the minimum distance $d^e_{tm}={\rm min}(d^e_{tmn})$ is sampled for this bead.
For this frame, the mean $d^e_t=<d^e_{tm}>$ of all the minimum NB distances is determined for time $t$.
The average equilibrium distance in all trajectories $d^e=<d^e_t>$ was defined as the equilibrium distance.
As AA and CG simulations were performed under the NPT ensemble at 0.1 MPa thereby the estimation of NB $\sigma$ was ranged in [0.89$d^e$, 1.00$d^e$] , which contains the stationary point of the LJ potential $2^{-\frac{1}{6}}$.
This range was extended to [0.89$d^e$, 1.11$d^e$] as the action space.

\begin{figure}[htbp]
    \centering
    \begin{subfigure}[b]{0.32\textwidth}
        \centering
        \includegraphics[width=0.96\textwidth]{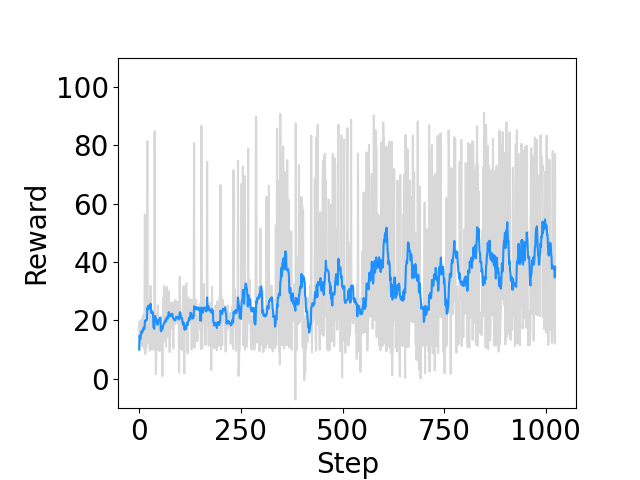}
        \subcaption{}
    \end{subfigure}
    \begin{subfigure}[b]{0.32\textwidth}
        \centering
        \includegraphics[width=0.96\textwidth]{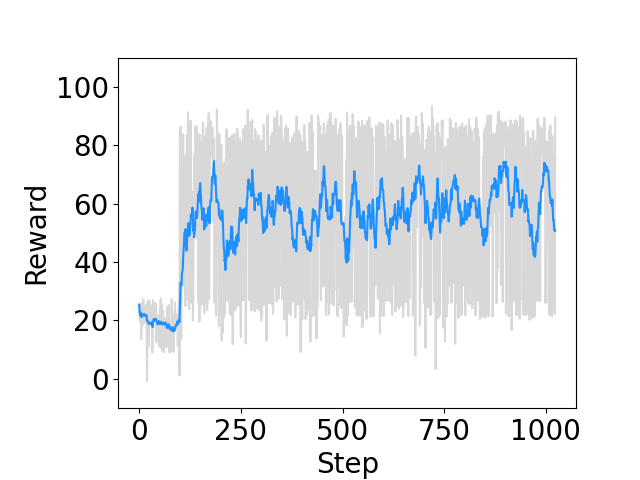}
        \subcaption{}
    \end{subfigure}
    \begin{subfigure}[b]{0.32\textwidth}
        \centering
        \includegraphics[width=0.96\textwidth]{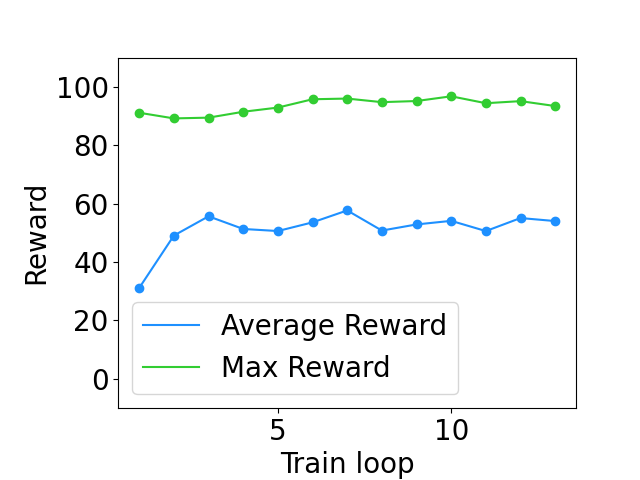}
        \subcaption{}
    \end{subfigure}
    \\
    \begin{subfigure}[b]{0.45\textwidth}
        \centering
        \includegraphics[width=0.96\textwidth]{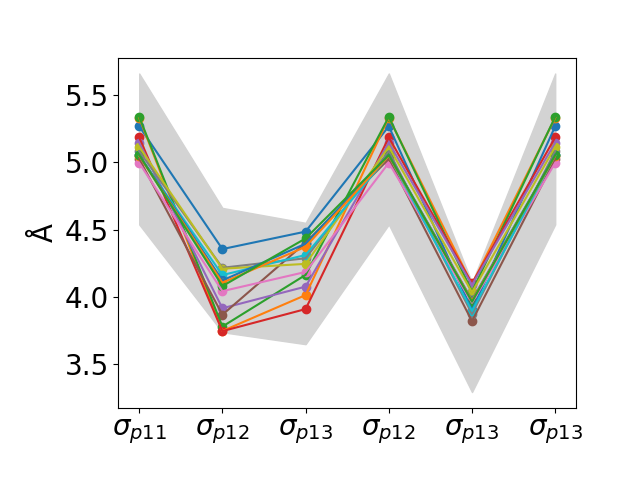}
        \subcaption{}
    \end{subfigure}
    \begin{subfigure}[b]{0.45\textwidth}
        \centering
        \includegraphics[width=0.96\textwidth]{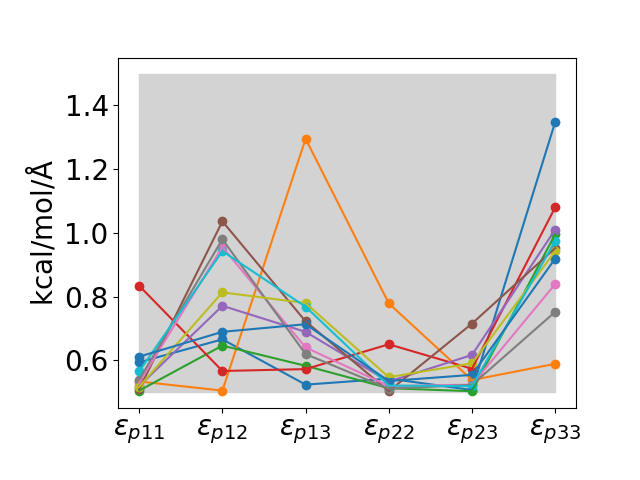}
        \subcaption{}
    \end{subfigure}
    \captionsetup{font=scriptsize}
    \caption{
    Training convergence and statistics-guided reduction.
    (a) Reward and its moving average in the first and (b) last training loops.
    The quantitative convergence criterion was the ratio of the maximum reward of the last training loop to that of the earlier training loops:
    $C_n=\frac{R^{max}_n}{R^{max}_{1:n-1}}$
    . The selectivity criterion was the highest reward: coefficients were chosen among those with the highest rewards from the training loops.
    Each training loop contained 1024 single-step training episodes.
    One trial took about 4 minutes and the overall training took approximately 40 days.
    The 4 minutes were consumed by the 75000 timesteps for the slant model stretch while other tasks were performed in parallel.
    The subsequent training loops with higher and more stable rewards were based on the previous training loops.
    (c) Average and maximum rewards for different training loops.
    The rewards confirmed the convergence.
    After training, the best coefficients for the different training loops were collected to perform statistics-guided reduction.
    The first reduction was inferred from the geometric characteristics of the CG models, and a constraint of $\sigma_{p11}=\sigma_{p22}=\sigma_{p33}$ was applied ($pmn$ represents a bead pair of CLm and CLn), which is also helpful for fragile behaviors in the vertical and horizontal directions.
    (d) The second reduction was in the distance coefficients using statistics.
    $\sigma$ except for $\sigma_{p23}$ oscillated around its equilibrium distance $d^e$ whereas an obvious increase of $\sigma_{p23}$ was observed.
    Then only $\sigma_{p11}$ and $\sigma_{p23}$ were kept as independent distance coefficients: $\sigma_{p12}=\frac{d^e_{p12}}{d^e_{p11}}\sigma_{p11}, \sigma_{p13}=\frac{d^e_{p13}}{d^e_{p11}}\sigma_{p11}$.
    (e) The third reduction in the energy coefficients was also statistically determined.
    After the first and second reduction, we observe that $\epsilon_{p23}$ is usually the smallest energy coefficients, which may correspond to the $\sigma{p23}$ enlargement.
    Then only $\epsilon_{p23}$ and $\epsilon_{p23}$ were kept as independent energy coefficient: $\epsilon_{p11}=\epsilon_{p22}=\epsilon_{p33}=\epsilon_{p12}=\epsilon_{p13}$.
    }
    \label{fig:convergence_and_statistics}
\end{figure}

The coefficients were trained using several training loops, each containing 1024 single-step training episodes.
When a training loop was completed, the information of the agent was saved and loaded for subsequent trainings.
Considering the time consumption, 4 minutes for one trial and 3 days for a training loop were spent.
Several computation tests were designed to start in parallel, and the 75000 timesteps of slant model stretch simulation were responsible for the 4 minutes.
The convergence criterion was defined as the relative error between the maximum reward of the latest training loop and the previous training loops.
For the $n$-th training loop , the convergence criterion $C_n$ was the relative advantages between the maximum reward $R^{max}_n$ of this training loop and all of the previous maximum reward $R^{max}_{1:n-1}$:
\begin{equation}C_n=\frac{R^{max}_n}{R^{max}_{1:n-1}}\end{equation}
. If $C_n$ was lower than 0.05 for 5 training loops, the training was considered convergent.
The coefficients were selected from the best coefficients (highest reward) among the different training loops.
In other words, the selectivity criterion yields the highest reward.
Referring to the first training loop, the last training loop achieved a higher reward much faster and more stably (Figure~\ref{fig:convergence_and_statistics}(a)).
The average and maximum rewards of the different training loops were compared (Figure~\ref{fig:convergence_and_statistics}(b)) to demonstrate the convergence of the training.
The component match degrees of the reward function with the learning steps are shown in Figure~\ref{fig:component_match_degrees_during_training}.

When the training was confirmed to be convergent, the coefficients with the highest reward in each training loop were extracted to determine important coefficients.
The first reduction in the NB coefficients was inspired by the laminar geometry:
\begin{equation}\sigma_{p11}=\sigma_{p22}=\sigma_{p33}\end{equation}
, where $pmn$ denotes the bead pair between the CLm and CLn.
The structures of the CG models with lower degrees of freedom were more orderly.
Thus, the equilibrium distances of the same types should be nearly identical to represent the laminar structure.
In other words, the equilibrium distance of CL1:CL1, CL2:CL2 and CL3:CL3 should be nearly the same, whereas in the mapped AA trajectories, they are not (Figure~\ref{fig:nonbonded_equilibrium_distance_distributions_equilibrium}, and Figure~\ref{fig:nonbonded_equilibrium_distance_distributions_stretch}).
The \textit{equilibrium distances} of the ideal crystal structure were selected as the estimated value of the distance coefficient ($\sigma$) for each pair.
These \textit{equilibrium distances} do not correspond to the well-relaxed structures or characteristic distances of the radial distribution function.
The first physics-informed restraint was also beneficial for reproducing fragile fractures in vertical and horizontal models.

The subsequent reductions in the NB coefficients were statistically guided.

Second reduction: Referring to $d^e$, only $\sigma_{p23}$ had an unexpected intention to increase (Figure~\ref{fig:convergence_and_statistics}(d)), maintaining $\sigma_{p11}$ and $\sigma_{p23}$ only:
\begin{equation}\sigma_{p12}=\frac{d^e_{p12}}{d^e_{p11}}\sigma_{p11}, \sigma_{p23}=\frac{d^e_{p23}}{d^e_{p11}}\sigma_{p11}, \sigma_{p13}=\frac{d^e_{p13}}{d^e_{p11}}\sigma_{p11}\end{equation}
.

Third reduction: $\epsilon_{p23}$ were usually the smallest energy coefficients in the system (Figure~\ref{fig:convergence_and_statistics}(e)), trying to keep only
$\epsilon_{p11}$ and $\epsilon_{p23}$ independent:
\begin{equation}\epsilon_{p11}=\epsilon_{p22}=\epsilon_{p33}=\epsilon_{p12}=\epsilon_{p13}\end{equation}
.

After these dimensionality reduction, only 9 independent coefficients were preserved: 3 BD $k$, 1 HB $\epsilon$, 1 HB $\sigma$, 2 LJ $\epsilon$ and 2 LJ $\sigma$.

The determined BD coefficients are listed in Table~\ref{tab:bonded_coefficients} and the NB coefficients are listed in Table~\ref{tab:nonbonded_coefficients} along with the BD geometry parameters $l^e$ and NB equilibrium distances $d^e$ under both equilibrium and stretch conditions.
All coefficients are given with four significant digits.

These reductions and resultant coefficients informed us one thing: the sampled average values from AA trajectories cannot be directly used as equilibrium points for CG interactions.
The laminar structure and coefficients of CNC is critical to be artificial, and the $\theta_0$ was also chosen of the \textit{equilibrium angle} in the ideal artificial model (the strict laminar model build by coordinate offset).
This estimated $\theta_0$ is beneficial but not strongly supported by molecular details.

\begin{table}[h]
    \centering
    \caption{Bonded coefficients}
    \label{tab:bonded_coefficients}
    \begin{tabular}{c c c c c c c c}
        \hline
        \hline
        {type}&{id}&{$k$(kcal/mol/$\rm\r{A}^2$)}&{$l_0$(\rm\r{A})}&{$l^e_{\rm AA,E}$(\rm\r{A})}&{$l^e_{\rm CG,E}$(\rm\r{A})}&{$l^e_{\rm AA,S}$(\rm\r{A})}&{$l^e_{\rm CG,S}$(\rm\r{A})}\\
        \hline
        \multirow{3}{*}{bond}&{1}&{64.06}&{5.270}&{5.270}&{5.258}&{5.262}&{5.263}\\
        &{2}&{82.42}&{2.493}&{2.493}&{2.473}&{2.493}&{2.479}\\
        &{3}&{112.3}&{2.066}&{2.066}&{2.061}&{2.066}&{2.065}\\
        \hline
        {}&{}&{$k$(kcal/mol)}&{$\theta_0(^\circ)$}&{$\theta^e_{\rm AA,E}(^\circ)$}&{$\theta^e_{\rm CG,E}(^\circ)$}&{$\theta^e_{\rm AA,S}(^\circ)$}&{$\theta^e_{\rm CG,S}(^\circ)$}\\
        \hline
        \multirow{6}{*}{angle}&{1}&{30.97}&{163.5}&{163.5}&{162.3}&{163.7}&{161.7}\\
        &{2}&{38.79}&{74.2}&{74.2}&{74.7}&{74.2}&{75.4}\\
        &{3}&{30.52}&{93.2}&{93.2}&{94.0}&{93.3}&{93.4}\\
        &{4}&{27.28}&{90.0}&{90.0}&{89.2}&{90.2}&{89.9}\\
        &{5}&{25.97}&{102.0}&{102.0}&{101.4}&{102.0}&{100.1}\\
        &{6}&{22.42}&{166.3}&{166.3}&{165.1}&{166.5}&{165.1}\\
        \hline
        {}&{}&{$k$(kcal/mol)}&{$\chi_0(^\circ)$}&{$\chi^e_{\rm AA,E}(^\circ)$}&{$\chi^e_{\rm CG,E}(^\circ)$}&{$\chi^e_{\rm AA,S}(^\circ)$}&{$\chi^e_{\rm CG,S}(^\circ)$}\\
        \hline
        \multirow{3}{*}{improper}&{1}&{0.1702}&{178.0}&{178.0}&{175.6}&{177.0}&{171.2}\\
        &{2}&{0.1555}&{2.3}&{2.3}&{2.1}&{3.2}&{3.5}\\
        &{3}&{0.1698}&{2.9}&{2.9}&{2.1}&{3.6}&{3.8}\\
        \hline
        \hline
    \end{tabular}
\end{table}

\begin{table}[h]
    \centering
    \caption{Nonbonded coefficients}
    \label{tab:nonbonded_coefficients}
    \begin{tabular}{c c c c c c c c}
        \hline
        \hline
        {type}&{pair}&{$\epsilon$(kcal/mol)}&{$\sigma$(\rm\r{A})}&{$d^e_{\rm AA,E}$(\rm\r{A})}&{$d^e_{\rm CG,E}$(\rm\r{A})}&{$d^e_{\rm AA,S}$(\rm\r{A})}&{$d^e_{\rm CG,S}$(\rm\r{A})}\\
        \hline
        \multirow{2}{*}{HB}&{2:3-1}&{1.404}&{6.017}\\
        &{3:2-1}&{1.404}&{6.017}\\
        \hline
        \multirow{6}{*}{LJ}&{1-1}&{0.9031}&{5.101}&{5.398}&{5.750}&{5.558}&{5.706}\\
        &{1-2}&{0.9031}&{4.201}&{4.769}&{4.473}&{4.840}&{4.584}\\
        &{1-3}&{0.9031}&{4.101}&{4.712}&{4.609}&{4.723}&{4.681}\\
        &{2-2}&{0.9031}&{5.101}&{4.636}&{5.260}&{4.544}&{5.347}\\
        &{2-3}&{0.5031}&{4.083}&{3.983}&{4.018}&{4.113}&{4.091}\\
        &{3-3}&{0.9031}&{5.101}&{4.528}&{5.271}&{4.488}&{5.356}\\
        \hline
        \hline
    \end{tabular}
\end{table}

\subsection{Bonded properties}\indent

For the BD interactions and validation of the BD coefficients, we considered the axial elastic modulus and polymer stiffness of a single cellulose chain, as listed in Table~\ref{tab:bonded_properties}.
All the values after $\pm$ represent the corresponding standard errors.
Considering the stability and computational expense (particularly for polymer stiffness), the AA references were only computed once.
Axial stretch was performed in an xyz-periodic cell with a deformation load along the chains, which is a basic mechanical property.
The simulation of polymer stiffness is a single chain centered in a larger cell.
For axial stretch, Poisson effect was presented in CG model and elastic modulus of AA and CG are 133.5 GPa and 130.3 GPa respectively, where reference modulus is 120-140 GPa\cite{diddens2008anisotropic,iwamoto2009elastic} from experiments and 110-200 GPa\cite{dri2013anisotropy,shishehbor2019effects,wu2014tensile} from simulations.
For polymer stiffness, persistence lengths of AA and CG range from 9.7-9.8~nm and 8.9-10.2~nm with reference data of 6-25~nm from experiments\cite{hilton2022single} and simulations\cite{kroon1997estimation}.
Notably, the polymer stiffness data were not stable for the AA simulations because of the intertwining in some cases.
Therefore, persistence lengths and end-to-end distances of different lengths (10, 20, and 30 residues) are listed.
When considering polymer stiffness, the joint O atoms between two cellulose residues in the AA models and CL1 beads on the backbone in the CG models were node particles to be considered.
The reference data for training were based on AA information only, and the mapped AA trajectories were compared with the CG trajectories after training.
In this study, all effective $k$ values were smaller than their estimations based on distributions (Table~\ref{tab:bonded_force_constants_without_rescale}), particularly for angles and dihedrals, to reproduce the polymer stiffness performance in a system with fewer degrees of freedom (Table~\ref{tab:bonded_properties_without_rescale}).

\begin{table}[h]
    \centering
    \caption{Bonded properties}
    \label{tab:bonded_properties}
    \begin{tabular}{c c c c c}
        \hline
        \hline
        \multicolumn{2}{c}{type}&{AA}&{CG}&{Error}\\
        \hline
        \multicolumn{2}{c}{Axial modulus (GPa)}&{133.53}&{130.42$\pm$0.01}&{2.33\%}\\
        \hline
        \multirow{3}{*}{\makecell{Persistence\\length (nm)}}&{10}&{9.755}&{9.020$\pm$0.024}&{7.53\%}\\
        &{20}&{9.815}&{10.185$\pm$0.079}&{3.77\%}\\
        &{30}&{9.716}&{11.221$\pm$0.168}&{15.49\%}\\
        \hline
        \multirow{3}{*}{\makecell{End-to-end\\distance (nm)}}&{10}&{4.523}&{4.529$\pm$0.000}&{0.13\%}\\
        &{20}&{9.355}&{9.198$\pm$0.004}&{1.68\%}\\
        &{30}&{13.536}&{13.575$\pm$0.017}&{0.29\%}\\
        \hline
        \hline
    \end{tabular}
\end{table}

\begin{table}[h]
    \centering
    \caption{Nonbonded properties}
    \label{tab:nonbonded_properties}
    \begin{tabular}{c c c c c}
        \hline
        \hline
        \multicolumn{2}{c}{type}&{AA}&{CG}&{Error}\\
        \hline
        \multirow{3}{*}{\makecell{Strength\\(MPa)}}&{V}&{1224.0$\pm$9.7}&{1191.6$\pm$6.0}&{2.65\%}\\
        &{H}&{638.2$\pm$4.8}&{653.4$\pm$1.1}&{2.38\%}\\
        &{S}&{467.0$\pm$3.7}&{509.2$\pm$4.6}&{9.04\%}\\
        \hline
        \multirow{3}{*}{\makecell{Toughness\\(MPa)}}&{V}&{111.0$\pm$2.5}&{117.2$\pm$3.2}&{5.59\%}\\
        &{H}&{77.9$\pm$2.8}&{78.2$\pm$1.3}&{0.39\%}\\
        &{S}&{190.1$\pm$4.6}&{167.0$\pm$3.9}&{12.15\%}\\
        \hline
        \multirow{2}{*}{\makecell{Direction\\angle}}&{R}&{2.29$\pm$0.00}&{2.33$\pm$0.00}&{1.75\%}\\
        &{F}&{1.92$\pm$0.01}&{2.02$\pm0.01$}&{5.21\%}\\
        \hline
        \hline
    \end{tabular}
\end{table}

\subsection{Nonbonded properties}\indent

\begin{figure}[htbp]
    \centering
    \scriptsize
    \begin{minipage}[b]{0.32\textwidth}
        \centering
        \includegraphics[width=0.35\textwidth]{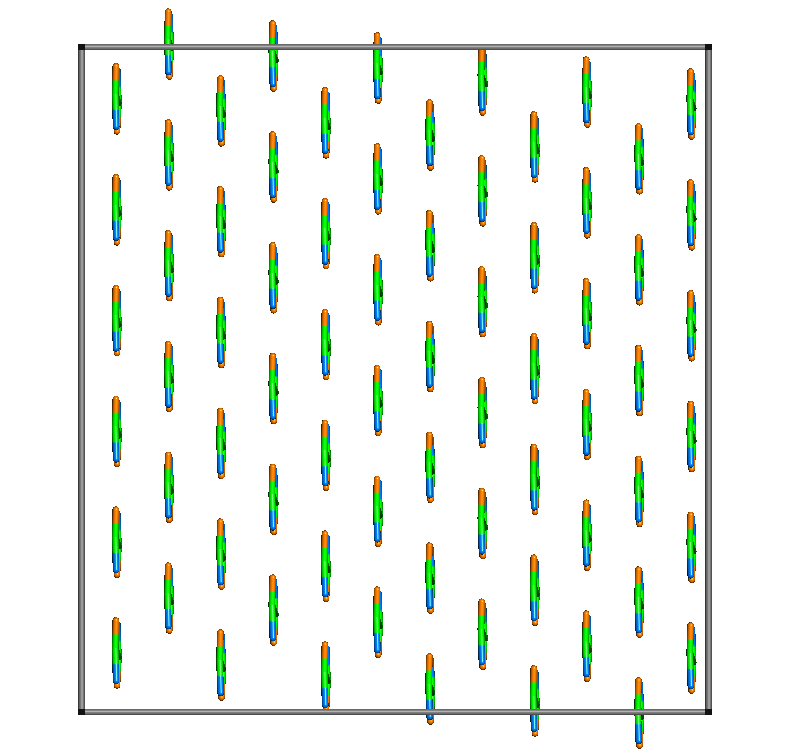}
        \includegraphics[width=0.35\textwidth]{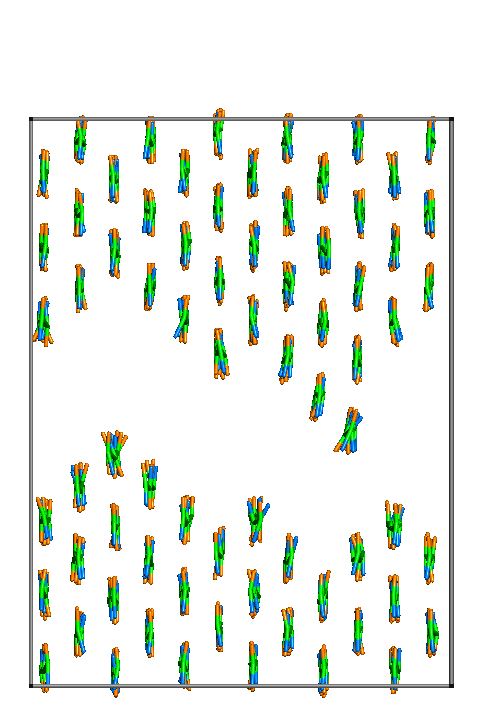}
        \\
        Vertical
    \end{minipage}
    \begin{minipage}[b]{0.32\textwidth}
        \centering
        \includegraphics[width=0.36\textwidth]{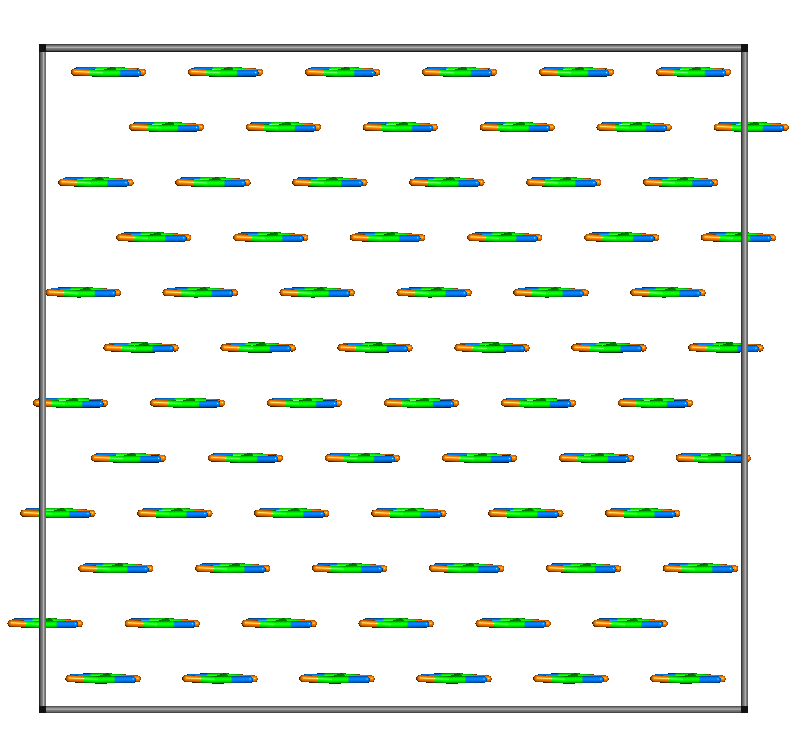}
        \includegraphics[width=0.36\textwidth]{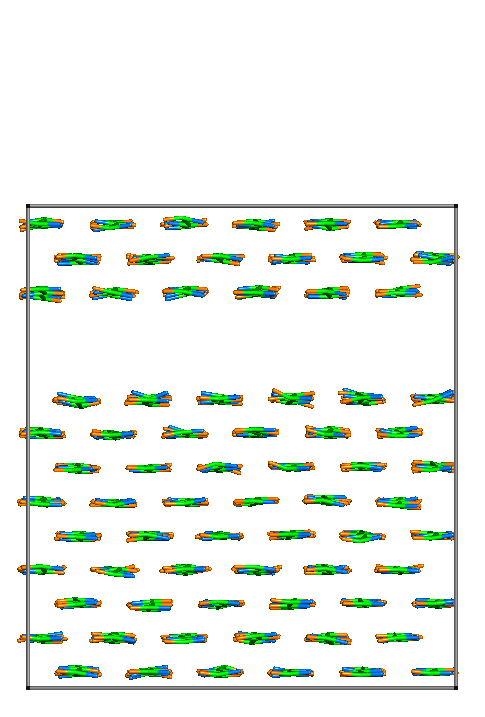}
        \\
        Horizontal
    \end{minipage}
    \begin{minipage}[b]{0.32\textwidth}
        \centering
        \includegraphics[width=0.48\textwidth]{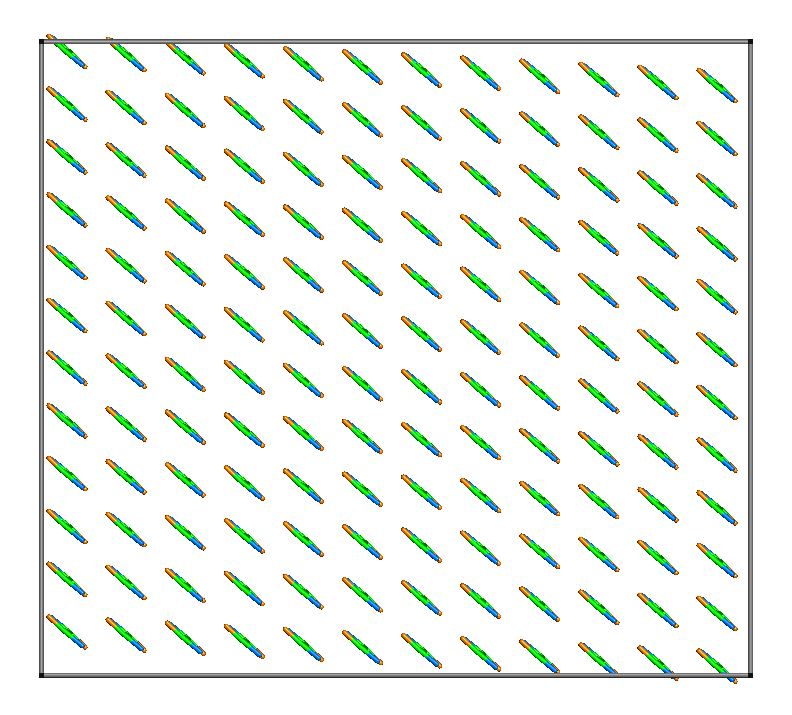}
        \includegraphics[width=0.48\textwidth]{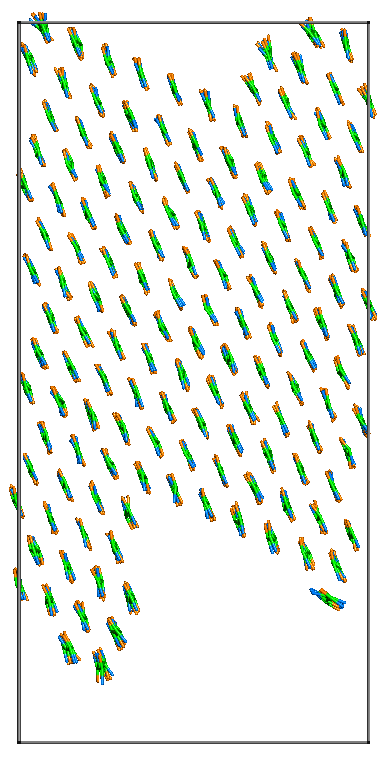}
        \\
        Slant
    \end{minipage}
    \captionsetup{font=scriptsize}
    \caption{
    Transverse behaviors of CG models in the characteristic directions.
    Important frictional sliding in the slant direction and fragile fractures in the vertical and horizontal directions were reproduced.
    }
    \label{fig:transverse_behaviors}
\end{figure}

The NB properties include the transverse strength and toughness, as listed in Table~\ref{tab:nonbonded_properties}.
All the values after $\pm$ are the corresponding standard errors.
The transverse behaviors of CG models are shown in Figure~\ref{fig:transverse_behaviors}.
The reference strengths of the vertical, horizontal and slant models are 0.9 GPa\cite{wu2014tensile}, 0.5 GPa\cite{wu2014tensile} and 0.7 GPa\cite{shishehbor2019effects}, which were well reproduced by both AA and CG simulations.
To reduce the effect of randomness, 100 replica simulations of transverse stretch were performed.
The strength and toughness distributions of the CG models were similar to those of the AA reference data.
Only the distributions of the horizontal CG models were narrower than those of their AA counterparts.
The average strength and toughness values were similar to those obtained from the AA data.
The frictional sliding of the slant models was quantitatively validated by the direction angle (Figure~\ref{fig:direction_angle_definition}).
These data validate the CG model for reproducing the transverse anisotropy.
Although all $\sigma$ values were larger than expected, the equilibrium distances in the CG models were only slightly larger than those in the AA models (less than 15\%, Table~\ref{tab:nonbonded_coefficients}).
Although the accuracy of the strength and toughness distributions is problematic owing to the representability problem of CG models, the full distributions of transverse strength and toughness are shown in Figure~\ref{fig:transverse_mechanical_property_distributions} for reference.
Other stretch curves collections are shown in Figure~\ref{fig:transverse_stretch_performance_collection}.

\subsection{Structural properties from equilibrium conditions}\indent

\begin{figure}[htbp]
    \centering
    \scriptsize
    \begin{subfigure}[b]{0.96\textwidth}
        \begin{minipage}[b]{0.32\textwidth}
            \centering
            \includegraphics[width=0.96\textwidth]{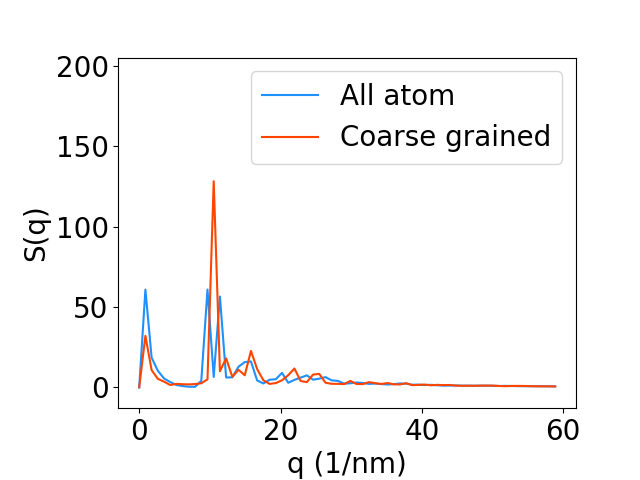}
            \\
            CL1
        \end{minipage}
        \begin{minipage}[b]{0.32\textwidth}
            \centering
            \includegraphics[width=0.96\textwidth]{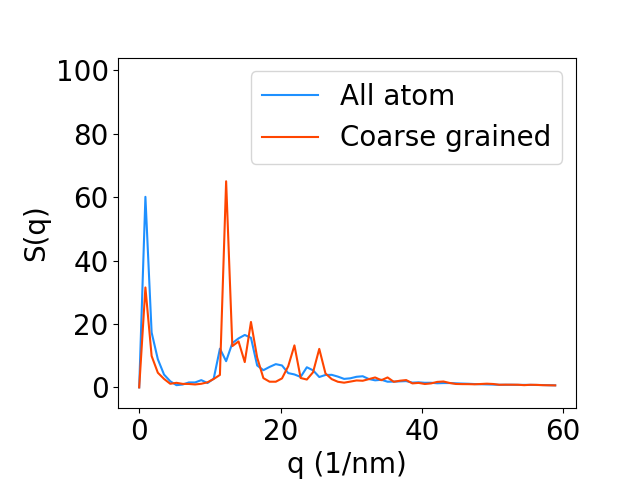}
            \\
            CL2
        \end{minipage}
        \begin{minipage}[b]{0.32\textwidth}
            \centering
            \includegraphics[width=0.96\textwidth]{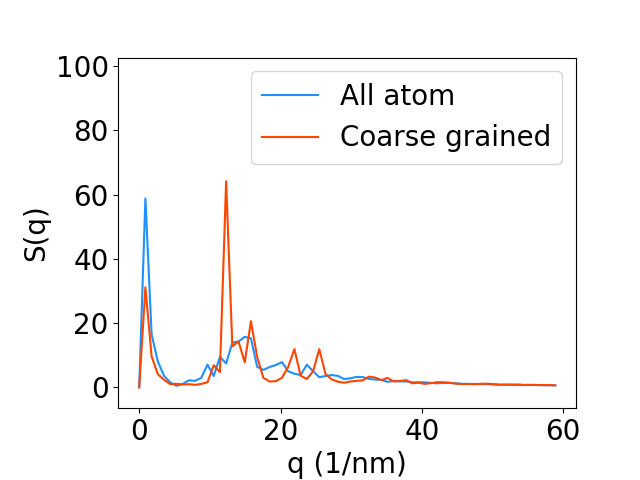}
            \\
            CL3
        \end{minipage}
        \subcaption{}
    \end{subfigure}
    \\
    \begin{subfigure}[b]{0.96\textwidth}
        \begin{minipage}[b]{0.32\textwidth}
            \centering
            \includegraphics[width=0.96\textwidth]{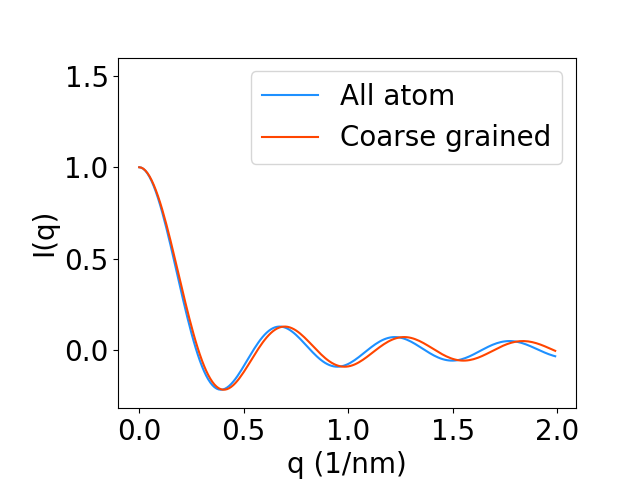}
            \\
            CL1
        \end{minipage}
        \begin{minipage}[b]{0.32\textwidth}
            \centering
            \includegraphics[width=0.96\textwidth]{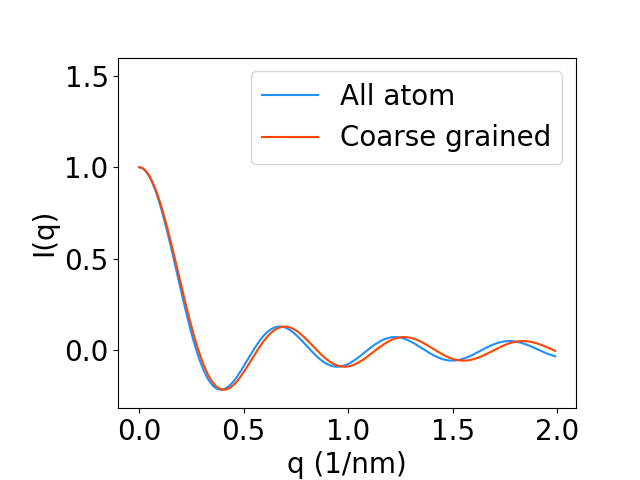}
            \\
            CL2
        \end{minipage}
        \begin{minipage}[b]{0.32\textwidth}
            \centering
            \includegraphics[width=0.96\textwidth]{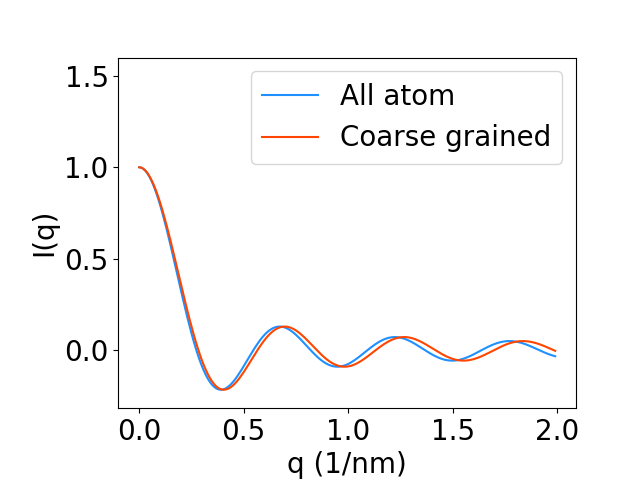}
            \\
            CL3
        \end{minipage}
        \subcaption{}
    \end{subfigure}
    \\
    \begin{subfigure}[b]{0.96\textwidth}
        \begin{minipage}[b]{0.32\textwidth}
            \centering
            \includegraphics[width=0.96\textwidth]{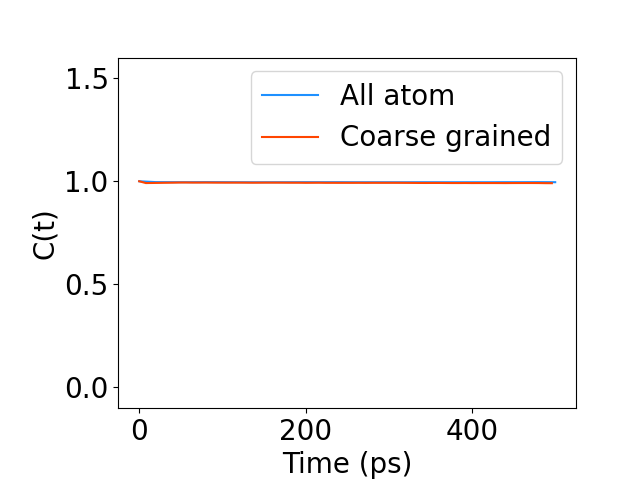}
            \\
            CL1
        \end{minipage}
        \begin{minipage}[b]{0.32\textwidth}
            \centering
            \includegraphics[width=0.96\textwidth]{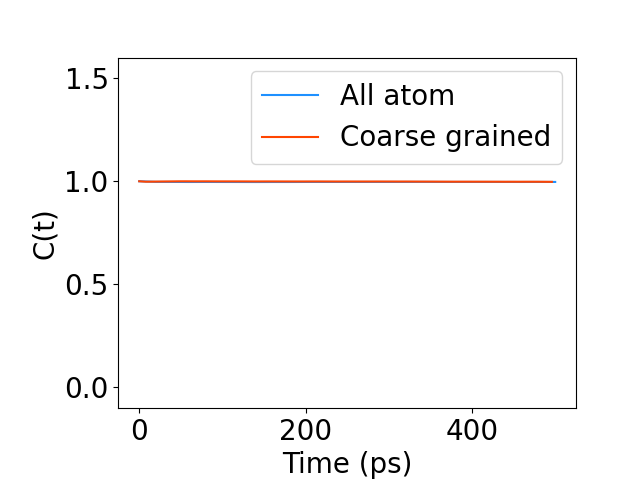}
            \\
            CL2
        \end{minipage}
        \begin{minipage}[b]{0.32\textwidth}
            \centering
            \includegraphics[width=0.96\textwidth]{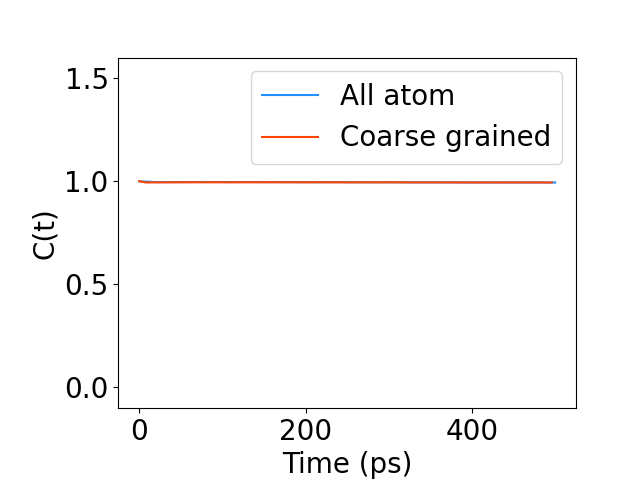}
            \\
            CL3
        \end{minipage}
        \subcaption{}
    \end{subfigure}
    \\
    \begin{subfigure}[b]{0.96\textwidth}
        \begin{minipage}[b]{0.32\textwidth}
            \centering
            \includegraphics[width=0.96\textwidth]{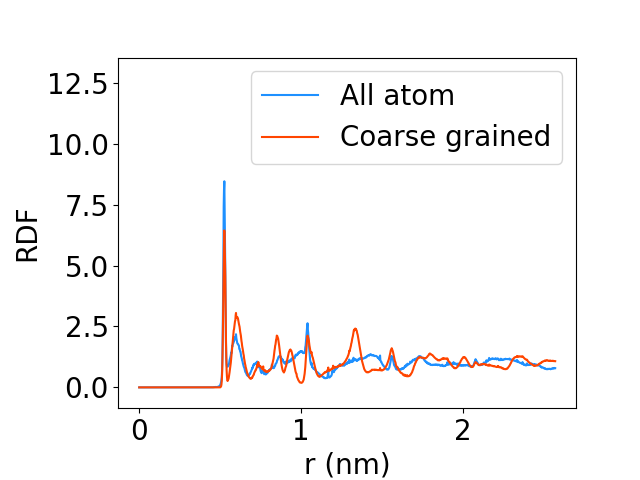}
            \\
            CL1:CL1 Whole
        \end{minipage}
        \begin{minipage}[b]{0.32\textwidth}
            \centering
            \includegraphics[width=0.96\textwidth]{structural_property_equilibrium_rdf_layer.png}
            \\
            CL1:CL1 HBonding layer
        \end{minipage}
        \begin{minipage}[b]{0.32\textwidth}
            \centering
            \includegraphics[width=0.96\textwidth]{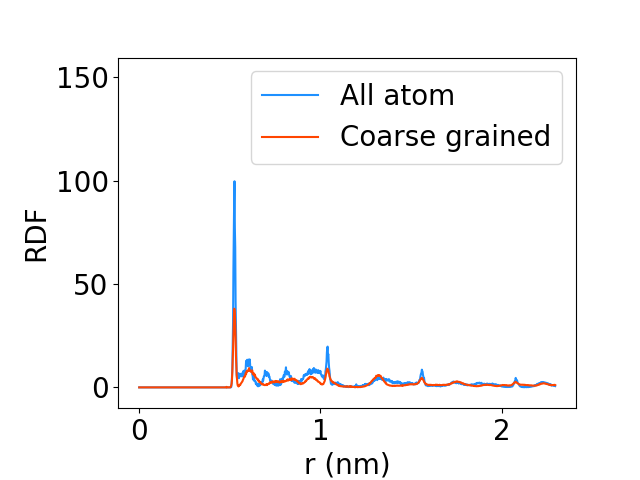}
            \\
            CL1:CL1 Inter layers
        \end{minipage}
        \subcaption{}
    \end{subfigure}
    \captionsetup{font=scriptsize}
    \caption{
    Structural properties of mapped AA and CG from equilibrium simulations of CNCs.
    (a) Static structure factor of beads.
    (b) Normalized scattering intensity of beads.
    (c) Angular correlation function of beads.
    They stressed the structural stability from a quantitative perspective.
    (d) RDF of CL1:CL1 pair from a whole CNC, a HBonding layer slice, and a perpendicular inter HBonding layer slice under equilibrium conditions.
    The slices were orthogonal decompositions of the structures to a certain extent, and contributed to the understanding of anisotropy.
    These data confirm the structural and anisotropic reproduction of the CG model under the equilibrium conditions.
    }
    \label{fig:structural_properties_equilibrium}
\end{figure}

\begin{figure}[htbp]
    \centering
    \scriptsize
    \begin{subfigure}[b]{0.96\textwidth}
        \begin{minipage}[b]{0.32\textwidth}
            \centering
            \includegraphics[width=0.96\textwidth]{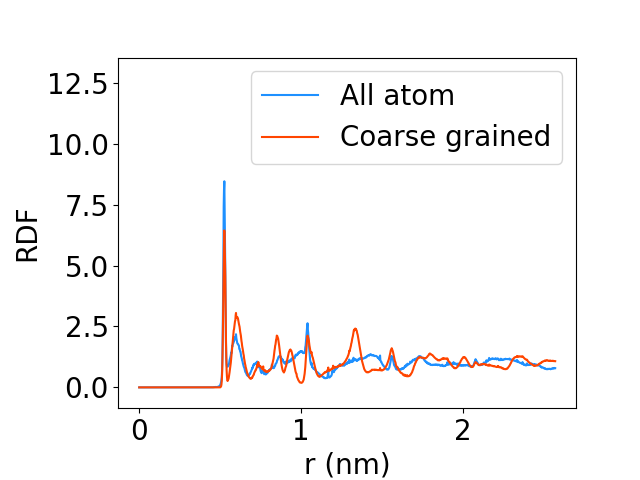}
            \\
            CL1:CL1
        \end{minipage}
        \begin{minipage}[b]{0.32\textwidth}
            \centering
            \includegraphics[width=0.96\textwidth]{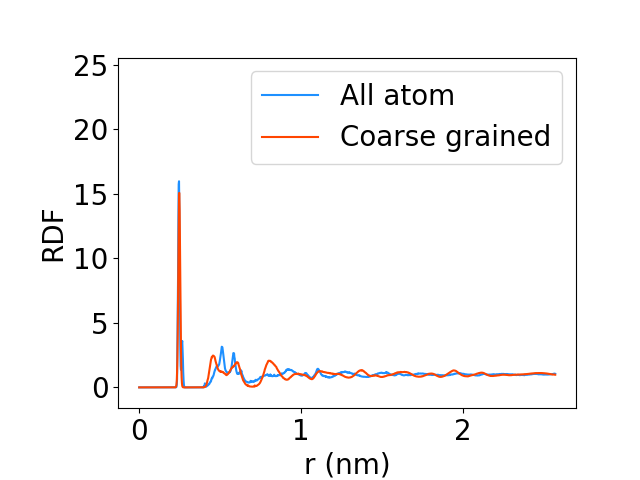}
            \\
            CL1:CL2
        \end{minipage}
        \begin{minipage}[b]{0.32\textwidth}
            \centering
            \includegraphics[width=0.96\textwidth]{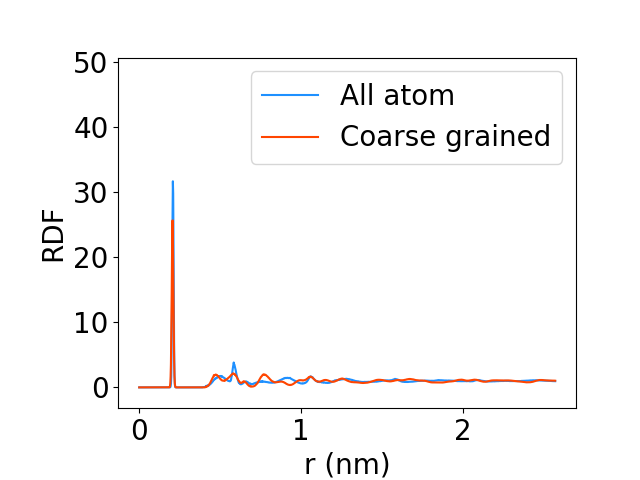}
            \\
            CL1:CL3
        \end{minipage}
        \\
        \begin{minipage}[b]{0.32\textwidth}
            \centering
            \includegraphics[width=0.96\textwidth]{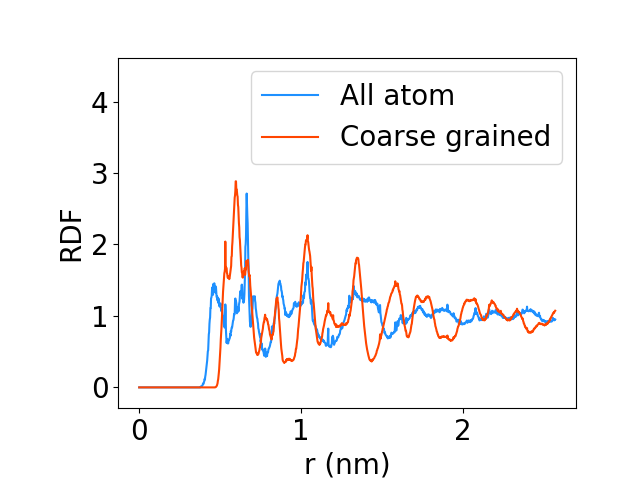}
            \\
            CL2:CL2
        \end{minipage}
        \begin{minipage}[b]{0.32\textwidth}
            \centering
            \includegraphics[width=0.96\textwidth]{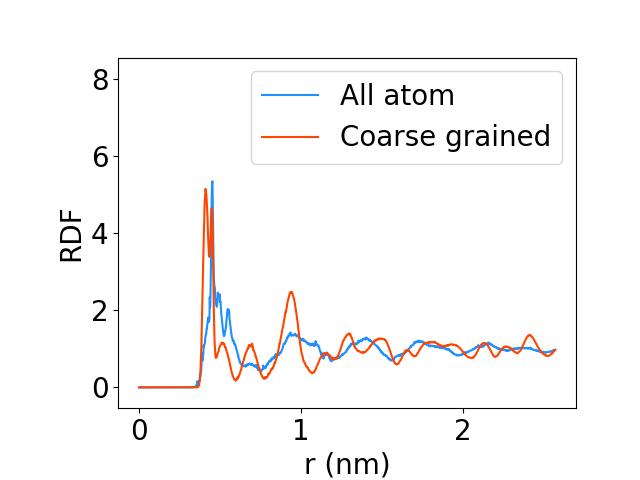}
            \\
            CL2:CL3
        \end{minipage}
        \begin{minipage}[b]{0.32\textwidth}
            \centering
            \includegraphics[width=0.96\textwidth]{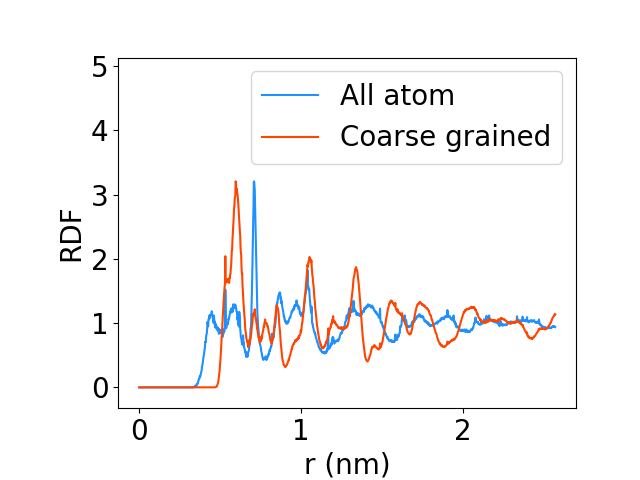}
            \\
            CL3:CL3
        \end{minipage}
        \subcaption{}
    \end{subfigure}
    \\
    \begin{subfigure}[b]{0.32\textwidth}
        \centering
        \includegraphics[width=0.96\textwidth]{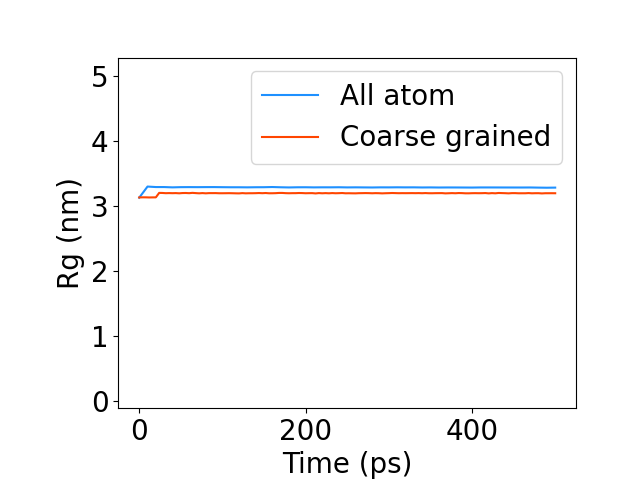}
        \subcaption{}
    \end{subfigure}
    \begin{subfigure}[b]{0.32\textwidth}
        \centering
        \includegraphics[width=0.96\textwidth]{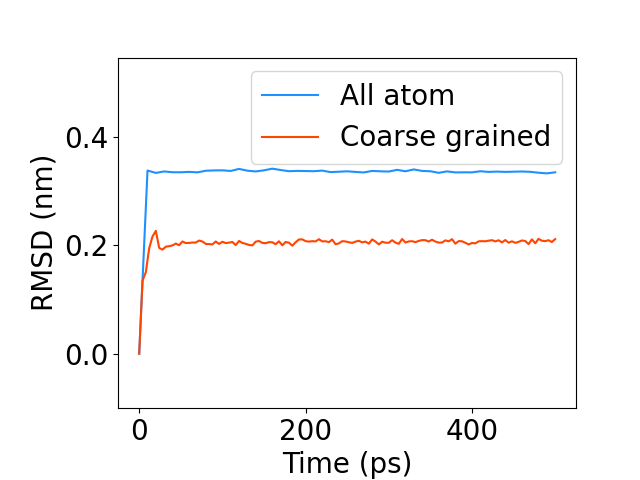}
        \subcaption{}
    \end{subfigure}
    \captionsetup{font=scriptsize}
    \caption{
    Additional structural properties of both mapped AA and CG from equilibrium simulations of CNCs.
    (a) RDF of all bead pairs.
    (b) Radius of gyration and (c) Root-mean-square deviation during the relaxations.
    These data further confirm the structural and anisotropic reproduction of the CG model under the equilibrium conditions.
    }
    \label{fig:additional_structural_properties_equilibrium}
\end{figure}

Structural properties including spatial correlations, BD geometry parameter distributions, and NB equilibrium distance distributions are quantitative measures for validating the CG models.
The equilibrium conditions refer to snapshots from equilibrium simulations of CNC crystals on both AA and CG scales.
The static structure factor (S(q)), normalized scattering intensity (I(q)), angular correlation function (C(t)), and radical distribution function (RDF) of all pairs, the CL1:CL1 pair from the whole laminar CNC, the CL1:CL1 pair within one HBonding layer, and the CL1:CL1 pair from one inter HBonding layer slice of both the mapped AA and CG equilibrium CNC trajectories are shown in Figure~\ref{fig:structural_properties_equilibrium} and Figure~\ref{fig:additional_structural_properties_equilibrium}.
The RDF of one layer and inter layer were computed from the sliced equilibrium structures, aiming to provide additional insights about anisotropy.
In addition, the radius of gyration (Rg), and root-mean-square deviation (RMSD) during the relaxation processes were shown in (Figure~\ref{fig:additional_structural_properties_equilibrium}(b)(c)).
S(q), I(q), C(t), RDF, Rg, and RMSD were computed using the saxs, sans, rdf, gyrate, and rms tools of GROMACS\cite{abraham2015gromacs,bjelkmar2010implementation}.
Full comparisons of the BD geometry parameter distributions and NB equilibrium distance distributions under the equilibrium conditions and stretch progresses are shown in Figure~\ref{fig:bonded_geometry_parameter_distributions_equilibrium}, Figure~\ref{fig:bonded_geometry_parameter_distributions_stretch}, Figure~\ref{fig:nonbonded_equilibrium_distance_distributions_equilibrium}, and Figure~\ref{fig:nonbonded_equilibrium_distance_distributions_stretch}.
These alignments for the spatial correlations confirm the effectiveness of this CG model, and the horizontal curve of C(t) quantitatively stresses structural stability.
Although the comparisons of these data under non-equilibrium conditions are not rigorous (and I(q) computation failed technically), the structural properties from the stretch processes are also included in Figure~\ref{fig:structural_properties_stretch} for reference.

\subsection{Baseline methods for coarse grained models}\indent

Except for the RL-parameterized CG potential, the baseline models were also tested in parallel.
The simulations we used for the AA and CG baselines were the slant model equilibrium without any external loads under the NPT ensemble, as we used for the RL-parameterized model.
The reference AA trajectory was simulated for 20~ns, using a timestep of 2~fs.
Bottom-up baselines (including IBI, RE, and FM) and Top-down baseline MARTINI 3 were compared to illustrate the necessity of this model.

Bottom-up methods were utilized via VOTCA tools\cite{ruhle2009versatile,ruhle2011hybrid,mashayak2015relative}.
RE was used to focus on NB interactions, and the BD interactions coefficients for the model were obtained from the IBI, whereas IBI and FM were independently trained.
The IBI and RE models were trained with 200 training steps respectively, where one training step contained 1000000 simulation timesteps of 1~fs (1~ns for each);
and the FM model was resolved on the full 20~ns mapped AA trajectories.
One trial simulation of IBI or FM took approximately 10 minutes, and 200 training steps took about 40 and 50 hours, respectively.
By monitoring the potential energy well at 0.5~nm for the CL1:CL1 pair and CL1-CL1 bond, we noticed that the potential energy continuously increased during training, which made it difficult to execute more training technically.
The top-down baseline was the MARTINI 3 cellulose potentials from the Polyply references\cite{grunewald2022martini,grunewald2022polyply}.

The FM to match the mean forces did not produce a stable structure (it required a very small timestep in the energy minimization and the distance between bonds exceeded the table range in the relaxation), and the IBI and RE models were checked under the NVT ensemble without velocity generalization because the velocity and pressure manipulations would also lead to unstable structures (the periodic chains rotated and translated).
Thus polymer stiffness simulation results were provided but were problematic because we were not sure about the effect of velocity generalization in these cases, while transverse stretch strength and toughness were also problematic without pressure control.
The MARTINI 3 cellulose potentials were better at preserving the parallel crystal structure than those of the bottom-up baselines and were simulated under the NPT ensemble.

The comparison focused on the axial elastic modulus and transverse mechanical properties, as listed in Table~\ref{tab:baselines_bonded_properties} and Table~\ref{tab:baselines_nonbonded_properties}, respectively.
The inferior performances of the bottom-up baselines were within expectations and aligned with the limitations described in the ``Introduction''.
It should be emphasized that all the polymer stiffness data were obtained from simulations with velocity generalization under the NVT ensemble, regardless of the NVT or NPT marks.
In addition to the numerical properties in the tables, the structures after relaxation and fracture (Figure~\ref{fig:baselines_structure_relaxation} and Figure~\ref{fig:baselines_structure_fracture}), together with the RDF for IBI and RE after training (Figure~\ref{fig:baselines_rdf_ibi} and Figure~\ref{fig:baselines_rdf_re}) are provided.
The results of these conventional baselines confirm the validation and superiority of this RL-parameterized model following an extended bottom-up approach, in which the well-designed mapping and interactions play pivotal roles.

\begin{table}[h]
    \centering
    \caption{Baselines bonded properties}
    \label{tab:baselines_bonded_properties}
    \begin{tabular}{c c c c c c c c}
        \hline
        \hline
        \multirow{2}{*}{type}&\multirow{2}{*}{\makecell{Axial\\Modulus (GPa)}}&\multicolumn{3}{c}{Persistence Length (nm)}&\multicolumn{3}{c}{End to end distance (nm)}\\
        \cline{3-8}
        &&{10}&{20}&{30}&{10}&{20}&{30}\\
        \hline
        {IBI(NVT)}&{1966.0}&{6.225}&{8.704}&{10.949}&{4.466}&{9.078}&{13.639}\\
        \hline
        {RE(NVT)}&{1967.4}&{7.625}&{7.820}&{12.855}&{4.491}&{9.022}&{13.759}\\
        \hline
        {MARTINI(NPT)}&{10.7}&{5.440}&{5.054}&{6.807}&{4.134}&{8.078}&{11.907}\\
        \hline
        {RL(NPT)}&{130.5}&{8.990}&{10.176}&{10.106}&{4.529}&{9.199}&{13.432}\\
        \hline
        {AA(NPT)}&{133.5}&{9.755}&{9.815}&{9.716}&{4.523}&{9.355}&{13.536}\\
        \hline
        &\multicolumn{7}{c}{Error}\\
        \hline
        {IBI(NVT)}&{1373\%}&{36\%}&{11\%}&{13\%}&{1\%}&{3\%}&{1\%}\\
        \hline
        {RE(NVT)}&{1374\%}&{22\%}&{20\%}&{32\%}&{1\%}&{1\%}&{2\%}\\
        \hline
        {MARTINI(NPT)}&{92\%}&{44\%}&{49\%}&{30\%}&{9\%}&{14\%}&{12\%}\\
        \hline
        {RL(NPT)}&{2\%}&{8\%}&{4\%}&{4\%}&{0\%}&{2\%}&{1\%}\\
        \hline
        \hline
    \end{tabular}
\end{table}

\begin{table}[h]
    \centering
    \caption{Baselines nonbonded properties}
    \label{tab:baselines_nonbonded_properties}
    \begin{tabular}{c c c c c c c}
        \hline
        \hline
        \multirow{2}{*}{type}&\multicolumn{3}{c}{Strength (MPa)}&\multicolumn{3}{c}{Toughness (GPa)}\\
        \cline{2-7}
        &{V}&{H}&{S}&{V}&{H}&{S}\\
        \hline
        {IBI(NVT)}&{1544.0}&{893.6}&{656.3}&{153.4}&{52.9}&{316.7}\\
        \hline
        {RE(NVT)}&{1337.9}&{552.2}&{869.8}&{160.7}&{82.0}&{352.9}\\
        \hline
        {MARTINI(NPT)}&{471.6}&{268.1}&{259.7}&{62.0}&{396.6}&{838.7}\\
        \hline
        {RL(NPT)}&{1191.6$\pm$6.0}&{653.4$\pm$1.1}&{509.2$\pm$4.6}&{117.2$\pm$3.2}&{78.2$\pm$1.3}&{167.0$\pm$3.9}\\
        \hline
        {AA(NPT)}&{1224.0$\pm$9.7}&{638.2$\pm$4.8}&{467.0$\pm$3.7}&{111.0$\pm$2.5}&{77.9$\pm$2.8}&{190.1$\pm$4.6}\\
        \hline
        &\multicolumn{6}{c}{Error}\\
        \hline
        {IBI(NVT)}&{26\%}&{40\%}&{41\%}&{38\%}&{32\%}&{67\%}\\
        \hline
        {RE(NVT)}&{13\%}&{13\%}&{86\%}&{45\%}&{5\%}&{86\%}\\
        \hline
        {MARTINI(NPT)}&{61\%}&{58\%}&{44\%}&{44\%}&{309\%}&{341\%}\\
        \hline
        {RL(NPT)}&{3\%}&{2\%}&{9\%}&{6\%}&{0\%}&{12\%}\\
        \hline
        \hline
    \end{tabular}
\end{table}

\subsection{Training without thresholds and training using other optimization methods}\indent

The advantages of the thresholds and min in the reward function were emphasized via parallel training without thresholds and min function, as shown in Figure~\ref{fig:convergence_and_statistics_without_threshold} and Figure~\ref{fig:component_match_degrees_during_training_without_threshold}.
The results from the parallel training showed that the thresholds did not harm the best performances and accelerated learning in the early stage.

On the other hand, the parameterization of the interaction potentials is similar to that of the hyperparameter optimization of artificial neural networks.
To further validate the effectiveness of RL, parallel trainings were performed using other optimization methods including the Tree-structured Parzen Estimator\cite{ozaki2020multiobjective} (TPE, an implementation of Bayesian Optimization) and Covariance Matrix Adaptation Evolution Strategy\cite{iruthayarajan2010covariance} (CMAES), which are widely applied for hyperparameter tuning.
CMAES and TPE were utilized by the implementation of OPTUNA\cite{akiba2019optuna}, and both the cases using reward functions with/without thresholds were trained (Figure~\ref{fig:reward_tpe_cmaes}).
The comparison emphasized the optimization capability of RL to addressed highly nonlinear potential parameterization problems and maintain critical stochasticity to compensate for insufficient sampling.
These results are not directly related to the CG model itself, but are important to demonstrate the superiority of the reward function form and degenerate RL approach, and are presented in the Supplementary Information.

\subsection{Additional inspections}\indent

Additional inspections of the mechanical properties of the CG model were performed to verify its transferability and generalization to other cases.

\subsubsection{Draw-out and Tear-apart}\indent

\begin{figure}[htbp]
    \centering
    \begin{subfigure}[b]{0.38\textwidth}
        \centering
        \scriptsize
        \begin{minipage}[b]{0.48\textwidth}
            \centering
            \includegraphics[width=0.7\textwidth]{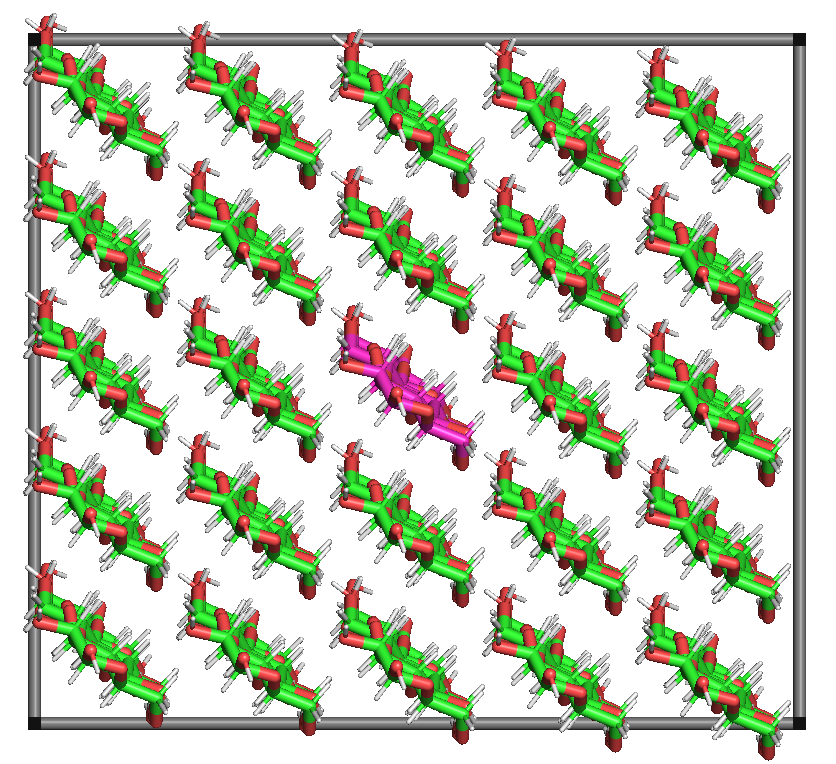}
            \includegraphics[width=0.9\textwidth]{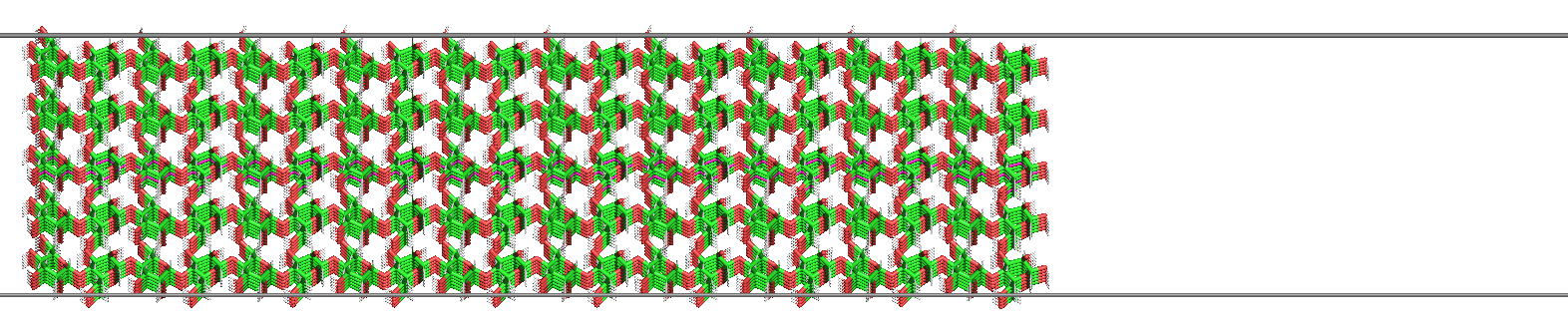}
            \includegraphics[width=0.9\textwidth]{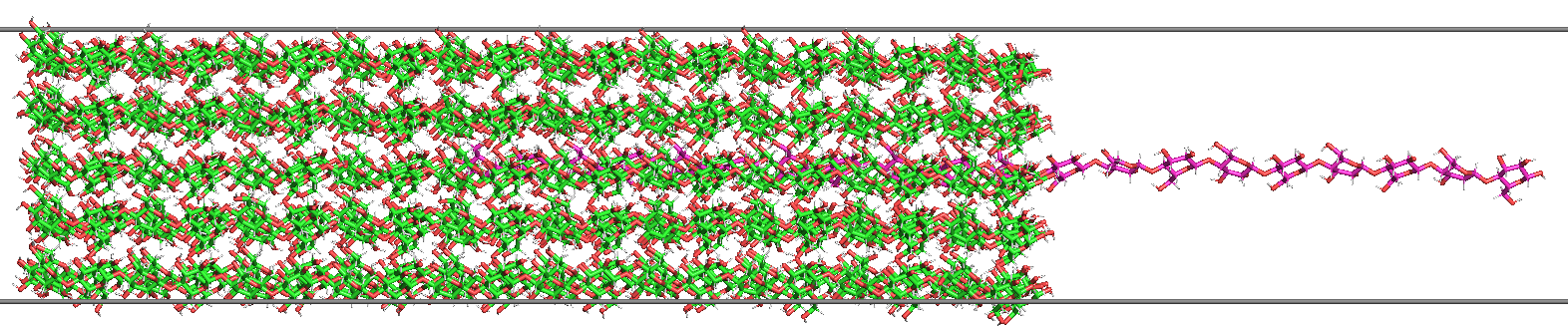}
            \\
            AA
        \end{minipage}
        \begin{minipage}[b]{0.48\textwidth}
            \centering
            \includegraphics[width=0.7\textwidth]{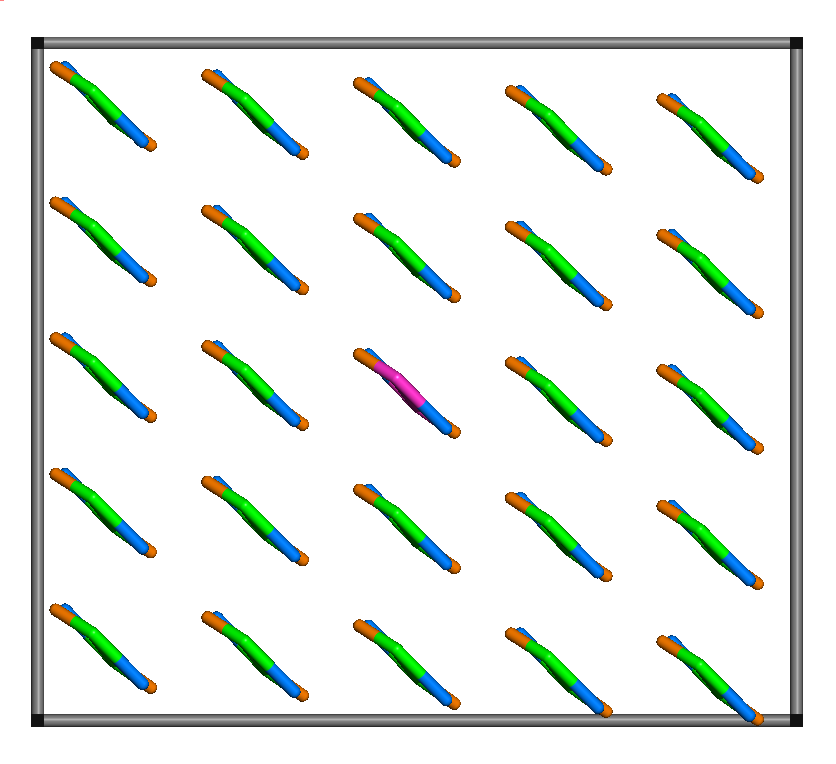}
            \includegraphics[width=0.9\textwidth]{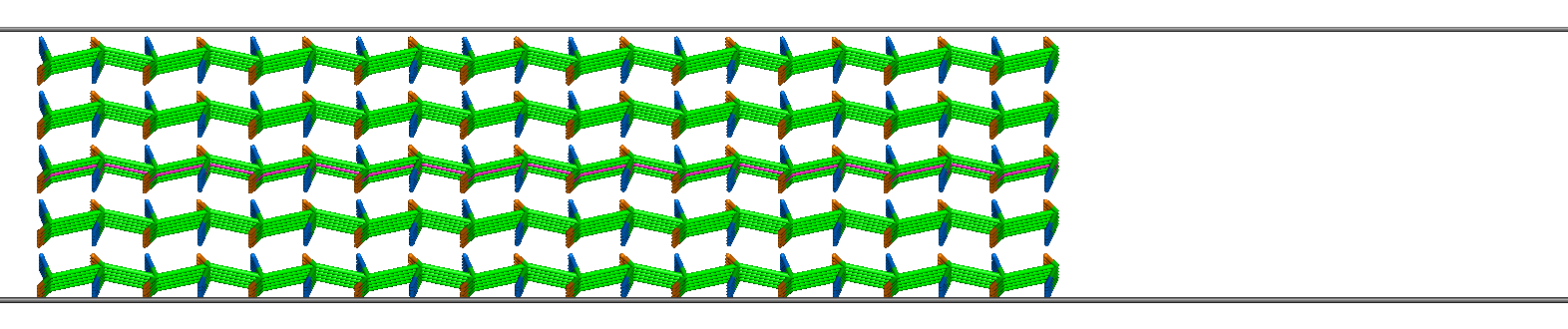}
            \includegraphics[width=0.9\textwidth]{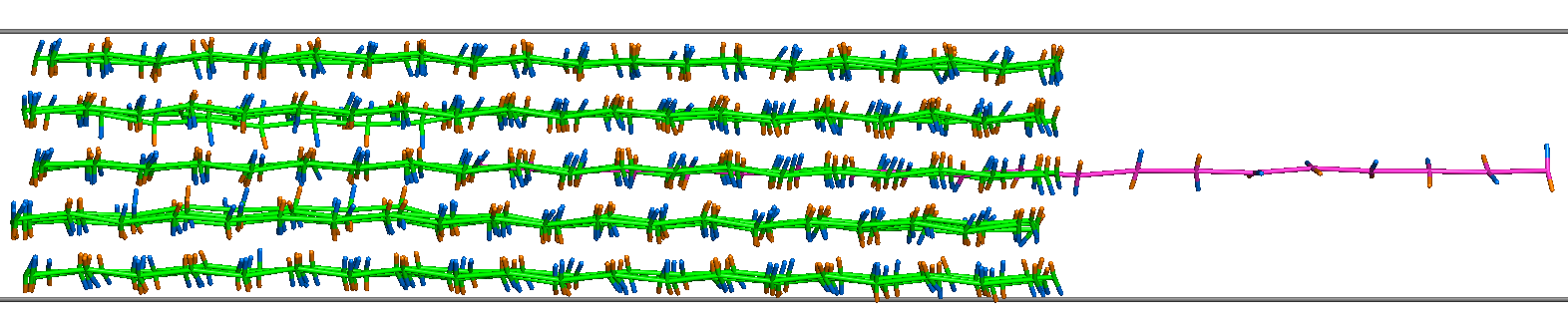}
            \\
            CG
        \end{minipage}
        \subcaption{}
    \end{subfigure}
    \begin{subfigure}[b]{0.58\textwidth}
        \centering
        \scriptsize
        \begin{minipage}[b]{0.48\textwidth}
            \centering
            \includegraphics[width=1.0\textwidth]{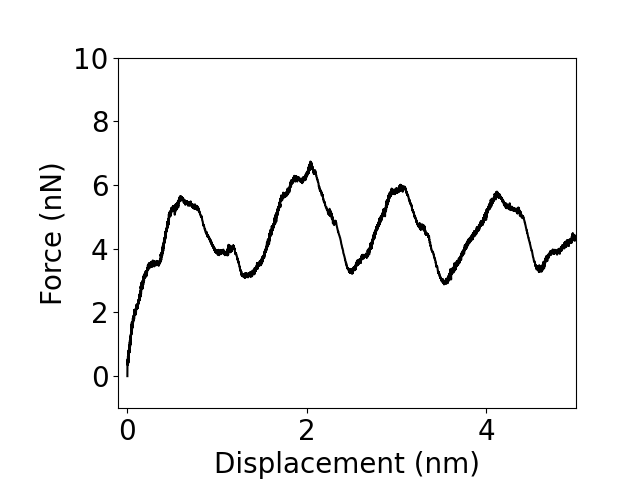}
            \\
            AA
        \end{minipage}
        \begin{minipage}[b]{0.48\textwidth}
            \centering
            \includegraphics[width=1.0\textwidth]{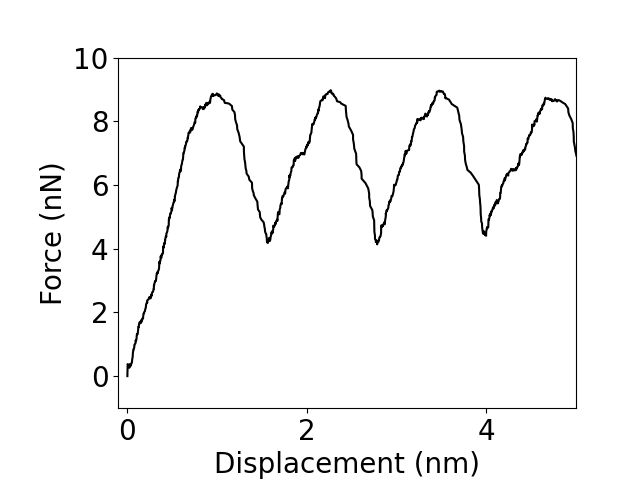}
            \\
            CG
        \end{minipage}
        \subcaption{}
    \end{subfigure}
    \\
    \begin{subfigure}[b]{0.66\textwidth}
        \centering
        \scriptsize
        \begin{minipage}[b]{0.48\textwidth}
            \centering
            \includegraphics[width=0.45\textwidth]{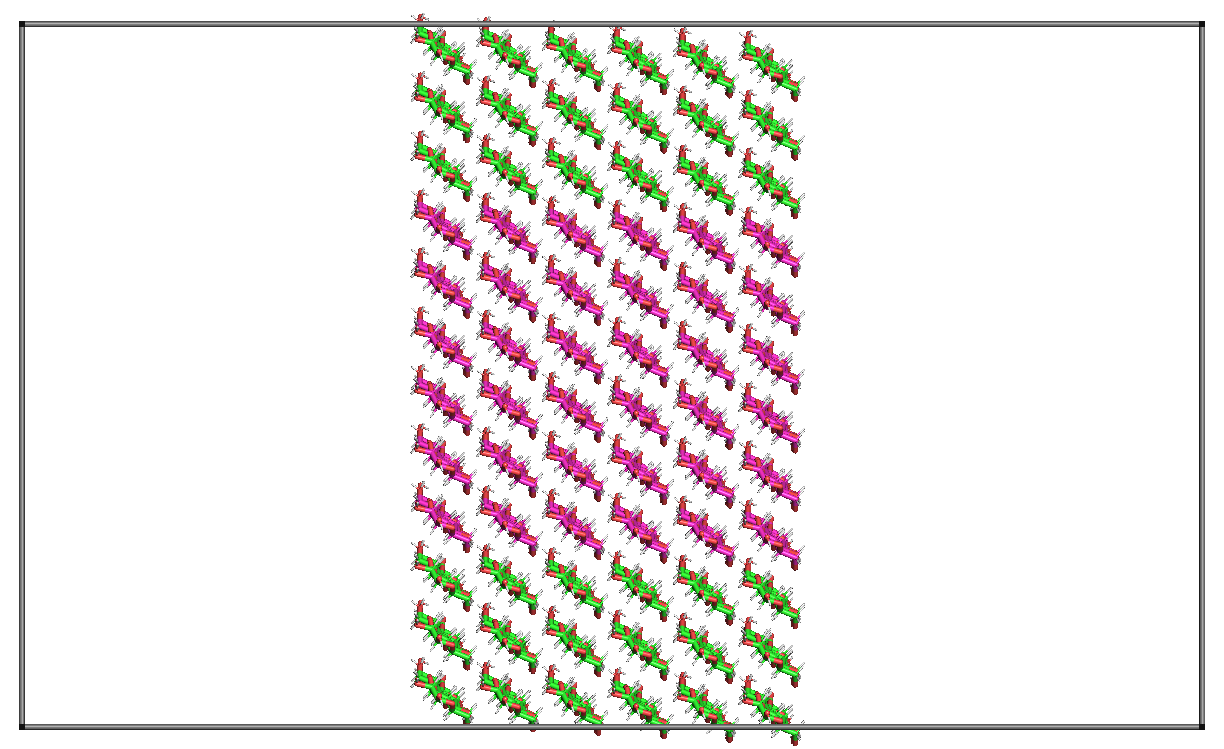}
            \includegraphics[width=0.45\textwidth]{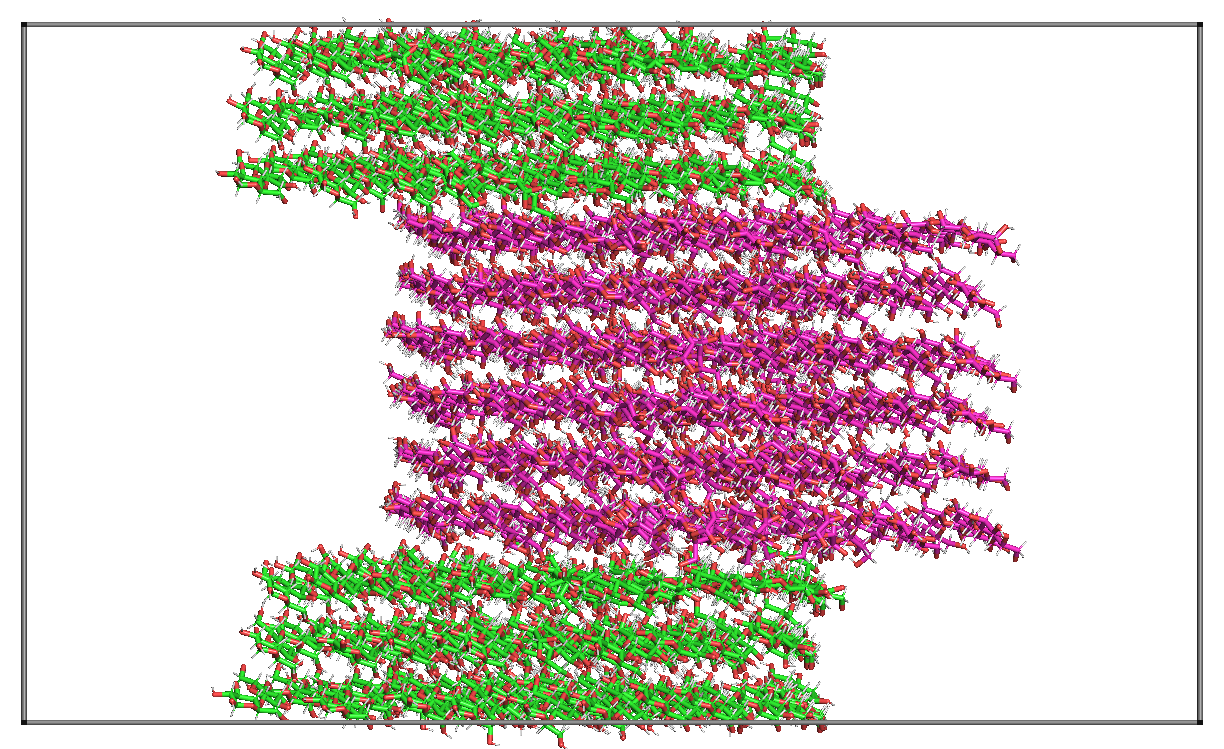}
            \\
            AA front view
            \\
            \includegraphics[width=0.45\textwidth]{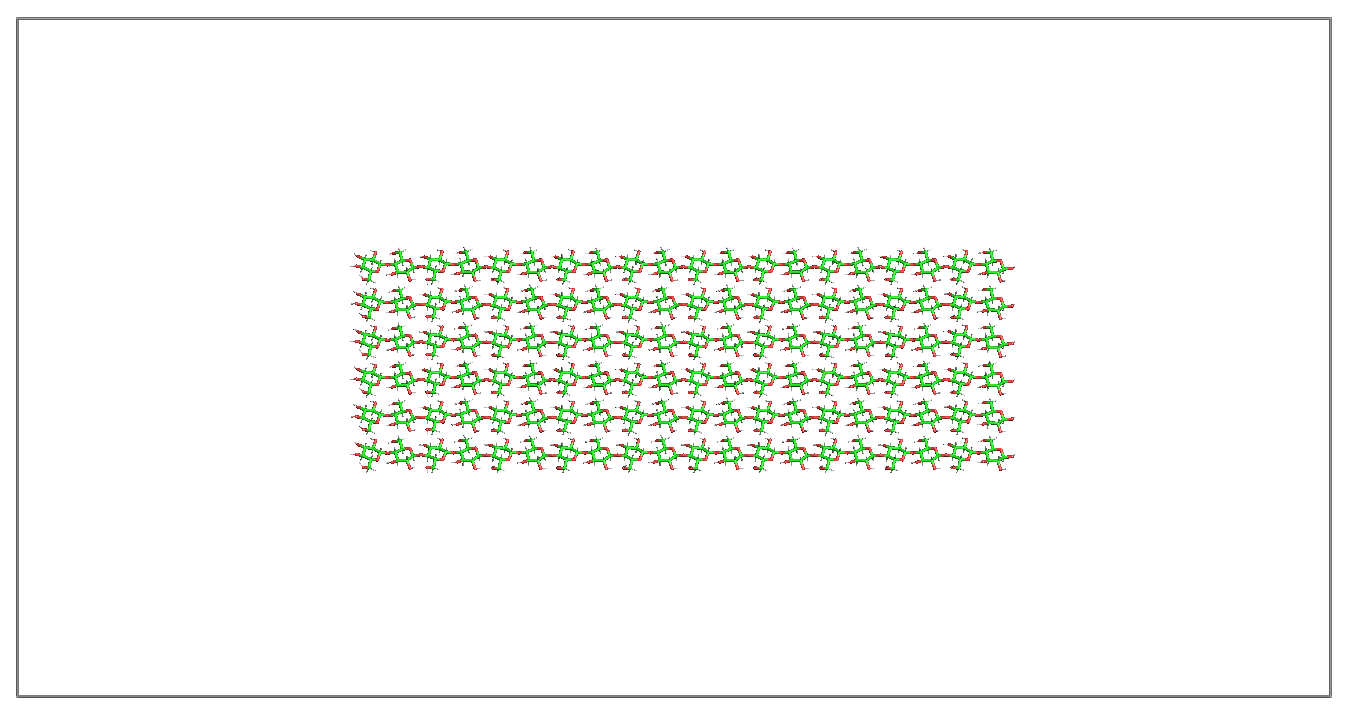}
            \includegraphics[width=0.45\textwidth]{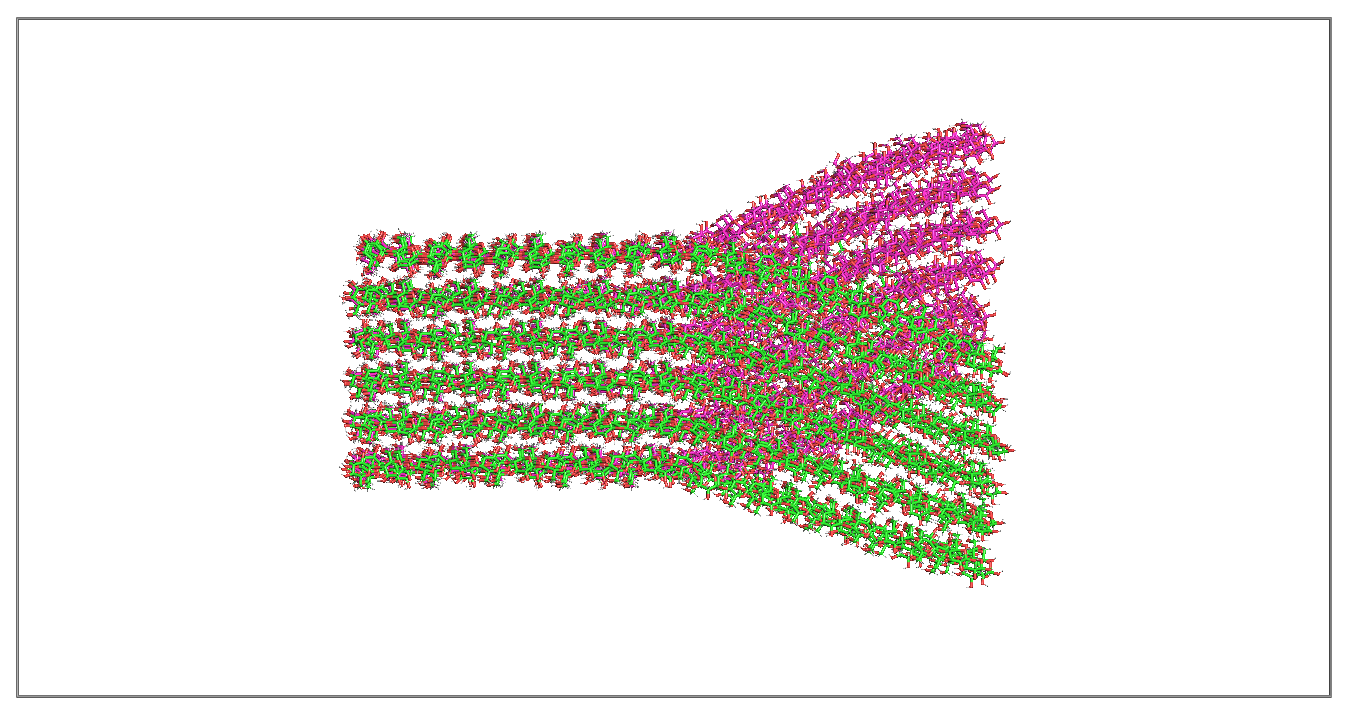}
            \\
            AA top view
        \end{minipage}
        \begin{minipage}[b]{0.48\textwidth}
            \centering
            \includegraphics[width=0.45\textwidth]{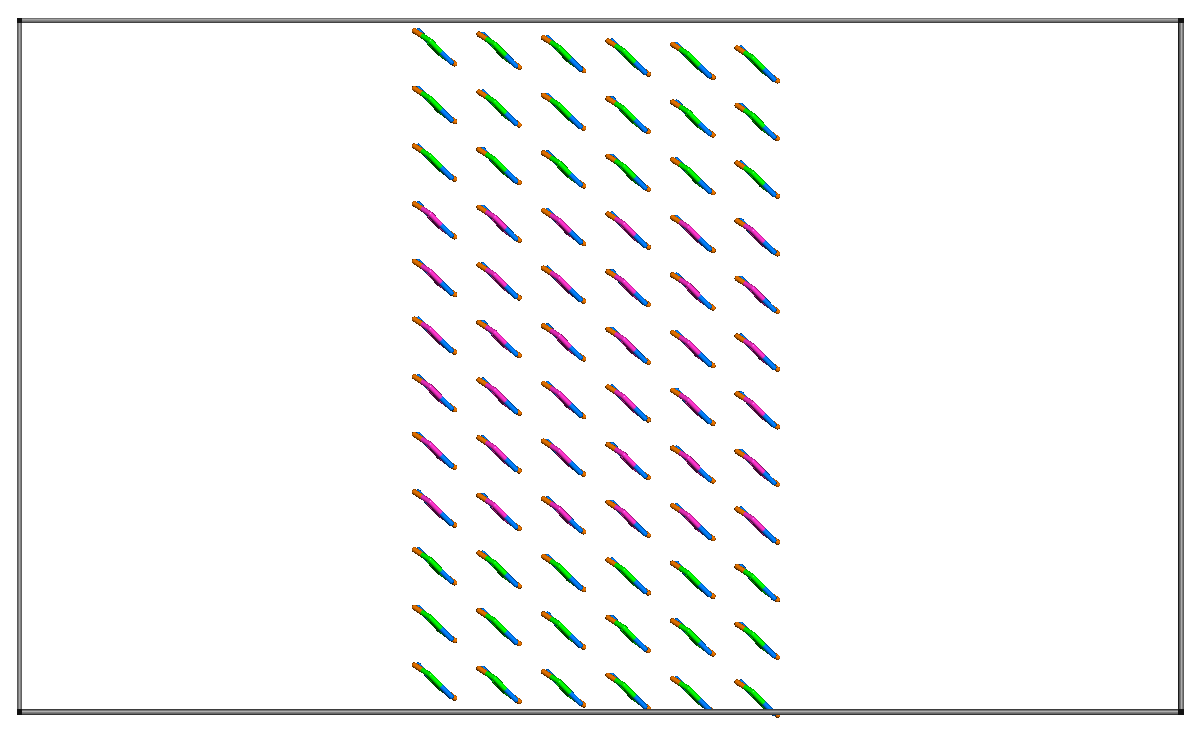}
            \includegraphics[width=0.45\textwidth]{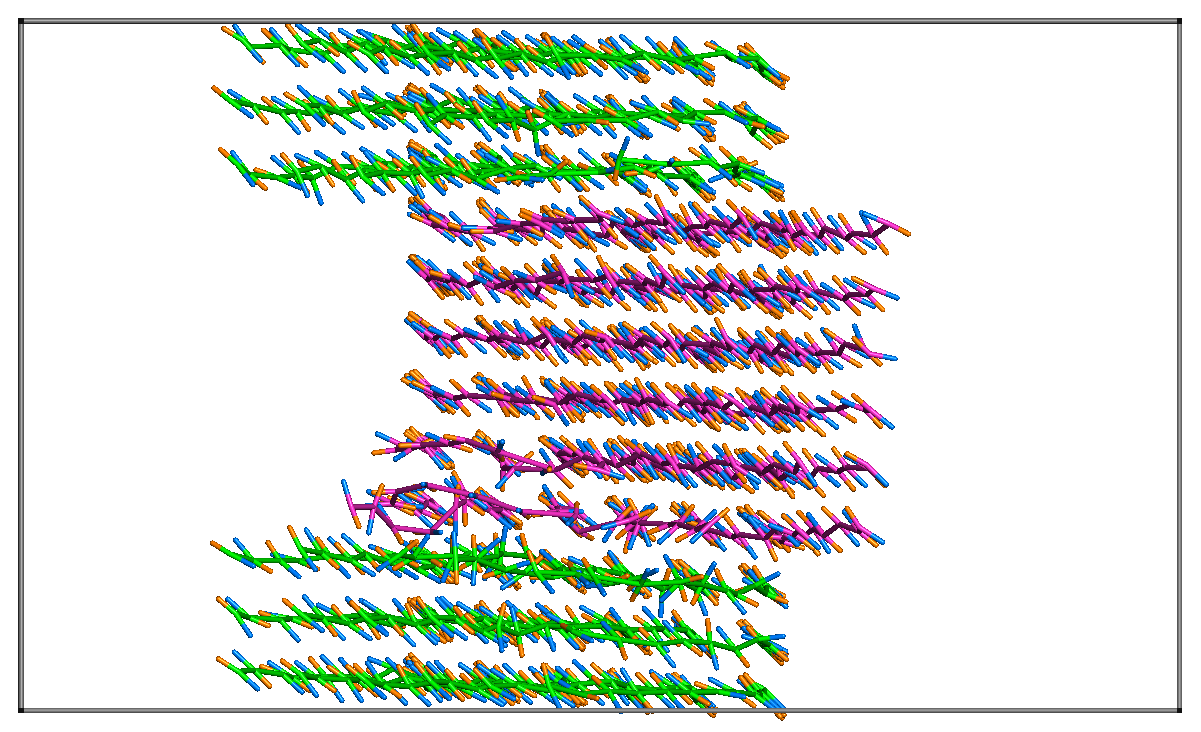}
            \\
            CG front view
            \\
            \includegraphics[width=0.45\textwidth]{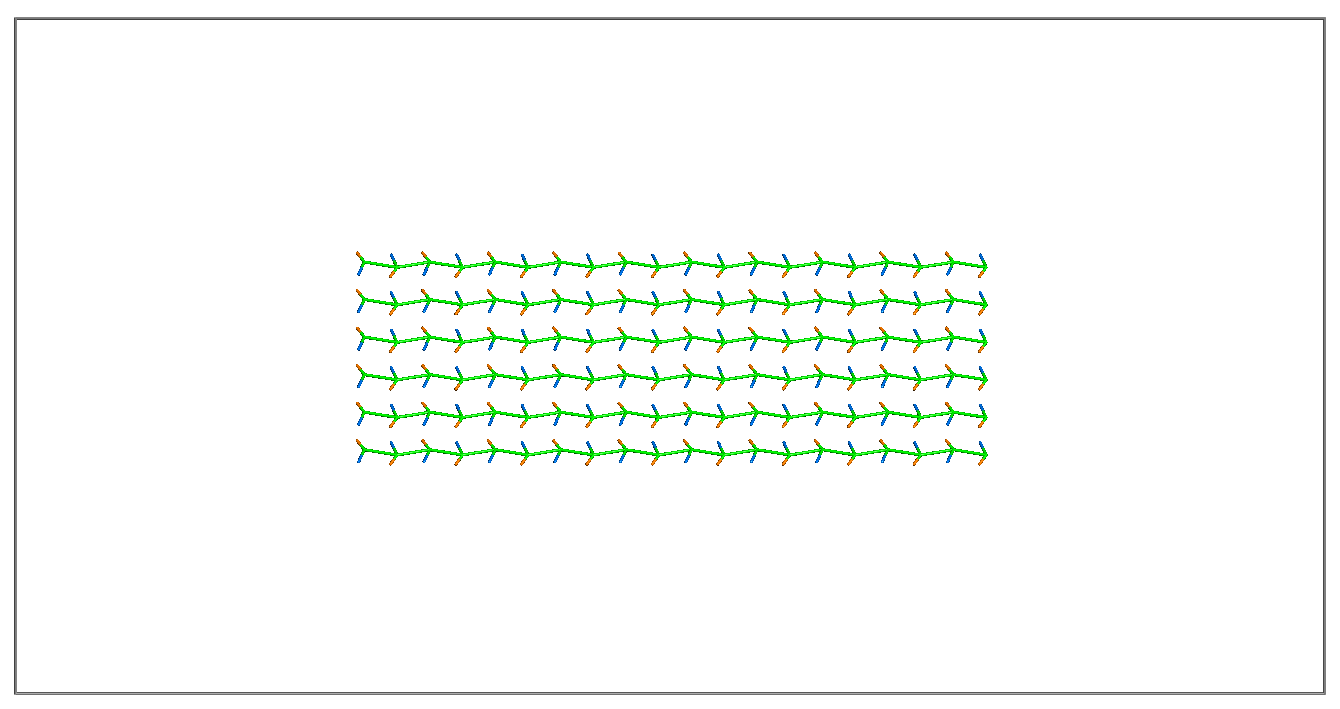}
            \includegraphics[width=0.45\textwidth]{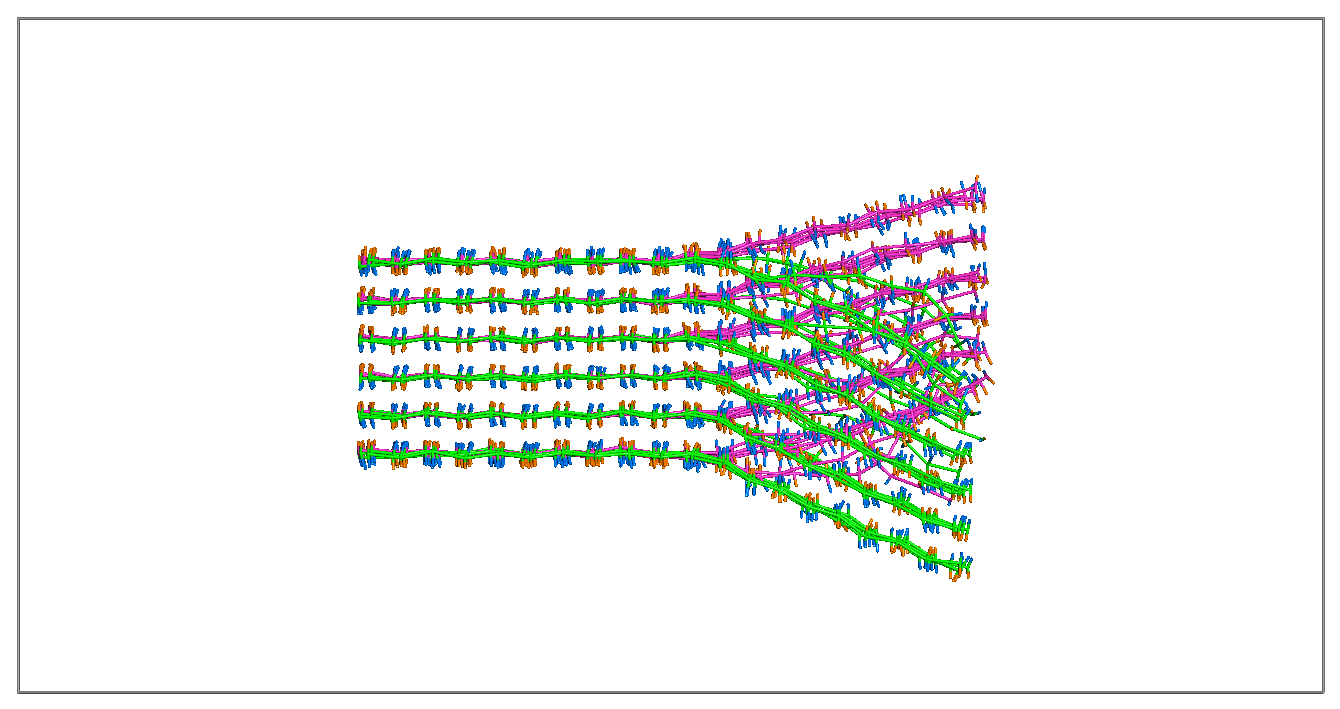}
            \\
            CG top view
        \end{minipage}
        \subcaption{}
    \end{subfigure}
    \begin{subfigure}[b]{0.30\textwidth}
        \centering
        \includegraphics[width=0.98\textwidth]{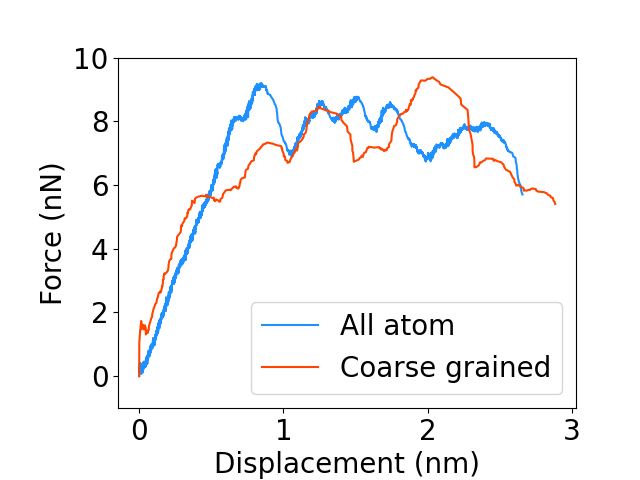}
        \subcaption{}
    \end{subfigure}
    \captionsetup{font=scriptsize}
    \caption{
    Draw-out and Tear-apart.
    (a) Draw-out and (b) force-displacement curves.
    (c) Tear-apart and (d) force-displacement curves.
    The transverse stretch resembled an opening fracture, whereas the in-plane and out-of-plane shear were examined using draw-out and tear-apart simulations, respectively.
    The draw-out and tear-apart tests were performed separately under the NPT and NVT ensembles by steered molecular dynamics with axial towing of the middle cellulose chain or perpendicular towing of the middle chains.
    The force-displacement curves of the draw-out presented periodic fading force curves that were ruled by their structure, whereas the tear-apart forces were incremental.
    The maximum force and frequency of the draw-out and tear-apart matched those of the AA references.
    Therefore, the CG model can quantitatively represent the in-plane and out-of-plane shear.
    }
    \label{fig:draw_out_and_tear_apart}
\end{figure}

Opening, in-plane shear, and out-of-plane shear are the three types of fracture.
The transverse stretch used in training corresponded to the type of opening fracture.
For the in-plane and out-of-plane shear, draw-out and tear-apart tests were performed using steered molecular dynamics (Figure~\ref{fig:draw_out_and_tear_apart}).
A draw-out test was performed in a periodic cell under the NPT ensemble, and the middle cellulose chain was pulled out of the CNC.
A tear-apart test towed the tips of the middle part of the cellulose chains perpendicular to their backbones under the NVT ensemble.
The structure and force-displacement curves are shown in Figure~\ref{fig:draw_out_and_tear_apart}.
Determined by the periodic structure and HBonds of cellulose, draw-out was a periodic process with force fading.
The tear-apart test is a type of force-increasing process, and the displacements and maximum forces of the AA and CG simulations are similar.
These results indicate that the CG model can accurately illustrate the three types of CNC fracture.

\subsubsection{Adhesion and Bending}\indent

\begin{figure}[htbp]
    \centering
    \begin{subfigure}[b]{0.96\textwidth}
        \centering
        \scriptsize
        \begin{minipage}[b]{0.48\textwidth}
            \centering
            \includegraphics[width=0.48\textwidth]{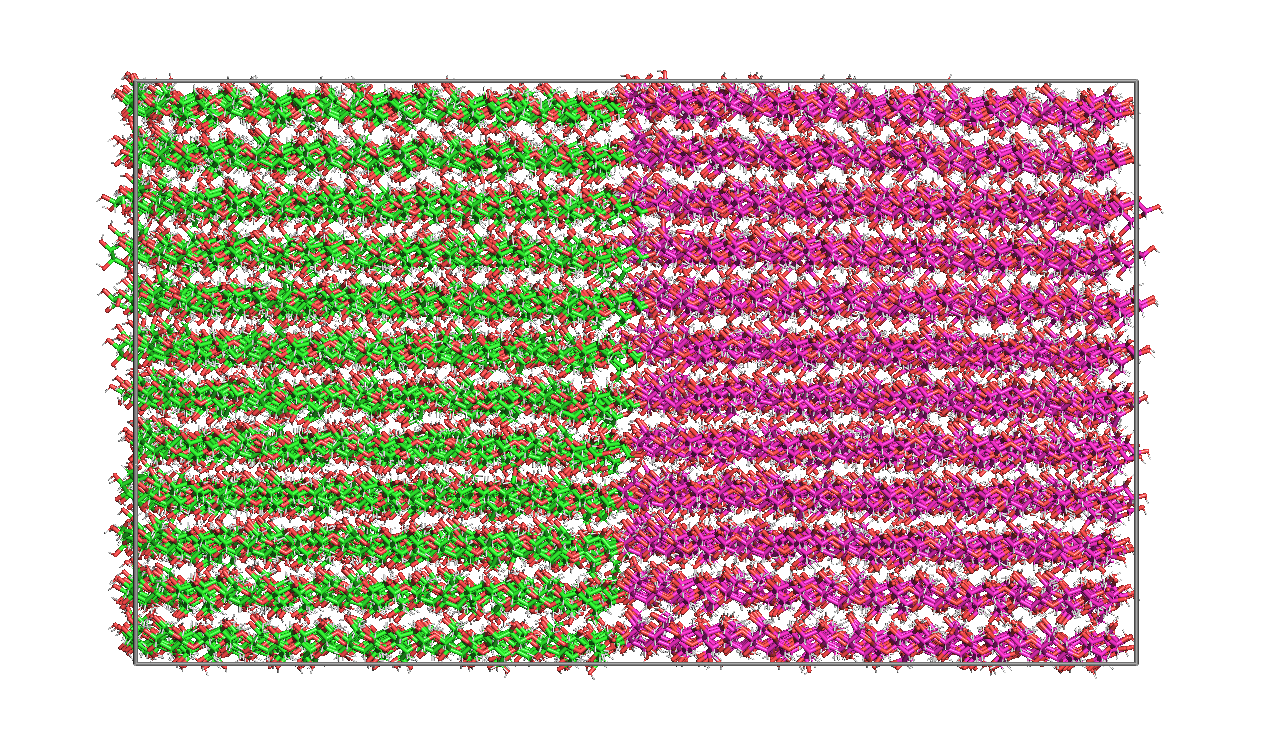}
            \includegraphics[width=0.48\textwidth]{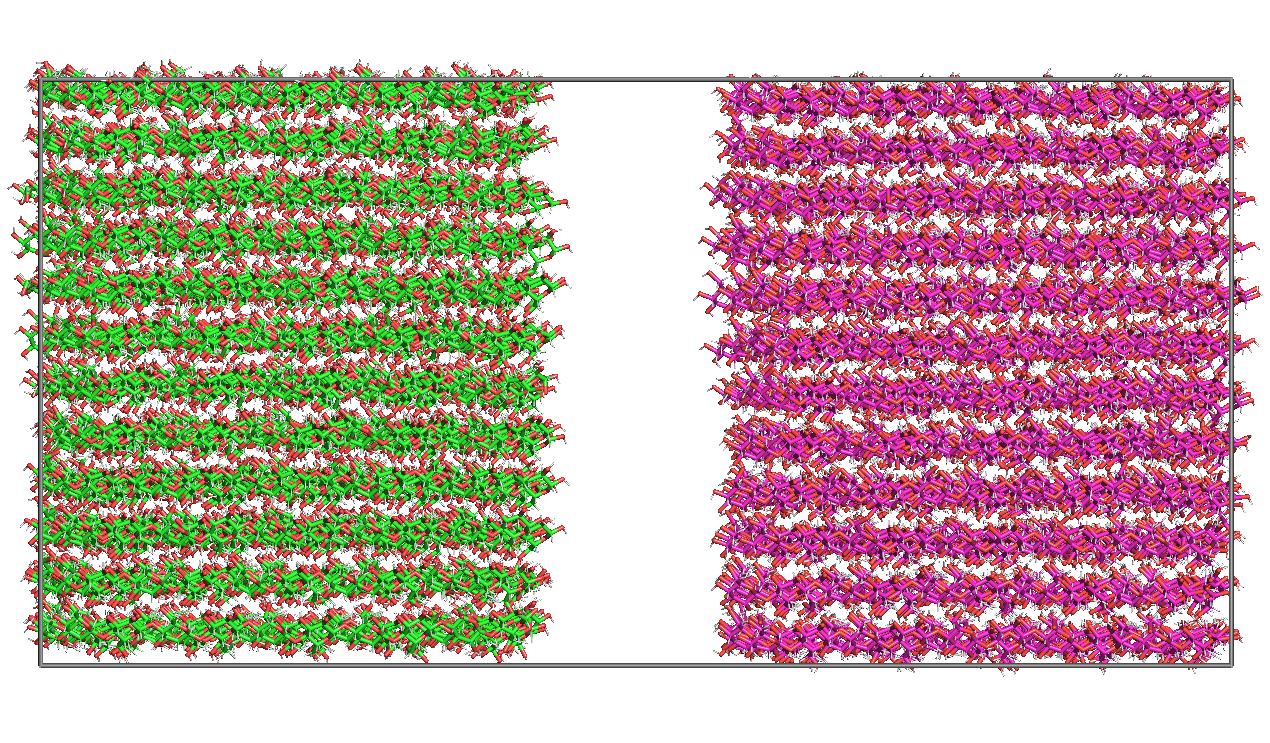}
            \\
            AA
            \\
            \includegraphics[width=0.48\textwidth]{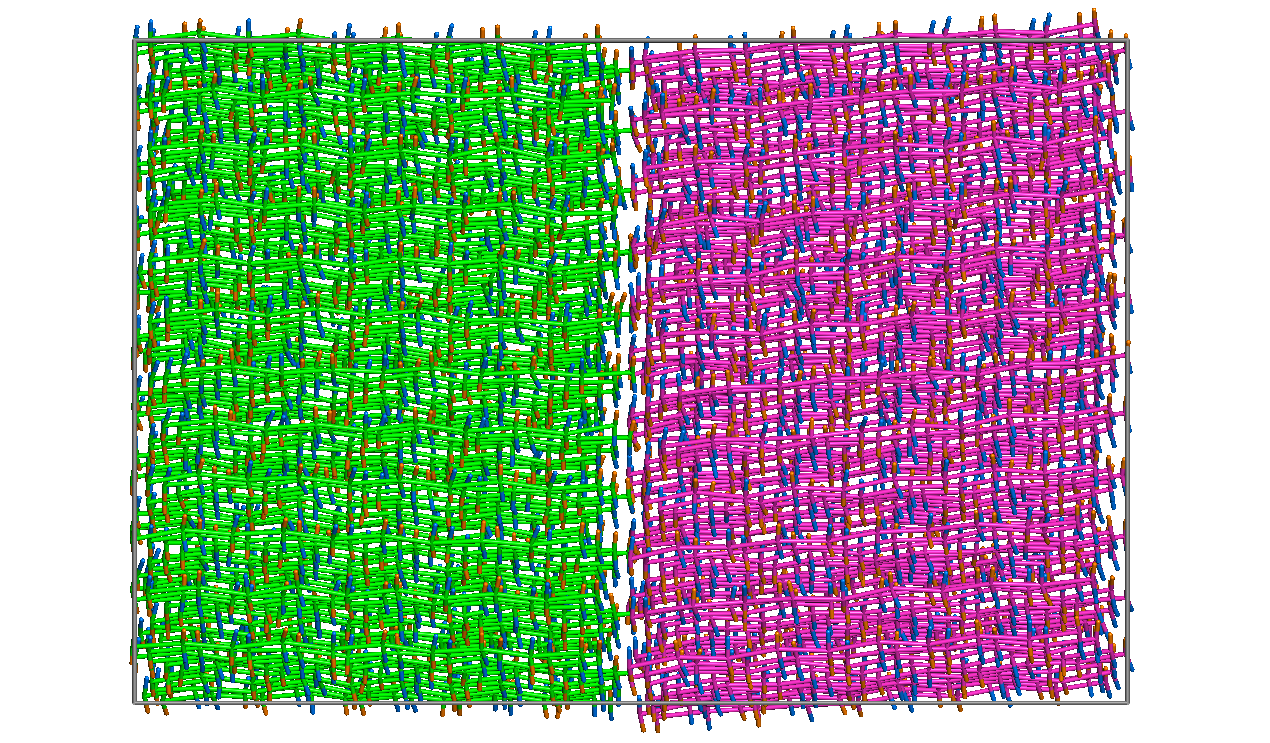}
            \includegraphics[width=0.48\textwidth]{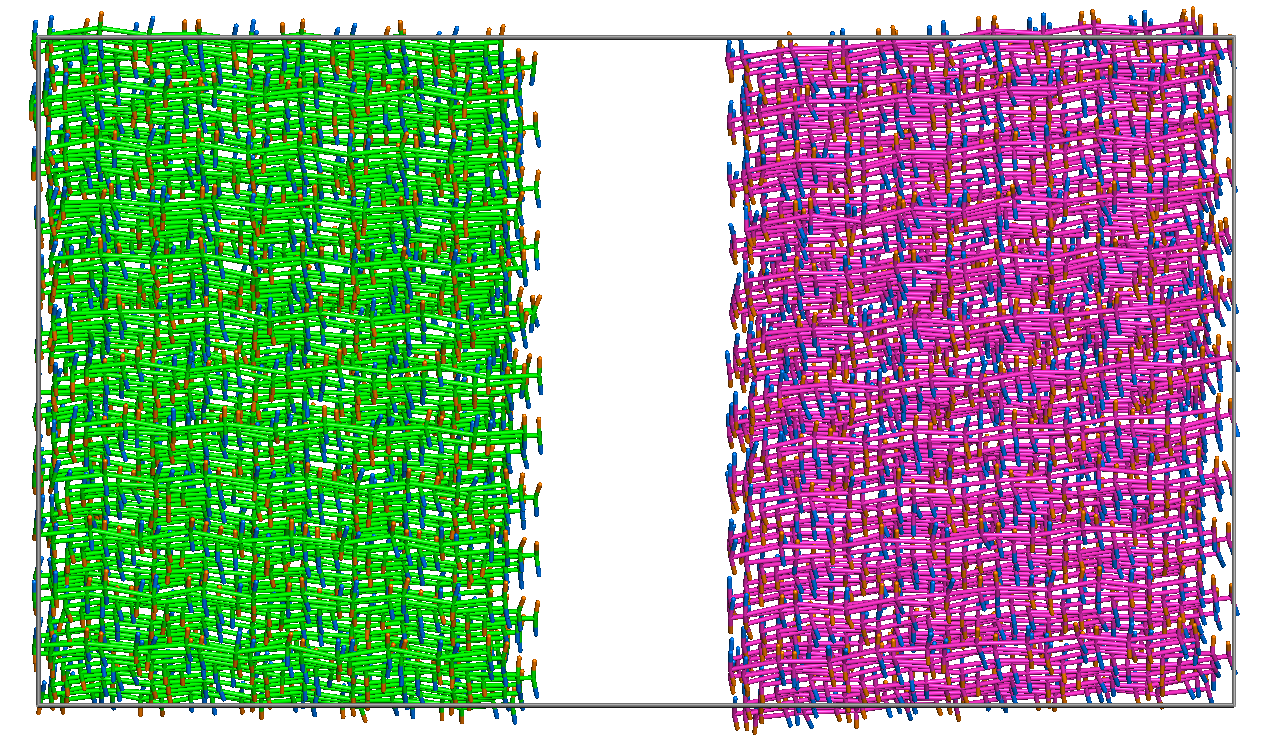}
            \\
            CG
        \end{minipage}
        \begin{minipage}[b]{0.48\textwidth}
            \centering
            \includegraphics[width=0.96\textwidth]{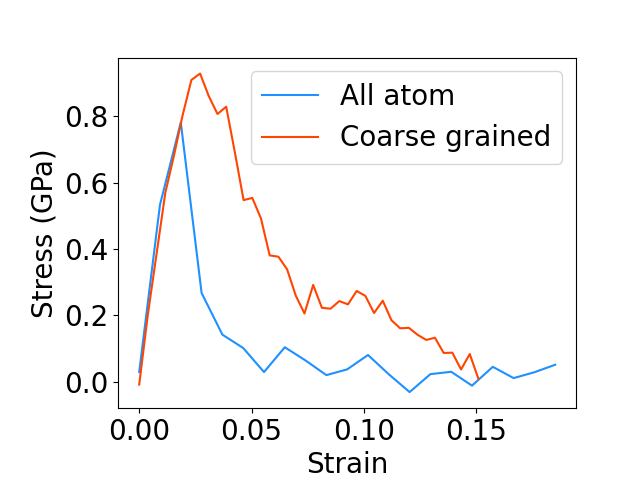}
        \end{minipage}
        \subcaption{}
    \end{subfigure}
    \\
    \begin{subfigure}[b]{0.96\textwidth}
        \centering
        \scriptsize
        \begin{minipage}[b]{0.48\textwidth}
            \centering
            \vspace{8.0pt}
            \includegraphics[width=0.48\textwidth]{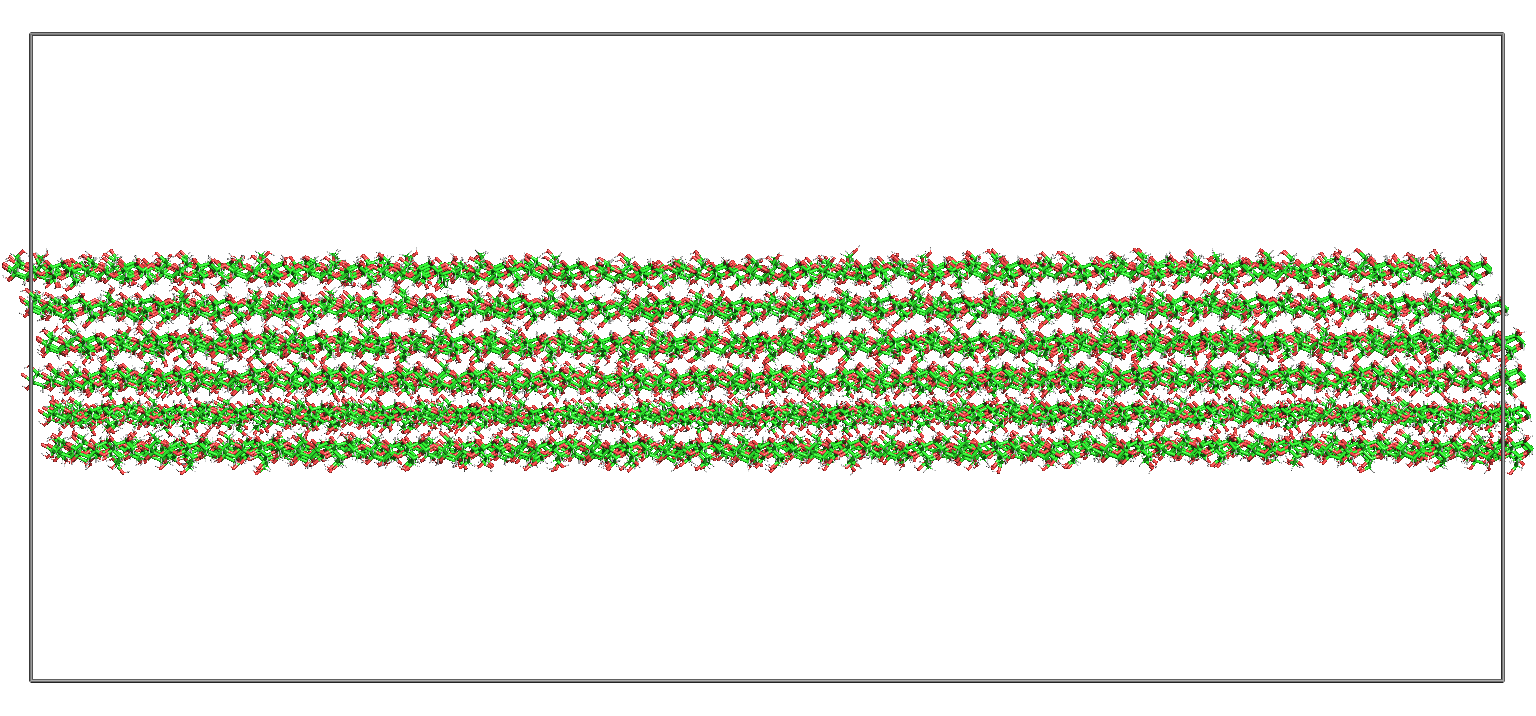}
            \includegraphics[width=0.48\textwidth]{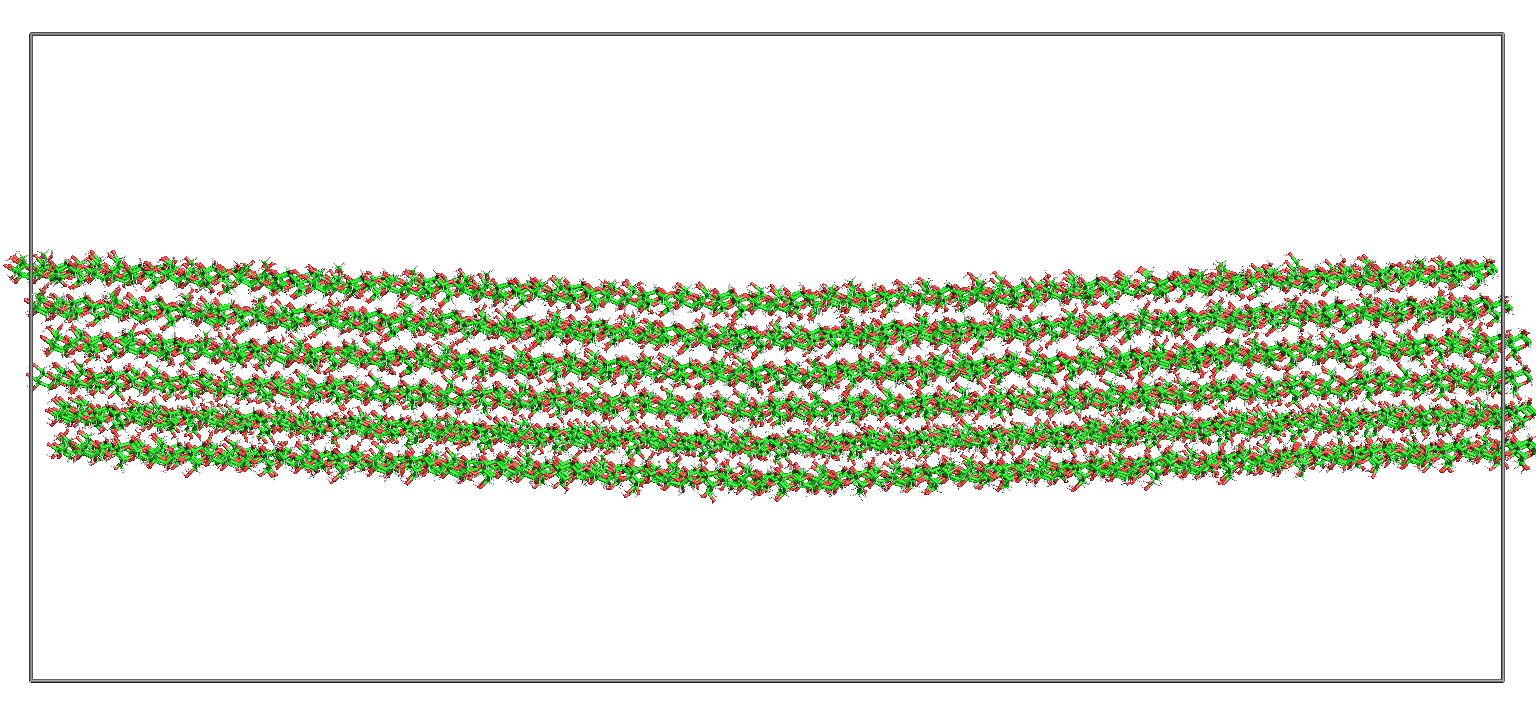}
            \\
            AA
            \\
            \includegraphics[width=0.48\textwidth]{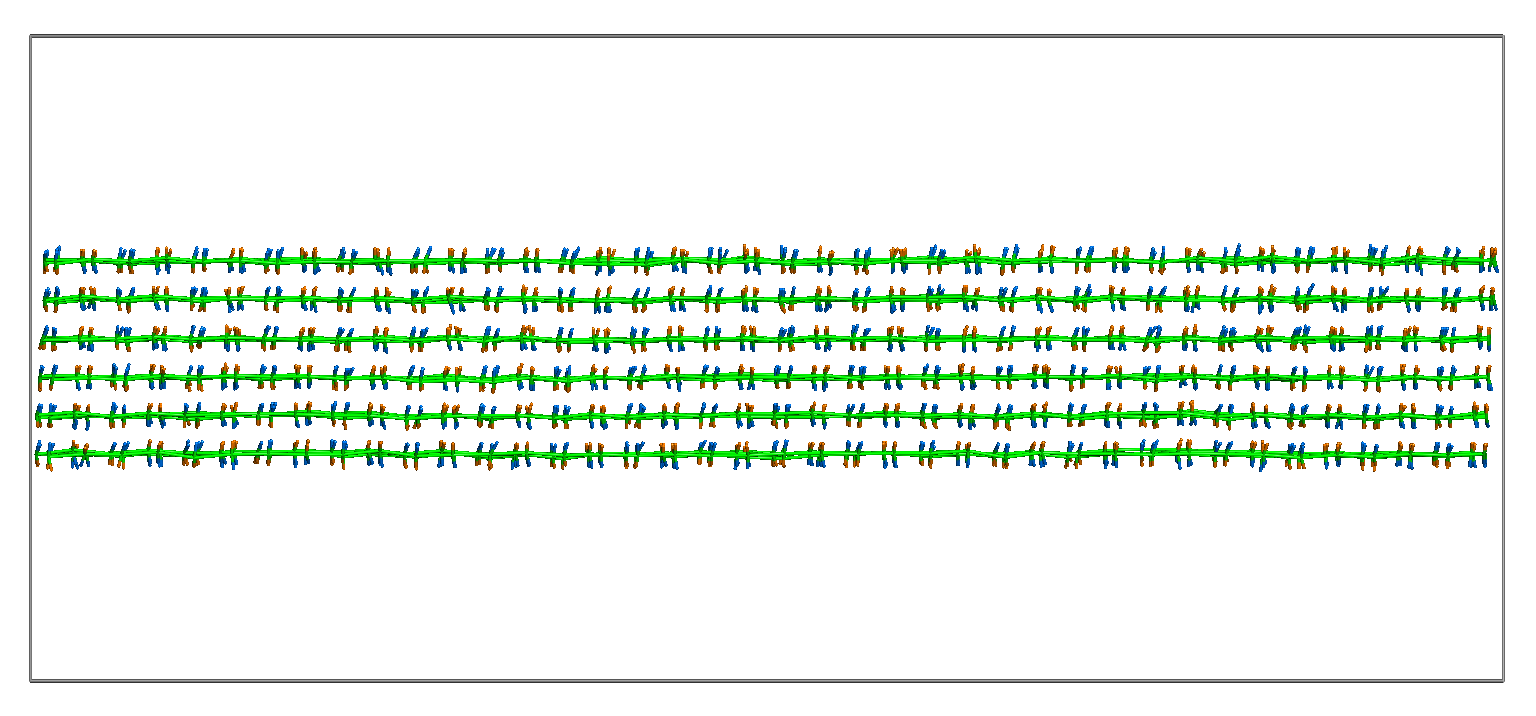}
            \includegraphics[width=0.48\textwidth]{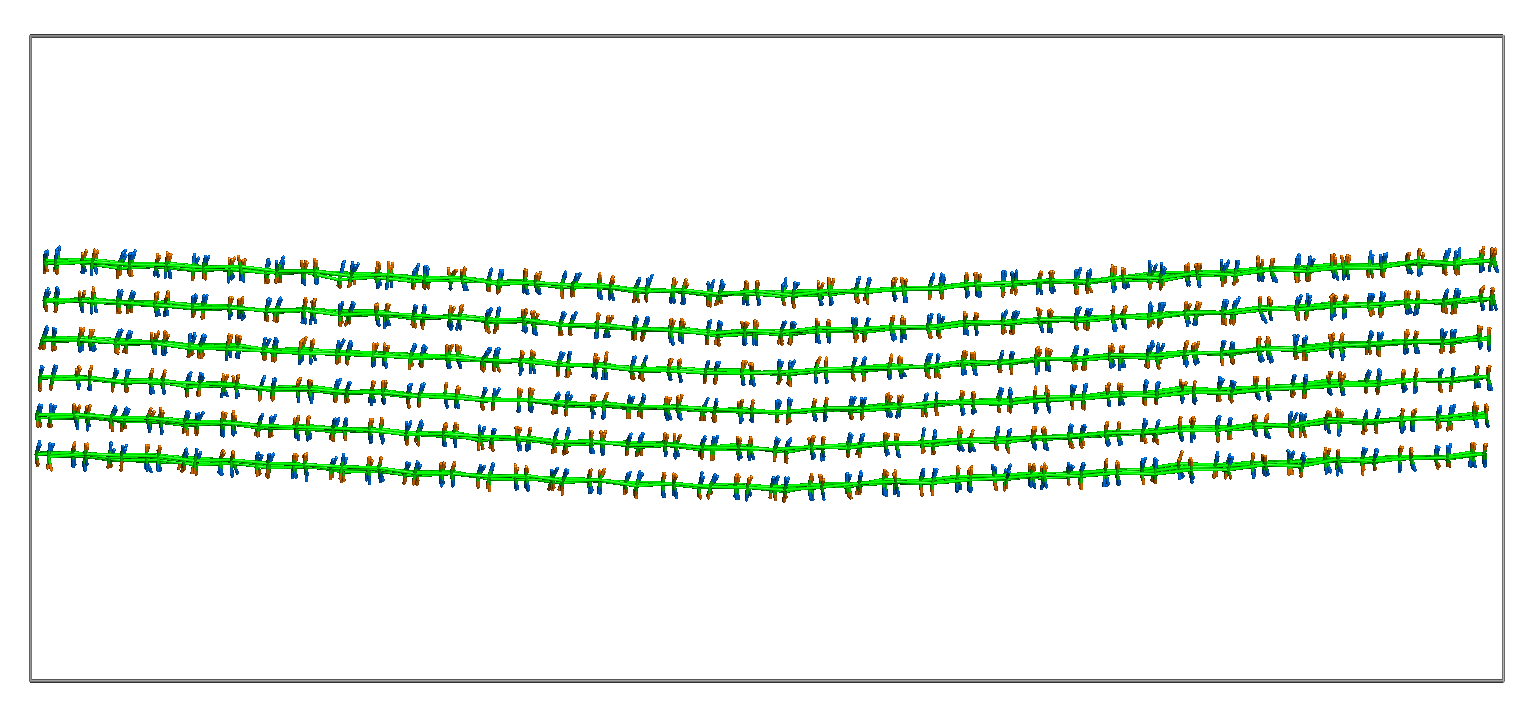}
            \\
            CG
        \end{minipage}
        \begin{minipage}[b]{0.48\textwidth}
            \centering
            \includegraphics[width=0.96\textwidth]{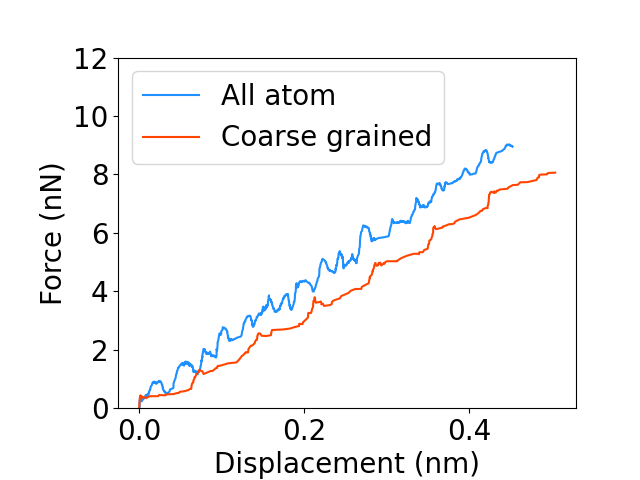}
        \end{minipage}
        \subcaption{}
    \end{subfigure}
    \captionsetup{font=scriptsize}
    \caption{
    Adhesion and Bending in aperiodic systems.
    (a) Adhesion and stress-strain curves.
    Two CNCs were placed tightly close along the chain direction and were clung by the interface adhesion energy.
    They were then torn away from one another.
    The adhesion energy of the CG model was larger.
    (b) Bending and force-displacement curves.
    A vertical load was applied at the center of the CNC bundle to induce bending.
    The force curves of adhesion and bending confirm the capabilities of the CG model in aperiodic systems.
    Unlike the axial modulus dominated by bonds, the bending is also affected by the angles and dihedrals.
    The slightly lower bending modulus was assumed to be caused by the rescaling of angle and dihedral force constants.
    }
    \label{fig:adhesion_and_bending}
\end{figure}

Although our CG model was designed for pure periodic systems, adhesion and bending loads were applied to verify its applicability to aperiodic systems (Figure~\ref{fig:adhesion_and_bending}).
An adhesion test was performed to determine the adhesion energy of nearby aperiodic CNC in the axial direction.
A vertical load was applied at the center point of the cellulose bundle for the bending test.
The adhesion energy of the CG model was evidently larger than that of AA, and the slightly lower force of the CG model in bending may result from the BD force constant rescaling of the angles and dihedrals.
The results confirmed that the CG model could also be used in aperiodic CNC systems with a slightly lower accuracy.

\subsubsection{Transverse arrangement}\indent

\begin{figure}[htbp]
    \centering
    \begin{subfigure}[b]{0.96\textwidth}
        \centering
        \scriptsize
        \begin{minipage}[b]{0.32\textwidth}
            \centering
            \includegraphics[width=0.96\textwidth]{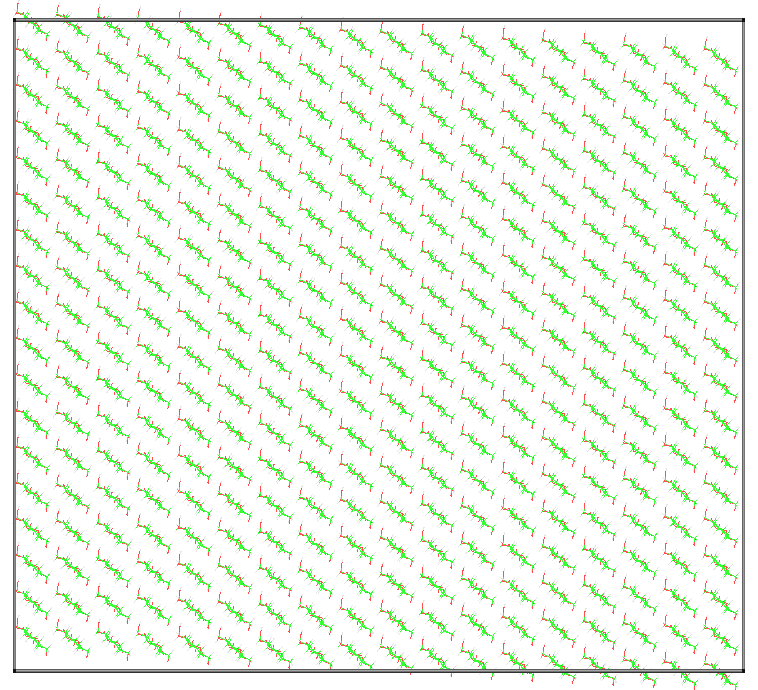}
            \\
            Crystal
        \end{minipage}
        \begin{minipage}[b]{0.32\textwidth}
            \centering
            \includegraphics[width=0.96\textwidth]{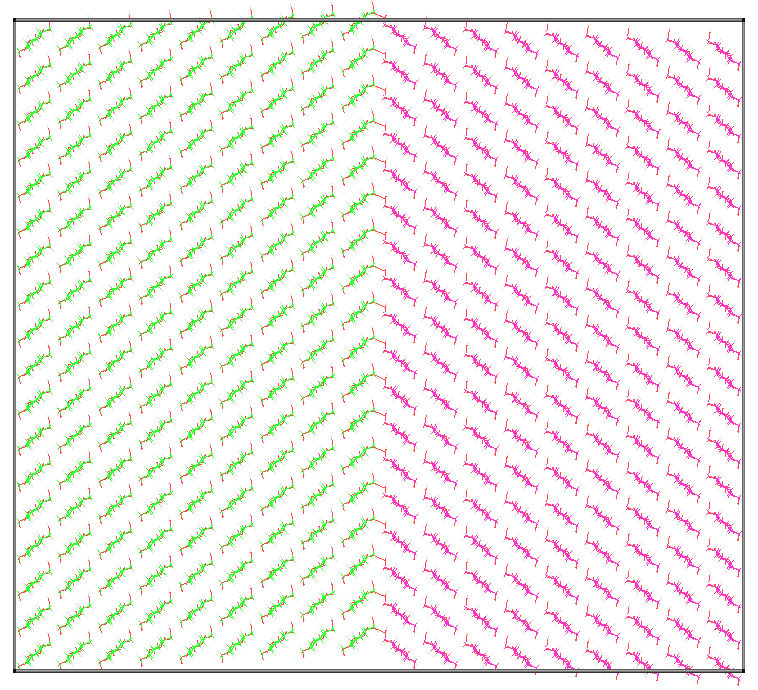}
            \\
            Symmetric
        \end{minipage}
        \begin{minipage}[b]{0.32\textwidth}
            \centering
            \includegraphics[width=0.96\textwidth]{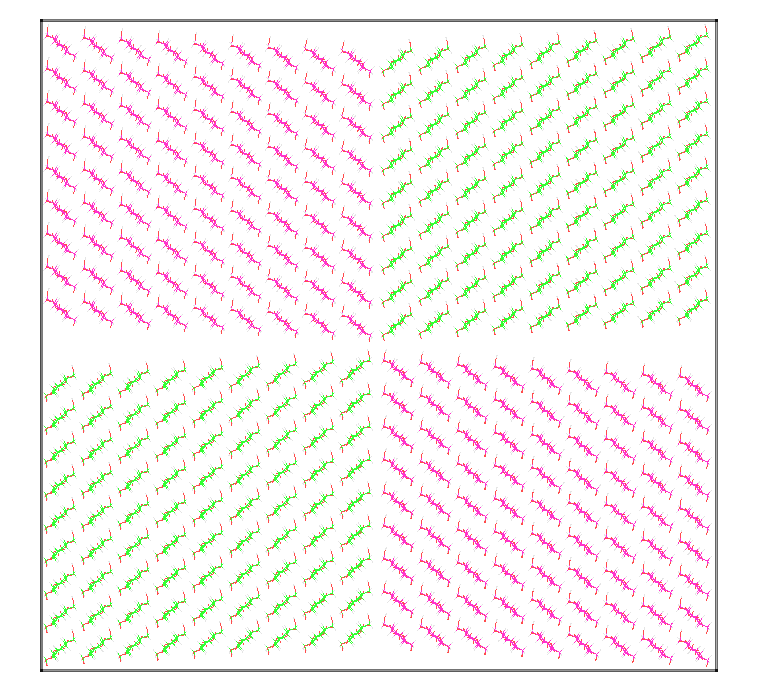}
            \\
            Cross
        \end{minipage}
        \subcaption{}
    \end{subfigure}
    \\
    \begin{subfigure}[b]{0.96\textwidth}
        \centering
        \scriptsize
        \begin{minipage}[b]{0.32\textwidth}
            \centering
            \includegraphics[width=0.48\textwidth]{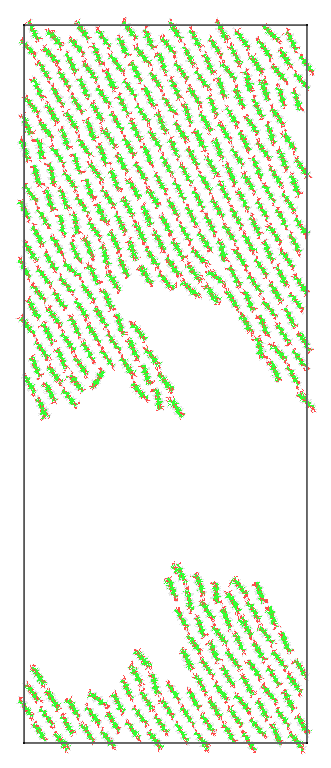}
            \includegraphics[width=0.48\textwidth]{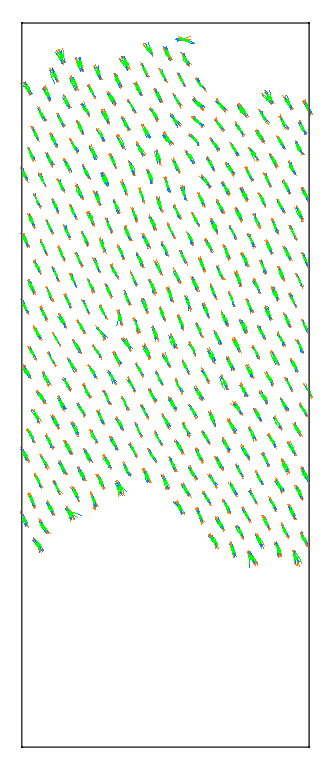}
            \\
            Crystal AA and CG
        \end{minipage}
        \begin{minipage}[b]{0.32\textwidth}
            \centering
            \includegraphics[width=0.48\textwidth]{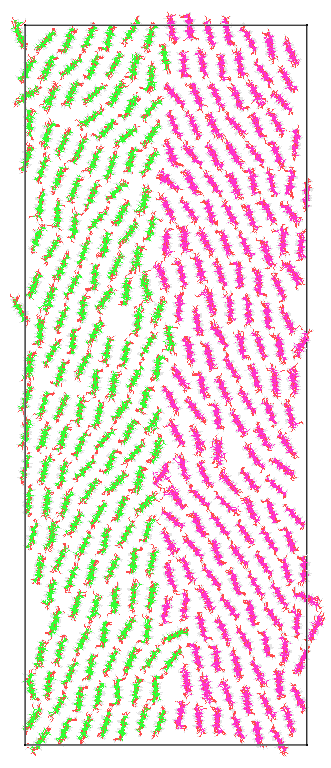}
            \includegraphics[width=0.48\textwidth]{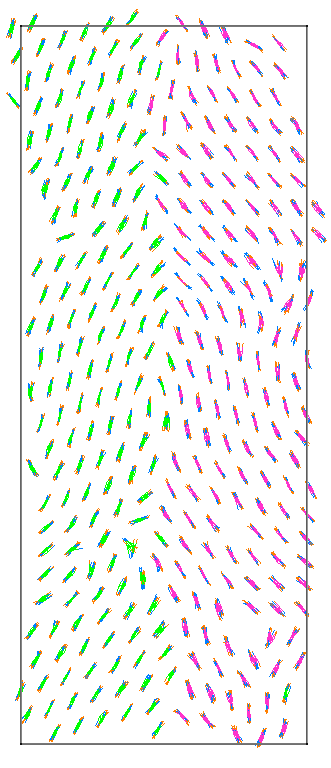}
            \\
            Symmetric AA and CG
        \end{minipage}
        \begin{minipage}[b]{0.32\textwidth}
            \centering
            \includegraphics[width=0.48\textwidth]{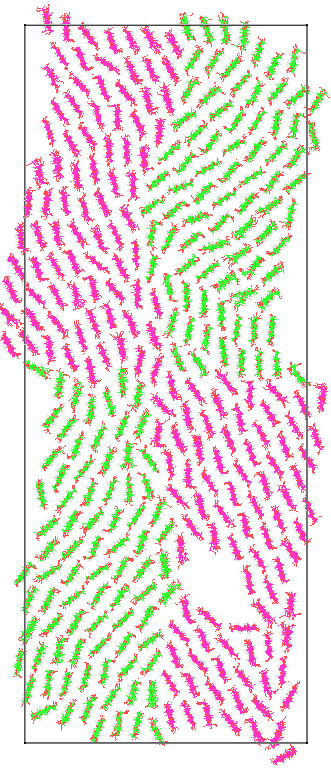}
            \includegraphics[width=0.48\textwidth]{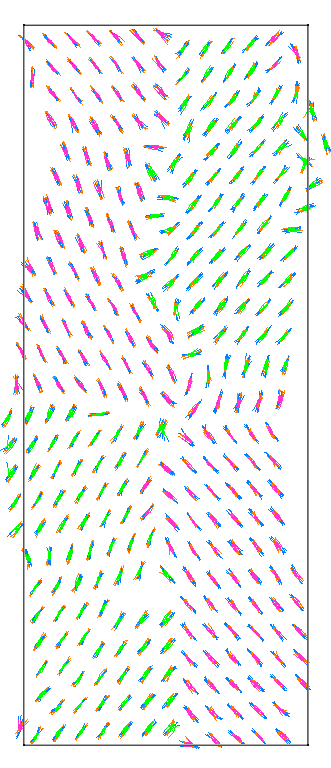}
            \\
            Cross AA and CG
        \end{minipage}
        \subcaption{}
    \end{subfigure}
    \\
    \begin{subfigure}[b]{0.96\textwidth}
        \centering
        \scriptsize
        \begin{minipage}[b]{0.32\textwidth}
            \centering
            \includegraphics[width=0.96\textwidth]{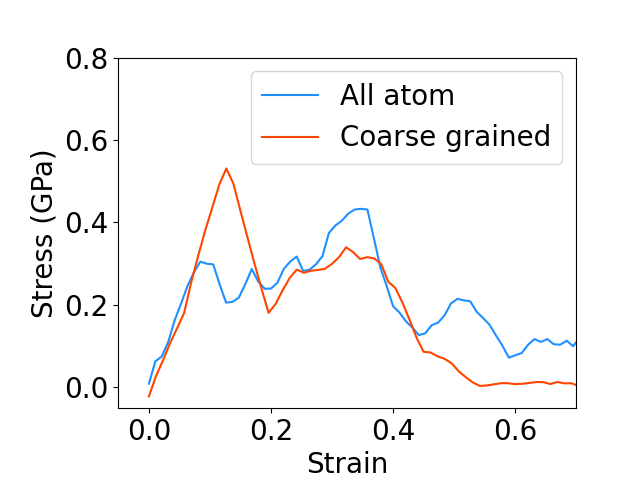}
            \\
            Crystal
        \end{minipage}
        \begin{minipage}[b]{0.32\textwidth}
            \centering
            \includegraphics[width=0.96\textwidth]{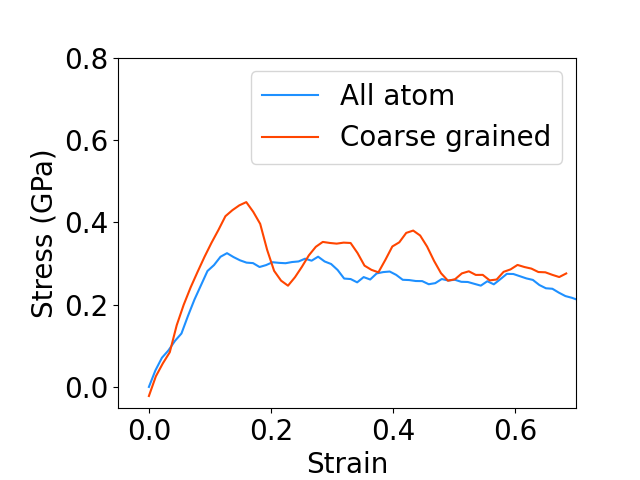}
            \\
            Symmetric
        \end{minipage}
        \begin{minipage}[b]{0.32\textwidth}
            \centering
            \includegraphics[width=0.96\textwidth]{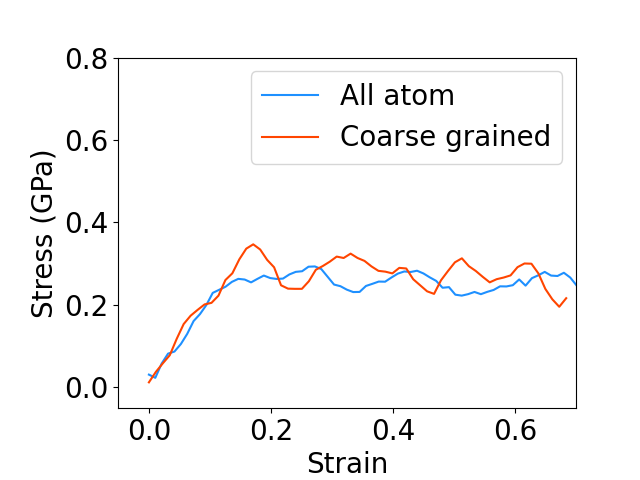}
            \\
            Cross
        \end{minipage}
        \subcaption{}
    \end{subfigure}
    \captionsetup{font=scriptsize}
    \caption{
    Transverse arrangement.
    (a) Transverse arrangement patterns.
    The crystal pattern was the same as that used for the potential parameterization.
    The symmetric pattern was inspired by the twin structure.
    The cross pattern was an X-shaped arrangement.
    (b) AA (left) and CG (right) structures under same vertical strain.
    In the symmetric and cross patterns, the global stretch was composed of the deformation of units, whose frictional sliding was hampered.
    In this case, local rotations became the dominant behavior, leading to better toughness and ductility.
    (c) Stress-strain curves of the three patterns.
    The transverse arrangements in the symmetric and cross patterns were better in ductility and could not be expressed by the previous CG models.
    These arrangements should be emphasized in future studies.
    }
    \label{fig:transverse_arrangement}
\end{figure}

Unlike previous BD directionality CG models\cite{fan2015coarse,shishehbor2019effects,ramezani2019mechanical,shishehbor2021influence}, the CG model was designed to present directionality via branched beads and HBond-like interactions.
This makes it possible to consider the transverse arrangements of CNCs with cross and symmetric patterns.
The results of the arrangement patterns confirmed that the fractures and performance differed significantly from those of the crystals, as shown in Figure~\ref{fig:transverse_arrangement}.
The symmetric pattern was inspired by the twin structure, and the results indicated that this pattern exhibited a better toughness than the crystal.
The cross pattern was symmetric in both directions.
In these two patterns, frictional sliding was impeded and transformed into local rotations, thereby achieving better toughness and ductility in the symmetric pattern.
These results emphasize the importance of the transverse arrangement, which cannot be described by previous CG models.

\subsubsection{Brick-and-mortar}\indent

\begin{figure}[htbp]
    \centering
    \begin{subfigure}[b]{0.64\textwidth}
        \centering
        \scriptsize
        \includegraphics[width=0.48\textwidth]{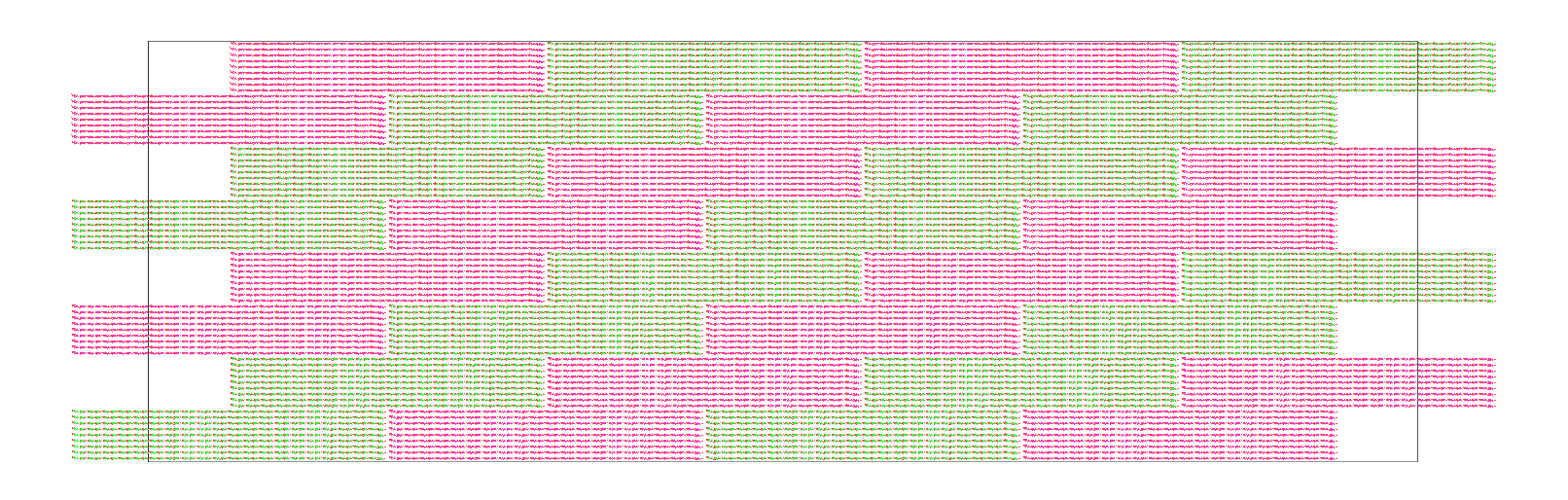}
        \includegraphics[width=0.48\textwidth]{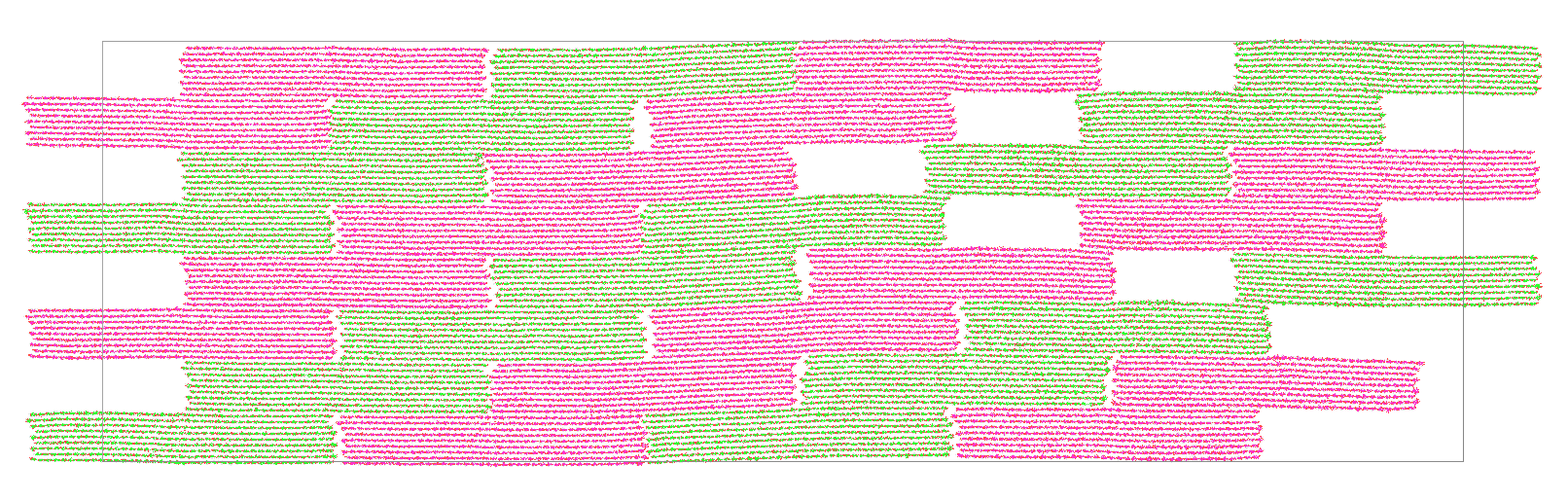}
        \\
        AA
        \\
        \includegraphics[width=0.48\textwidth]{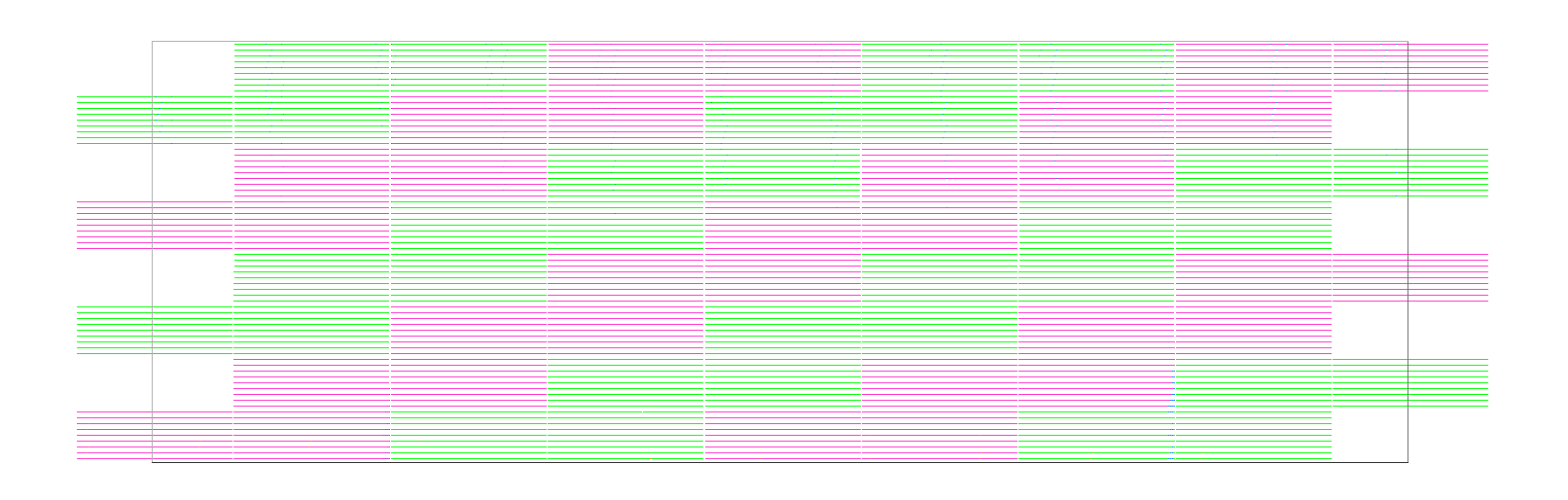}
        \includegraphics[width=0.48\textwidth]{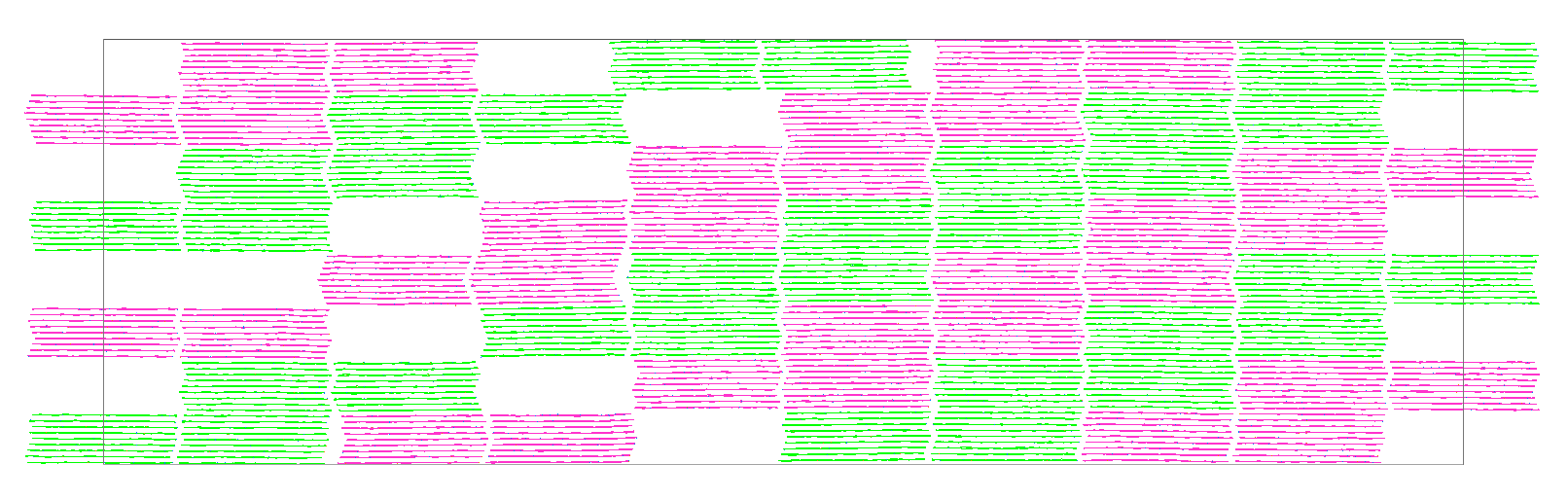}
        \\
        CG
        \\
        \subcaption{}
    \end{subfigure}
    \begin{subfigure}[b]{0.32\textwidth}
        \centering
        \includegraphics[width=0.96\textwidth]{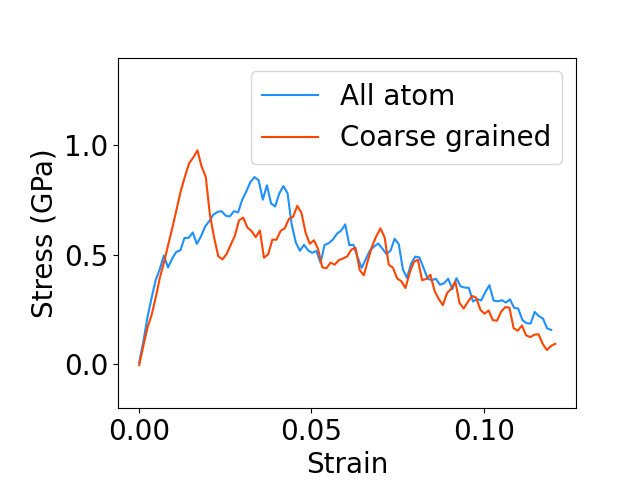}
        \subcaption{}
    \end{subfigure}
    \\
    \begin{subfigure}[b]{0.48\textwidth}
        \centering
        \includegraphics[width=0.96\textwidth]{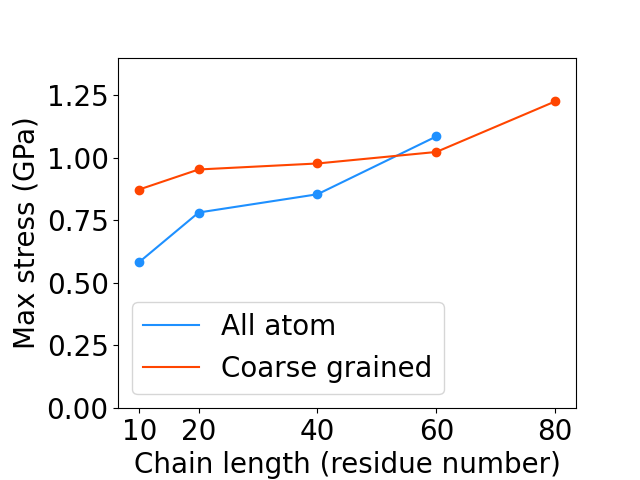}
        \subcaption{}
    \end{subfigure}
    \begin{subfigure}[b]{0.48\textwidth}
        \centering
        \includegraphics[width=0.96\textwidth]{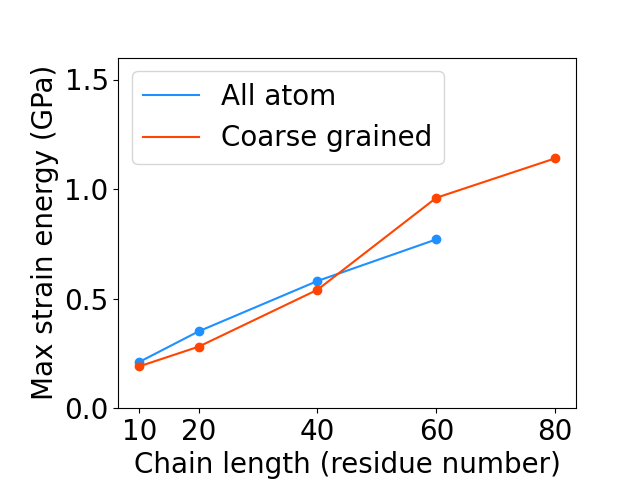}
        \subcaption{}
    \end{subfigure}
    \captionsetup{font=scriptsize}
    \caption{
    Brick-and-mortar.
    (a) Brick-and-mortar models and fractures.
    (b) Stress-strain curves when the elementary cellulose chain length was 40 residues.
    (c) Strength and (d) toughness curves of different elementary chain lengths.
    Highly aligned CNCs can assemble into brick-and-mortar structures \cite{li2004nanoscale,natarajan2018bioinspired} in the axial direction, and this staggered structure can improve the strength and toughness.
    The brick-and-mortar stretch was also a CNC-level in-plane shear, and was heavily influenced by the length of the elementary cellulose chains.
    The CG model accurately describes the structural behavior of the brick-and-mortar.
    The brick-and-mortar fracture was composed of major local slips and decentralized minor slips, and the toughness is positively correlated with the overlapping length.
    }
    \label{fig:brick_and_mortar}
\end{figure}

According to previous studies, highly aligned CNCs can be assembled into staggered brick-and-mortar\cite{li2004nanoscale,natarajan2018bioinspired} and chiral nematic helicoidal structures\cite{natarajan2018bioinspired}.
The brick-and-mortar structure can increase the stiffness, strength, and toughness, and this test can be considered as a mesoscopic in-plane shear, as shown in Figure~\ref{fig:brick_and_mortar}.
In these simulations, each CNC brick was composed of 36 parallel cellulose chains and 4$\times$8 CNCs were assembled.
The elementary cellulose chains contained 10, 20, 40, 60, or 80 cellulose residues.
There are 1.5 million atoms in the AA system if the elementary brick is 60 cellulose residues long.
Longer cellulose bricks were simulated using the CG model only. 
Local slips and small decentralized slips appeared as frictional fracture behaviors for the staggered structures with different brick lengths.
Stretch toughness and overlapping length also demonstrated a positive correlation.

\subsubsection{Computational efficiency}\indent

\begin{figure}[htbp]
    \centering
    \begin{subfigure}[b]{0.48\textwidth}
        \centering
        \includegraphics[width=0.96\textwidth]{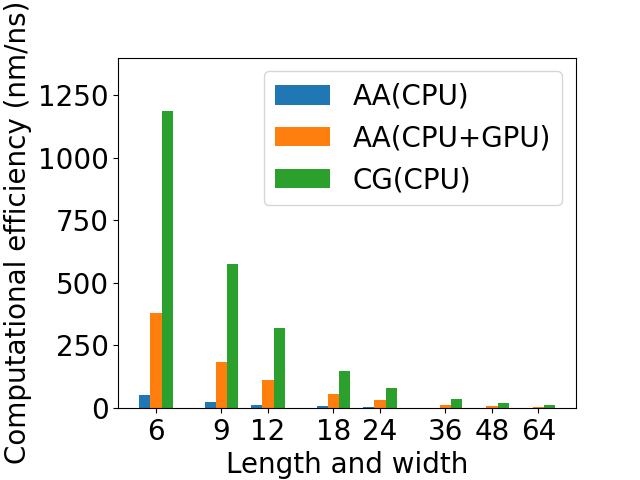}
        \subcaption{}
    \end{subfigure}
    \begin{subfigure}[b]{0.48\textwidth}
        \centering
        \includegraphics[width=0.96\textwidth]{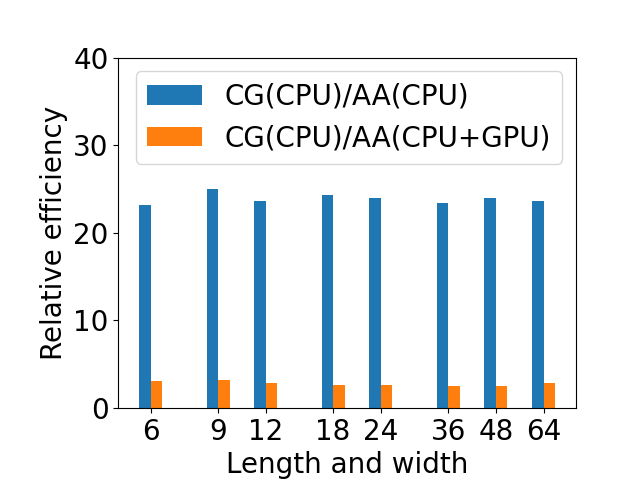}
        \subcaption{}
    \end{subfigure}
    \captionsetup{font=scriptsize}
    \caption{
    Computational efficiency.
    (a) Referring to the AA simulations with the same amount of CPU resources, the CG model was 20 times faster in the CNC equilibrium simulations of different transverse sizes.
    (b) The efficiency of the CG model was still triple when additional exclusive GPU resources were utilized for the AA.
    However, its computational efficiency has not yet been fully proven owing to the lack of GPU acceleration.
    }
    \label{fig:computational_efficiency}
\end{figure}

The computational efficiencies of the AA and CG models were compared using equilibrium simulations.
A sequence of square periodic CNCs (the word square indicates that the row and column numbers of the transverse section of the CNCs were the same) equilibrium simulations of different sizes were used, in which the atom number of the AA models ranged from seven kilos to one million.
The benchmark simulations were performed using central processing units (CPU) and graphical processing units (GPU).
The HB interaction does not support hardware acceleration, and the AA simulations using only the same amount of CPU resources are shown in Figure~\ref{fig:computational_efficiency}.
For the AA and CG simulations using only the CPU, six cores were allocated exclusively to each task.
Six cores and an additional exclusive GPU were allocated for the accelerated AA simulations using a GPU.
The results confirmed that the CG model was 20 times faster than its AA counterpart when using the same resources.
Even when referring to the AA models with additional GPU acceleration, the efficiencies of the CG models were tripled.
When GPU-accelerated HB interaction is implemented and tested in the future, its potential can be fully exploited.

\section*{Discussion}
\addtocounter{section}{1}
\setcounter{subsection}{0}
\addcontentsline{toc}{section}{\protect\numberline{}Discussion}\indent

Cellulose materials are usually composed of long and thin cellulose nanofibrils, and previous CG models of cellulose have emphasized their axial properties\cite{qin2017optimizing,ray2021mechanics}.
However, CNCs are anisotropic and much shorter, and their transverse properties should not be disparaged, which is important for the future development of advanced cellulose materials.
Some exhibit transverse anisotropy on the CG scale using anisotropic particles\cite{rolland2020new,nguyen2022systematic} or additional artificial bonds\cite{fan2015coarse,ramezani2019mechanical,shishehbor2019effects,shishehbor2021influence}, which are still incapable of fully representing the transverse anisotropy.

The challenges and main characteristics of the CG model include the flat residues and hydrogen bond interactions.
This approximate CG model could reveal the laminar structures and frictional sliding, which are two critical properties for the transverse anisotropy of CNCs, and the previous CG models cannot reflect these accurately.
Except for the HBond interactions, only the harmonic and Lennard-Jones potentials were included in the BD and NB interactions.
This model also confirms the potential parameterization capability of RL: only a discriminant reward function was required, which was further emphasized via the comparisons with TPE and CMAES.
The reward function for RL includes the matching degree of the BD properties (axial modulus and polymer stiffness) and the NB properties (transverse strength and toughness).
RL may be more widely applied to potential parameterizations in the future.

In the context of CG modeling, this model followed a bottom-up approach starting from molecular details but using strength and toughness as optimization targets, rather than the potential of mean force or other structural distributions.
It chose simple analytical function forms, which is also similar to the top-down approach to a certain extent.
However, the key of this model is still molecular details: its mapping and topology restore the molecular shape and frictions, and the interaction functions implement anisotropic HBonding.
All of these enable this model to explore both the equilibrium conditions and the dynamic properties with quantitative accuracy.
On the other hand, this model preserved the necessary molecular structures, and may be suitable for AA reconstruction based upon CG models.

However, several problems are encountered in the CG model.
The first is the inherent representability problem of CG modeling\cite{dannenhoffer2019compatible,dunn2016van,jin2019understanding,johnson2007representability,wagner2016representability}.
Although the transverse strength and toughness are important target properties for RL parameterization, there is one critical problem: the strength and toughness data and distribution comparisons between AA and CG models are deficient in the solid physics background.
Fortunately, the structural properties are well aligned as an offset, otherwise the basic physical consistency of this model is questionable.
Although it can explore both the equilibrium and non-equilibrium properties with quantitative accuracy, it must be described as an approximate model.
The second is the rescaling of the angle and the dihedral interaction force constants, which is also a problem of representability resulting from reductions.
The CG model, which has significantly fewer degrees of freedom, must compensate for the polymer stiffness properties by decreasing the force constants (Table~\ref{tab:bonded_properties_without_rescale}).
The previous CG model addressing the persistence length did not require a precise match (40~nm)\cite{goundla2014coarse}.
Some studies have also estimated the persistence length of a bundle of cellulose chains instead of a single free chain\cite{glass2012reach}.
These smaller force constants may cause problems for other properties involving the angles and dihedrals.
For example, a lower stress was observed during the bending test.

NB equilibrium distances are also problematic.
The equilibrium distances of the CG model were larger than those of their AA counterparts (Figure~\ref{fig:nonbonded_equilibrium_distance_distributions_equilibrium} and Figure~\ref{fig:nonbonded_equilibrium_distance_distributions_stretch}), and the enlargement of the CL2-CL3 NB distance coefficients was determined using RL (Figure~\ref{fig:convergence_and_statistics}).
This means that the distance coefficients and equilibrium distances do not correspond to the well-relaxed AA equilibrium distances or radial distribution function.
In this study, the equilibrium distance of a pair was defined as the average minimum distance.
Nearly all minimum distances of pairs are from beads of nearby HBonding layers; only the minimum distances of CL2-CL3 are from beads within the same HBonding layer.
Therefore, we infer that the equilibrium distances from nearby HBonding plane pairs are appropriate estimations of $\sigma$ values for the LJ potentials because the dominant NB interactions are HBonds in the HBonding layers.
An overall enlargement of the NB distance coefficients is assumed to be required by the increased $\sigma_{p23}$ to retain the crystal structure.
Moreover, the enlargement of $\sigma_{p23}$ may be an assimilation of $\sigma$ because the equilibrium distance between CL2 and CL3 is the smallest.
The phenomenon of frictional sliding may be presented more easily if all $\sigma$ values are close to each other.
This also influenced the accuracy of the $\theta_0$ in the HB potential.
In the first reduction, the distance coefficients between pairs of the same type were artificially constrained to be the same because the laminar structure and coefficients are critical for the CG model.
To align with the artificial constraints, $\theta_0$ in the HB potential was determined by the ideal model, which was constructed using coordinate offset and was ideally parallel.
Therefore, the $\theta_0$ is also an approximation without a strong theoretical background.

Other minor problems also exist.
Our model may not perform well in aperiodic cellulose systems (as in the adhesion test).
And the pressure curve of the CG model under the NPT ensemble exhibited a larger oscillation amplitude (Figure~\ref{fig:timestep_validation_thermodyanmics}).
Otherwise, some molecular details were ignored in the CG model.
For instance, the rotational degrees of freedom of the hydrogen atoms on the branch hydroxyls were neglected in harmonic BD potentials (Figure~\ref{fig:secondary_summits_of_the_boned_geometry_parameter_distributions}).
Moreover, one technical problem temporarily restrained the application potential of the CG model: 
the HBond-like pair style does not support hardware acceleration.
Therefore, the computational efficiency of the CG model has not yet been fully realized.

As a pioneering work that introduced HBonds, the CG model of CNC could promote further researches.
The development of cellulose CG models has mainly focused on the native cellulose.
However, solvent environments, particularly humidity, are key factors for cellulose materials\cite{goundla2014coarse,petridis2014hydration,hou2020strengthening,guo2023molecular} and are not well represented in cellulose CG models\cite{ray2021mechanics}.
Chemical surface modifications and interactions with other molecules such as hemicellulose or lignin are also critical for enhancing the performance of cellulose materials.
These problems are strongly dependent on the hydroxyl groups of cellulose residues.
Regarding the computational resources and expertise, these interactions were not considered in this study.
With the help of HBond-like interactions, later CG models could attempt to present more properties of the cellulose.
We hope that this model will facilitate future development of CG models.

As an application of RL for potential parameterization, our methods and implementations are not heavily dependent on cellulose, and may help in general CG model development.
In this CG model, much prior knowledge, including pre-defined mapping and interaction potential expressions, is exploited to reduce the difficulty and enhance the physical explainability.
However, the resultant effect of prior knowledge remains unclear.
Fully automatic optimization may achieve models with better performances.
Transfer Learning and Imitation Learning methods may also be applied in the future to further accelerate RL in the potential parameterization applications.

\section*{Conclusions}
\addtocounter{section}{1}
\setcounter{subsection}{0}
\addcontentsline{toc}{section}{\protect\numberline{}Conclusions}\indent

In this study, we introduce a novel CG cellulose model with explicit branch beads and HBond-like interactions that can reproduce the anisotropy of CNCs.
The strength and toughness in the three characteristic directions of the orthotropic transverse section were quantitatively reproduced without anisotropic particles or additional bonds.
In this pioneering study, degenerate RL was applied as a nonlinear optimizer to perform direct analytical potential parameterization, and its physical meaning was enhanced by integrating the Boltzmann Inversion method.
This model synthesizes physical knowledge, traditional modeling, and Machine Learning methods.
It aids in understanding the importance of hydrogen bonds and the transverse anisotropy.
By using an appropriate reward function based on the target properties and statistics-guided reduction, the RL agent can determine the appropriate coefficients without consuming considerable computational resources.
This model may also contribute to the general CG modeling in its own manner.
It is an approximate model that combines bottom-up and top-down approaches, considering both the equilibrium and non-equilibrium properties.
Rather than targeting traditional mapped structural distributions, this CG model extends the bottom-up approach by implementing anisotropic interactions and hydrogen bonding and incorporating mechanical properties.
Another apparent characteristic is the simple analytical function forms supported by the optimized mapping and topology, which quantitatively explore the dynamic properties.
A model that is not too coarse may also contribute to multiscale simulations that require backmapping and reconstruction.
More specifically, the target anisotropic properties considered were the BD properties (axial elastic modulus and polymer stiffness) and NB properties (transverse stretch strength and toughness), all of which were accurately matched with errors below 16\%.
Several additional inspections were performed after the training, including in-plane shear, out-of-plane shear, adhesion energy, bending, transverse arrangement, and brick-and-mortar structures.
The transferability and generalization of this CG model were proven through additional inspections.
These results emphasize that transverse anisotropy is important for transverse arrangements, which has not been quantitatively captured in previous CG models.
Two transverse arrangement patterns with improved toughness and ductility were introduced.
We found a positive correlation between the overlapping length and toughness of the brick-and-mortar structures.
Data on transverse arrangements and brick-and-mortar structures may aid advanced material design in the future.
This model may also aid in addressing interactions between cellulose and other polar molecules on a mesoscopic scale.
This study emphasizes that Machine Learning can also be applied to construct effective analytical potentials that are better in terms of explainability, transferability, and kinetics.
Although our results and comparisons with TPE and CMAES demonstrate the efficiency of RL, many refinements remain to be explored, particularly for molecular systems.
For example, the necessity of inspiration from human knowledge and traditional methods such as Boltzmann Inversion is still unclear.
Finally, our basic parameterization approach is general for the applications, implying that our workflow may also be applicable to other molecular systems.
Transfer Learning and Imitation Learning methods may also be integrated with RL to solve similar problems in the future.

\section*{Conflict of Interest declaration}
\addtocounter{section}{1}
\setcounter{subsection}{0}
\addcontentsline{toc}{section}{\protect\numberline{}Conflict of Interest declaration}\indent

The author declares no competing financial interest.

\section*{Data availability}
\addtocounter{section}{1}
\setcounter{subsection}{0}
\addcontentsline{toc}{section}{\protect\numberline{}Data availability}\indent

The modified hbond/dreiding/lj/angleoffset was merged into the official LAMMPS EXTRA-MOLECULE at

$\textrm{https://github.com/lammps/lammps}$

\noindent
. The codes for potential parameterization training are available at

$\textrm{https://github.com/EiPiFun/rl-cll-cg}$

\noindent
. The training data, models, simulation codes, processed data, plot scripts, and latest manuscript are available at

$\textrm{https://github.com/EiPiFun/rl-cll-cg-data}$

\noindent
. The raw data, pre-processing codes, and post-processing codes are available upon requests.

\section*{Acknowledgments}
\addtocounter{section}{1}
\setcounter{subsection}{0}
\addcontentsline{toc}{section}{\protect\numberline{}Acknowledgments}\indent

Xu Dong thanks the Department of Engineering Mechanics of Zhejiang University for the funding and computational resources.

\section*{Author Contributions}
\addtocounter{section}{1}
\setcounter{subsection}{0}
\addcontentsline{toc}{section}{\protect\numberline{}Author Contributions}\indent

Xu Dong: Conceptualization, Data curation, Formal analysis, Investigation, Methodology, Project administration, Resources, Software, Validation, Visualization, Writing - original draft, Writing - review \& editing.

A professor from the Department of Engineering Mechanics of Zhejiang University provided some additional Methodology, Resources, and Writing - review \& editing helps and chose to be anonymous.

\linespread{1.0}\footnotesize\bibliography{cll_cg_references}
\addtocounter{section}{1}
\setcounter{subsection}{0}
\addcontentsline{toc}{section}{\protect\numberline{}References}

\setcounter{section}{0}
\setcounter{subsection}{0}

\setcounter{table}{0}
\setcounter{figure}{0}
\setcounter{equation}{0}

\renewcommand{\thesection}{S\arabic{section}}
\renewcommand{\thesubsection}{S\arabic{subsection}}

\renewcommand{\thetable}{S\arabic{table}}
\renewcommand{\thefigure}{S\arabic{figure}}
\renewcommand{\theequation}{S\arabic{equation}}

\section*{Supplementary Information}
\addtocounter{section}{1}
\setcounter{subsection}{0}
\addcontentsline{toc}{section}{\protect\numberline{}Supplementary Information}\indent

\normalsize The following subsections and data are included in the Supplementary Information.
\\\indent

Force and energy curve of LJ and HB potentials

Sensitivity and Ablation analysis for potential formula

All atom transverse stretch performance at varying stretch speed

Thermodynamics properties using a 12~fs timestep

Component match degrees of training loops

Bonded properties without rescale

Transverse strength and toughness distributions

Transverse stretch curves and distributions

Bonded geometry parameter distributions

Nonbonded equilibrium distance distributions

Structural properties from stretch processes

Baseline method structural properties for coarse grained models

Training using reward function without thresholds

Training using other optimization methods

Secondary summits of the bonded geometry parameter distributions
\\\indent

\refstepcounter{table}\label{tab:sensitivity_and_ablation_check_equilibrium_thermodynamics}\tablename\quad\thetable,\quad
\refstepcounter{table}\label{tab:sensitivity_and_ablation_check_nonbonded_properties}\tablename\quad\thetable,\quad
\refstepcounter{table}\label{tab:bonded_force_constants_without_rescale}\tablename\quad\thetable,\quad
\refstepcounter{table}\label{tab:bonded_properties_without_rescale}\tablename\quad\thetable

\refstepcounter{figure}\label{fig:nonbonded_energy_curve}\figurename\quad\thefigure,\quad
\refstepcounter{figure}\label{fig:sensitivity_and_ablation_check_structure_relaxation}\figurename\quad\thefigure,\quad
\refstepcounter{figure}\label{fig:aa_transverse_stretch_performance_with_speed}\figurename\quad\thefigure,\quad
\refstepcounter{figure}\label{fig:aa_transverse_stretch_curve_collection}\figurename\quad\thefigure,\quad
\refstepcounter{figure}\label{fig:direction_angle_definition}\figurename\quad\thefigure,\quad
\refstepcounter{figure}\label{fig:timestep_validation_thermodyanmics}\figurename\quad\thefigure,\quad
\refstepcounter{figure}\label{fig:component_match_degrees_during_training}\figurename\quad\thefigure,\quad
\refstepcounter{figure}\label{fig:transverse_mechanical_property_distributions}\figurename\quad\thefigure,\quad

\refstepcounter{figure}\label{fig:transverse_stretch_performance_collection}\figurename\quad\thefigure,\quad
\refstepcounter{figure}\label{fig:bonded_geometry_parameter_distributions_equilibrium}\figurename\quad\thefigure,\quad
\refstepcounter{figure}\label{fig:bonded_geometry_parameter_distributions_stretch}\figurename\quad\thefigure,\quad
\refstepcounter{figure}\label{fig:nonbonded_equilibrium_distance_distributions_equilibrium}\figurename\quad\thefigure,\quad
\refstepcounter{figure}\label{fig:nonbonded_equilibrium_distance_distributions_stretch}\figurename\quad\thefigure,\quad
\refstepcounter{figure}\label{fig:structural_properties_stretch}\figurename\quad\thefigure,\quad
\refstepcounter{figure}\label{fig:baselines_structure_relaxation}\figurename\quad\thefigure,\quad

\refstepcounter{figure}\label{fig:baselines_structure_fracture}\figurename\quad\thefigure,\quad
\refstepcounter{figure}\label{fig:baselines_rdf_ibi}\figurename\quad\thefigure,\quad
\refstepcounter{figure}\label{fig:baselines_rdf_re}\figurename\quad\thefigure,\quad
\refstepcounter{figure}\label{fig:convergence_and_statistics_without_threshold}\figurename\quad\thefigure,\quad
\refstepcounter{figure}\label{fig:component_match_degrees_during_training_without_threshold}\figurename\quad\thefigure,\quad
\refstepcounter{figure}\label{fig:reward_tpe_cmaes}\figurename\quad\thefigure,\quad
\refstepcounter{figure}\label{fig:secondary_summits_of_the_boned_geometry_parameter_distributions}\figurename\quad\thefigure

\end{document}


\begin{center}
\LARGE Analytical coarse grained potential parameterization by\\
Reinforcement Learning for anisotropic cellulose\\
\end{center}

\begin{center}
\renewcommand{\thefootnote}{\fnsymbol{footnote}}
Xu Dong$^{1,}$\footnote{Corresponding author, Email: donx@zuaa.zju.edu.cn}\\
$^{1}$\textit{Department of Engineering Mechanics, Zhejiang University, Hangzhou 310027, China}
\end{center}

\begin{center}
Keywords: Reinforcement Learning, Boltzmann Inversion, coarse grained, cellulose, anisotropy, hydrogen bonds
\end{center}

\setcounter{section}{0}
\setcounter{subsection}{0}

\setcounter{table}{0}
\setcounter{figure}{0}
\setcounter{equation}{0}

\renewcommand{\thesection}{S\arabic{section}}
\renewcommand{\thesubsection}{S\arabic{subsection}}

\renewcommand{\thetable}{S\arabic{table}}
\renewcommand{\thefigure}{S\arabic{figure}}
\renewcommand{\theequation}{S\arabic{equation}}

\section*{Supplementary Information}
\addtocounter{section}{1}
\setcounter{subsection}{0}
\addcontentsline{toc}{section}{\protect\numberline{}Supplementary Information}\indent

\subsection*{Force and energy curve of LJ and HB potentials}\indent
\addcontentsline{toc}{subsection}{\protect\numberline{}Force and energy curve of LJ and HB potentials}\indent

\begin{figure}[htbp]
    \centering
    \begin{subfigure}[b]{0.96\textwidth}
        \centering
        \begin{minipage}[b]{0.48\textwidth}
            \centering
            \includegraphics[width=0.96\textwidth]{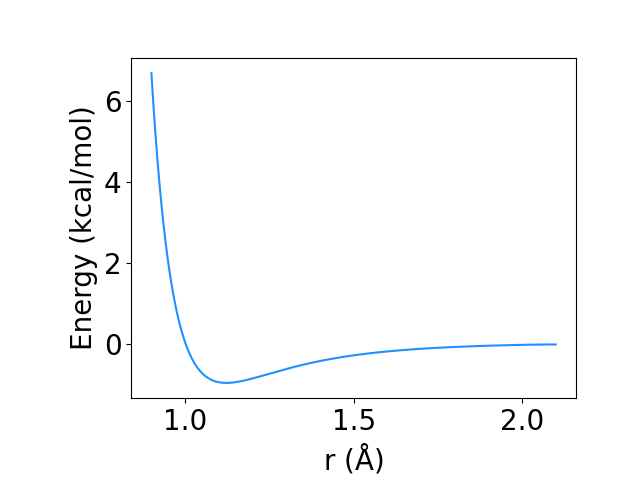}
            \\
            LJ
        \end{minipage}
        \begin{minipage}[b]{0.48\textwidth}
            \centering
            \includegraphics[width=0.96\textwidth]{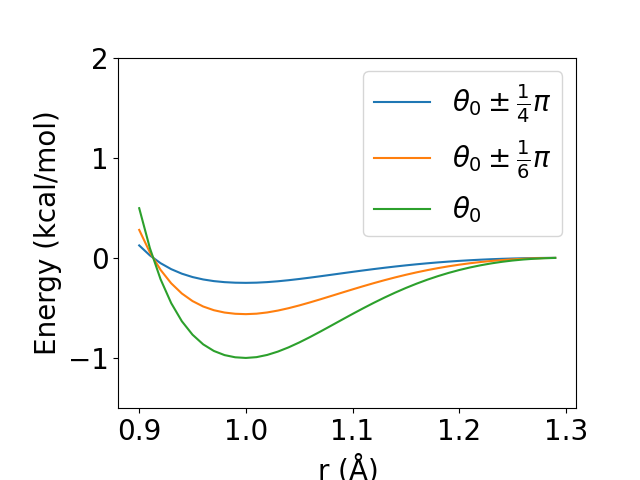}
            \\
            HB
        \end{minipage}
        \subcaption{}
    \end{subfigure}
    \\
    \begin{subfigure}[b]{0.96\textwidth}
        \centering
        \begin{minipage}[b]{0.48\textwidth}
            \centering
            \includegraphics[width=0.96\textwidth]{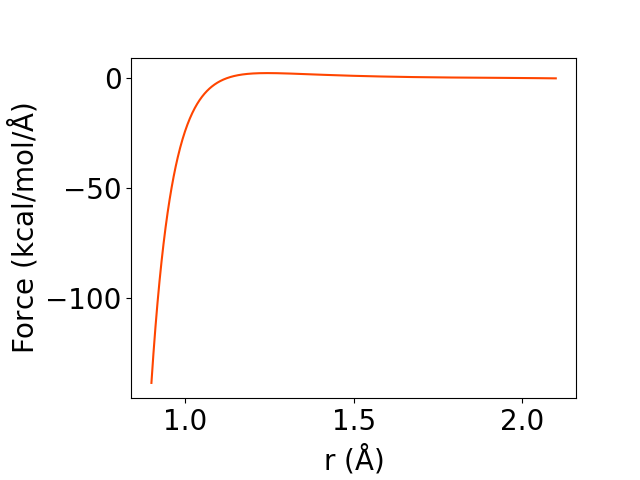}
            \\
            LJ
        \end{minipage}
        \begin{minipage}[b]{0.48\textwidth}
            \centering
            \includegraphics[width=0.96\textwidth]{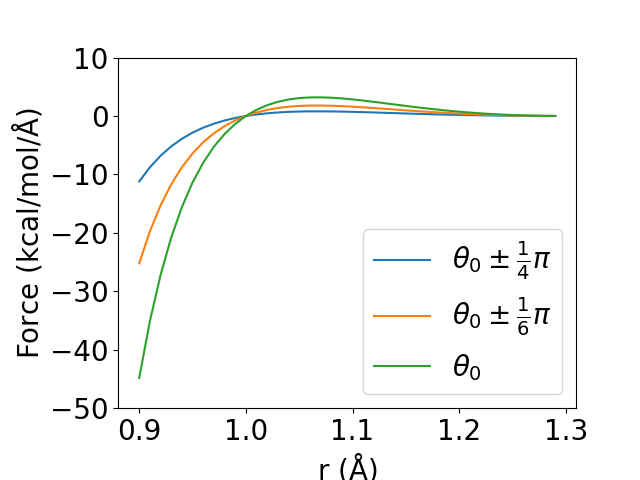}
            \\
            HB
        \end{minipage}
        \subcaption{}
    \end{subfigure}
    \captionsetup{font=scriptsize}
    \caption{
    NB energy and force curves.
    (a) LJ and HB energy curves.
    (b) LJ and HB force curves.
    Both the LJ and HB potentials are based on Lennard-Jones interactions, and they follow the 12-6 and 12-10 formulas respectively.
    The key difference is that the HB potential is scaled by the power of the cosine of the A:H-D angle, while the energy shift and force smoothing of the LJ potential confirm zero energy and force at the cutoff distance.
    Although modified by smoothing, the energy and force are always continuous between $r_{in}$ and $r_{cut}$.
    These curves are defined using $\epsilon=1.0$ and $\sigma=1.0$.
    More specifically, $r_{in}=2.0$ and $r_{cut}=2.1$ for the LJ potential, and $r_{in}=2.0$ and $r_{cut}=2.1$ for the HB potential.
    }
    \label{fig:nonbonded_energy_curve}
\end{figure}

The LJ potential used in this study, defined with different Lennard-Jones style functions, applies an energy shift and force smoothing to ensure that both energy and force vanish at the cutoff distance.
In contrast, the HB potential is also influenced by the A:H-D angle.
The smoothing operation does not influence the continuity of the potentials between $r_{in}$ and $r_{cut}$.
The force and energy curves for both potentials are shown in Figure~\ref{fig:nonbonded_energy_curve}, with $\epsilon=1$ and $\sigma=1$ applied in both cases.

\clearpage

\subsection*{Sensitivity and Ablation analysis for potential formula}\indent
\addcontentsline{toc}{subsection}{\protect\numberline{}Sensitivity and Ablation analysis for potential formula}\indent

\begin{figure}[htbp]
    \centering    
    \begin{subfigure}[b]{0.72\textwidth}
        \centering
        \scriptsize
        \begin{minipage}[b]{0.32\textwidth}
            \centering
            \includegraphics[width=0.96\textwidth]{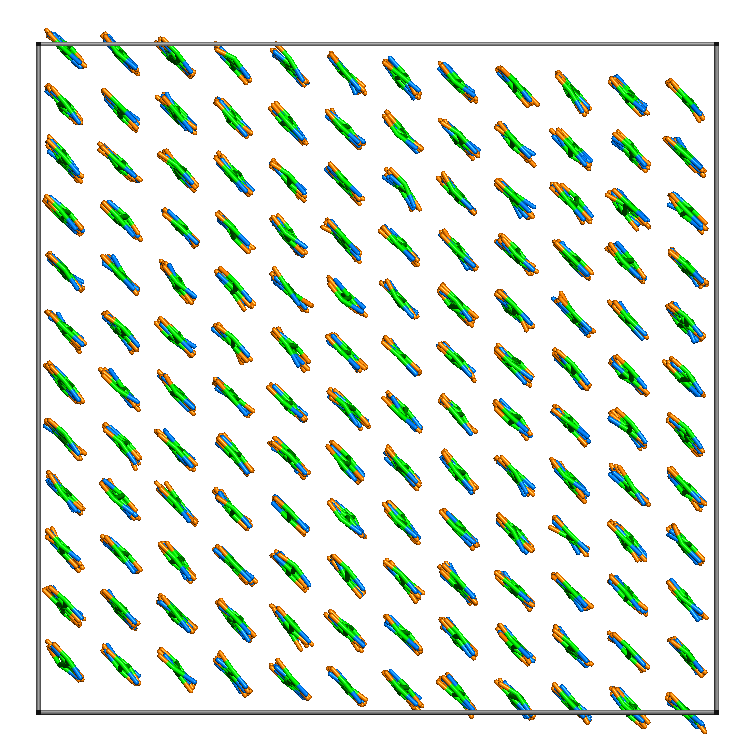}
            \\
            0.5x $\epsilon$
        \end{minipage}
        \begin{minipage}[b]{0.32\textwidth}
            \centering
            \includegraphics[width=0.96\textwidth]{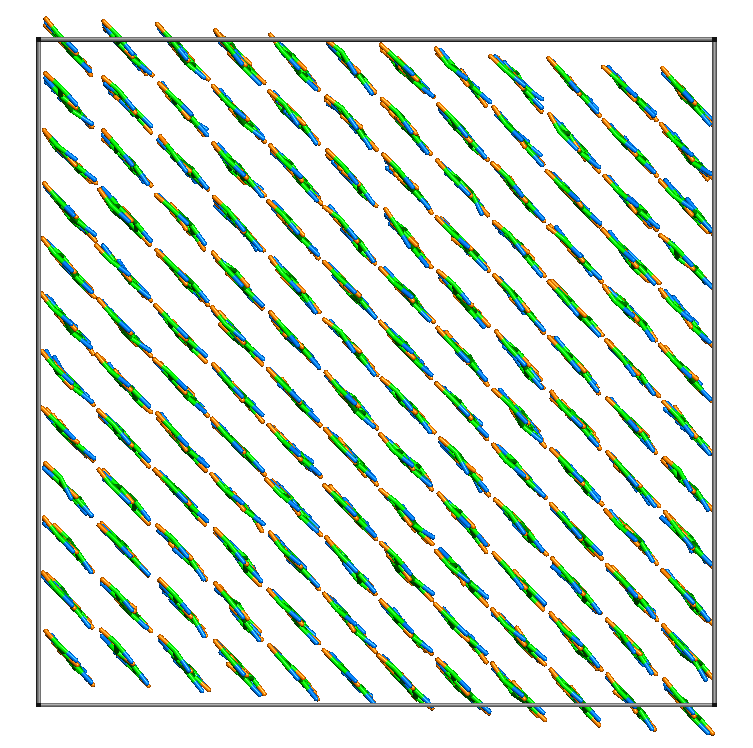}
            \\
            2.0x $\epsilon$
        \end{minipage}
        \begin{minipage}[b]{0.32\textwidth}
            \centering
            \includegraphics[width=0.96\textwidth]{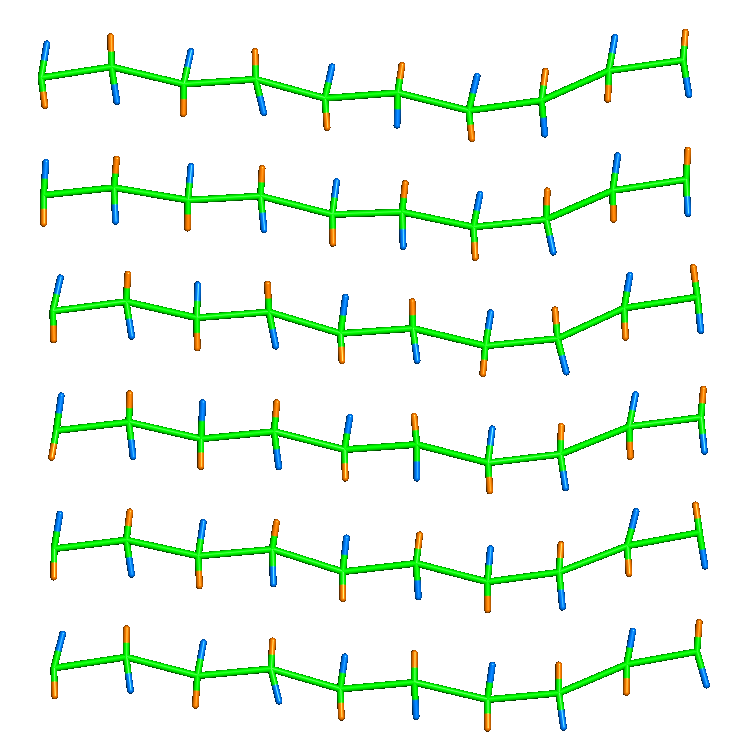}
            \\
            2.0x $\epsilon$ one layer
        \end{minipage}
        \\
        \begin{minipage}[b]{0.32\textwidth}
            \centering
            \includegraphics[width=0.96\textwidth]{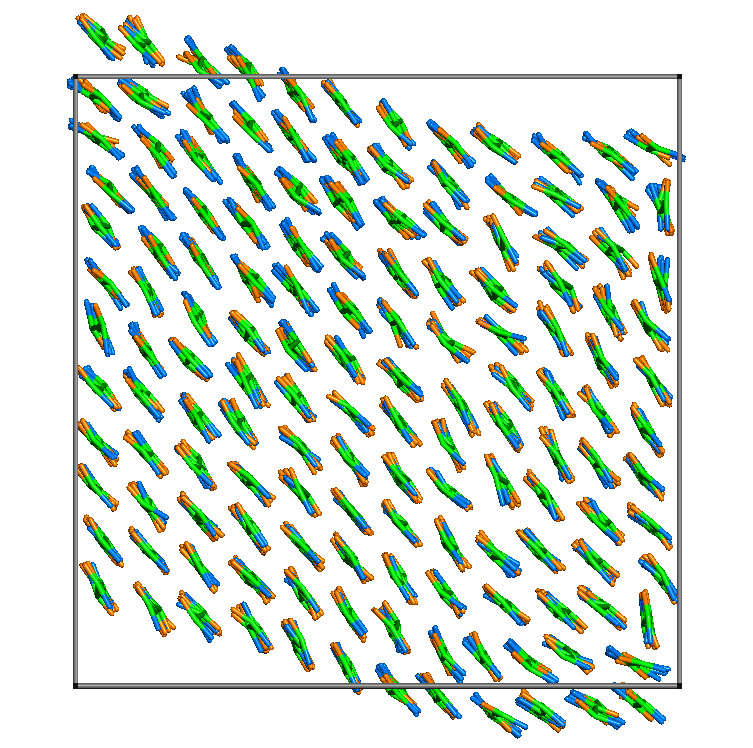}
            \\
            0.8x $\sigma$
        \end{minipage}
        \begin{minipage}[b]{0.32\textwidth}
            \centering
            \includegraphics[width=0.96\textwidth]{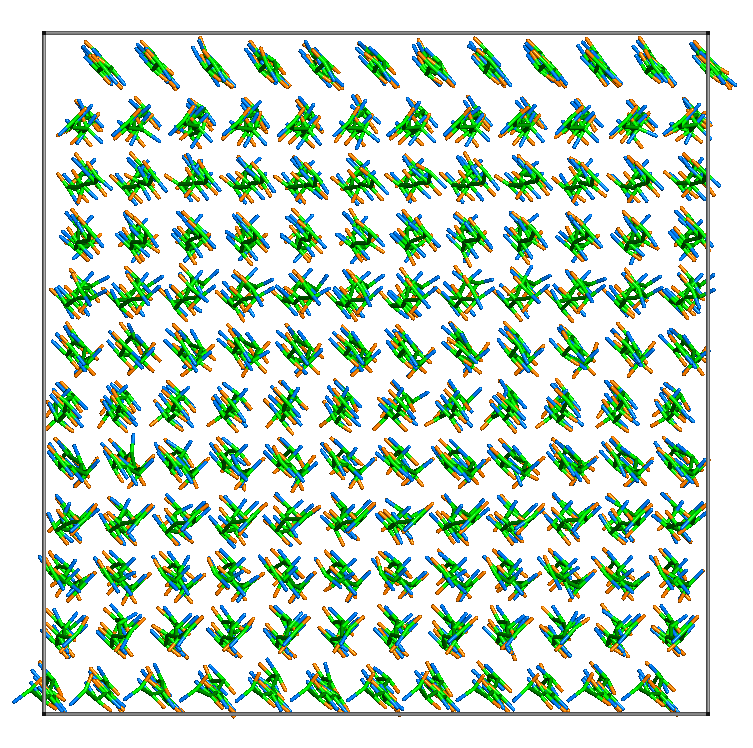}
            \\
            1.2x $\sigma$
        \end{minipage}
        \begin{minipage}[b]{0.32\textwidth}
            \centering
            \includegraphics[width=0.96\textwidth]{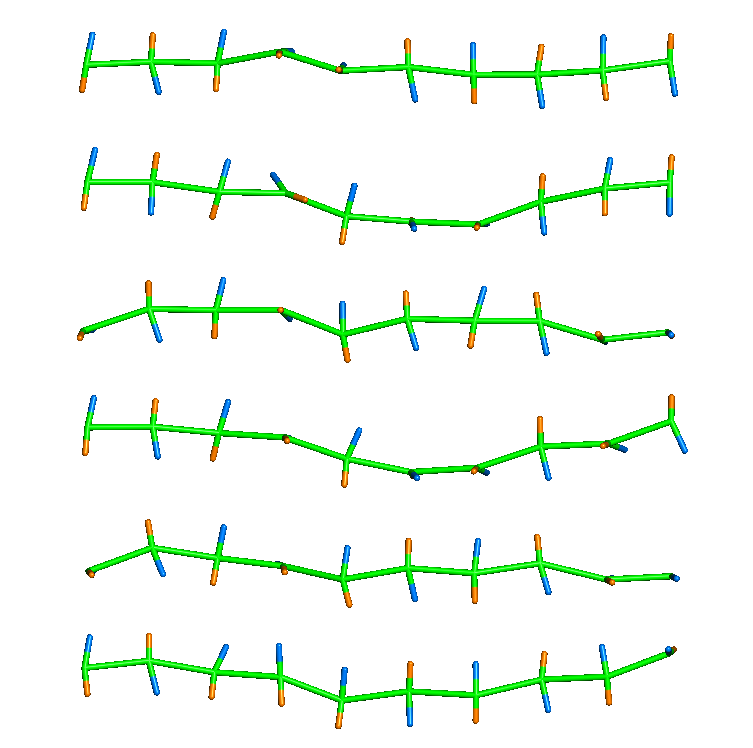}
            \\
            1.2x $\sigma$ one layer
        \end{minipage}
        \\
        \begin{minipage}[b]{0.32\textwidth}
            \centering
            \includegraphics[width=0.96\textwidth]{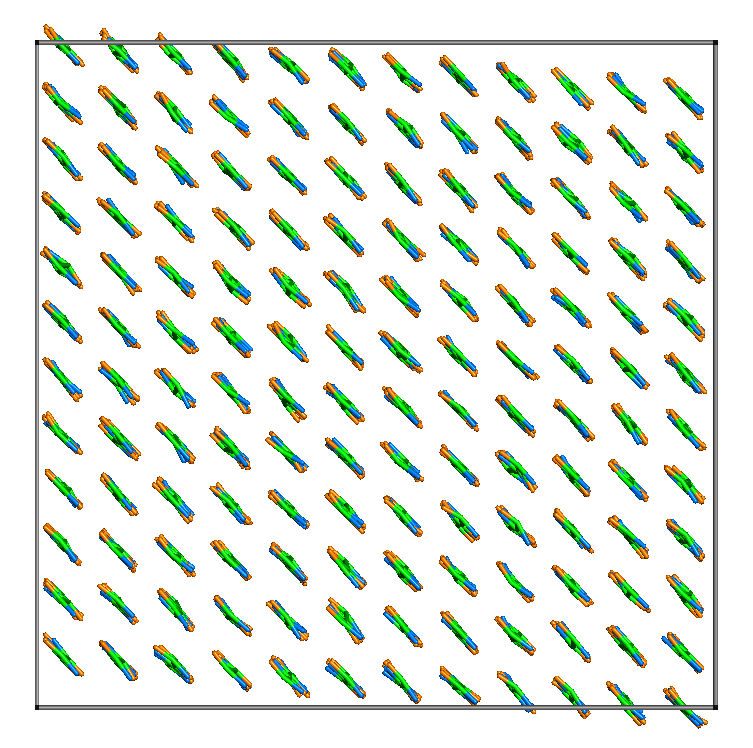}
            \\
            ${\rm cos}^2$
        \end{minipage}
        \begin{minipage}[b]{0.32\textwidth}
            \centering
            \includegraphics[width=0.96\textwidth]{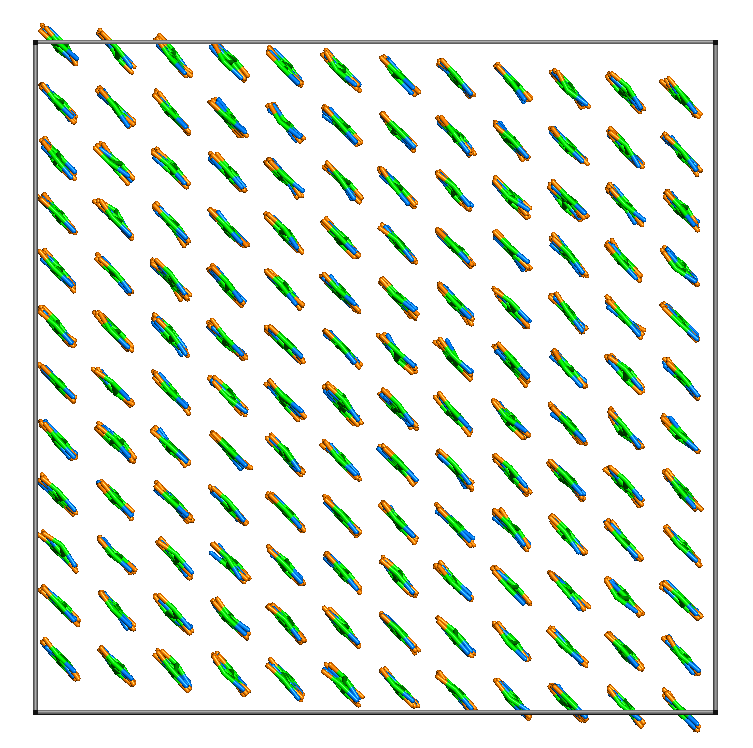}
            \\
            ${\rm cos}^6$
        \end{minipage}
        \begin{minipage}[b]{0.32\textwidth}
            \centering
            \includegraphics[width=0.96\textwidth]{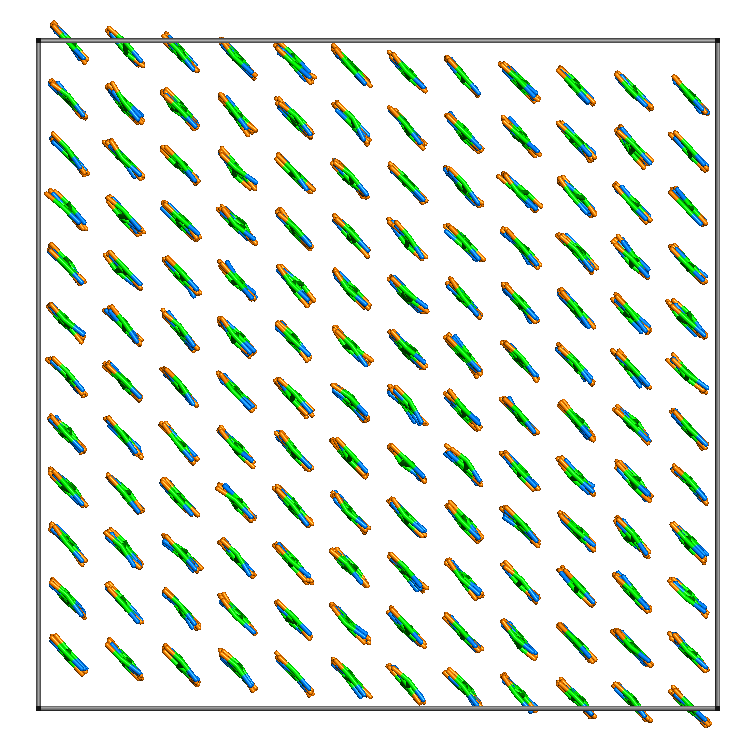}
            \\
            ${\rm cos}^8$
        \end{minipage}
        \\
        \begin{minipage}[b]{0.32\textwidth}
            \centering
            \includegraphics[width=0.96\textwidth]{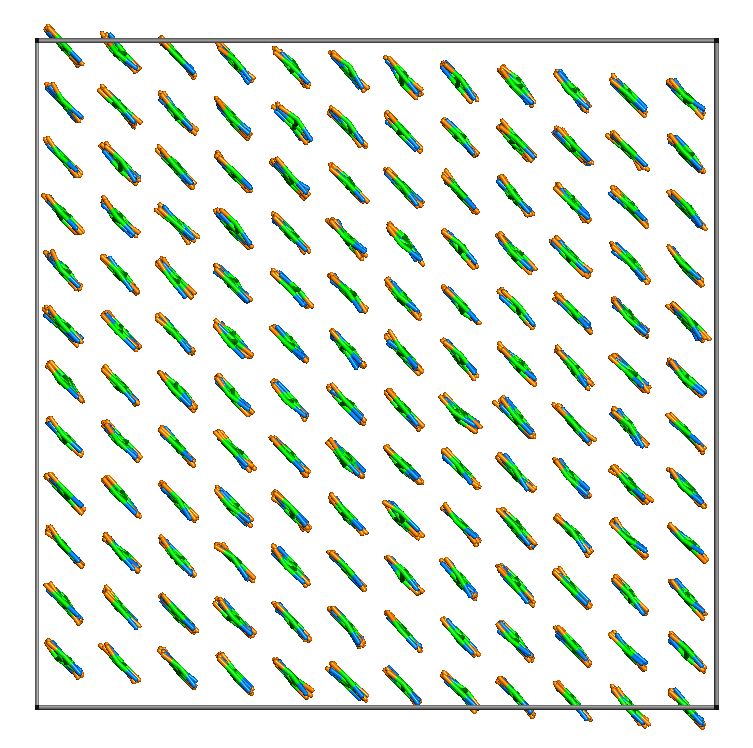}
            \\
            $\theta_0=180.0$
        \end{minipage}
        \begin{minipage}[b]{0.32\textwidth}
            \centering
            \includegraphics[width=0.96\textwidth]{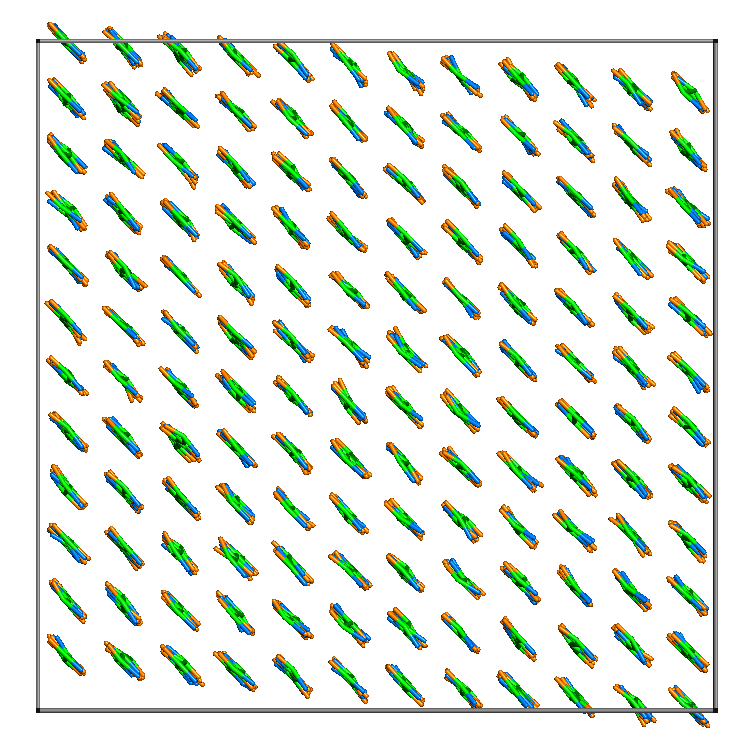}
            \\
            Without HB
        \end{minipage}
        \begin{minipage}[b]{0.32\textwidth}
            \centering
            \includegraphics[width=0.96\textwidth]{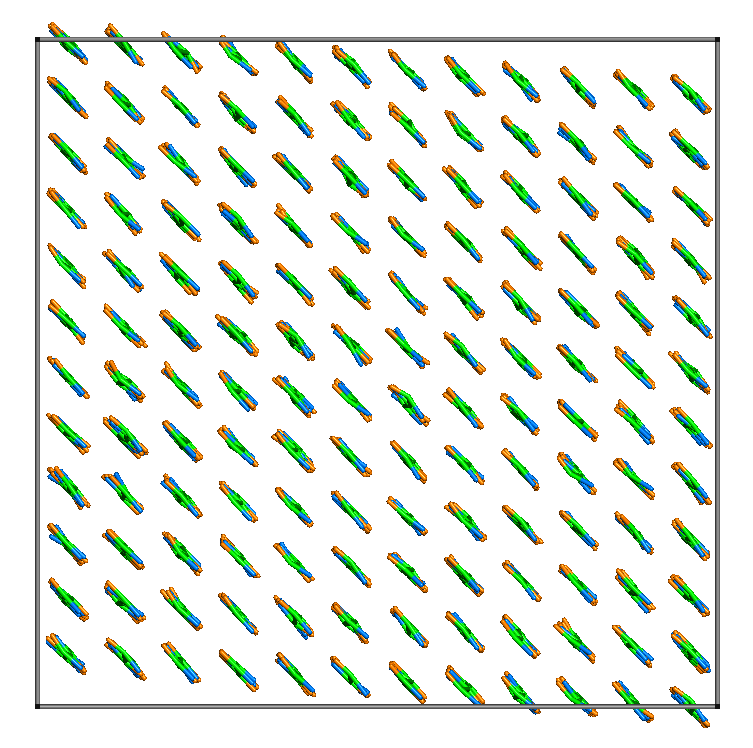}
            \\
            CG
        \end{minipage}
        \subcaption{}
    \end{subfigure}
    \captionsetup{font=scriptsize}
    \caption{
    Structures after relaxation of sensitivity and ablation checks.
    The application of the Lennard-Jones potential at the CG level requires sensitivity and ablation analyses to ensure validity.
    For the sensitivity check, NB energy coefficients $\epsilon$ were scaled to be half or double, NB distance coefficients $\sigma$ were scaled to be 0.8 times or 1.2 times;
    and the exponent of the cosine of the HB potential was changed to 2, 6, 8 instead of 4.
    For the ablation check, $\theta_0$ in the HB potential was reset to $\pi$ or the HB potential was fully removed.
    For all of these options and the unchanged CG reference (denoted as CG, using the same coefficients in the paper), the structures after relaxation/before stretch were all provided.
    Relaxation structures demonstrated the robustness of NB potential function forms, but large energy coefficients or unmatched distance coefficients caused unstable structures, as the one layer figure showed.
    }
    \label{fig:sensitivity_and_ablation_check_structure_relaxation}
\end{figure}

The selection of the Lennard-Jones potential in this model lacked a strong physics background.
It was primarily chosen for its simplicity and the ease of stabilizing the geometric structure based on its stationary point.
Consequently, the sensitivity and ablation checks for these potential forms are necessary.
Using the optimized coefficients from the main paper and locking all BD potential coefficients, variations were tested: half/double energy coefficients (while distance coefficients were kept) and 0.8/1.2 times distance coefficients (while energy coefficients were kept).
These tests involved 100 replica simulations.
Furthermore, although we followed the original literature that used an exponent of 4 for the cosine term, exponents of 2, 6, and 8 were also evaluated (while all other coefficients were retained).
Ablation tests included resetting $\theta_0$ to $\pi$ and the complete removal of the HB potential.
All the sensitivity and ablation data are listed in Table~\ref{tab:sensitivity_and_ablation_check_equilibrium_thermodynamics} and Table~\ref{tab:sensitivity_and_ablation_check_nonbonded_properties}, respectively.
The corresponding structures after relaxation further confirm the robustness of the potentials, as shown in Figure \ref{fig:sensitivity_and_ablation_check_structure_relaxation}.
These data indicate that the potential function forms are not significantly sensitive to coefficients; the results for significantly deviated coefficients showed quantitative rather than qualitative changes, and the simulations remained stable.

\begin{table}[h]
    \centering
    \caption{Sensitivity and ablation check equilibrium thermodynamics}
    \label{tab:sensitivity_and_ablation_check_equilibrium_thermodynamics}
    \begin{tabular}{c c c c}
        \hline
        \hline
        \multicolumn{2}{c}{type}&{Temperature (T)}&{Pressure YY (MPa)}\\
        \hline
        \multirow{2}{*}{$\epsilon$}&{0.5x}&{300.2$\pm$0.1}&{18.4$\pm$0.7}\\
        &{2.0x}&{300.3$\pm$0.1}&{13.4$\pm$0.8}\\
        \hline
        \multirow{2}{*}{$\sigma$}&{0.8x}&{300.2$\pm$0.1}&{86.0$\pm$1.2}\\
        &{1.2x}&{300.7$\pm$0.1}&{-4.2$\pm$0.5}\\
        \hline
        \multirow{3}{*}{${\rm cos}^{n}$}&{2}&{300.2$\pm$0.1}&{6.2$\pm$0.7}\\
        &{6}&{300.3$\pm$0.1}&{10.0$\pm$0.8}\\
        &{8}&{300.2$\pm$0.1}&{2.5$\pm$0.8}\\
        \hline
        {$\theta_0$}&{180.0}&{300.0$\pm$0.1}&{25.1$\pm$0.7}\\
        \hline
        {HB}&{no}&{300.0$\pm$0.1}&{-13.6$\pm$0.7}\\
        \hline
        \multicolumn{2}{c}{CG}&{300.2$\pm$0.1}&{11.9$\pm$0.7}\\
        \hline
        \multicolumn{2}{c}{AA}&{299.9$\pm$0.0}&{-0.8$\pm$1.2}\\
        \hline
        \hline
    \end{tabular}
\end{table}

\begin{table}[h]
    \centering
    \caption{Sensitivity and ablation check nonbonded properties}
    \label{tab:sensitivity_and_ablation_check_nonbonded_properties}
    \begin{tabular}{c c c c c c c c}
        \hline
        \hline
        \multicolumn{2}{c}{\multirow{2}{*}{type}}&\multicolumn{3}{c}{Strength (MPa)}&\multicolumn{3}{c}{Toughness (GPa)}\\
        \cline{3-8}
        &&{V}&{H}&{S}&{V}&{H}&{S}\\
        \hline
        \multirow{2}{*}{$\epsilon$}&{0.5x}&{522.0$\pm$4.8}&{252.0$\pm$0.7}&{223.6$\pm$3.0}&{53.3$\pm$1.5}&{38.5$\pm$0.7}&{70.7$\pm$2.1}\\
        &{2.0x}&{2550.2$\pm$6.2}&{1520.0$\pm$4.4}&{1165.9$\pm$9.7}&{228.6$\pm$5.6}&{406.4$\pm$7.7}&{347.5$\pm$8.1}\\
        \hline
        \multirow{2}{*}{$\sigma$}&{0.8x}&{447.2$\pm$9.5}&{30.6$\pm$0.7}&{387.8$\pm$4.9}&{89.1$\pm$3.3}&{1.3$\pm$0.2}&{146.2$\pm$5.6}\\
        &{1.2x}&{654.8$\pm$21.5}&{568.6$\pm$14.7}&{500.1$\pm$10.9}&{100.0$\pm$3.3}&{116.1$\pm$3.3}&{155.3$\pm$5.6}\\
        \hline
        \multirow{3}{*}{${\rm cos}^{n}$}&{2}&{1244.1$\pm$4.7}&{617.0$\pm$0.9}&{498.3$\pm$4.7}&{123.5$\pm$3.1}&{109.8$\pm$3.7}&{164.5$\pm$4.2}\\
        &{6}&{1177.8$\pm$7.3}&{672.7$\pm$1.1}&{512.7$\pm$4.0}&{111.7$\pm$2.9}&{77.2$\pm$0.5}&{159.9$\pm$4.0}\\
        &{8}&{1178.5$\pm$6.9}&{686.5$\pm$1.0}&{511.7$\pm$5.2}&{121.2$\pm$3.4}&{77.6$\pm$0.4}&{155.5$\pm$4.1}\\
        \hline
        {$\theta_0$}&{180.0}&{1168.9$\pm$7.5}&{637.4$\pm$1.0}&{506.7$\pm$4.2}&{112.6$\pm$3.1}&{84.7$\pm$2.0}&{151.2$\pm$4.0}\\
        \hline
        {HB}&{no}&{718.7$\pm$4.6}&{520.4$\pm$2.0}&{357.3$\pm$2.8}&{98.7$\pm$3.1}&{141.5$\pm$1.5}&{98.7$\pm$3.4}\\
        \hline
        \multicolumn{2}{c}{CG}&{1191.9$\pm$6.0}&{653.5$\pm$1.1}&{509.8$\pm$4.6}&{117.0$\pm$3.2}&{78.1$\pm$1.2}&{166.9$\pm$3.8}\\
        \hline
        \multicolumn{2}{c}{AA}&{1224.0$\pm$9.7}&{638.2$\pm$4.2}&{467.0$\pm$3.7}&{111.0$\pm$2.5}&{77.9$\pm$2.8}&{190.1$\pm$4.6}\\
        \hline
        &&\multicolumn{6}{c}{Error}\\
        \hline
        \multirow{2}{*}{$\epsilon$}&{0.5x}&{57\%}&{61\%}&{52\%}&{52\%}&{51\%}&{63\%}\\
        &{2.0x}&{108\%}&{38\%}&{150\%}&{106\%}&{422\%}&{83\%}\\
        \hline
        \multirow{2}{*}{$\sigma$}&{0.8x}&{63\%}&{95\%}&{17\%}&{20\%}&{98\%}&{23\%}\\
        &{1.2x}&{47\%}&{11\%}&{7\%}&{10\%}&{49\%}&{18\%}\\
        \hline
        \multirow{3}{*}{${\rm cos}^{n}$}&{2}&{2\%}&{3\%}&{7\%}&{11\%}&{41\%}&{13\%}\\
        &{6}&{4\%}&{5\%}&{10\%}&{7\%}&{1\%}&{16\%}\\
        &{8}&{4\%}&{8\%}&{10\%}&{9\%}&{0\%}&{18\%}\\
        \hline
        {$\theta_0$}&{180.0}&{5\%}&{0\%}&{9\%}&{1\%}&{9\%}&{20\%}\\
        \hline
        {HB}&{no}&{41\%}&{18\%}&{23\%}&{11\%}&{82\%}&{48\%}\\
        \hline
        \multicolumn{2}{c}{CG}&{3\%}&{2\%}&{9\%}&{5\%}&{0\%}&{12.2\%}\\
        \hline
        \hline
    \end{tabular}
\end{table}

\clearpage

\subsection*{All atom transverse stretch performance at varying stretch speed}\indent
\addcontentsline{toc}{subsection}{\protect\numberline{}All atom transverse stretch performance at varying stretch speed}\indent

\begin{figure}[htbp]
    \centering
    \begin{subfigure}[b]{0.32\textwidth}
        \centering
        \includegraphics[width=0.96\textwidth]{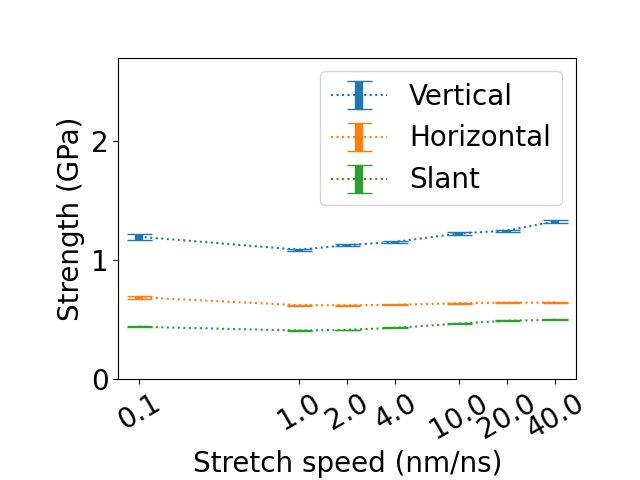}
        \subcaption{}
    \end{subfigure}
    \begin{subfigure}[b]{0.32\textwidth}
        \centering
        \includegraphics[width=0.96\textwidth]{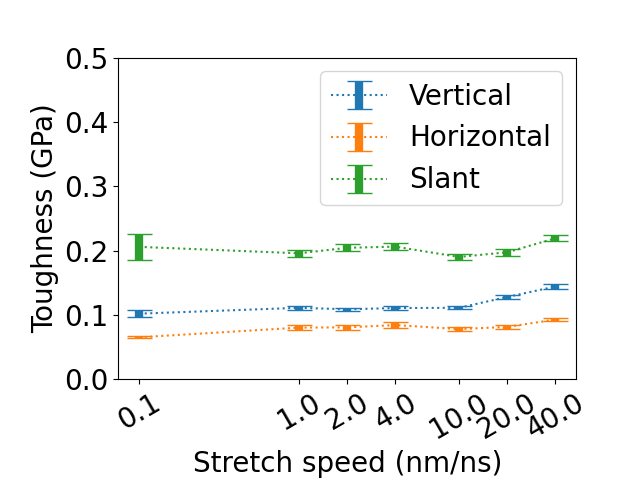}
        \subcaption{}
    \end{subfigure}
    \begin{subfigure}[b]{0.32\textwidth}
        \centering
        \includegraphics[width=0.96\textwidth]{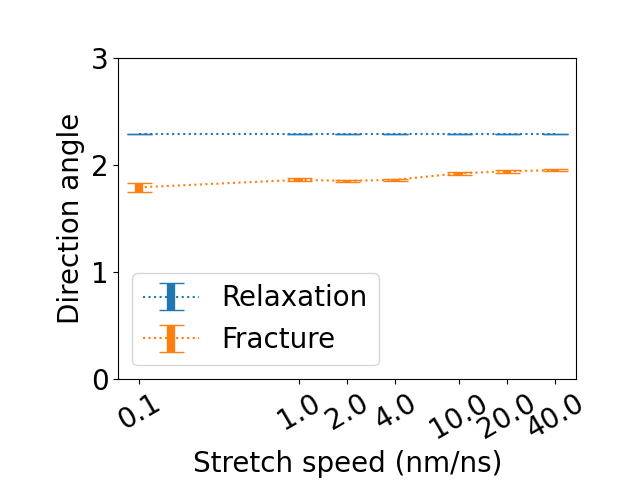}
        \subcaption{}
    \end{subfigure}
    \captionsetup{font=scriptsize}
    \caption{
    AA transverse stretch performance with respect to speed.
    Transverse (a) strength, (b) toughness and (c) direction angle with standard errors at varying stretch speeds are presented.
    100 replica transverse stretches at stretch speeds of 1.0, 2.0, 4.0, 10.0, 20.0 and 40.0~nm/ns were summarized with averages and standard errors.
    Only 10 replica at 0.1~nm/ns were enforced to control computation costs.
    All data except for toughness of vertical models at 40.0~nm/ns were within 15\% relative deviations referenced to average values at 10.0~nm/ns.
    10.0~nm/ns was chosen for parameterization finally.
    }
    \label{fig:aa_transverse_stretch_performance_with_speed}
\end{figure}

\begin{figure}[htbp]
    \centering    
    \begin{subfigure}[b]{0.96\textwidth}
        \centering
        \scriptsize
        \begin{minipage}[b]{0.32\textwidth}
            \centering
            \includegraphics[width=0.96\textwidth]{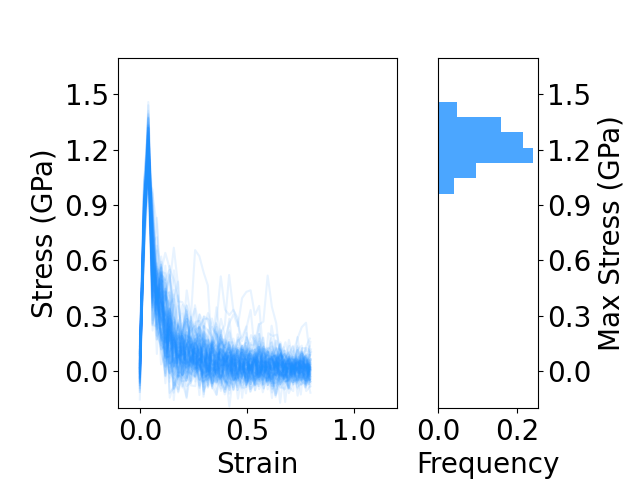}
            \\
            Vertical
        \end{minipage}
        \begin{minipage}[b]{0.32\textwidth}
            \centering
            \includegraphics[width=0.96\textwidth]{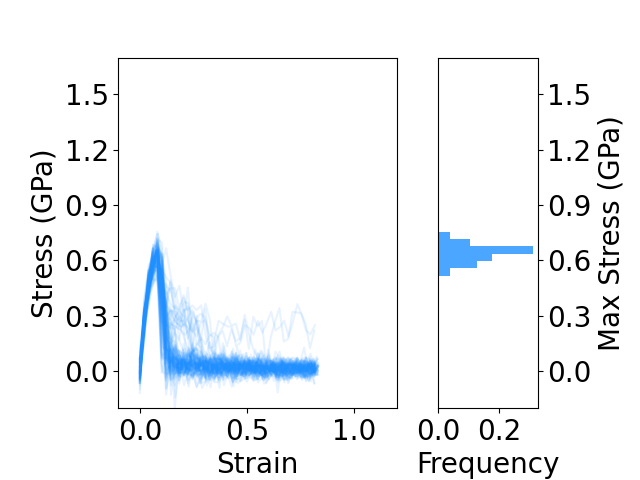}
            \\
            Horizontal
        \end{minipage}
        \begin{minipage}[b]{0.32\textwidth}
            \centering
            \includegraphics[width=0.96\textwidth]{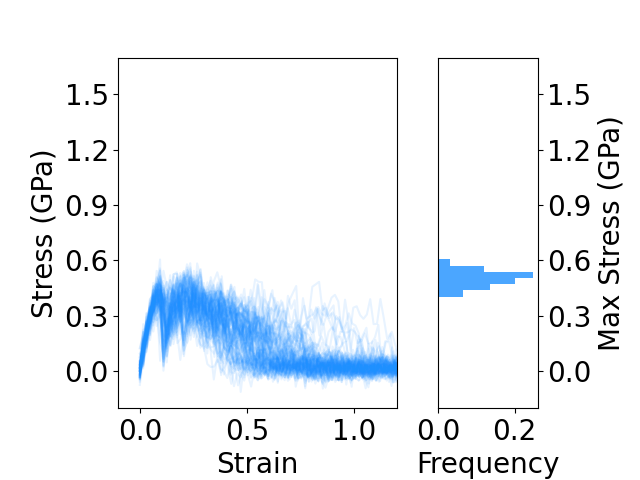}
            \\
            Slant
        \end{minipage}
    \subcaption{}
    \end{subfigure}
    \\
    \begin{subfigure}[b]{0.96\textwidth}
        \centering
        \scriptsize
        \begin{minipage}[b]{0.32\textwidth}
            \centering
            \includegraphics[width=0.96\textwidth]{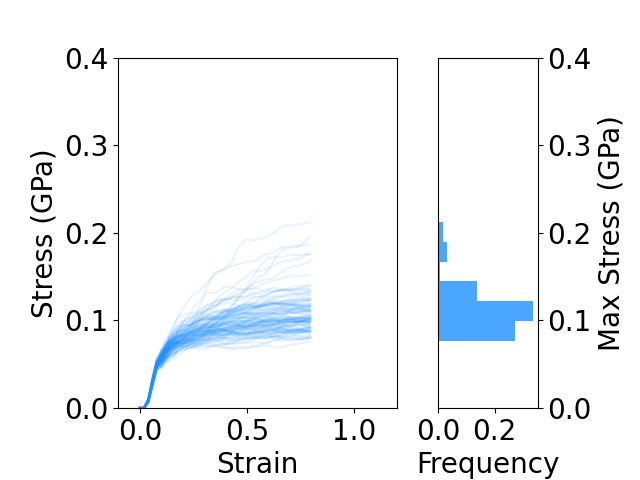}
            \\
            Vertical
        \end{minipage}
        \begin{minipage}[b]{0.32\textwidth}
            \centering
            \includegraphics[width=0.96\textwidth]{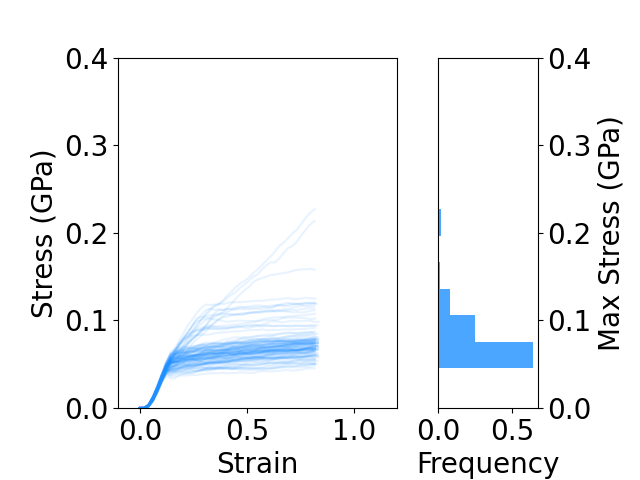}
            \\
            Horizontal
        \end{minipage}
        \begin{minipage}[b]{0.32\textwidth}
            \centering
            \includegraphics[width=0.96\textwidth]{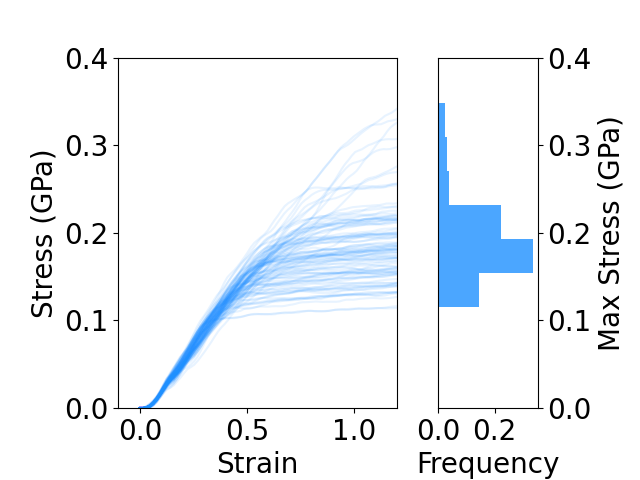}
            \\
            Slant
        \end{minipage}
        \subcaption{}
    \end{subfigure}
    \captionsetup{font=scriptsize}
    \caption{
    AA stretch curve collections.
    (a) Stress-strain and (b) strain energy collection of AA transverse stretches in characteristic directions at 10.0~nm/ns.
    Each collection contains 100 curves, and the strength or toughness distribution is enclosed on the right.
    }
    \label{fig:aa_transverse_stretch_curve_collection}
\end{figure}

\begin{figure}[htbp]
    \centering
    \begin{subfigure}[b]{0.24\textwidth}
        \includegraphics[width=0.96\textwidth]{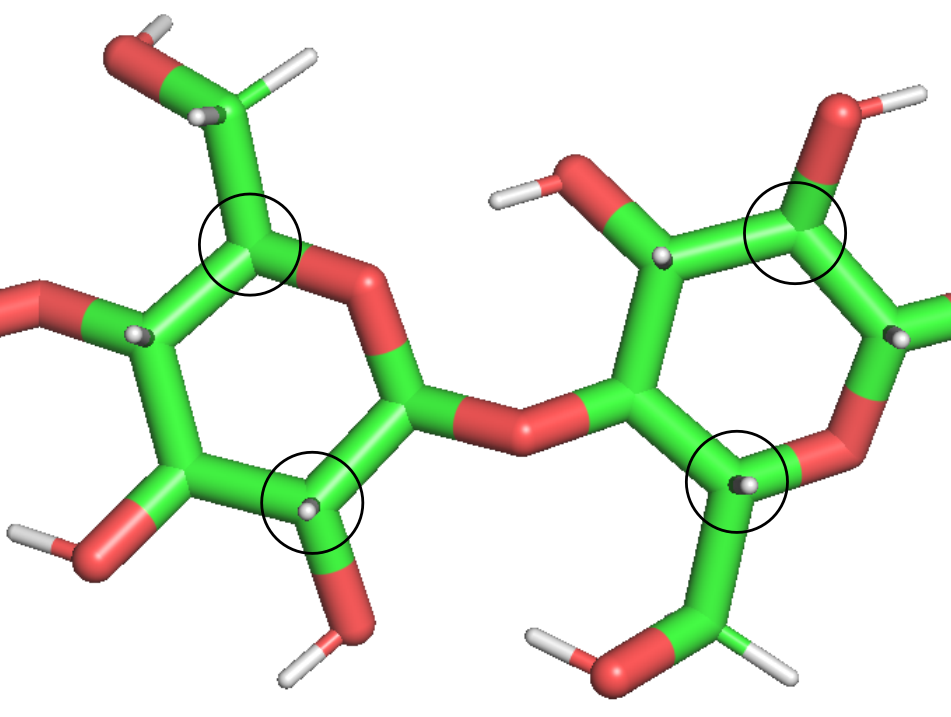}
        \subcaption{}
    \end{subfigure}
    \begin{subfigure}[b]{0.16\textwidth}
        \includegraphics[width=0.96\textwidth]{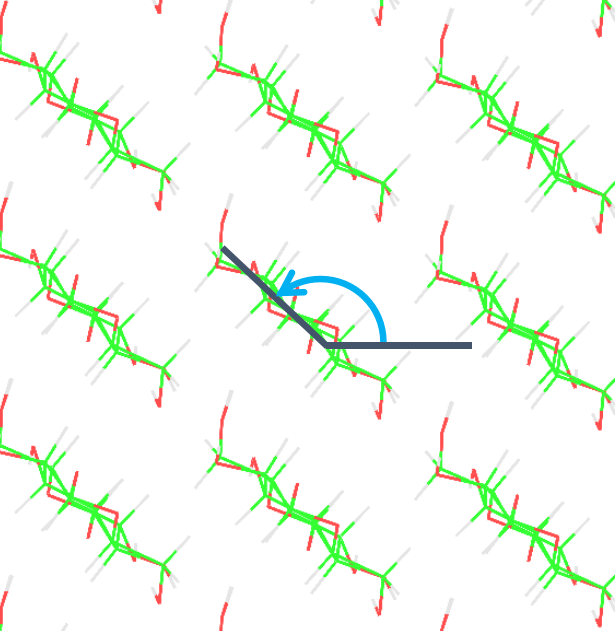}
        \subcaption{}
    \end{subfigure}
    \captionsetup{font=scriptsize}
    \caption{
    Definition of direction angle.
    (a) Carbon atoms considered for the direction angle.
    (b) The projection angle of the considered carbon atoms was defined as the direction angle for the AA models, whereas the CL1 and CL2 beads were used for the CG models.
    The direction angle was defined to quantify the frictional sliding of the slant models.
    }
    \label{fig:direction_angle_definition}
\end{figure}

For the slant models, frictional sliding was quantitatively determined by the ``direction angle'', which was defined as the projection angle of the edge carbon atoms on the ring (Figure~\ref{fig:direction_angle_definition}).
Before selecting 10.0~nm/ns as the stretch speed for potential parameterization, the transverse stretch performancs of AA models were evaluated across three characteristic directions at 0.1, 1.0, 2.0, 4.0, 10.0, 20.0, and 40.0~nm/ns (Figure~\ref{fig:aa_transverse_stretch_performance_with_speed}).
Except for 0.1~nm/ns (at which speed 10 replica simulations are executed to save resources), 100 replica transverse stretch performance data were collected.
These results confirm that the stretch speed is not a critical factor, because the performance deviations at other speeds remained largely within 15\% of the values recorded at 10.0~nm/ns.
Only the toughness data of the vertical model at 40.0~nm/ns was beyond 15\% and 10.0~nm/ns was chosen.
Stress-strain curves with maximum stress distribution and strain energy curves with maximum strain energy distribution at 10.0~nm/ns are also provided in Figure~\ref{fig:aa_transverse_stretch_curve_collection}.

\newpage

\subsection*{Thermodynamics properties using a 12 fs timestep}\indent
\addcontentsline{toc}{subsection}{\protect\numberline{}Thermodynamics properties using a 12~fs timestep}\indent

\begin{figure}[htbp]
    \center
    \begin{subfigure}[b]{0.32\textwidth}
        \includegraphics[width=0.96\textwidth]{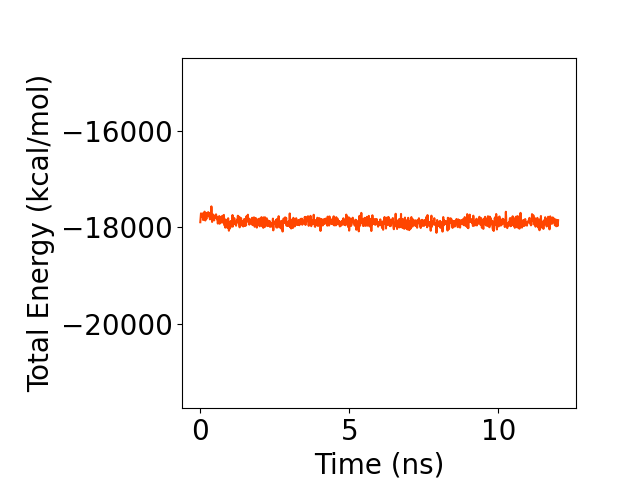}
        \subcaption{}
    \end{subfigure}
    \begin{subfigure}[b]{0.32\textwidth}
        \includegraphics[width=0.96\textwidth]{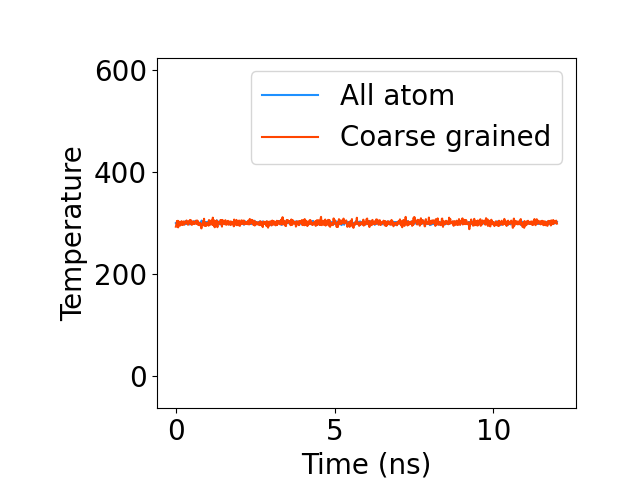}
        \subcaption{}
    \end{subfigure}
    \begin{subfigure}[b]{0.32\textwidth}
        \includegraphics[width=0.96\textwidth]{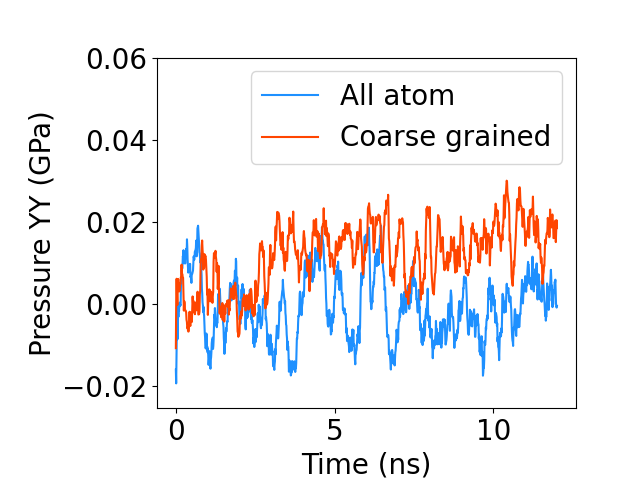}
        \subcaption{}
    \end{subfigure}
    \captionsetup{font=scriptsize}
    \caption{
    Thermodynamics properties from CNC equilibrium simulations of both AA and CG.
    (a) Total energy curve from CG equilibrium simulation under NVE ensemble.
    The stable total energy demonstrated the validation of a 12~fs timestep.
    (b) Temperature and (c) pressure curves from the 12~ns slant model equilibrium simulations using timestep of 2~fs for AA and 12~fs for CG.
    The comparison between the AA and CG models confirmed the suitability of timestep and stability of thermodyanmics.
    Compared with the AA model, the CG model remained thermodynamically stable even with a much larger timestep, while larger deviations were within acceptable limits.
    }
    \label{fig:timestep_validation_thermodyanmics}
\end{figure}

The timestep size is a critical parameter for CG models.
While a small timestep ensures system stability, it sacrifices the computational efficiency.
A 4~fs timestep was utilized for simulations of axial stretch and polymer stiffness, whereas a larger 12~fs timestep was applied for the more computationally expensive transverse stretch simulations.
The 12~fs value was determined via bisection of 10 and 15~fs to ensure the stability.
To validate this choice, equilibrium simulations without external loads were performed using a CNC slant model under NVE (for CG) and NPT (for both AA and CG, as in the case of transverse stretch) ensembles.
The timestep used for the AA was 2~fs.
The total energy of the system from the NVE CG trajectories, together with the temperature and pressure in the vertical direction (Pressure YY) from the NPT ensemble, are shown in Figure~\ref{fig:timestep_validation_thermodyanmics}.
As the figures show, the total energy was stable, and the amplitudes of the temperature and pressure of the CG model were slightly larger but also stable.

\subsection*{Component match degrees of training loops}\indent
\addcontentsline{toc}{subsection}{\protect\numberline{}Component match degrees of training loops}\indent

\begin{figure}[htbp]
    \centering    
    \begin{subfigure}[b]{0.96\textwidth}
        \centering
        \scriptsize
        \begin{minipage}[b]{0.32\textwidth}
            \centering
            \includegraphics[width=0.96\textwidth]{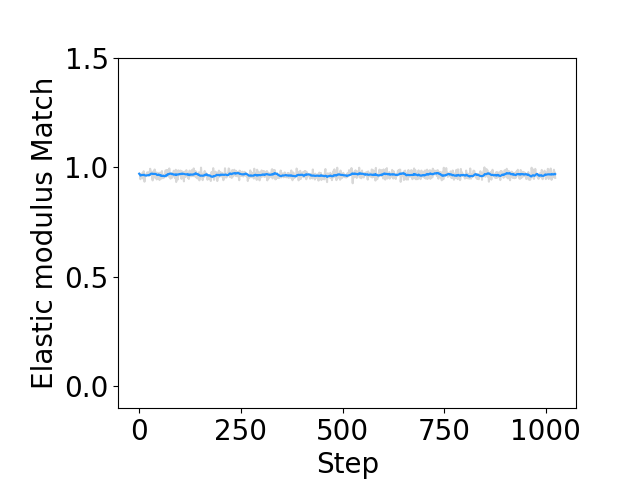}
            \includegraphics[width=0.96\textwidth]{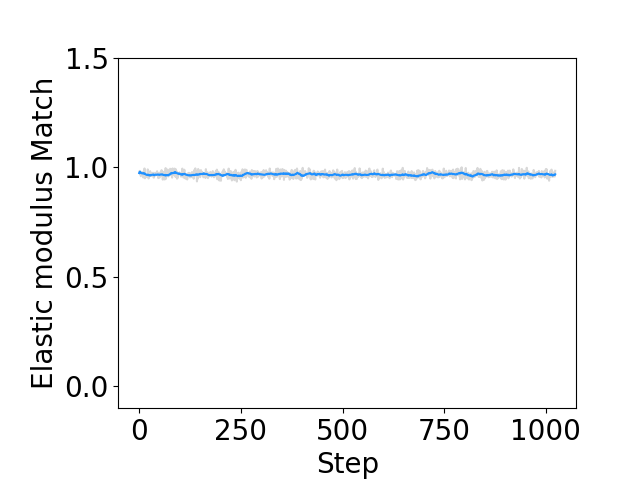}
            \\
            Axial modulus
        \end{minipage}
        \begin{minipage}[b]{0.32\textwidth}
            \centering
            \includegraphics[width=0.96\textwidth]{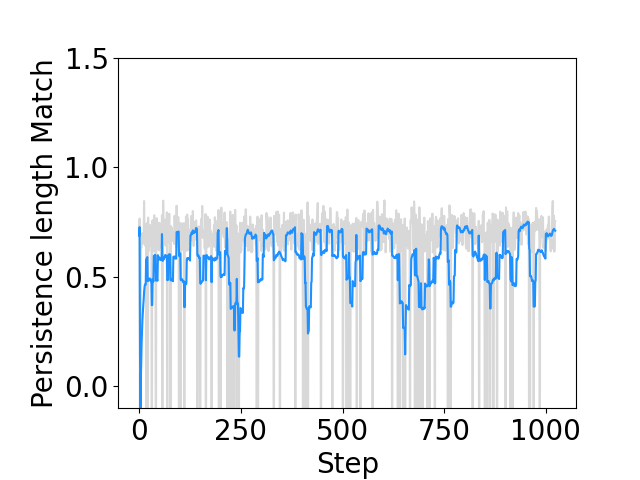}
            \includegraphics[width=0.96\textwidth]{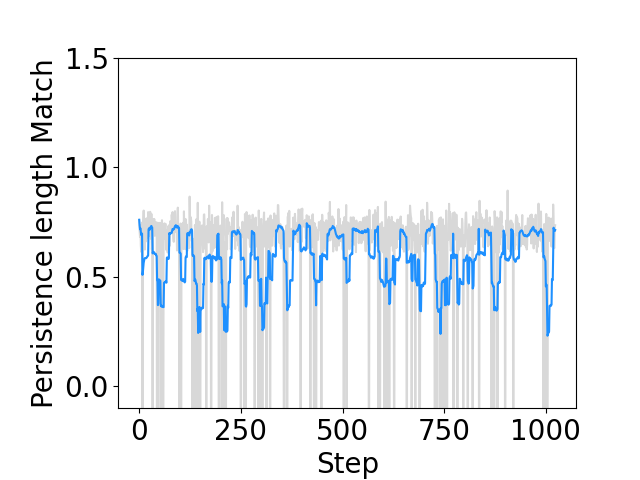}
            \\
            Persistence length
        \end{minipage}
        \begin{minipage}[b]{0.32\textwidth}
            \centering
            \includegraphics[width=0.96\textwidth]{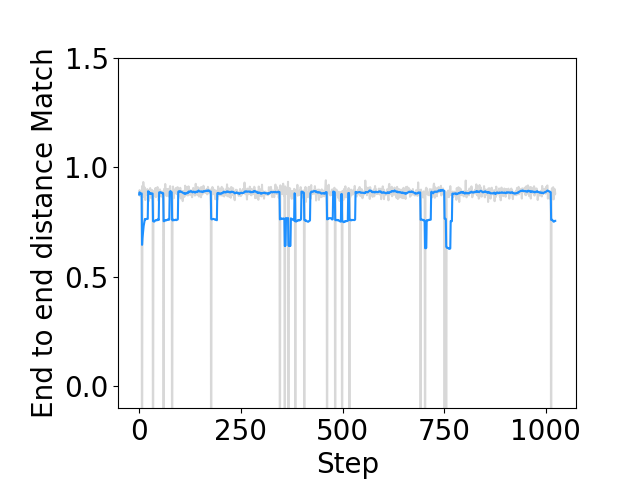}
            \includegraphics[width=0.96\textwidth]{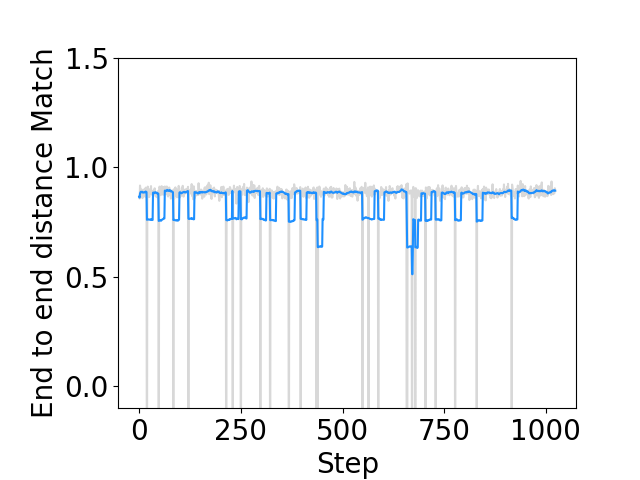}
            \\
            End to end distance
        \end{minipage}
        \\
        \begin{minipage}[b]{0.32\textwidth}
            \centering
            \includegraphics[width=0.96\textwidth]{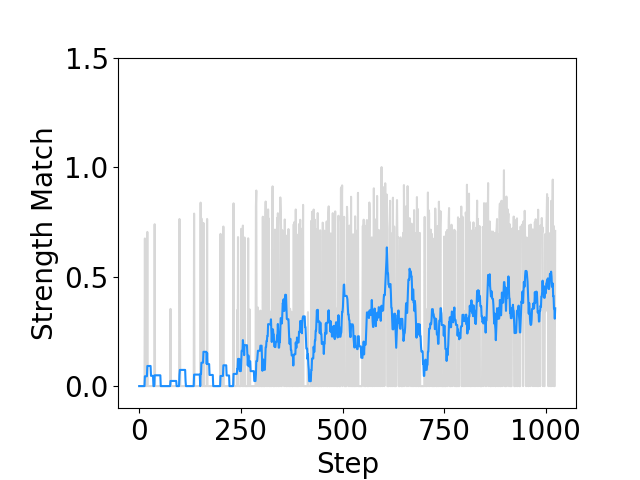}
            \includegraphics[width=0.96\textwidth]{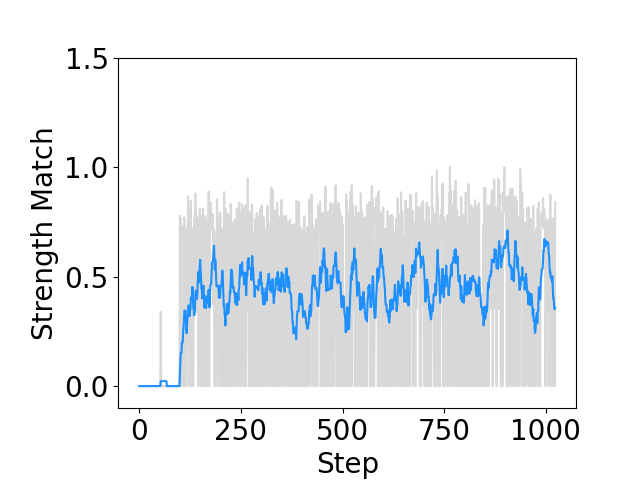}
            \\
            Transverse strength
        \end{minipage}
        \begin{minipage}[b]{0.32\textwidth}
            \centering
            \includegraphics[width=0.96\textwidth]{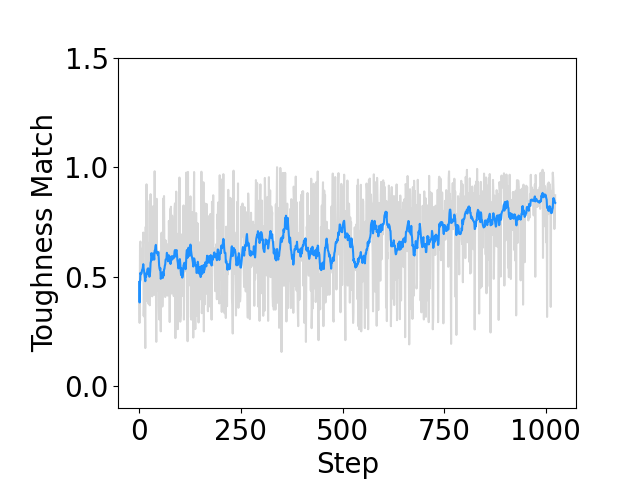}
            \includegraphics[width=0.96\textwidth]{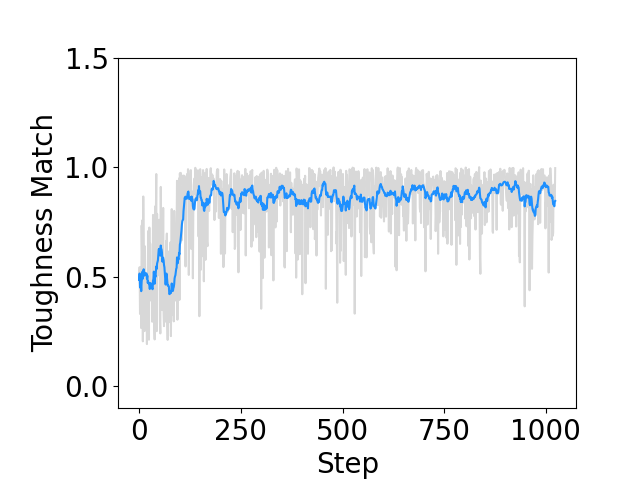}
            \\
            Transverse toughness
        \end{minipage}
    \end{subfigure}
    \captionsetup{font=scriptsize}
    \caption{
    Component match degrees of the reward function in the first (upper) and last (lower) training loops.
    The match degrees of the transverse strength and toughness confirmed the optimization by RL.
    The optimization for axial modulus and polymer stiffness was selectivity in fact.
    Thus, the polymer stiffness of the three chain lengths was computed to ensure validation, together with a specific strict match degree function.
    }
    \label{fig:component_match_degrees_during_training}
\end{figure}

The reward function for the RL training was composed of the match degrees and their thresholds and weights.
To describe the learning process during training, the component match degree curves with the training steps in both the first and last trainings loops are shown in Figure~\ref{fig:component_match_degrees_during_training}.
As included in the table of BD and NB properties, the component match degrees were of the elastic modulus, persistence length, end to end distance, transverse strength, and toughness.
The match degree of persistence length or end to end distance was the minimal value of the three chain lengths (10, 20, 30 residues long),
and the match degree of transverse strength or toughness was the minimal value of the three characteristic directions.

For each marked column, the upper image was from the first training loop whereas the lower image was from the last training loop.
As the figures show, the axial modulus was well matched even at the very beginning; and the transverse mechanical properties were obviously improved after training.
However, the match degree of transverse toughness was more reliable and higher in the last training loop, which illustrated the function of min for $M_s$ and $M_t$.

Therefore, the only problem was the polymer stiffness; there was no evidence of obvious improvement after training.
This was not the problem of computation convergence, as a 4~fs timestep was chosen and the polymer stiffness value stability was tested by extending the simulations steps to four times more.
As emphasized in the BD properties, rescaling of the angle and dihedral force constants was critical to compensate for the lower degrees of freedom while maintaining polymer stiffness.
This confirmed the difficulty of quantitatively reproducing polymer stiffness in the CG model.
Thus, the optimization of polymer stiffness is practically a selection process.

Three different chain lengths were considered during the training to harden the validation for all three chain lengths.
The match degree function for the polymer stiffness was slightly different:
\begin{equation}m=\frac{1}{1+\sqrt{|\frac{y_1-y_0}{y_0}|}}\end{equation}
. The square root function further lowers the match degrees for low values to enhance selectivity.

\subsection*{Bonded properties without rescale}\indent
\addcontentsline{toc}{subsection}{\protect\numberline{}Bonded properties without rescale}\indent

In this study, the BD force constants were rescaled to smaller values, referring to the estimations using the Boltzmann Inversion method (Table~\ref{tab:bonded_force_constants_without_rescale}).
The BD properties corresponding to the unrescaled BD force constants are listed in Table~\ref{tab:bonded_properties_without_rescale}, where the persistence length data are unacceptable.
The degrees of freedom of the CG models are significantly smaller than those of their AA models, and we assume that the smaller BD force constants for the angles and dihedrals are critical for reproducing polymer stiffness performance.

\begin{table}[h]
    \centering
    \caption{Bonded force constants without rescale}
    \label{tab:bonded_force_constants_without_rescale}
    \begin{tabular}{c c c c}
        \hline
        \hline
        {type}&{id}&{Optimized}&{Unrescaled}\\
        \hline
        {}&{}&\multicolumn{2}{c}{$k$(kcal/mol/$\rm\r{A}^2$)}\\
        \hline
        \multirow{3}{*}{bond}&{1}&{64.06}&{110.3}\\
        &{2}&{82.42}&{141.9}\\
        &{3}&{112.3}&{193.3}\\
        \hline
        {}&{}&\multicolumn{2}{c}{$k$(kcal/mol)}\\
        \hline
        \multirow{6}{*}{angle}&{1}&{30.97}&{274.5}\\
        &{2}&{38.79}&{343.7}\\
        &{3}&{30.52}&{270.5}\\
        &{4}&{27.28}&{241.8}\\
        &{5}&{25.97}&{230.2}\\
        &{6}&{22.42}&{198.7}\\
        \hline
        {}&{}&\multicolumn{2}{c}{$k$(kcal/mol)}\\
        \hline
        \multirow{3}{*}{improper}&{1}&{0.1702}&{21.57}\\
        &{2}&{0.1555}&{19.71}\\
        &{3}&{0.1698}&{21.53}\\
        \hline
        \hline
    \end{tabular}
\end{table}

\begin{table}[h]
    \centering
    \caption{Bonded properties without rescale}
    \label{tab:bonded_properties_without_rescale}
    \begin{tabular}{c c c c c}
        \hline
        \hline
        \multicolumn{2}{c}{type}&{AA}&{CG}&{Error}\\
        \hline
        \multicolumn{2}{c}{Axial modulus (GPa)}&{133.5}&{233.8}&{75.13\%}\\
        \hline
        \multirow{3}{*}{\makecell{Persistence\\length (nm)}}&{10}&{9.755}&{306.1}&{3137.87\%}\\
        &{20}&{9.815}&{310.5}&{3063.53\%}\\
        &{30}&{9.716}&{298.6}&{2973.33\%}\\
        \hline
        \multirow{3}{*}{\makecell{End-to-end\\distance (nm)}}&{10}&{4.523}&{4.689}&{3.67\%}\\
        &{20}&{9.355}&{9.888}&{5.70\%}\\
        &{30}&{13.536}&{15.069}&{11.33\%}\\
        \hline
        \hline
    \end{tabular}
\end{table}

\newpage

\subsection*{Transverse strength and toughness distributions}\indent
\addcontentsline{toc}{subsection}{\protect\numberline{}Transverse strength and toughness distributions}\indent

\begin{figure}[htbp]
    \centering
    \scriptsize
    \begin{subfigure}[b]{0.96\textwidth}
        \centering
        \scriptsize
        \begin{minipage}[b]{0.32\textwidth}
            \centering
            \includegraphics[width=0.96\textwidth]{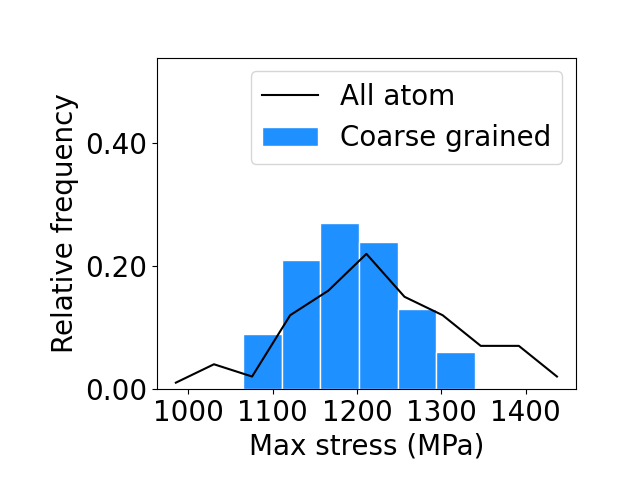}
            \\
            Vertical
        \end{minipage}
        \begin{minipage}[b]{0.32\textwidth}
            \centering
            \includegraphics[width=0.96\textwidth]{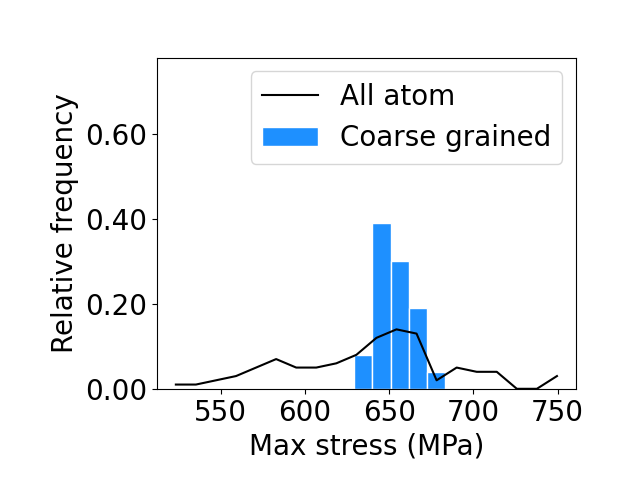}
            \\
            Horizontal
        \end{minipage}
        \begin{minipage}[b]{0.32\textwidth}
            \centering
            \includegraphics[width=0.96\textwidth]{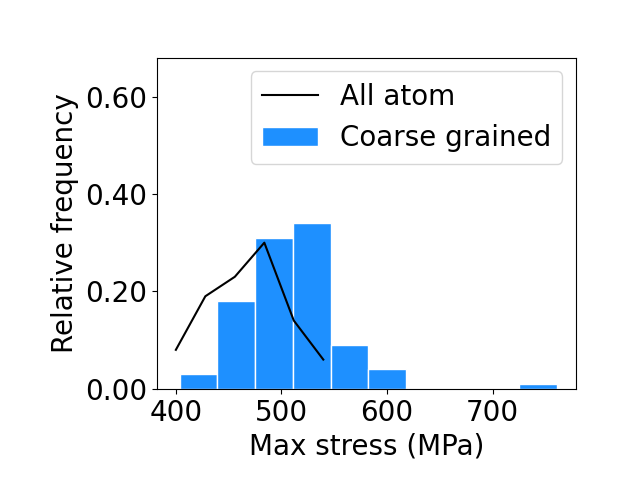}
            \\
            Slant
        \end{minipage}
        \subcaption{}
    \end{subfigure}
    \\
    \begin{subfigure}[b]{0.96\textwidth}
        \centering
        \scriptsize
        \begin{minipage}[b]{0.32\textwidth}
            \centering
            \includegraphics[width=0.96\textwidth]{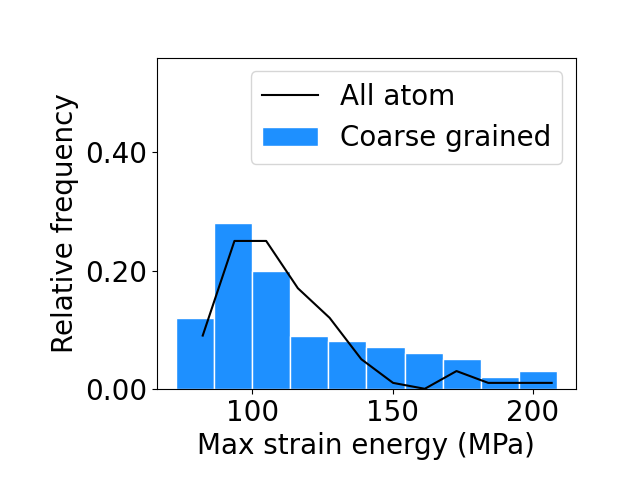}
            \\
            Vertical
        \end{minipage}
        \begin{minipage}[b]{0.32\textwidth}
            \centering
            \includegraphics[width=0.96\textwidth]{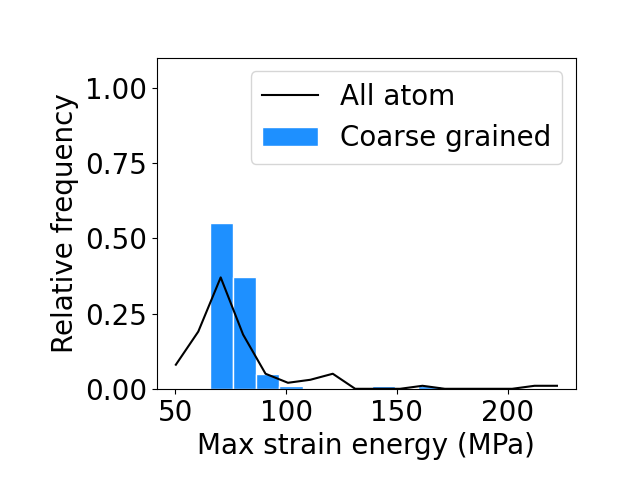}
            \\
            Horizontal
        \end{minipage}
        \begin{minipage}[b]{0.32\textwidth}
            \centering
            \includegraphics[width=0.96\textwidth]{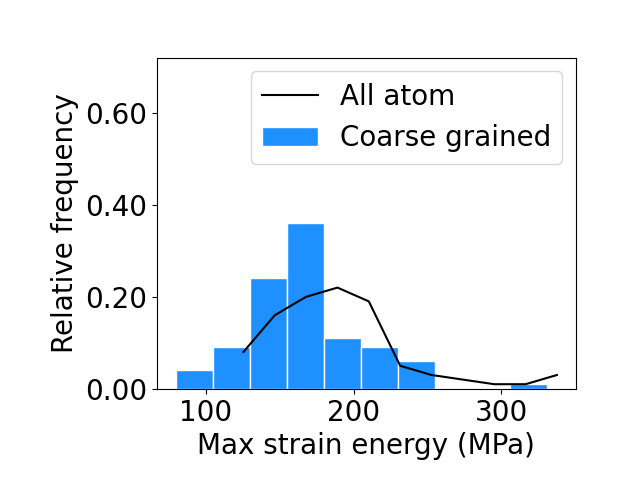}
            \\
            Slant
        \end{minipage}
        \subcaption{}
    \end{subfigure}
    \captionsetup{font=scriptsize}
    \caption{
    Transverse (a) strength and (b) toughness distributions.
    A total of 100 replica simulations were performed with different random seeds for both AA and CG models in three directions.
    For the vertical and slant models, the strength and toughness histograms were reproduced consistently, as their mean values and standard errors were similar.
    Despite the accuracy of the mean values, the horizontal CG models provided narrower distributions, which was within expectation.
    }
    \label{fig:transverse_mechanical_property_distributions}
\end{figure}

The NB properties include the transverse strength and toughness, as shown in Figure~\ref{fig:transverse_mechanical_property_distributions}.
Although the accuracy of the strength and toughness distributions is problematic owing to the representability problem of CG models, the full distributions of the transverse strength and toughness are shown in Figure~\ref{fig:transverse_mechanical_property_distributions} for reference.

\subsection*{Transverse stretch curves and distributions}\indent
\addcontentsline{toc}{subsection}{\protect\numberline{}Transverse stretch curves and distributions}\indent

\begin{figure}[htbp]
    \centering
    \begin{subfigure}[b]{0.96\textwidth}
        \centering
        \scriptsize
        \begin{minipage}[b]{0.32\textwidth}
            \centering
            \includegraphics[width=0.96\textwidth]{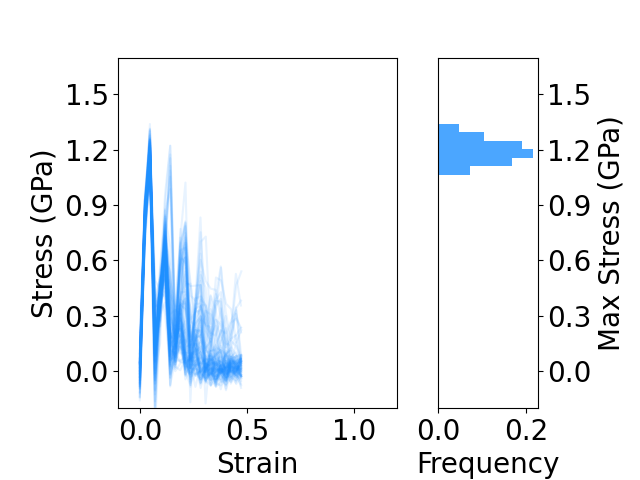}
            \\
            Vertical
        \end{minipage}
        \begin{minipage}[b]{0.32\textwidth}
            \centering
            \includegraphics[width=0.96\textwidth]{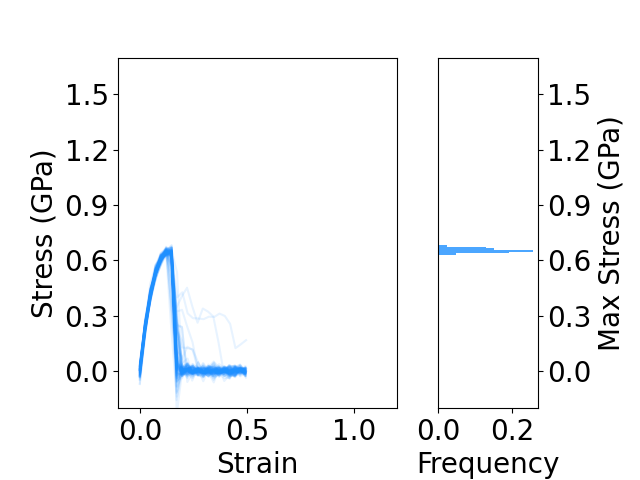}
            \\
            Horizontal
        \end{minipage}
        \begin{minipage}[b]{0.32\textwidth}
            \centering
            \includegraphics[width=0.96\textwidth]{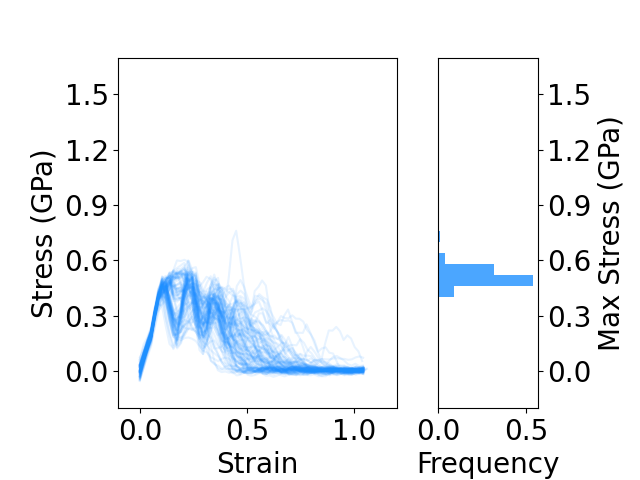}
            \\
            Slant
        \end{minipage}
    \subcaption{}
    \end{subfigure}
    \\
    \begin{subfigure}[b]{0.96\textwidth}
        \centering
        \scriptsize
        \begin{minipage}[b]{0.32\textwidth}
            \centering
            \includegraphics[width=0.96\textwidth]{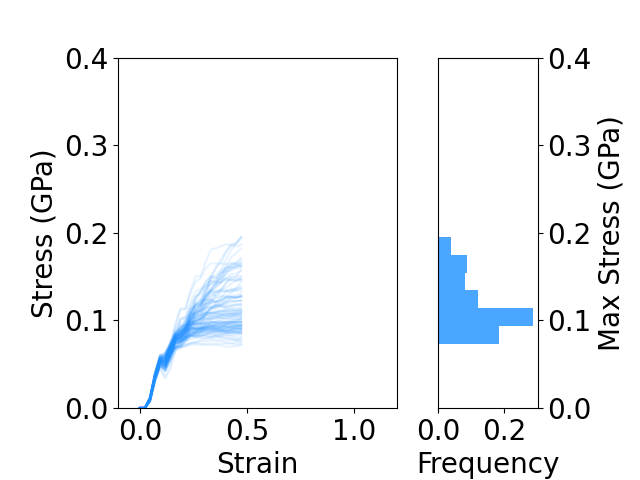}
            \\
            Vertical
        \end{minipage}
        \begin{minipage}[b]{0.32\textwidth}
            \centering
            \includegraphics[width=0.96\textwidth]{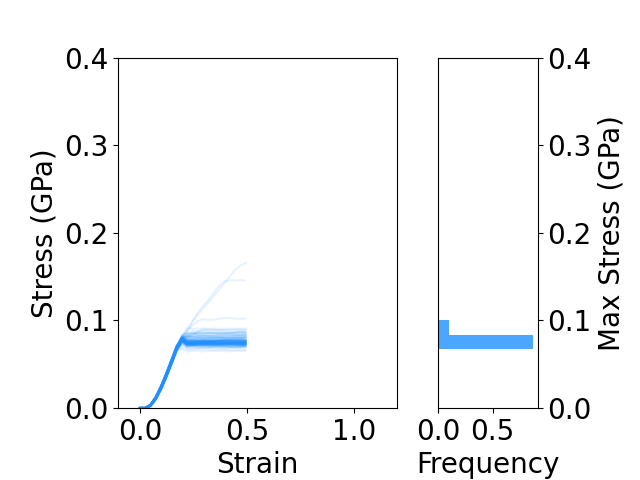}
            \\
            Horizontal
        \end{minipage}
        \begin{minipage}[b]{0.32\textwidth}
            \centering
            \includegraphics[width=0.96\textwidth]{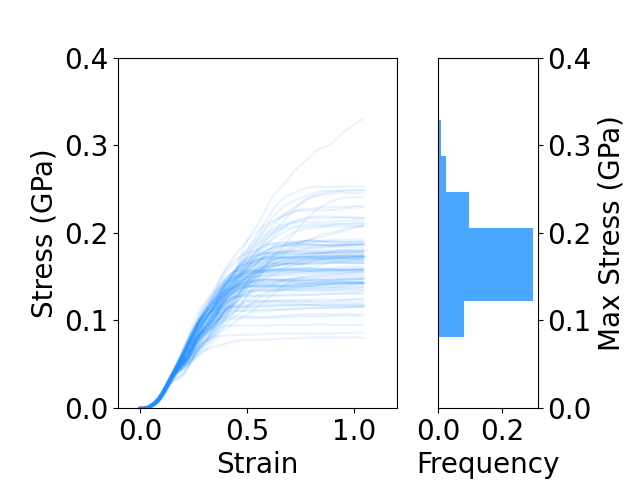}
            \\
            Slant
        \end{minipage}
        \subcaption{}
    \end{subfigure}
    \\
    \begin{subfigure}[b]{0.96\textwidth}
        \centering
        \scriptsize
        \begin{minipage}[b]{0.32\textwidth}
            \centering
            \includegraphics[width=0.96\textwidth]{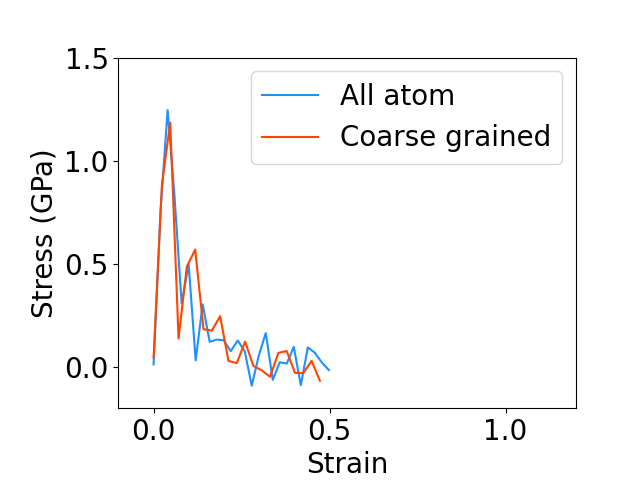}
            \\
            Vertical
        \end{minipage}
        \begin{minipage}[b]{0.32\textwidth}
            \centering
            \includegraphics[width=0.96\textwidth]{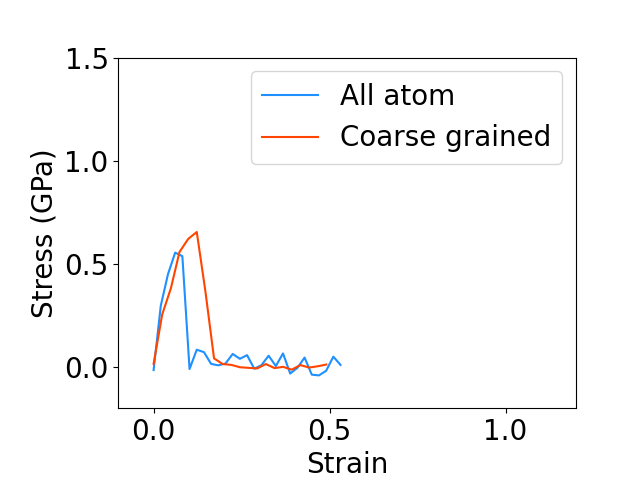}
            \\
            Horizontal
        \end{minipage}
        \\
        \centering
        \scriptsize
        \begin{minipage}[b]{0.32\textwidth}
            \centering
            \includegraphics[width=0.96\textwidth]{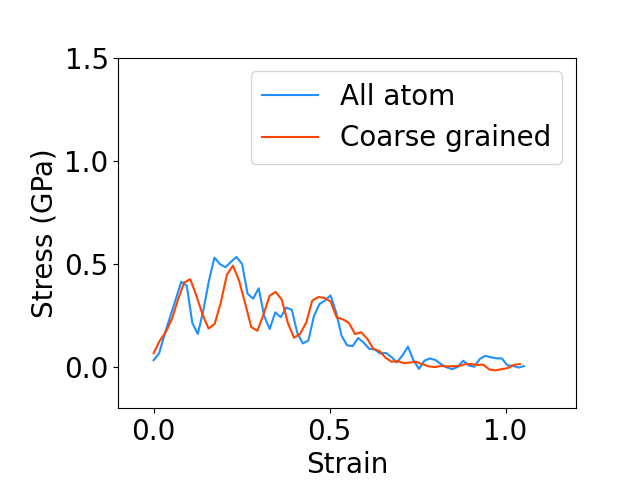}
            \\
            Slant
        \end{minipage}
        \begin{minipage}[b]{0.32\textwidth}
            \centering
            \includegraphics[width=0.96\textwidth]{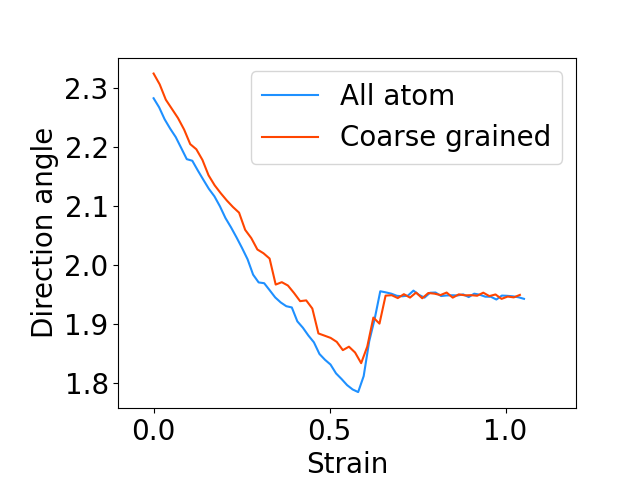}
            \\
            Direction angle
        \end{minipage}
    \end{subfigure}
    \captionsetup{font=scriptsize}
    \caption{
    CG stretch curve collections.
    CG transverse stretch (a) stress-strain and (b) strain energy collections in characteristic directions at 10.0~nm/ns.
    Each collection contains 100 curves, and the strength or toughness distribution is enclosed on the right.
    The two summits of the stress-strain curves for the vertical models were caused by horizontal dislocations.
    This implied that the fractured HBonding layers rotated and reconnected to their nearby layers and broke again.
    (c) Examples of transverse stress-strain curves in the characteristic directions.
    An example of the direction angle of the slant model is also included to quantitatively describe the frictional sliding.
    The fragility and ductility of the different models are validated via the example curves.
    }
    \label{fig:transverse_stretch_performance_collection}
\end{figure}

Transverse stretch curves and distributions of 100 replica simulations were used to elucidate the reproduced transverse anisotropy (Figure~\ref{fig:transverse_stretch_performance_collection}).
The direction angle of the slant model is also shown to describe the frictional sliding.
These results confirm that the fragility of the vertical and horizontal models and the toughness of the slant models are all represented in the CG model.
Two stress summits were observed in the stress-strain curves of the vertical model, which were caused by the horizontal dislocations.
When the HBonding layers broke in the vertical models, they rotated and reconnected, resulting in two fractures and two stress summits.

\subsection*{Bonded geometry parameter distributions}\indent
\addcontentsline{toc}{subsection}{\protect\numberline{}Bonded geometry parameter distributions}\indent

\begin{figure}[htbp]
    \centering
    \begin{subfigure}[b]{0.96\textwidth}
        \centering
        \scriptsize
        \begin{minipage}[b]{0.32\textwidth}
            \centering
            \includegraphics[width=0.96\textwidth]{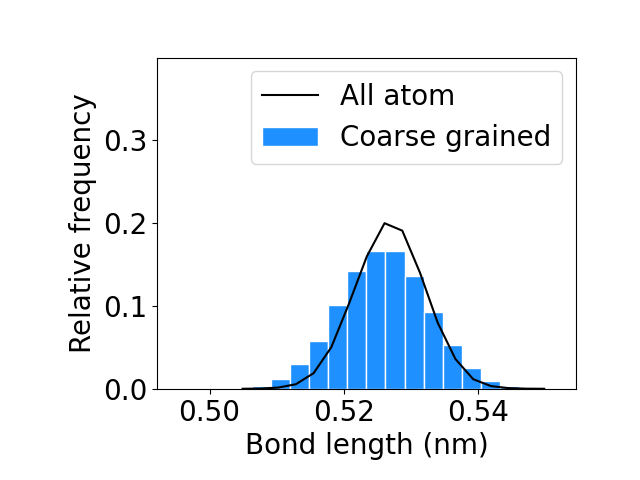}
            \\
            CL1-CL1
        \end{minipage}
        \begin{minipage}[b]{0.32\textwidth}
            \centering
            \includegraphics[width=0.96\textwidth]{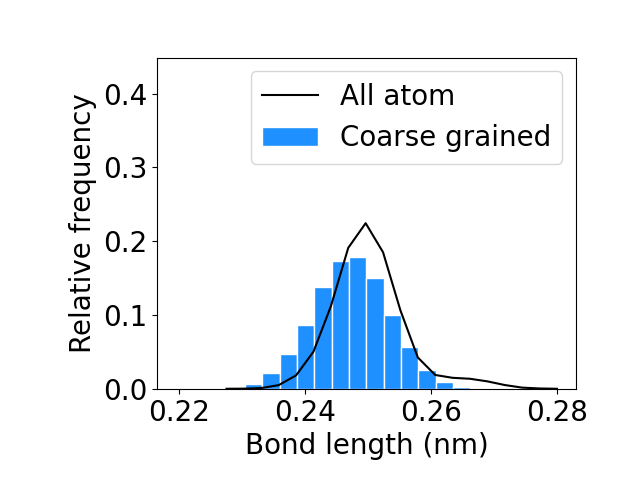}
            \\
            CL1-CL2
        \end{minipage}
        \begin{minipage}[b]{0.32\textwidth}
            \centering
            \includegraphics[width=0.96\textwidth]{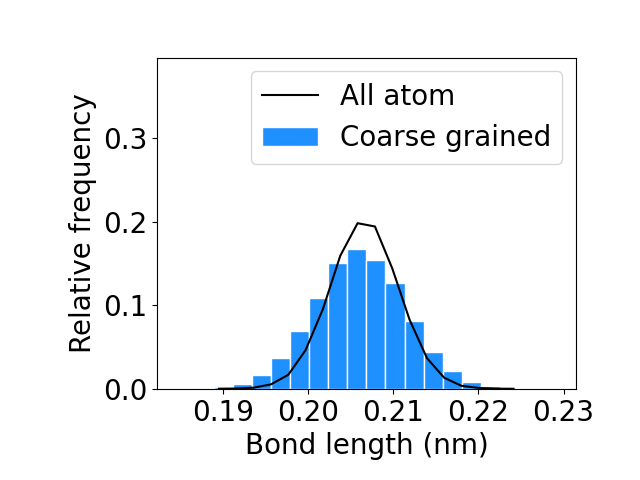}
            \\
            CL1-CL3
        \end{minipage}
        \subcaption{}
    \end{subfigure}
    \\
    \begin{subfigure}[b]{0.96\textwidth}
        \centering
        \scriptsize
        \begin{minipage}[b]{0.32\textwidth}
            \centering
            \includegraphics[width=0.96\textwidth]{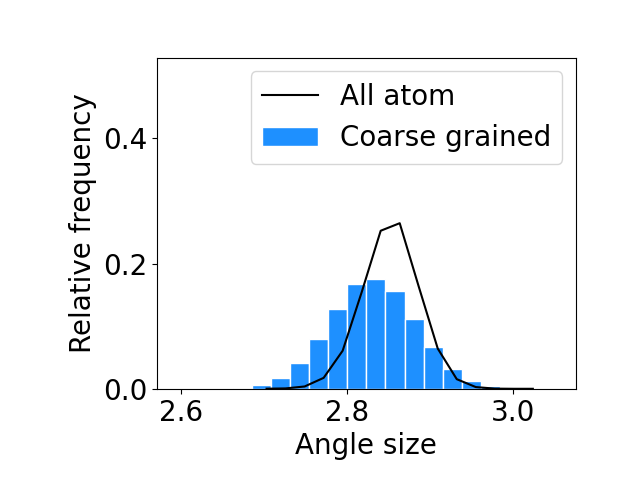}
            \\
            CL1-CL1-CL1
        \end{minipage}
        \begin{minipage}[b]{0.32\textwidth}
            \centering
            \includegraphics[width=0.96\textwidth]{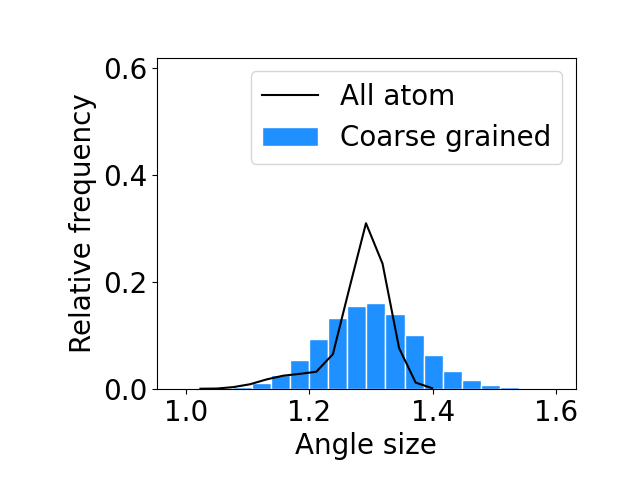}
            \\
            CL2-CL1-CL1\_1
        \end{minipage}
        \begin{minipage}[b]{0.32\textwidth}
            \centering
            \includegraphics[width=0.96\textwidth]{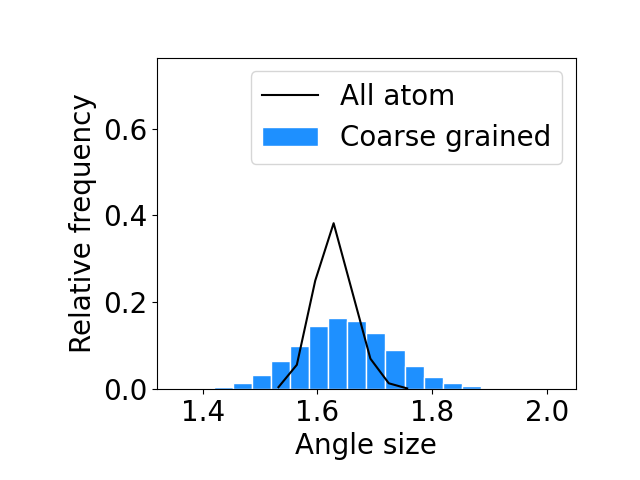}
            \\
            CL3-CL1-CL1\_1
        \end{minipage}
        \\
        \begin{minipage}[b]{0.32\textwidth}
            \centering
            \includegraphics[width=0.96\textwidth]{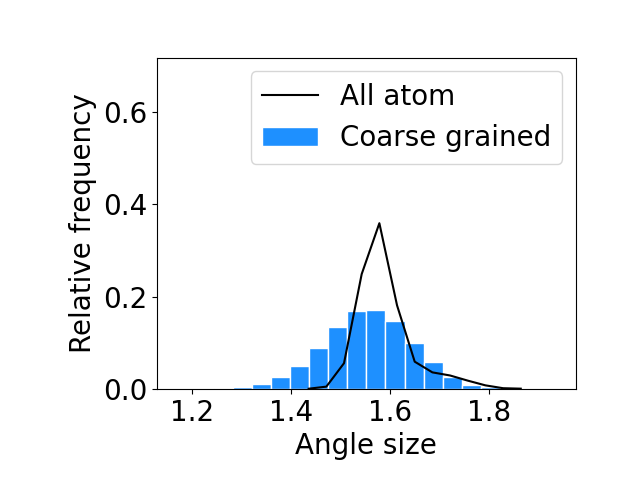}
            \\
            CL2-CL1-CL1\_2
        \end{minipage}
        \begin{minipage}[b]{0.32\textwidth}
            \centering
            \includegraphics[width=0.96\textwidth]{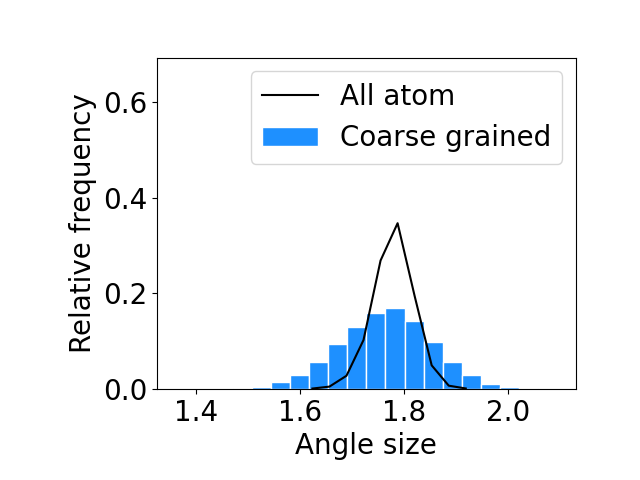}
            \\
            CL3-CL1-CL1\_2
        \end{minipage}
        \begin{minipage}[b]{0.32\textwidth}
            \centering
            \includegraphics[width=0.96\textwidth]{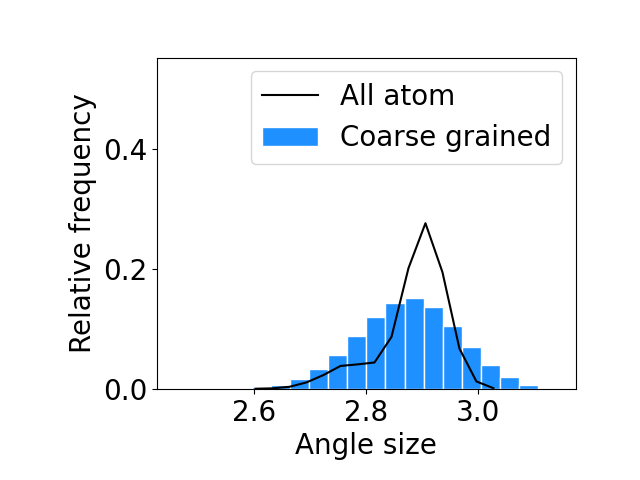}
            \\
            CL2-CL1-CL3
        \end{minipage}
        \subcaption{}
    \end{subfigure}
    \\
    \begin{subfigure}[b]{0.96\textwidth}
        \centering
        \scriptsize
        \begin{minipage}[b]{0.32\textwidth}
            \centering
            \includegraphics[width=0.96\textwidth]{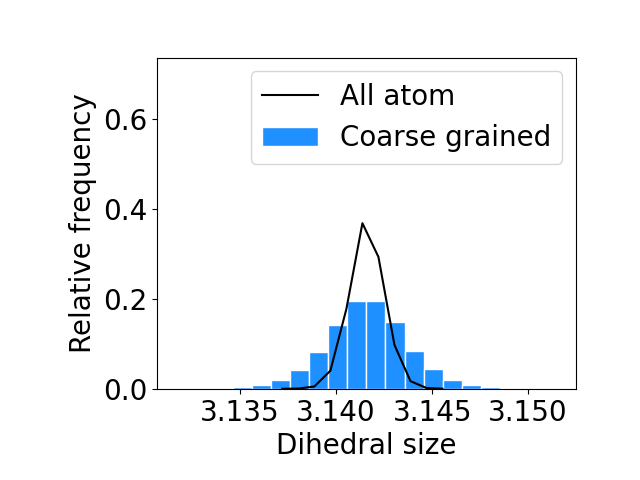}
            \\
            CL1-CL1-CL1-CL1
        \end{minipage}
        \begin{minipage}[b]{0.32\textwidth}
            \centering
            \includegraphics[width=0.96\textwidth]{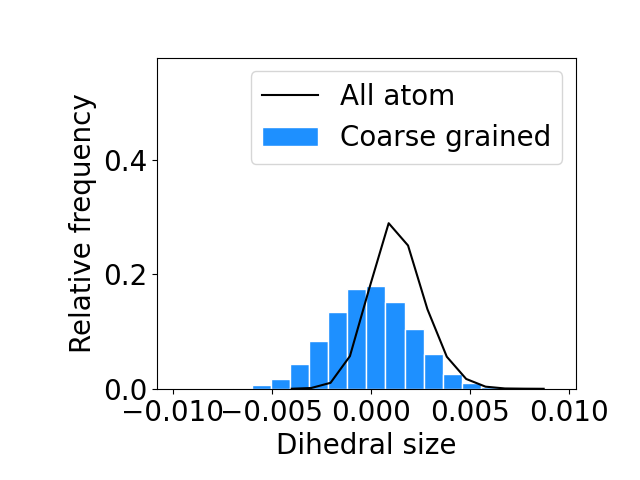}
            \\
            CL2-CL1-CL1-CL3
        \end{minipage}
        \begin{minipage}[b]{0.32\textwidth}
            \centering
            \includegraphics[width=0.96\textwidth]{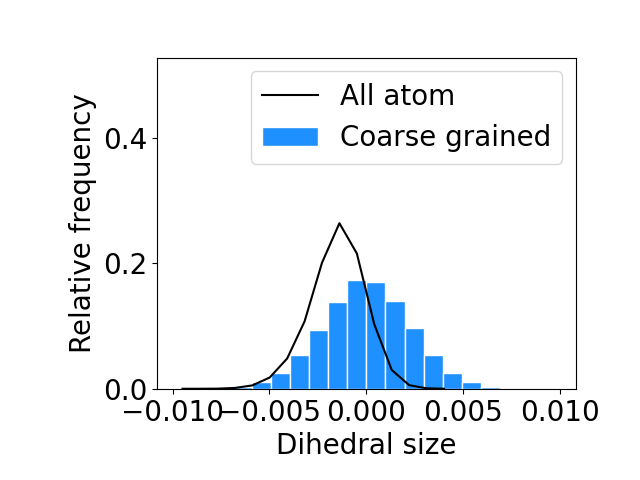}
            \\
            CL3-CL1-CL1-CL2
        \end{minipage}
        \subcaption{}
    \end{subfigure}
    \captionsetup{font=scriptsize}
    \caption{
    BD geometry parameter distributions of equilibrium simulations.
    Distributions of (a) bond length, (b) angle size, and (c) dihedral size.
    These data were derived from the equilibrium CNC trajectories of the CG models and mapped AA trajectories.
    The BD geometry parameters were well preserved in the CG model (with errors of less than 2\%).
    Owing to the smaller force constants, the distribution dispersions were larger than those of the AA reference as a compromise for reproducing the polymer stiffness.
    }
    \label{fig:bonded_geometry_parameter_distributions_equilibrium}
\end{figure}

\begin{figure}[htbp]
    \centering
    \begin{subfigure}[b]{0.96\textwidth}
        \centering
        \scriptsize
        \begin{minipage}[b]{0.32\textwidth}
            \centering
            \includegraphics[width=0.96\textwidth]{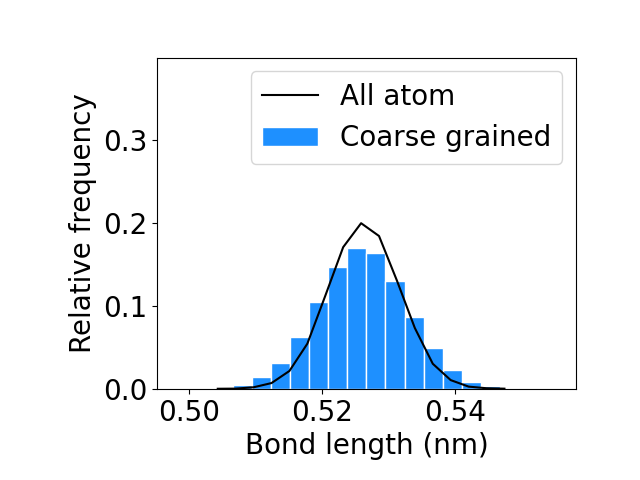}
            \\
            CL1-CL1
        \end{minipage}
        \begin{minipage}[b]{0.32\textwidth}
            \centering
            \includegraphics[width=0.96\textwidth]{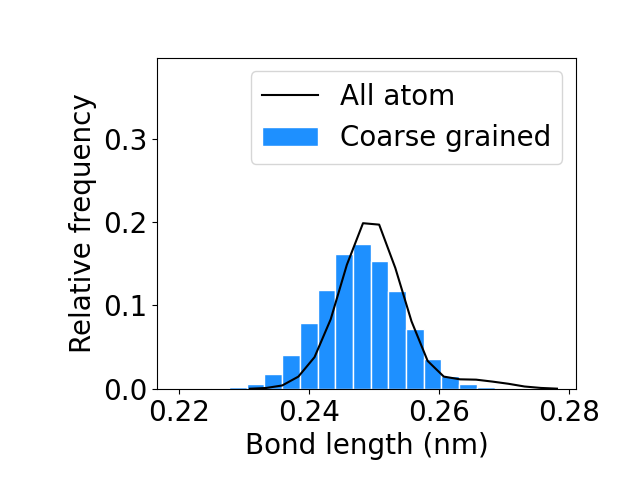}
            \\
            CL1-CL2
        \end{minipage}
        \begin{minipage}[b]{0.32\textwidth}
            \centering
            \includegraphics[width=0.96\textwidth]{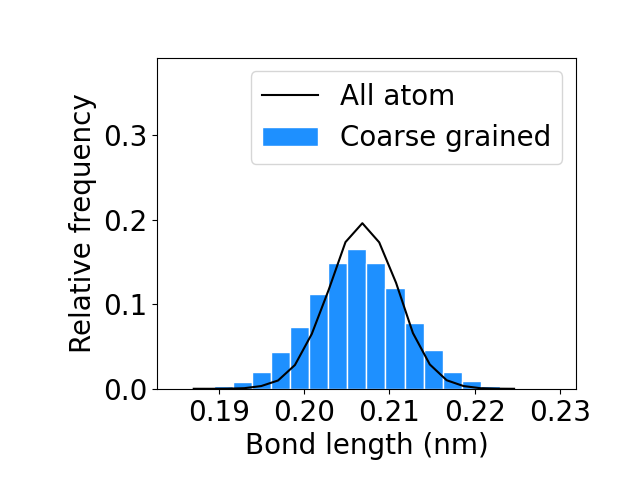}
            \\
            CL1-CL3
        \end{minipage}
        \subcaption{}
    \end{subfigure}
    \\
    \begin{subfigure}[b]{0.96\textwidth}
        \centering
        \scriptsize
        \begin{minipage}[b]{0.32\textwidth}
            \centering
            \includegraphics[width=0.96\textwidth]{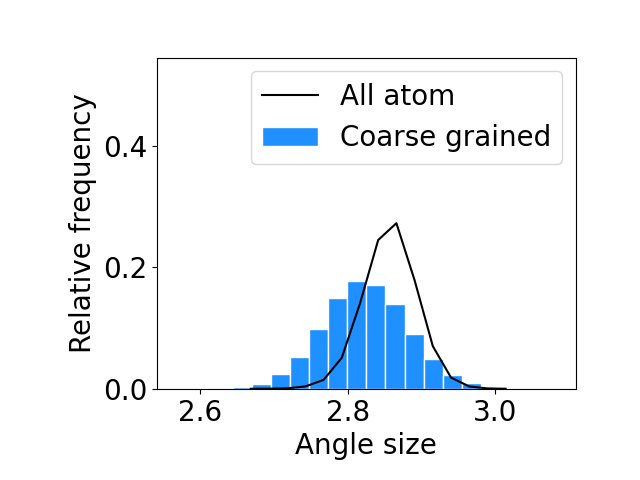}
            \\
            CL1-CL1-CL1
        \end{minipage}
        \begin{minipage}[b]{0.32\textwidth}
            \centering
            \includegraphics[width=0.96\textwidth]{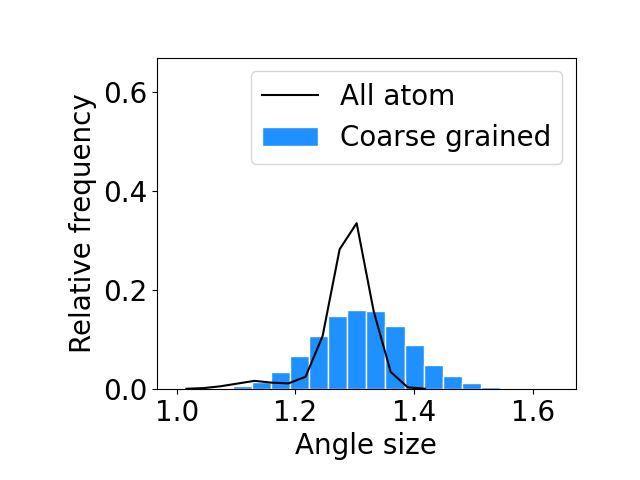}
            \\
            CL2-CL1-CL1\_1
        \end{minipage}
        \begin{minipage}[b]{0.32\textwidth}
            \centering
            \includegraphics[width=0.96\textwidth]{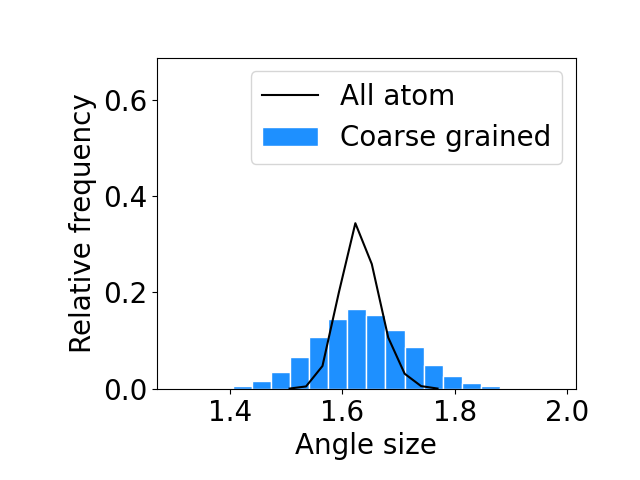}
            \\
            CL3-CL1-CL1\_1
        \end{minipage}
        \\
        \begin{minipage}[b]{0.32\textwidth}
            \centering
            \includegraphics[width=0.96\textwidth]{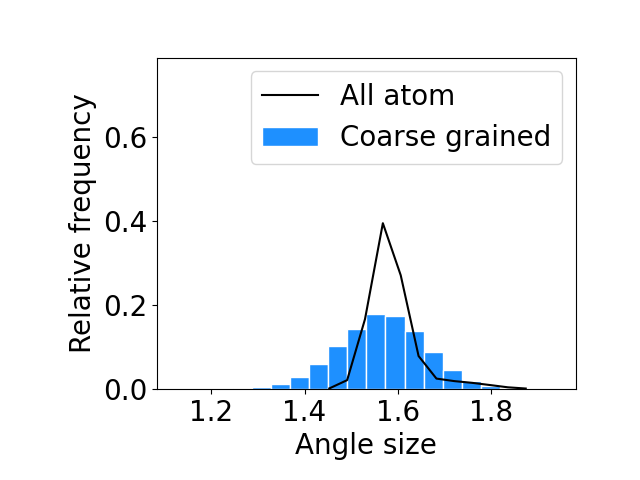}
            \\
            CL2-CL1-CL1\_2
        \end{minipage}
        \begin{minipage}[b]{0.32\textwidth}
            \centering
            \includegraphics[width=0.96\textwidth]{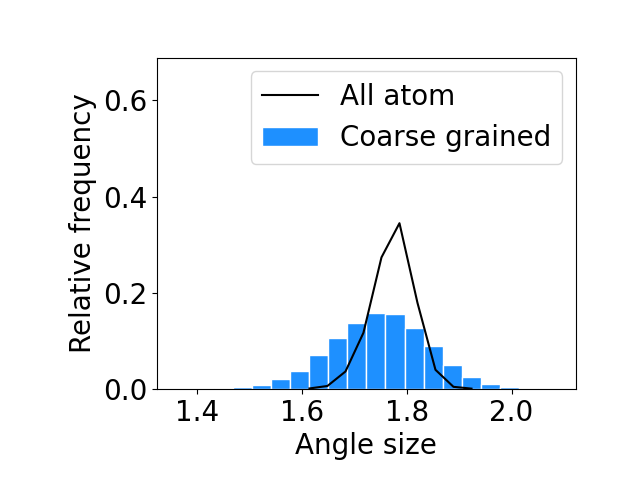}
            \\
            CL3-CL1-CL1\_2
        \end{minipage}
        \begin{minipage}[b]{0.32\textwidth}
            \centering
            \includegraphics[width=0.96\textwidth]{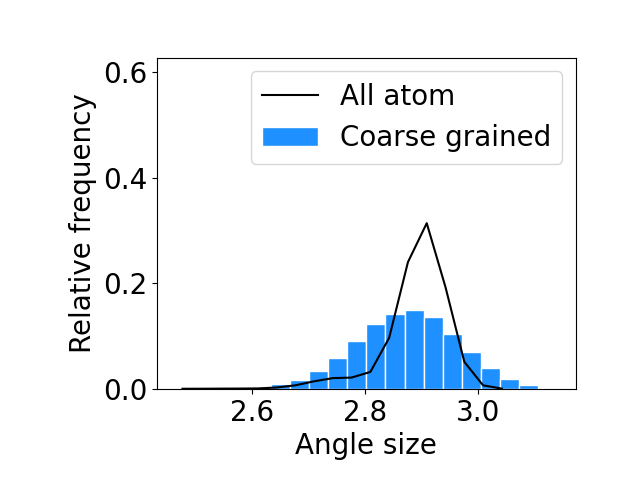}
            \\
            CL2-CL1-CL3
        \end{minipage}
        \subcaption{}
    \end{subfigure}
    \\
    \begin{subfigure}[b]{0.96\textwidth}
        \centering
        \scriptsize
        \begin{minipage}[b]{0.32\textwidth}
            \centering
            \includegraphics[width=0.96\textwidth]{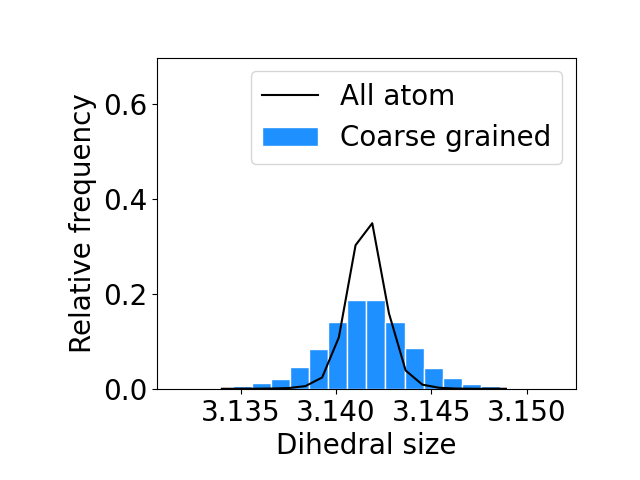}
            \\
            CL1-CL1-CL1-CL1
        \end{minipage}
        \begin{minipage}[b]{0.32\textwidth}
            \centering
            \includegraphics[width=0.96\textwidth]{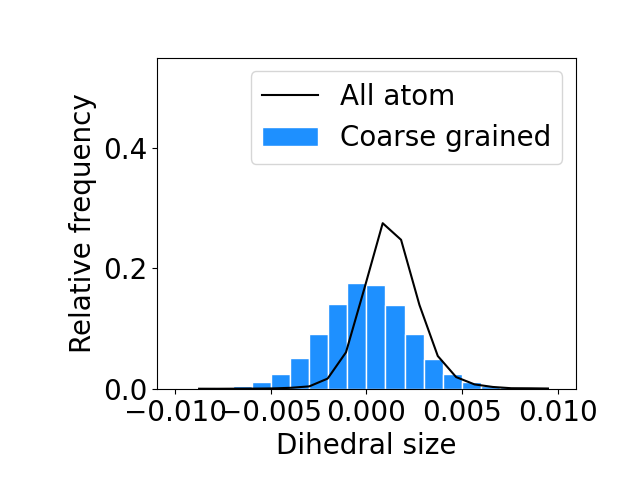}
            \\
            CL2-CL1-CL1-CL3
        \end{minipage}
        \begin{minipage}[b]{0.32\textwidth}
            \centering
            \includegraphics[width=0.96\textwidth]{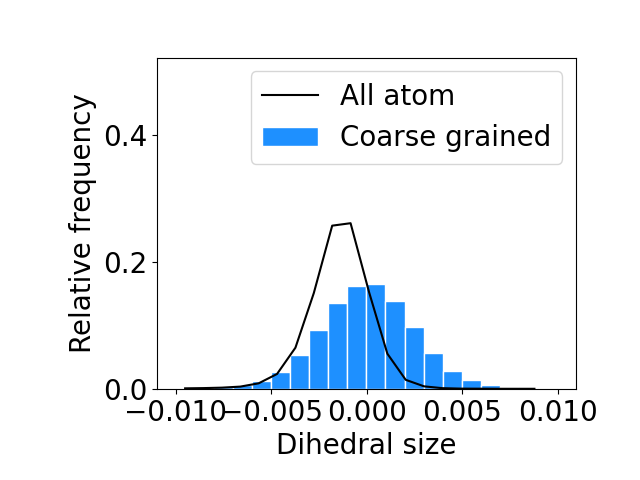}
            \\
            CL3-CL1-CL1-CL2
        \end{minipage}
        \subcaption{}
    \end{subfigure}
    \captionsetup{font=scriptsize}
    \caption{
    BD geometry parameter distributions of stretch processes.
    Distributions of (a) bond length, (b) angle size, and (c) dihedral size.
    These data were derived from the stretch CNC trajectories of the CG models and mapped AA trajectories.
    The distributions of AA and CG were similar to those of the equilibrium data.
    }
    \label{fig:bonded_geometry_parameter_distributions_stretch}
\end{figure}

After parameterization, the BD geometry parameters were sampled from the CNC equilibrium and stretch trajectories from the CG model and the mapped AA trajectories (Figure~\ref{fig:bonded_geometry_parameter_distributions_equilibrium} and Figure~\ref{fig:bonded_geometry_parameter_distributions_stretch}).
The BD geometry parameters of the CG models were similar to those of the AA reference data.
However, the distributions of the CG models exhibited larger dispersions owing to the rescaled BD force constants.

\newpage

\subsection*{Nonbonded equilibrium distance distributions}\indent
\addcontentsline{toc}{subsection}{\protect\numberline{}Nonbonded equilibrium distance distributions}\indent

\begin{figure}[htbp]
    \centering
    \scriptsize
    \begin{minipage}[b]{0.32\textwidth}
        \centering
        \includegraphics[width=0.96\textwidth]{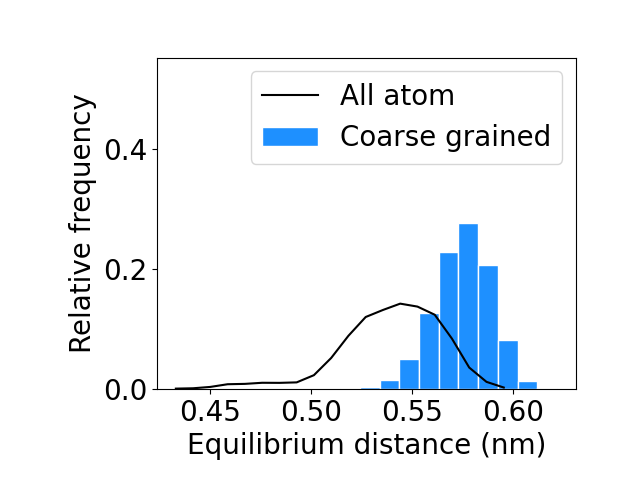}
        \\
        CL1:CL1
    \end{minipage}
    \begin{minipage}[b]{0.32\textwidth}
        \centering
        \includegraphics[width=0.96\textwidth]{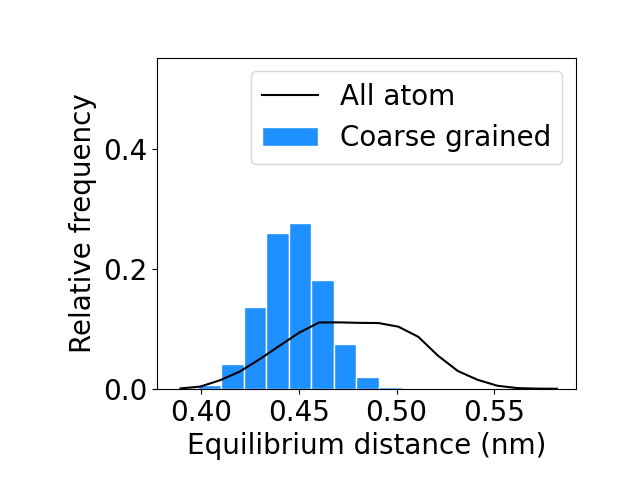}
        \\
        CL1:CL2
    \end{minipage}
    \begin{minipage}[b]{0.32\textwidth}
        \centering
        \includegraphics[width=0.96\textwidth]{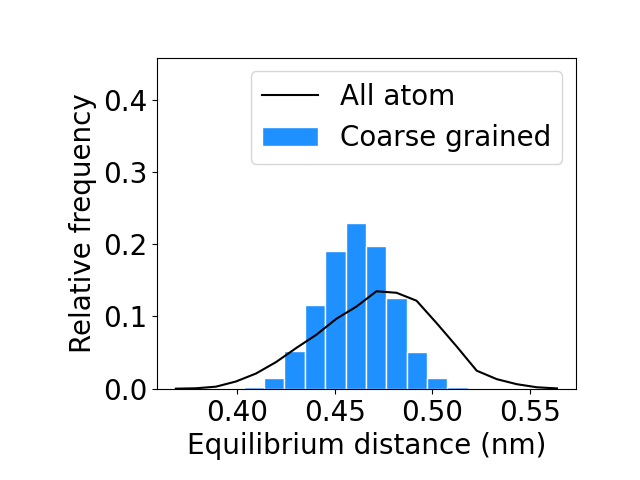}
        \\
        CL1:CL3
    \end{minipage}
    \\
    \begin{minipage}[b]{0.32\textwidth}
        \centering
        \includegraphics[width=0.96\textwidth]{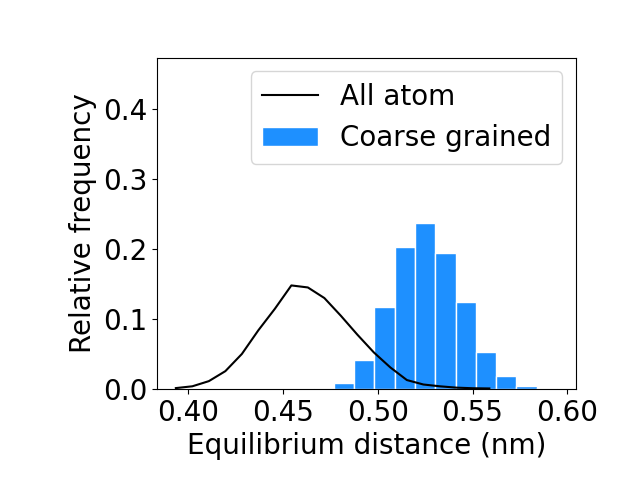}
        \\
        CL2:CL2
    \end{minipage}
    \begin{minipage}[b]{0.32\textwidth}
        \centering
        \includegraphics[width=0.96\textwidth]{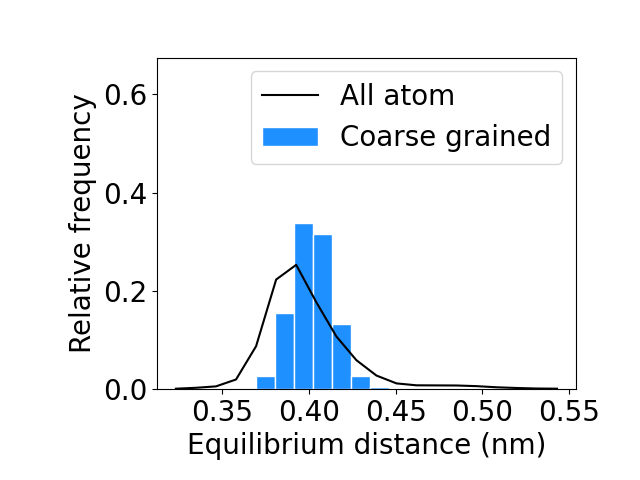}
        \\
        CL2:CL3
    \end{minipage}
    \begin{minipage}[b]{0.32\textwidth}
        \centering
        \includegraphics[width=0.96\textwidth]{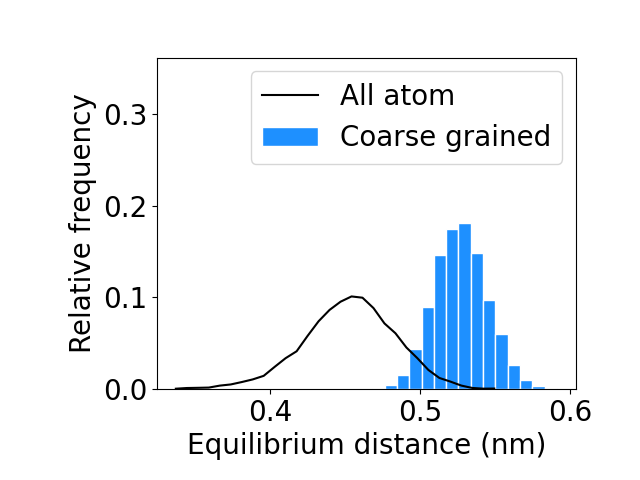}
        \\
        CL3:CL3
    \end{minipage}
    \captionsetup{font=scriptsize}
    \caption{
    NB equilibrium distance distributions of equilibrium simulations.
    These data were derived from the equilibrium CNC trajectories of the CG models and mapped AA trajectories.
    Corresponding to the unexpectedly larger NB distance coefficients, the equilibrium distances were often larger than those of the baselines to an acceptable extent with errors of less than 15\%.
    However, the resultant equilibrium distances of CL2:CL3 were not larger than those of the AA reference data, which elucidates the complexity of the multiple particle models and aids in understanding the increase in $\sigma_{p23}$.
    }
    \label{fig:nonbonded_equilibrium_distance_distributions_equilibrium}
\end{figure}

\begin{figure}[htbp]
    \centering
    \scriptsize
    \begin{minipage}[b]{0.32\textwidth}
        \centering
        \includegraphics[width=0.96\textwidth]{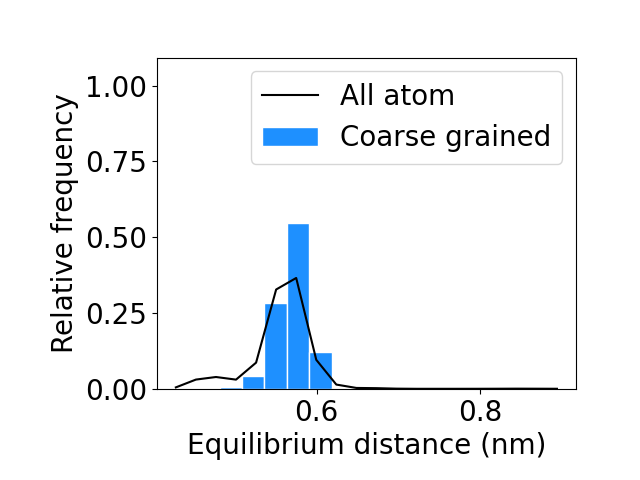}
        \\
        CL1:CL1
    \end{minipage}
    \begin{minipage}[b]{0.32\textwidth}
        \centering
        \includegraphics[width=0.96\textwidth]{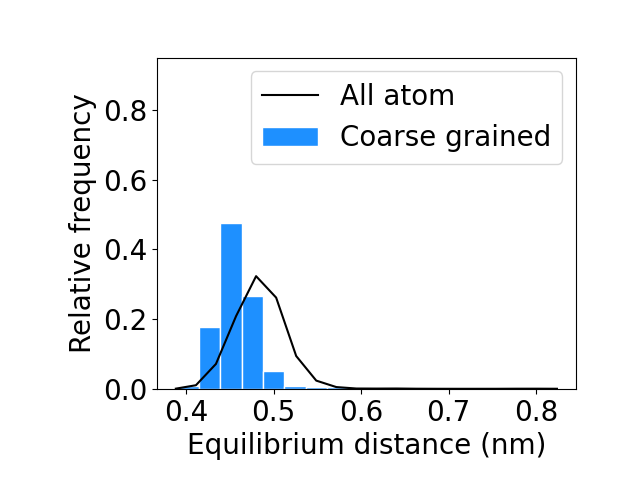}
        \\
        CL1:CL2
    \end{minipage}
    \begin{minipage}[b]{0.32\textwidth}
        \centering
        \includegraphics[width=0.96\textwidth]{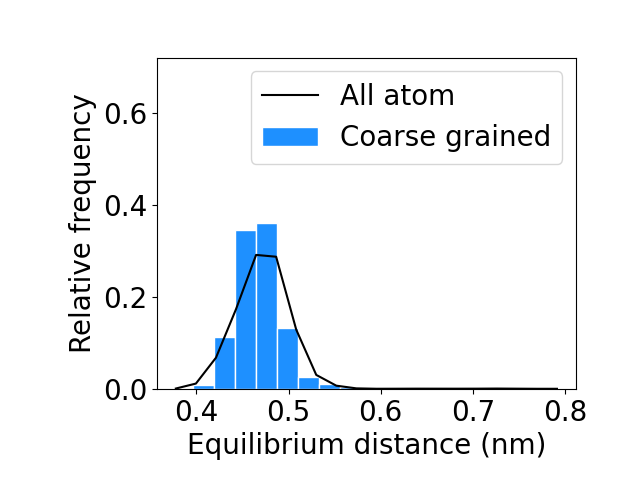}
        \\
        CL1:CL3
    \end{minipage}
    \\
    \begin{minipage}[b]{0.32\textwidth}
        \centering
        \includegraphics[width=0.96\textwidth]{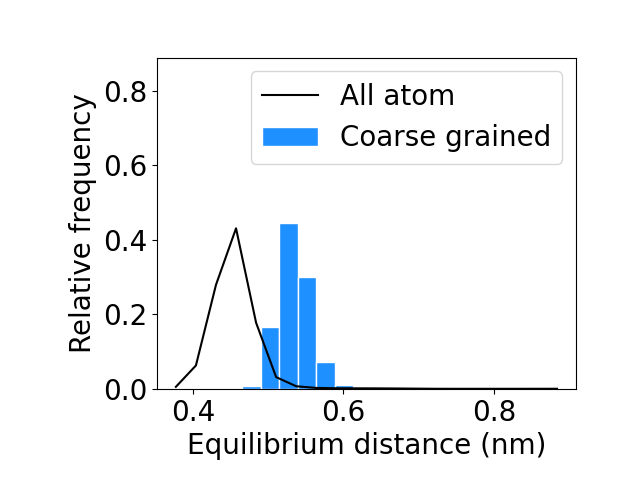}
        \\
        CL2:CL2
    \end{minipage}
    \begin{minipage}[b]{0.32\textwidth}
        \centering
        \includegraphics[width=0.96\textwidth]{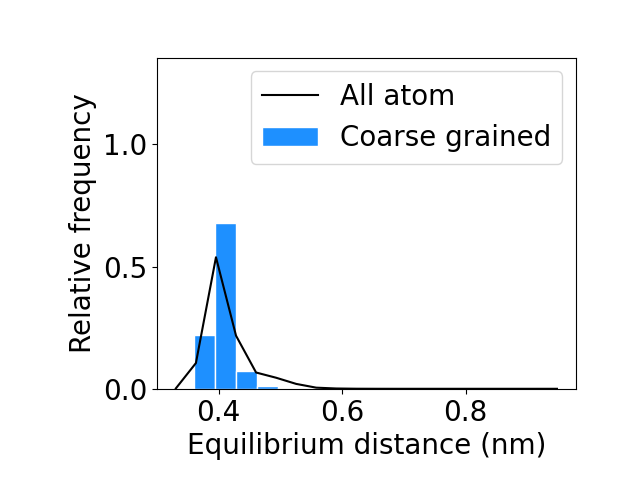}
        \\
        CL2:CL3
    \end{minipage}
    \begin{minipage}[b]{0.32\textwidth}
        \centering
        \includegraphics[width=0.96\textwidth]{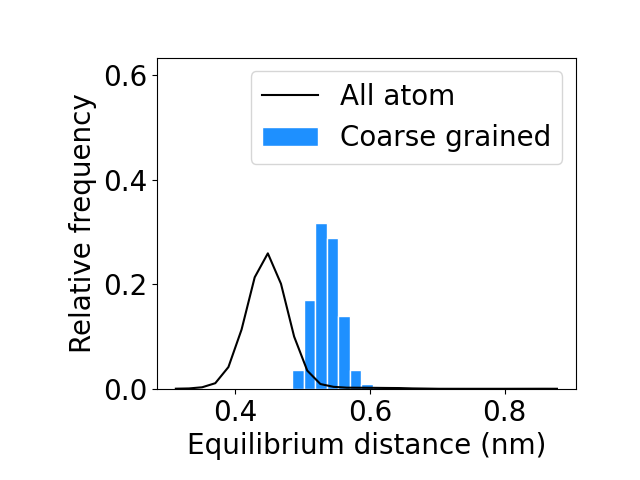}
        \\
        CL3:CL3
    \end{minipage}
    \captionsetup{font=scriptsize}
    \caption{
    NB equilibrium distance distributions of stretch processes.
    These data were derived from the stretch CNC trajectories of the CG models and mapped AA trajectories.
    The distributions of AA and CG were similar to the equilibrium data.
    }
    \label{fig:nonbonded_equilibrium_distance_distributions_stretch}
\end{figure}

After parameterization, NB equilibrium distances were sampled from the CNC equilibrium and stretch trajectories using the CG model and the mapped AA trajectories, as shown in Figure~\ref{fig:nonbonded_equilibrium_distance_distributions_equilibrium} and Figure~\ref{fig:nonbonded_equilibrium_distance_distributions_stretch}.
As mentioned in the Results/Coefficients section, NB distance coefficients are usually larger than the estimated range [0.89$d^e$, 1.00$d^e$] after parameterization by RL.
However, the equilibrium distance of the CL2:CL3 pair was not significantly larger, which emphasized the complexity of this CG model and demonstrates the reasonability of increase in $\sigma_{p23}$.
The errors in the equilibrium distance distributions are acceptable and less than 15\%.

\subsection*{Structural properties from stretch processes}\indent
\addcontentsline{toc}{subsection}{\protect\numberline{}Structural properties from stretch processes}\indent

\begin{figure}[htbp]
    \centering
    \begin{subfigure}[b]{0.96\textwidth}
        \begin{minipage}[b]{0.32\textwidth}
            \centering
            \includegraphics[width=0.96\textwidth]{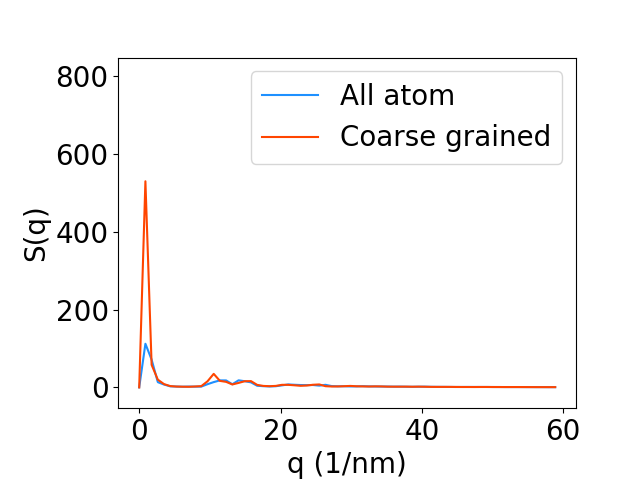}
            \\
            CL1
        \end{minipage}
        \begin{minipage}[b]{0.32\textwidth}
            \centering
            \includegraphics[width=0.96\textwidth]{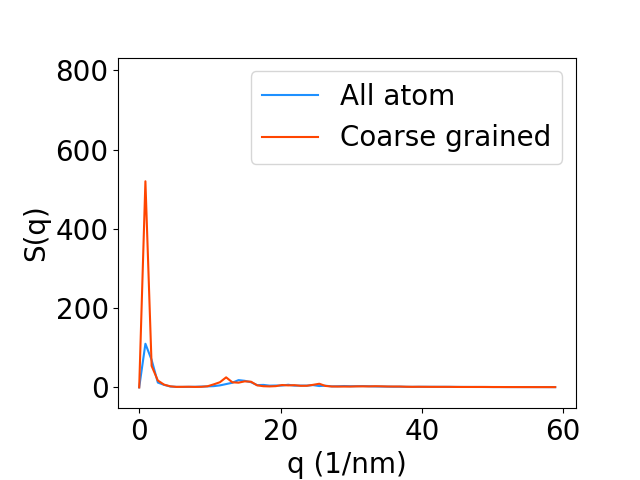}
            \\
            CL2
        \end{minipage}
        \begin{minipage}[b]{0.32\textwidth}
            \centering
            \includegraphics[width=0.96\textwidth]{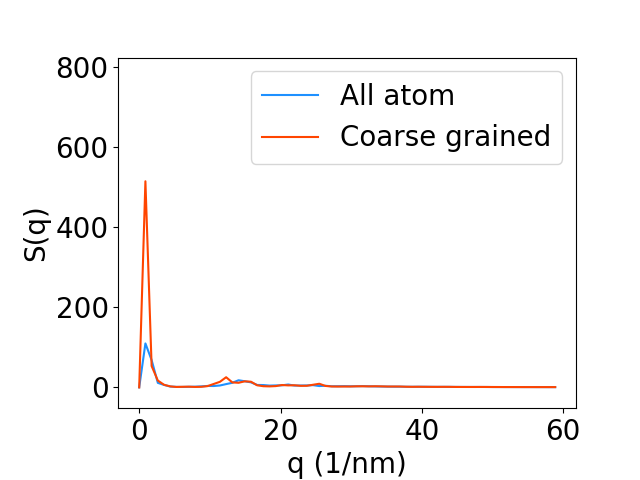}
            \\
            CL3
        \end{minipage}
        \subcaption{}
    \end{subfigure}
    \\
    \begin{subfigure}[b]{0.96\textwidth}
        \begin{minipage}[b]{0.32\textwidth}
            \centering
            \includegraphics[width=0.96\textwidth]{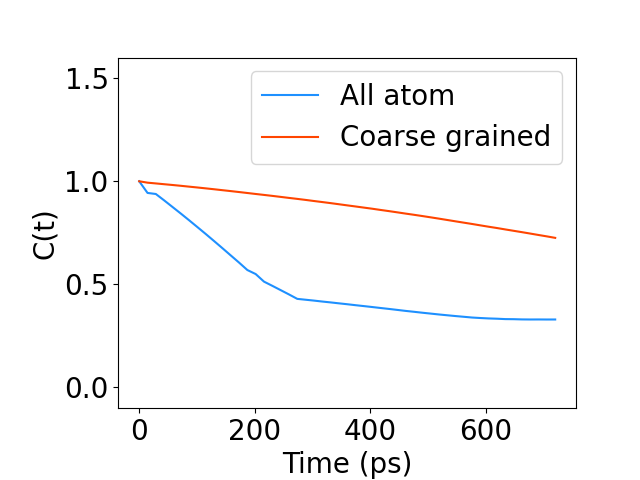}
            \\
            CL1
        \end{minipage}
        \begin{minipage}[b]{0.32\textwidth}
            \centering
            \includegraphics[width=0.96\textwidth]{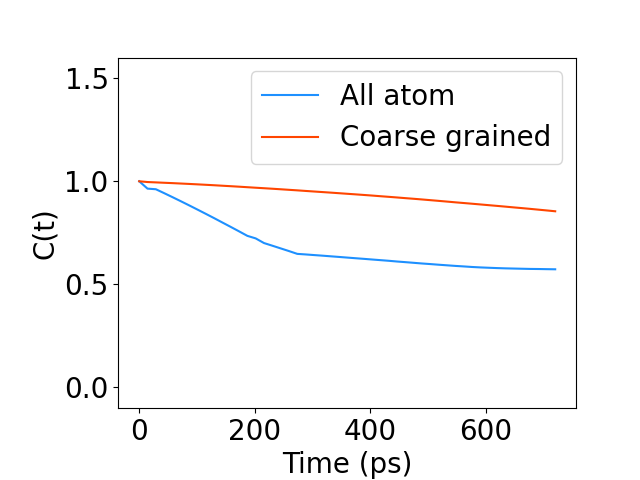}
            \\
            CL2
        \end{minipage}
        \begin{minipage}[b]{0.32\textwidth}
            \centering
            \includegraphics[width=0.96\textwidth]{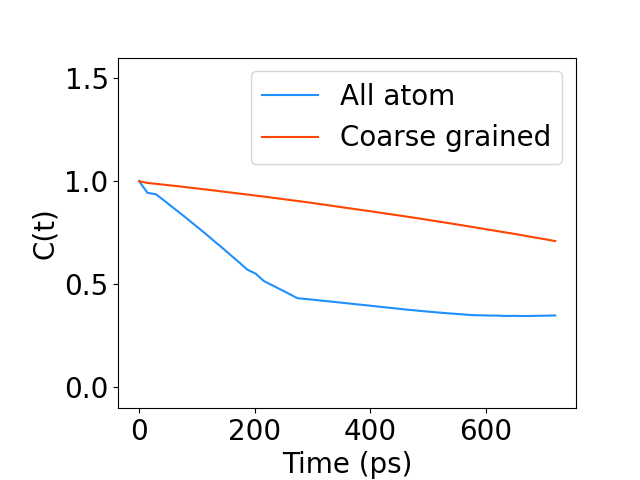}
            \\
            CL3
        \end{minipage}
        \subcaption{}
    \end{subfigure}
    \\
    \begin{subfigure}[b]{0.96\textwidth}
        \begin{minipage}[b]{0.32\textwidth}
            \centering
            \includegraphics[width=0.96\textwidth]{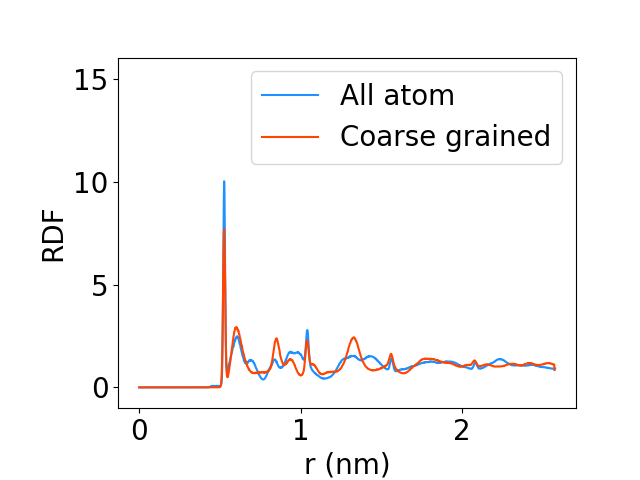}
            \\
            CL1:CL1
        \end{minipage}
        \begin{minipage}[b]{0.32\textwidth}
            \centering
            \includegraphics[width=0.96\textwidth]{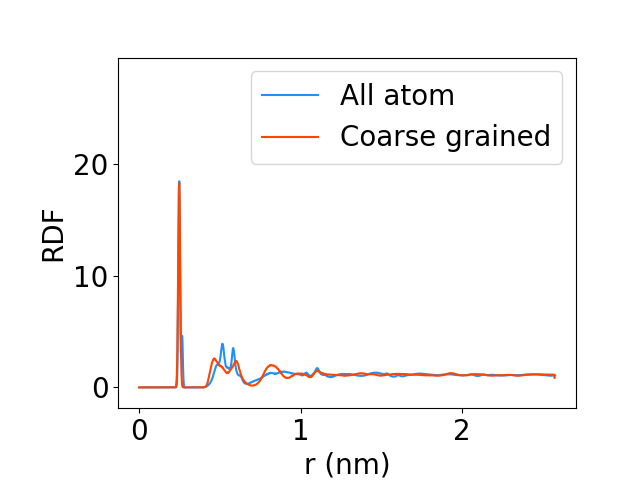}
            \\
            CL1:CL2
        \end{minipage}
        \begin{minipage}[b]{0.32\textwidth}
            \centering
            \includegraphics[width=0.96\textwidth]{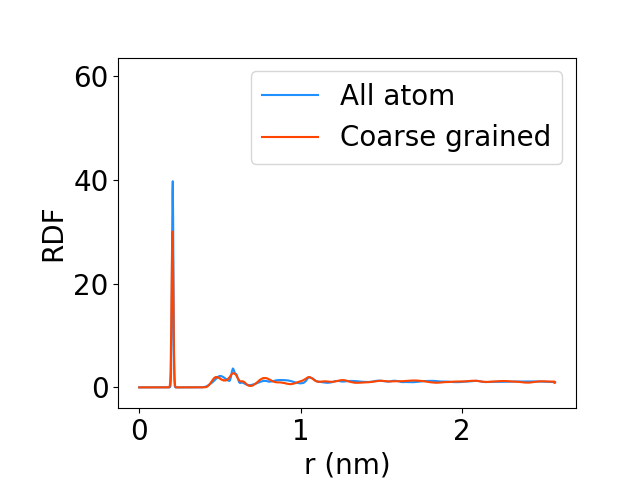}
            \\
            CL1:CL3
        \end{minipage}
        \subcaption{}
    \end{subfigure}
    \\
    \begin{subfigure}[b]{0.96\textwidth}
        \begin{minipage}[b]{0.32\textwidth}
            \centering
            \includegraphics[width=0.96\textwidth]{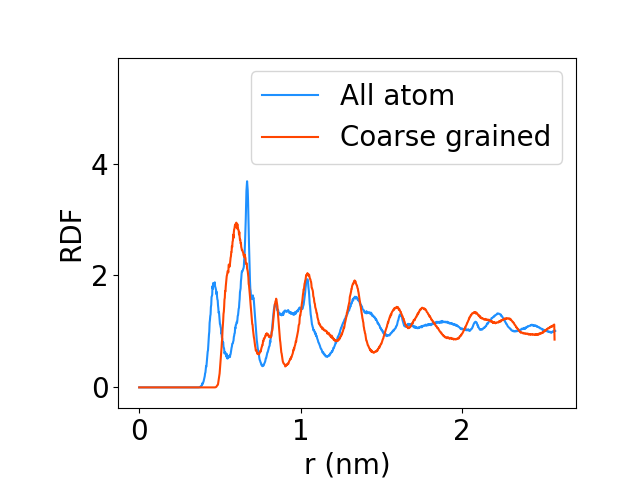}
            \\
            CL2:CL2
        \end{minipage}
        \begin{minipage}[b]{0.32\textwidth}
            \centering
            \includegraphics[width=0.96\textwidth]{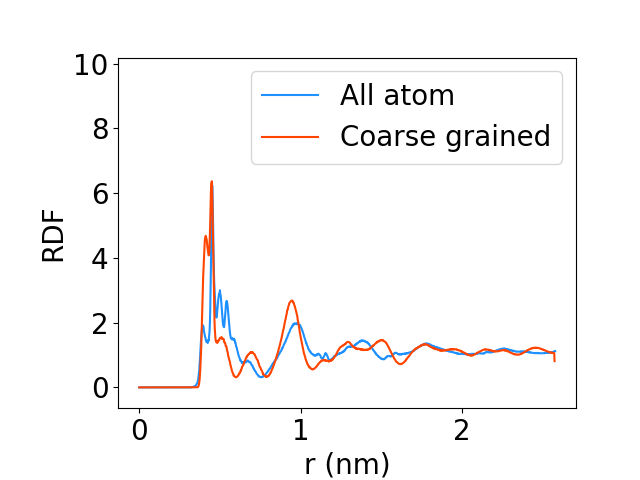}
            \\
            CL2:CL3
        \end{minipage}
        \begin{minipage}[b]{0.32\textwidth}
            \centering
            \includegraphics[width=0.96\textwidth]{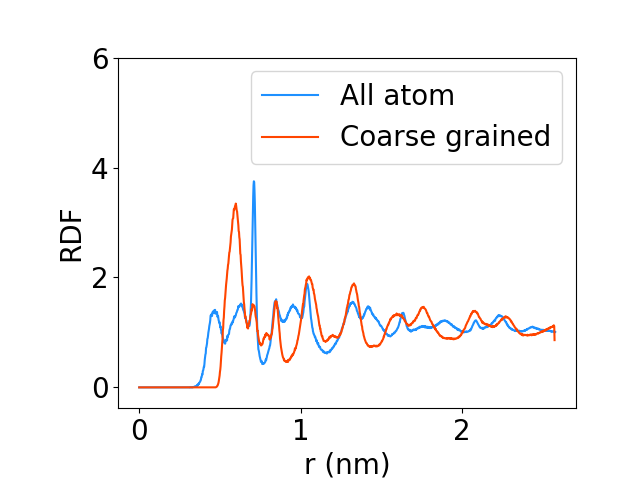}
            \\
            CL3:CL3
        \end{minipage}
        \subcaption{}
    \end{subfigure}
    \captionsetup{font=scriptsize}
    \caption{
    Structural properties of mapped AA and CG from stretch processes of CNCs.
    (a) Static structure factor of beads.
    (b) Angular correlation function of beads.
    Owing to the inherent reduction in the degrees of freedom, the structure of the CG model was obviously more orderly and the C(t) during stretch processes was higher.
    They provided additional information to demonstrate the structural similarities between CG and AA models.
    }
    \label{fig:structural_properties_stretch}
\end{figure}

Although the comparisons of these data from non-equilibrium conditions are not rigorous (and the normalized scattering intensity I(q) computation failed technically), the structural properties from the stretch processes are also presented for reference.
The static structure factor (S(q)), angular correlation function (C(t)), and radical distribution function (RDF) of all pairs of CG trajectories during the stretch processes were compared with those of the mapped AA trajectories, as shown in Figure~\ref{fig:structural_properties_stretch}.
These results help demonstrate the effectiveness of the CG model.

\clearpage

\subsection*{Baseline method structural properties for coarse grained models}\indent
\addcontentsline{toc}{subsection}{\protect\numberline{}Baseline method structural properties for coarse grained models}\indent

Except for the RL-parameterized CG potential, the baseline models were also tested in parallel.
The simulations we used for the AA and CG baselines were the slant model equilibrium without any external loads under the NPT ensemble, as we used for the RL-parameterized model.
The reference AA trajectory was simulated for 20~ns, using a timestep of 2~fs.
Bottom-up baselines (including IBI, RE, and FM) and Top-down baseline MARTINI 3 were compared to illustrate the necessity of this model.

\begin{figure}[htbp]
    \centering    
    \begin{subfigure}[b]{0.72\textwidth}
        \centering
        \scriptsize
        \begin{minipage}[b]{0.32\textwidth}
            \centering
            \includegraphics[width=0.60\textwidth]{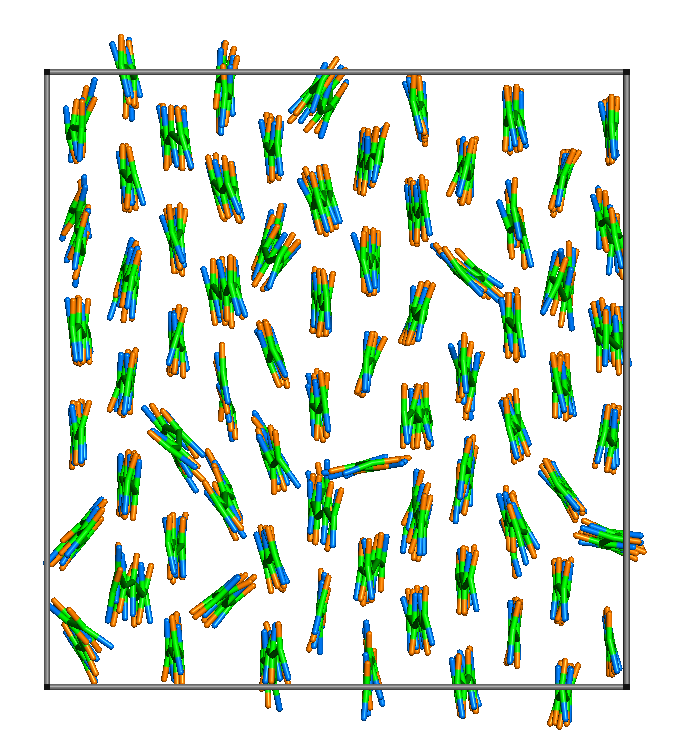}
            \\
            Vertical
        \end{minipage}
        \begin{minipage}[b]{0.32\textwidth}
            \centering
            \includegraphics[width=0.72\textwidth]{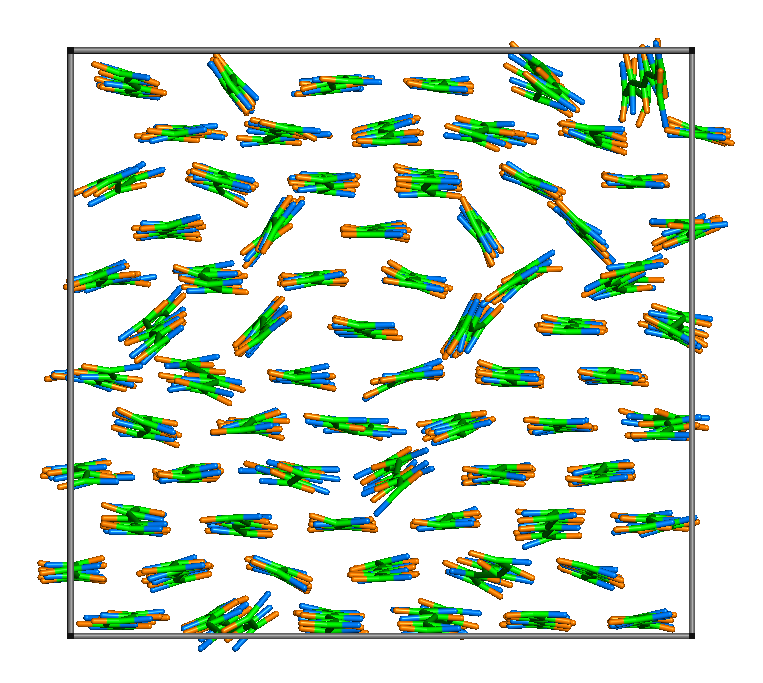}
            \\
            Horizontal
        \end{minipage}
        \begin{minipage}[b]{0.32\textwidth}
            \centering
            \includegraphics[width=0.96\textwidth]{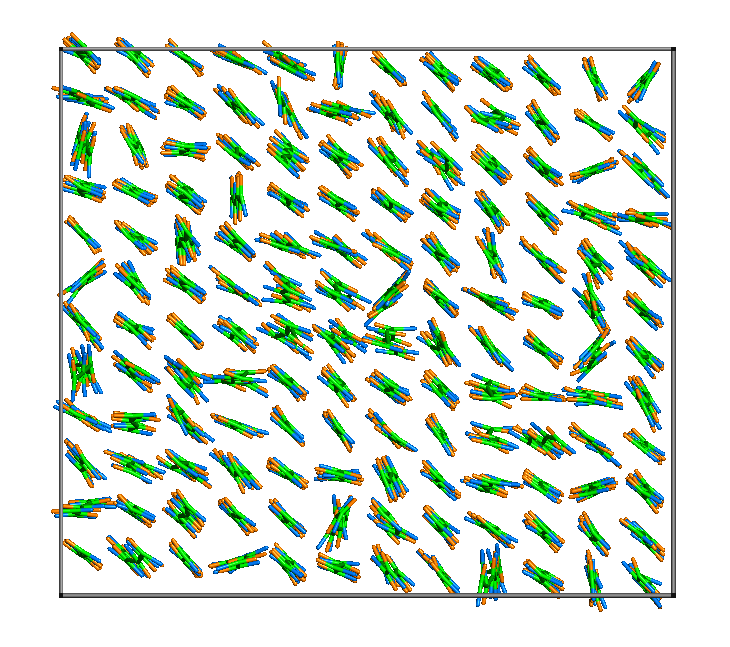}
            \\
            Slant
        \end{minipage}
        \subcaption{IBI}
    \end{subfigure}
    \begin{subfigure}[b]{0.72\textwidth}
        \centering
        \scriptsize
        \begin{minipage}[b]{0.32\textwidth}
            \centering
            \includegraphics[width=0.60\textwidth]{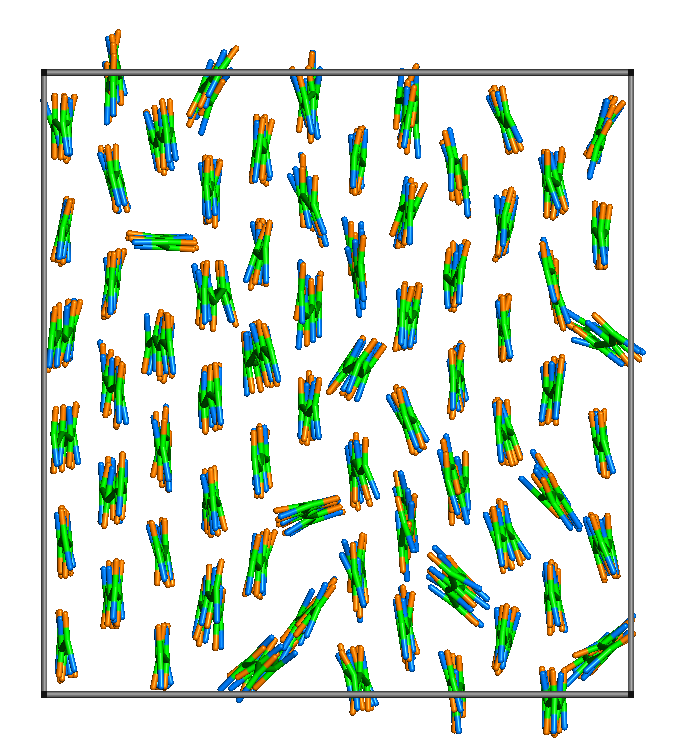}
            \\
            Vertical
        \end{minipage}
        \begin{minipage}[b]{0.32\textwidth}
            \centering
            \includegraphics[width=0.72\textwidth]{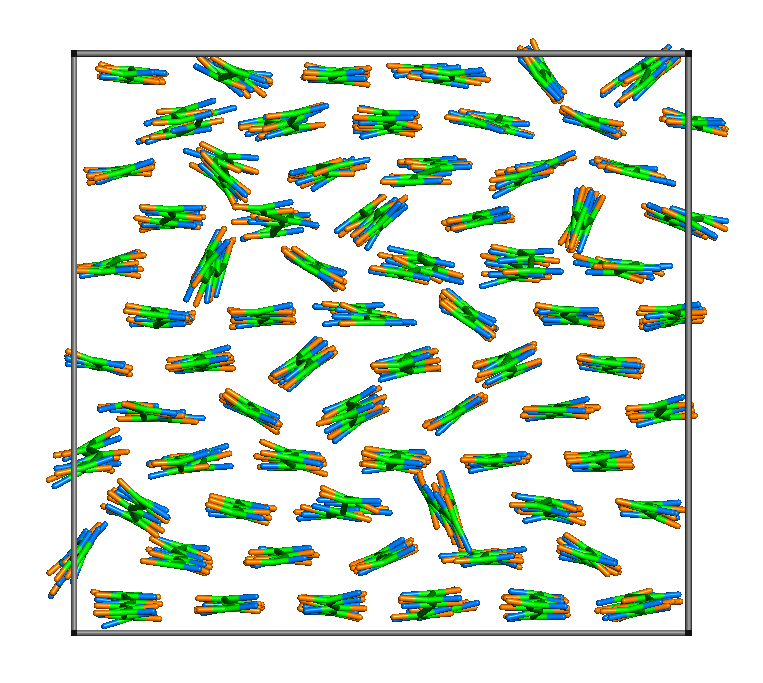}
            \\
            Horizontal
        \end{minipage}
        \begin{minipage}[b]{0.32\textwidth}
            \centering
            \includegraphics[width=0.96\textwidth]{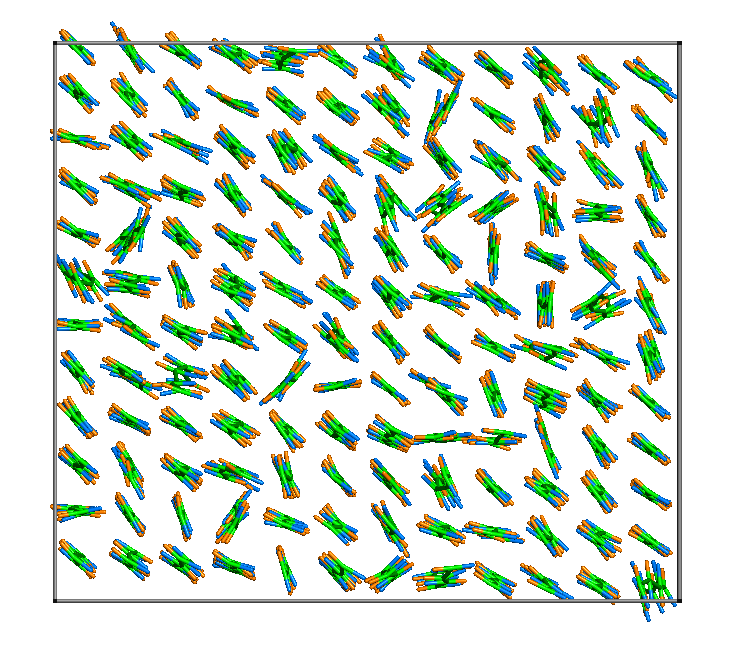}
            \\
            Slant
        \end{minipage}
        \subcaption{RE}
    \end{subfigure}
    \begin{subfigure}[b]{0.72\textwidth}
        \centering
        \scriptsize
        \begin{minipage}[b]{0.32\textwidth}
            \centering
            \includegraphics[width=0.60\textwidth]{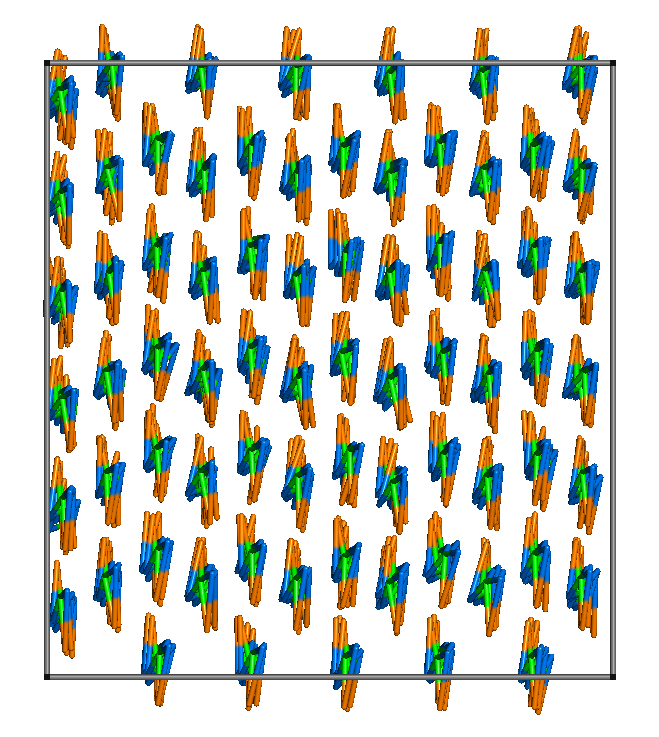}
            \\
            Vertical
        \end{minipage}
        \begin{minipage}[b]{0.32\textwidth}
            \centering
            \includegraphics[width=0.72\textwidth]{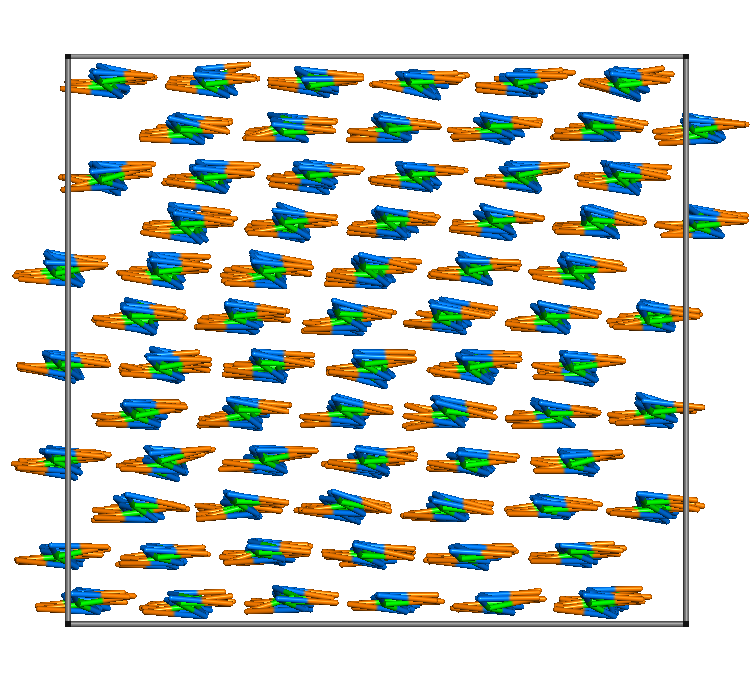}
            \\
            Horizontal
        \end{minipage}
        \begin{minipage}[b]{0.32\textwidth}
            \centering
            \includegraphics[width=0.96\textwidth]{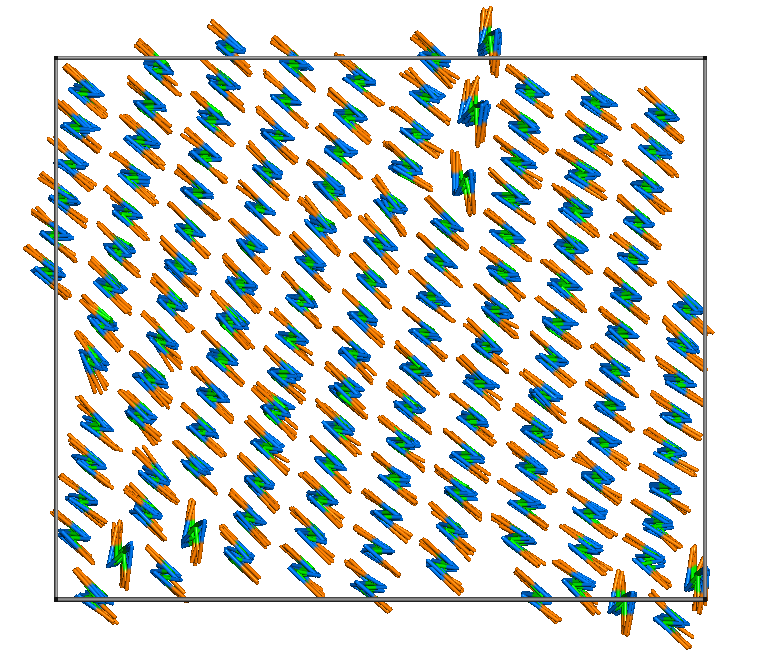}
            \\
            Slant
        \end{minipage}
        \subcaption{MARTINI}
    \end{subfigure}
    \captionsetup{font=scriptsize}
    \caption{
    Baselines relaxation structures of CG models by (a) IBI, (b) RE, and (c) MARTINI models in characteristic directions.
    The IBI and RE models were simulated under the NVT ensemble without velocity generalization, and only the MARTINI models were controlled by the NPT ensemble.
    However, the mode of IBI and RE cannot preserve laminar structures, as well as the MARTINI model.
    }
    \label{fig:baselines_structure_relaxation}
\end{figure}

\begin{figure}[htbp]
    \centering    
    \begin{subfigure}[b]{0.72\textwidth}
        \centering
        \scriptsize
        \begin{minipage}[b]{0.32\textwidth}
            \centering
            \includegraphics[width=0.72\textwidth]{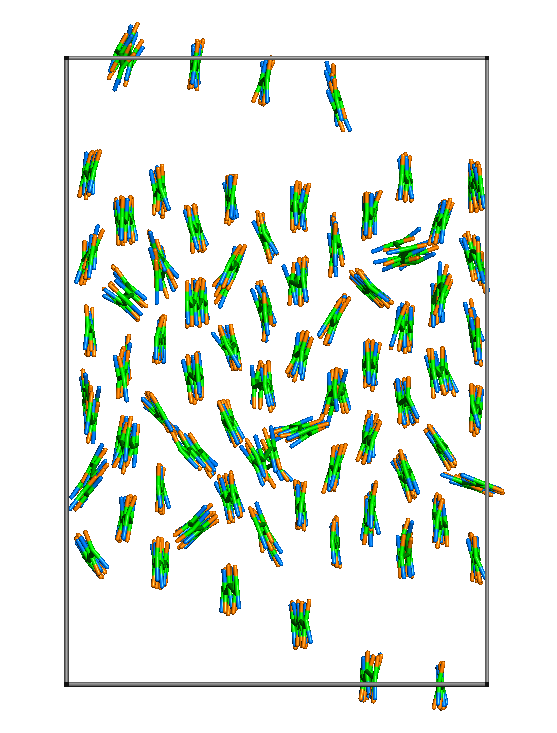}
            \\
            Vertical
        \end{minipage}
        \begin{minipage}[b]{0.32\textwidth}
            \centering
            \includegraphics[width=0.72\textwidth]{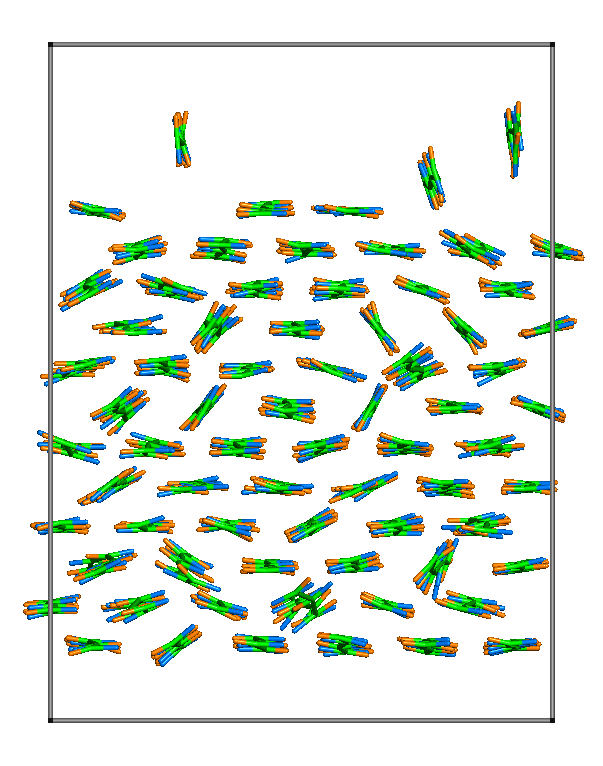}
            \\
            Horizontal
        \end{minipage}
        \begin{minipage}[b]{0.32\textwidth}
            \centering
            \includegraphics[width=0.96\textwidth]{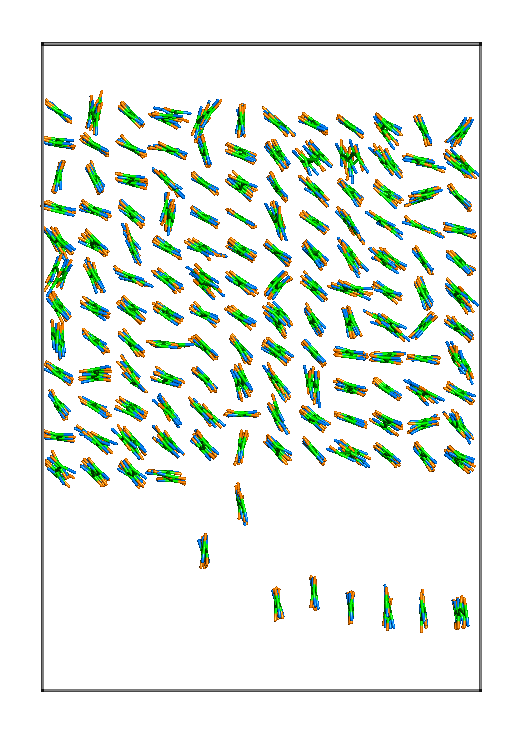}
            \\
            Slant
        \end{minipage}
        \subcaption{IBI}
    \end{subfigure}
    \begin{subfigure}[b]{0.72\textwidth}
        \centering
        \scriptsize
        \begin{minipage}[b]{0.32\textwidth}
            \centering
            \includegraphics[width=0.72\textwidth]{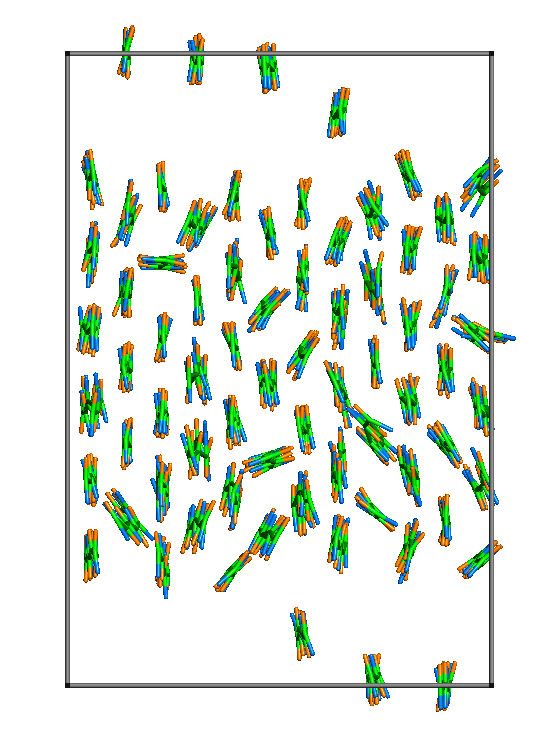}
            \\
            Vertical
        \end{minipage}
        \begin{minipage}[b]{0.32\textwidth}
            \centering
            \includegraphics[width=0.72\textwidth]{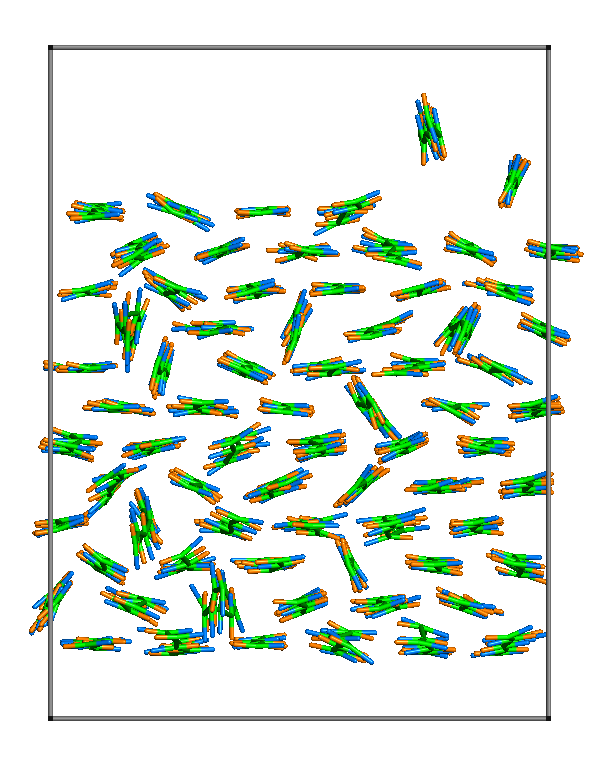}
            \\
            Horizontal
        \end{minipage}
        \begin{minipage}[b]{0.32\textwidth}
            \centering
            \includegraphics[width=0.96\textwidth]{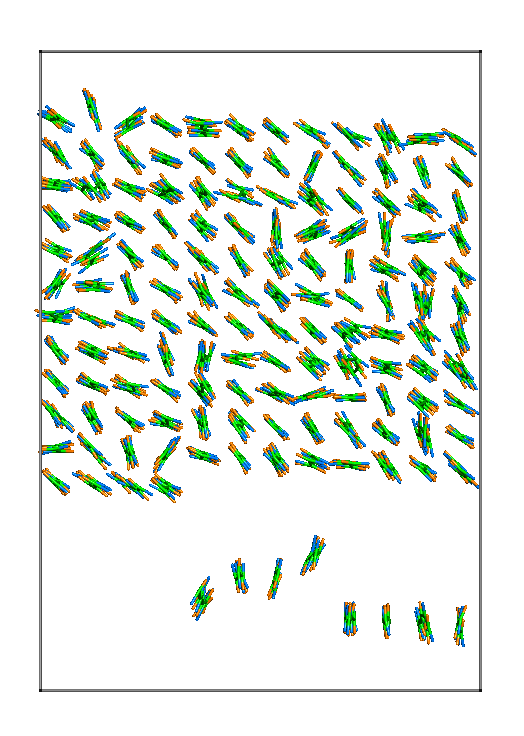}
            \\
            Slant
        \end{minipage}
        \subcaption{RE}
    \end{subfigure}
    \begin{subfigure}[b]{0.72\textwidth}
        \centering
        \scriptsize
        \begin{minipage}[b]{0.32\textwidth}
            \centering
            \includegraphics[width=0.72\textwidth]{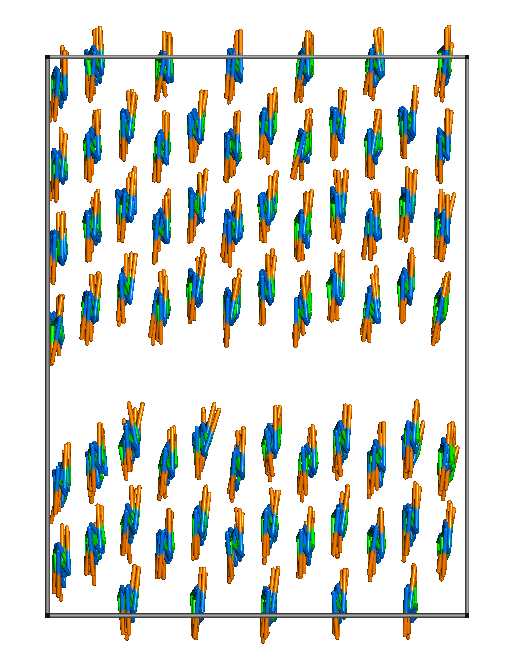}
            \\
            Vertical
        \end{minipage}
        \begin{minipage}[b]{0.32\textwidth}
            \centering
            \includegraphics[width=0.48\textwidth]{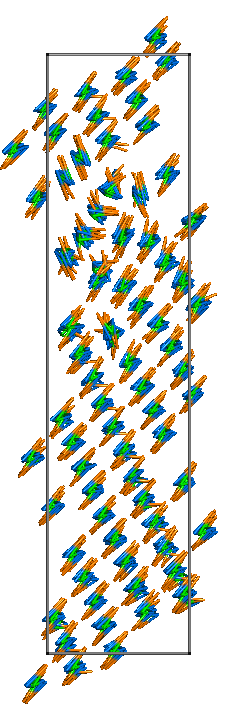}
            \\
            Horizontal
        \end{minipage}
        \begin{minipage}[b]{0.32\textwidth}
            \centering
            \includegraphics[width=0.48\textwidth]{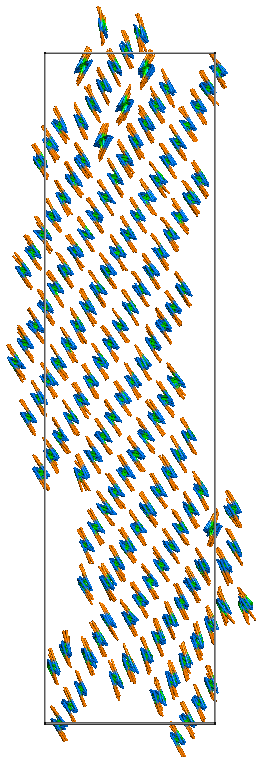}
            \\
            Slant
        \end{minipage}
        \subcaption{MARTINI}
    \end{subfigure}
    \captionsetup{font=scriptsize}
    \caption{
    Baselines fracture behaviors of CG models by (a) IBI, (b) RE, and (c) MARTINI models in characteristic directions.
    The IBI and RE models were simulated under the NVT ensemble without velocity generalization, and only the MARTINI models were controlled by the NPT ensemble.
    However, the IBI and RE models cannot reproduce the friction sliding behavior in the slant directions.
    The stretch result in the horizontal direction also emphasizes the weak NB interactions of the MARTINI 3 cellulose potential.
    For composing, the snapshots in the horizontal and slant directions were not even fractures because their models were stretched to form a long, thin strip and broke.
    }
    \label{fig:baselines_structure_fracture}
\end{figure}

\begin{figure}[htbp]
    \centering
    \scriptsize
    \begin{minipage}[b]{0.32\textwidth}
        \centering
        \includegraphics[width=0.96\textwidth]{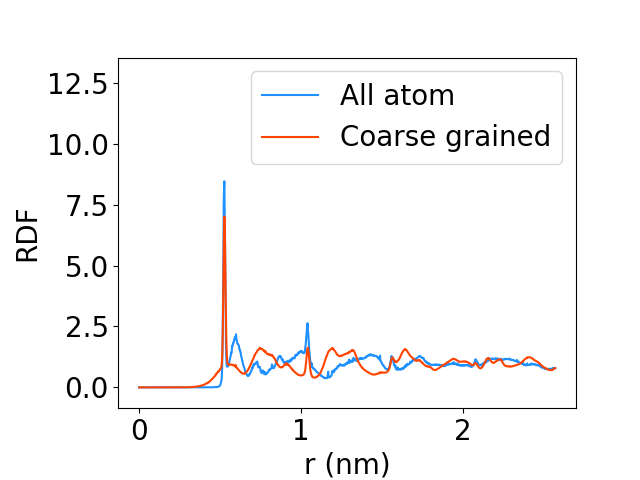}
        \\
        CL1:CL1
    \end{minipage}
    \begin{minipage}[b]{0.32\textwidth}
        \centering
        \includegraphics[width=0.96\textwidth]{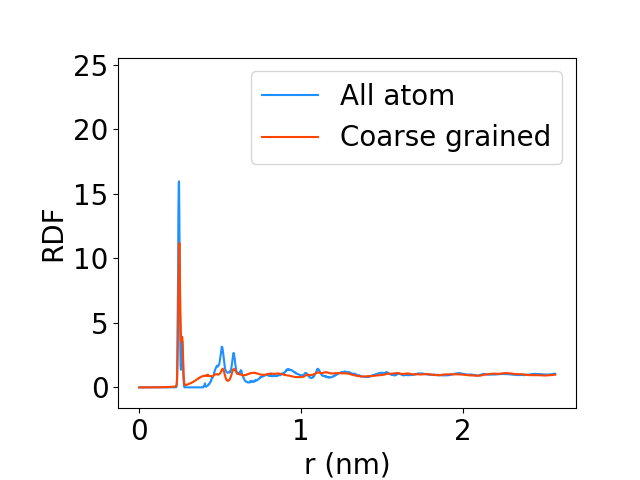}
        \\
        CL1:CL2
    \end{minipage}
    \begin{minipage}[b]{0.32\textwidth}
        \centering
        \includegraphics[width=0.96\textwidth]{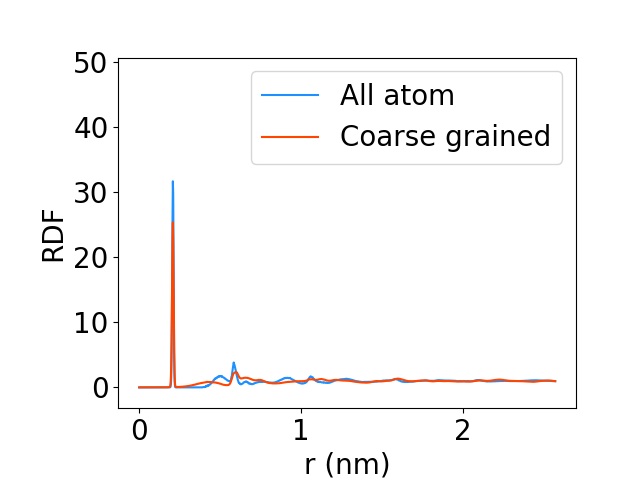}
        \\
        CL1:CL3
    \end{minipage}
    \\
    \begin{minipage}[b]{0.32\textwidth}
        \centering
        \includegraphics[width=0.96\textwidth]{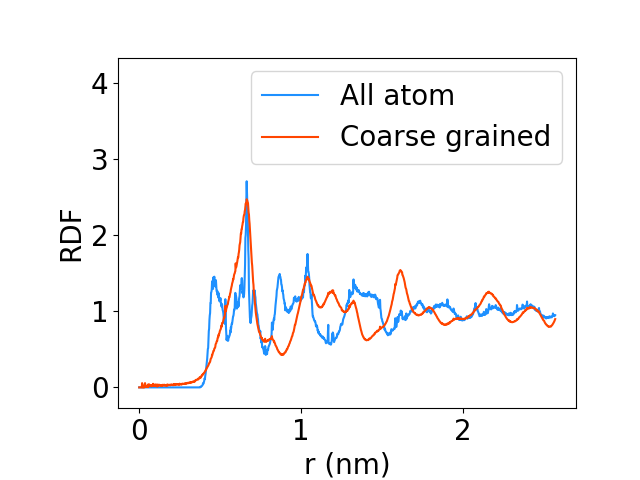}
        \\
        CL2:CL2
    \end{minipage}
    \begin{minipage}[b]{0.32\textwidth}
        \centering
        \includegraphics[width=0.96\textwidth]{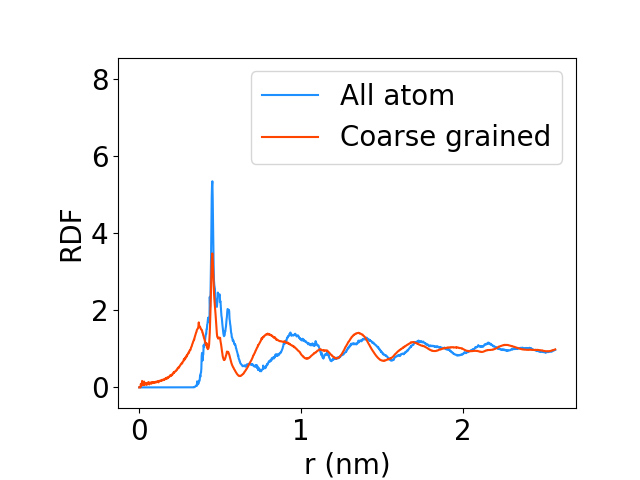}
        \\
        CL2:CL3
    \end{minipage}
    \begin{minipage}[b]{0.32\textwidth}
        \centering
        \includegraphics[width=0.96\textwidth]{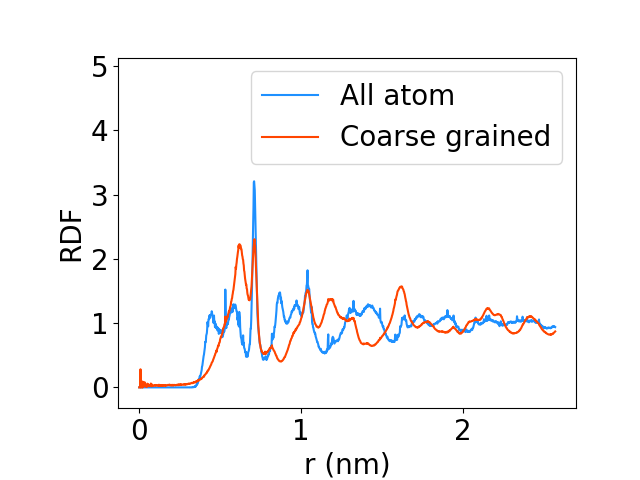}
        \\
        CL3:CL3
    \end{minipage}
    \captionsetup{font=scriptsize}
    \caption{
    RDF of both AA and baseline IBI CG from the equilibrium simulations of CNCs.
    Although the performances and structures are not well reproduced, the RDF data are relatively accurate.
    These data may illustrate the insufficient representability of RDF and conventional IBI methods.
    }
    \label{fig:baselines_rdf_ibi}
\end{figure}

\begin{figure}[htbp]
    \centering
    \scriptsize
    \begin{subfigure}[b]{0.96\textwidth}
        \begin{minipage}[b]{0.32\textwidth}
            \centering
            \includegraphics[width=0.96\textwidth]{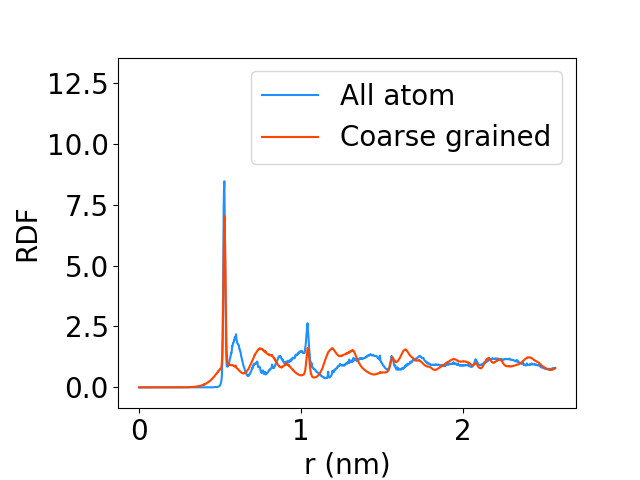}
            \\
            CL1:CL1
        \end{minipage}
        \begin{minipage}[b]{0.32\textwidth}
            \centering
            \includegraphics[width=0.96\textwidth]{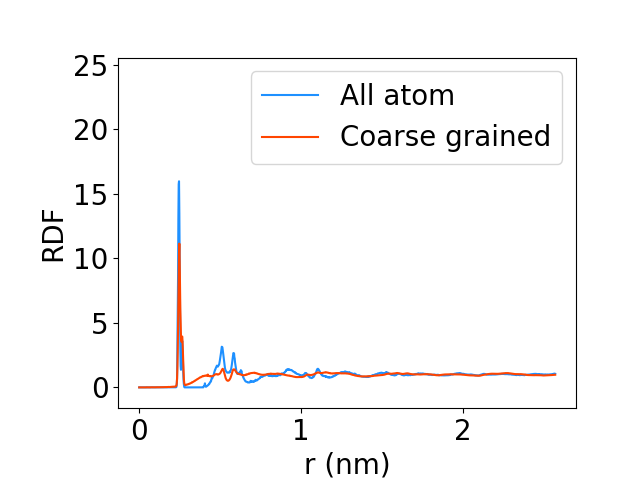}
            \\
            CL1:CL2
        \end{minipage}
        \begin{minipage}[b]{0.32\textwidth}
            \centering
            \includegraphics[width=0.96\textwidth]{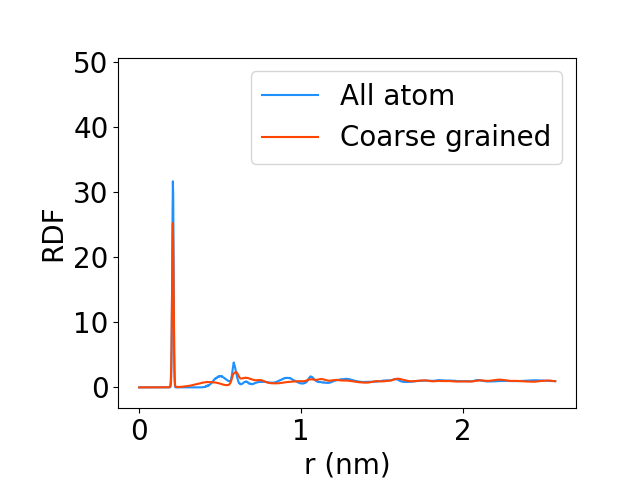}
            \\
            CL1:CL3
        \end{minipage}
        \subcaption{}
    \end{subfigure}
    \\
    \begin{subfigure}[b]{0.96\textwidth}
        \begin{minipage}[b]{0.32\textwidth}
            \centering
            \includegraphics[width=0.96\textwidth]{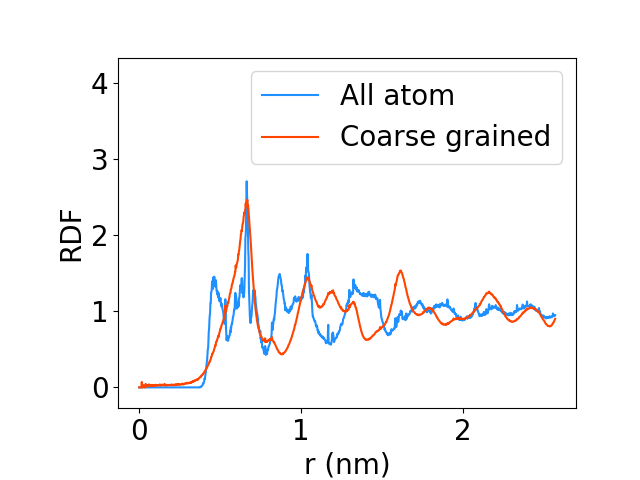}
            \\
            CL2:CL2
        \end{minipage}
        \begin{minipage}[b]{0.32\textwidth}
            \centering
            \includegraphics[width=0.96\textwidth]{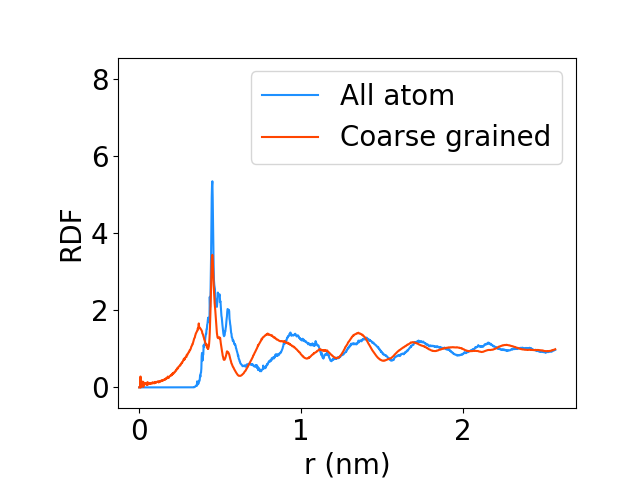}
            \\
            CL2:CL3
        \end{minipage}
        \begin{minipage}[b]{0.32\textwidth}
            \centering
            \includegraphics[width=0.96\textwidth]{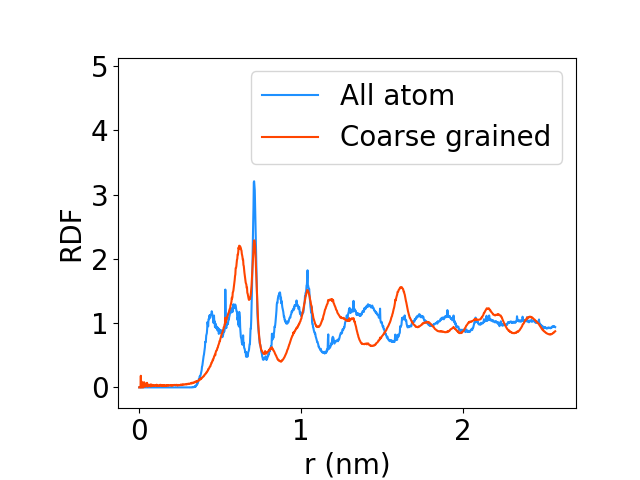}
            \\
            CL3:CL3
        \end{minipage}
        \subcaption{}
    \end{subfigure}
    \captionsetup{font=scriptsize}
    \caption{
    RDF of both AA and baseline RE CG from the equilibrium simulations of CNCs.
    Although the performances and structures are not well reproduced, the RDF data are relatively accurate.
    These data may illustrate the insufficient representability of RDF and conventional RE methods.
    }
    \label{fig:baselines_rdf_re}
\end{figure}

In addition to the numerical properties in the tables, the structures after relaxation and fracture are shown in Figure~\ref{fig:baselines_structure_relaxation} and Figure~\ref{fig:baselines_structure_fracture}, respectively.
Although the data performance and potential well did not converge to the reference, the RDF of both IBI and RE were relatively better reproduced, as shown in Figure~\ref{fig:baselines_rdf_ibi} and Figure~\ref{fig:baselines_rdf_re}.
Owing to its weak NB interactions and high fracture strain and toughness, the fracture structures shown for MARTINI in the horizontal and slant characteristic directions were not fractured because they were stripped like ribbons before fracture.
This may emphasize the fact that the mapping and properties considered in this work concentrates much on the molecular details, and may be difficult for traditional metrics to measure and reproduce.
The results of these conventional baselines confirm the validation and superiority of this RL-parameterized model following an extended bottom-up approach, in which the well-designed mapping and interactions play pivotal roles.

\clearpage

\subsection*{Training using reward function without thresholds}\indent
\addcontentsline{toc}{subsection}{\protect\numberline{}Training using reward function without thresholds}\indent

As mentioned in the main paper, the reward function was defined as the sum of the match degrees.
However, thresholds were introduced to increase discrimination, and when considering the match degree of strength and toughness, minimal values were considered.
We performed a parallel training using a reward function without thresholds and a minimal cut to better elucidate the importance and robustness of thresholds:
\begin{equation}M=\frac{1}{1+|\frac{y_1-y_0}{y_0}|}\end{equation}
, where $y_0$ and $y_1$ are the reference and sampled values, respectively.
The reward function is the sum of a series of matching degrees, except for the transverse strength $M_s$ and toughness $M_t$,
\begin{equation}R=\sum_{}M_i+w\cdot\frac{M_s+M_t}{2}\end{equation}
.

\begin{figure}[htbp]
    \centering
    \begin{subfigure}[b]{0.32\textwidth}
        \centering
        \includegraphics[width=0.96\textwidth]{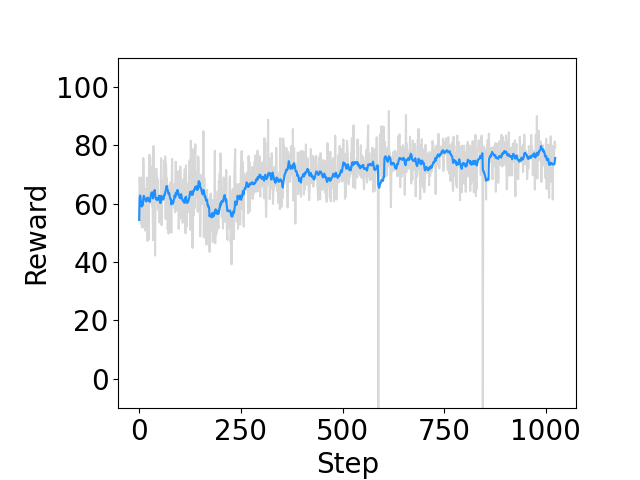}
        \subcaption{}
    \end{subfigure}
    \begin{subfigure}[b]{0.32\textwidth}
        \centering
        \includegraphics[width=0.96\textwidth]{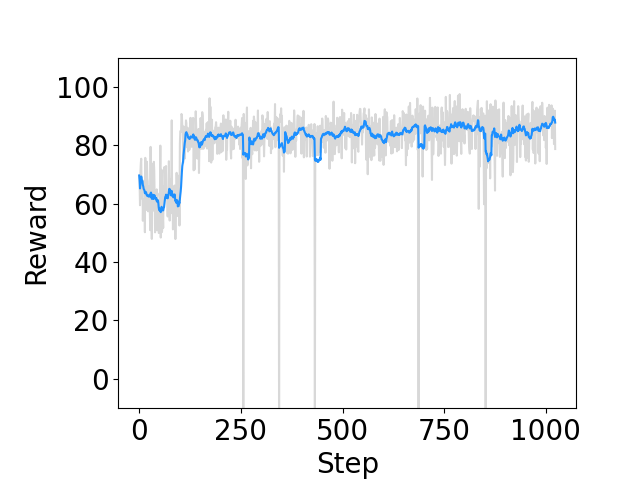}
        \subcaption{}
    \end{subfigure}
    \begin{subfigure}[b]{0.32\textwidth}
        \centering
        \includegraphics[width=0.96\textwidth]{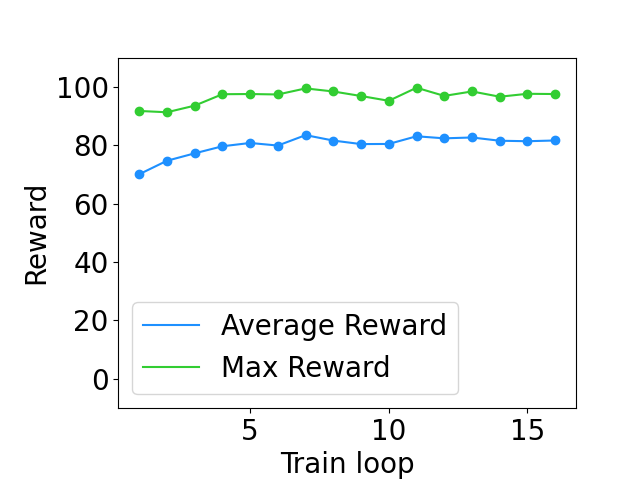}
        \subcaption{}
    \end{subfigure}
    \\
    \begin{subfigure}[b]{0.32\textwidth}
        \centering
        \includegraphics[width=0.96\textwidth]{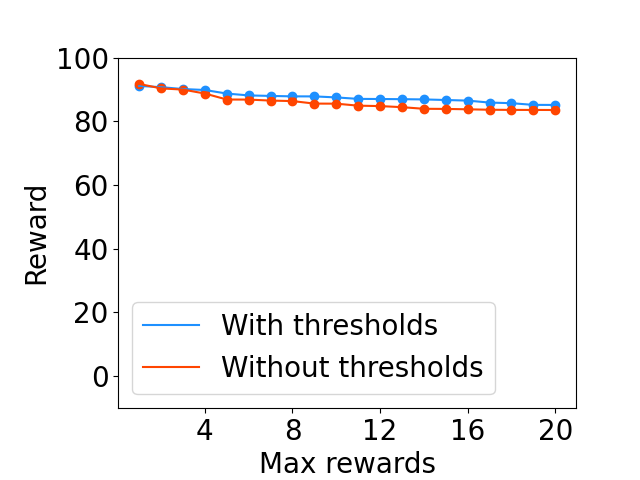}
        \subcaption{}
    \end{subfigure}
    \begin{subfigure}[b]{0.32\textwidth}
        \centering
        \includegraphics[width=0.96\textwidth]{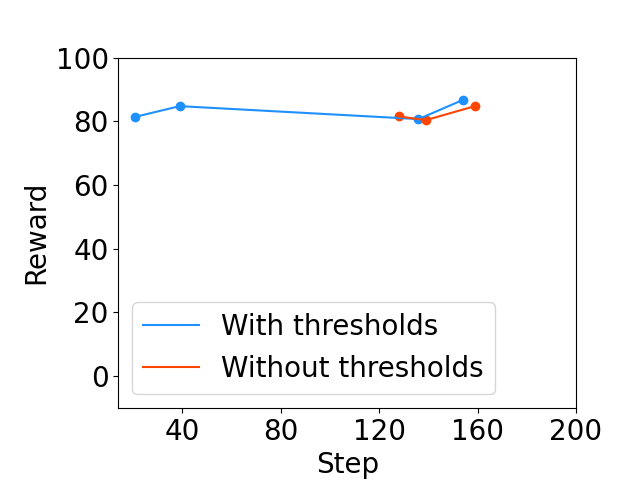}
        \subcaption{}
    \end{subfigure}
    \captionsetup{font=scriptsize}
    \caption{
    Training convergence and statistics without thresholds.
    (a) Reward and its moving average in the first and (b) last training loops.
    Each training loop contained 1024 training steps.
    The subsequent training loops with higher and more stable rewards, were based on the former training loops.
    (c) Average and maximum rewards for different training loops.
    The average and maximum rewards confirmed the convergence.
    (d) Maximum 20 rewards in the first training loop with and without thresholds.
    (e) Rewards higher than 80.0 with steps, from the beginning 200 steps of first training loops with and without thresholds.
    These data were obtained from parallel training which only changed the reward functions to be thresholds-less.
    The maximum rewards of the training loops without thresholds were only slightly higher owing to the removal of min for $M_s$ and $M_t$, which confirms the robustness of the thresholds.
    In addition, the maximum 20 rewards and the high rewards from the starting 200 steps further proved the advantages of thresholds: hitting coefficients with high rewards faster during the early training.
    This was not critical but really helpful during the early development stage.
    }
    \label{fig:convergence_and_statistics_without_threshold}
\end{figure}

\begin{figure}[htbp]
    \centering    
    \begin{subfigure}[b]{0.96\textwidth}
        \centering
        \scriptsize
        \begin{minipage}[b]{0.32\textwidth}
            \centering
            \includegraphics[width=0.96\textwidth]{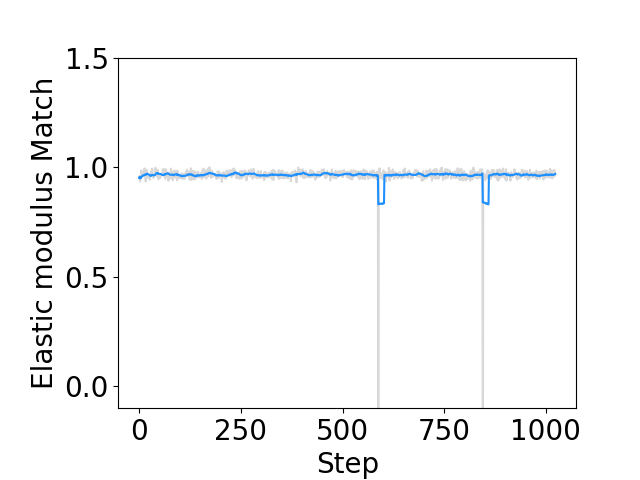}
            \includegraphics[width=0.96\textwidth]{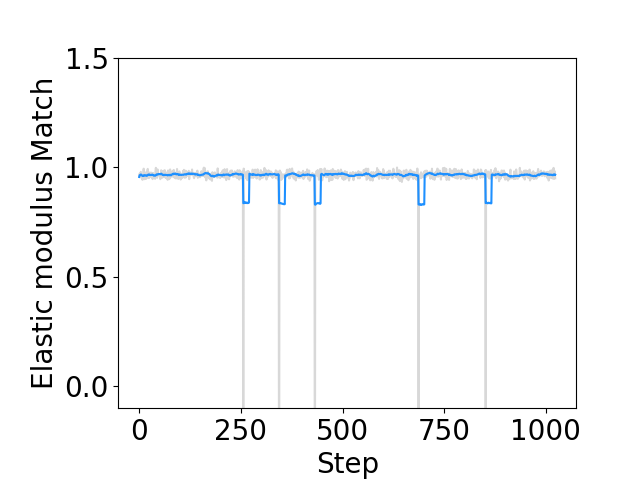}
            \\
            Axial modulus
        \end{minipage}
        \begin{minipage}[b]{0.32\textwidth}
            \centering
            \includegraphics[width=0.96\textwidth]{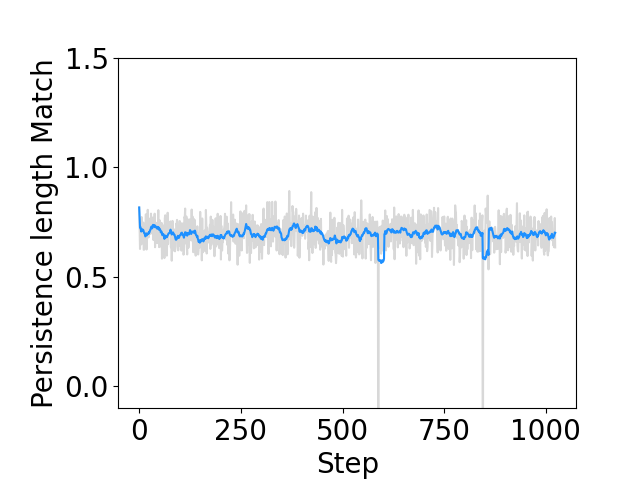}
            \includegraphics[width=0.96\textwidth]{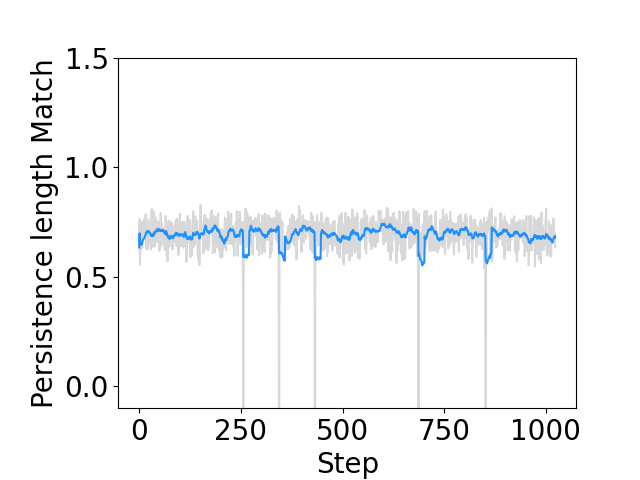}
            \\
            Persistence length
        \end{minipage}
        \begin{minipage}[b]{0.32\textwidth}
            \centering
            \includegraphics[width=0.96\textwidth]{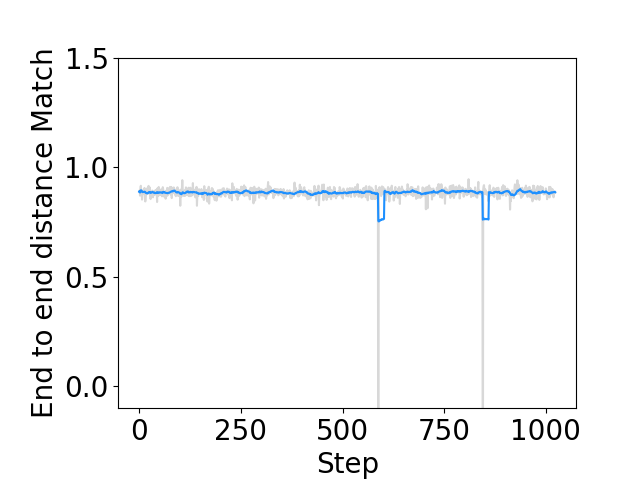}
            \includegraphics[width=0.96\textwidth]{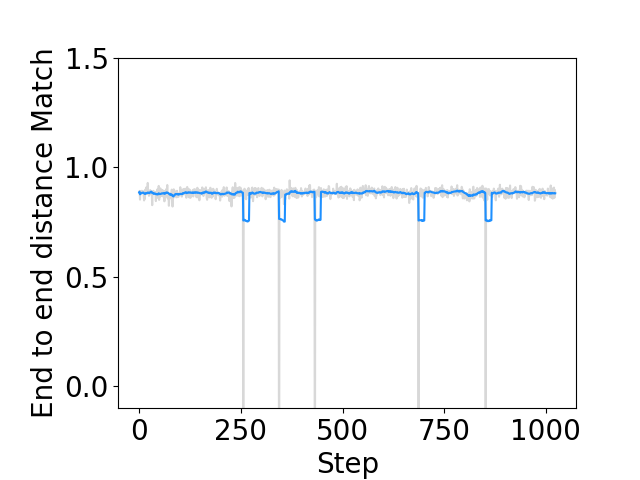}
            \\
            End to end distance
        \end{minipage}
        \\
        \begin{minipage}[b]{0.32\textwidth}
            \centering
            \includegraphics[width=0.96\textwidth]{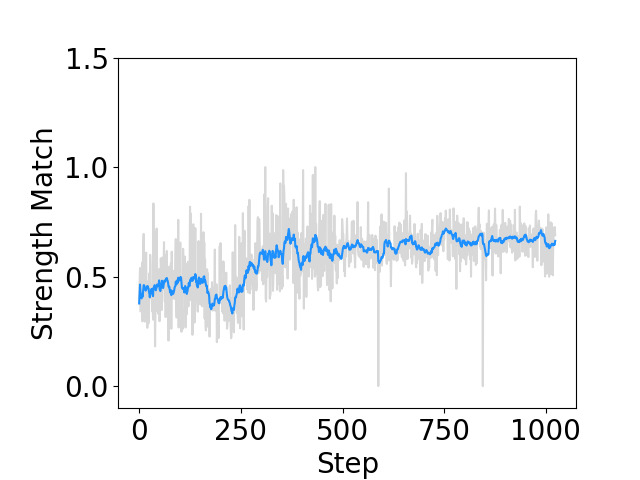}
            \includegraphics[width=0.96\textwidth]{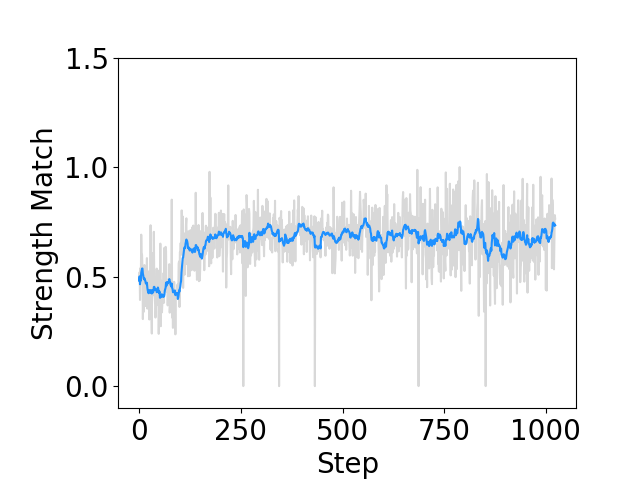}
            \\
            Transverse strength
        \end{minipage}
        \begin{minipage}[b]{0.32\textwidth}
            \centering
            \includegraphics[width=0.96\textwidth]{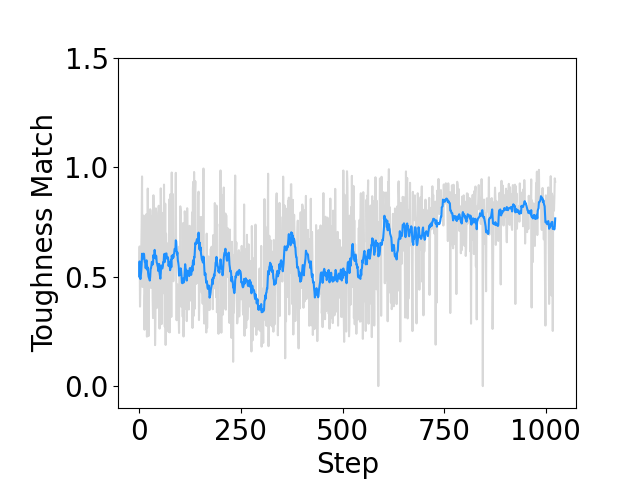}
            \includegraphics[width=0.96\textwidth]{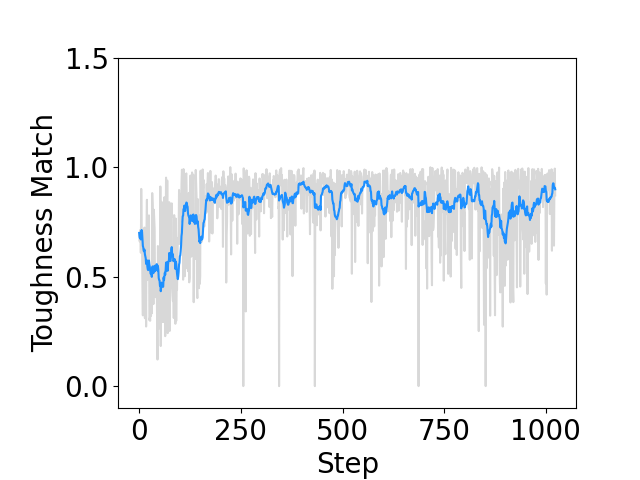}
            \\
            Transverse toughness
        \end{minipage}
    \end{subfigure}
    \captionsetup{font=scriptsize}
    \caption{
    Component match degrees of reward function in the first (upper) and last (lower) training loop without thresholds.
    The average match degrees were  higher than those of their counterparts with thresholds.
    When focusing on the best match degrees, these data did not outperform, which emphasizes the robustness of convergence.
    }
    \label{fig:component_match_degrees_during_training_without_threshold}
\end{figure}

Anything else was the same as the training in the main paper.
The training convergence and component match degrees are shown in Figure~\ref{fig:convergence_and_statistics_without_threshold} and Figure~\ref{fig:component_match_degrees_during_training_without_threshold}.
Training without thresholds resulted in a higher average reward; however, the critical maximum reward values remained very similar.
As shown by the component match degrees, the slightly higher maximum reward is mainly a result of the reward function form: it replaced ${\rm min}(M_s, M_t)$ with $\frac{1}{2}(M_s+M_t)$.
These data first confirms the robustness of the thresholds.

The real advantage of the reward function with thresholds was emphasized by the higher chance of hitting coefficients with high rewards, particularly at the beginning of the training.
This is illustrated by the 20 maximum reward cases for the first training loop (1024 training steps) using the reward function with and without thresholds (Figure~\ref{fig:convergence_and_statistics_without_threshold}(d)).

If we further concentrate on the starting 200 steps and target at the step whose reward was higher than 80 (Figure~\ref{fig:convergence_and_statistics_without_threshold}(e)), we will notice something more important:
When with threshold, although the average reward was definitely lower, it could hit good performance coefficients faster.
This is particularly helpful in the early stages of debugging and development.

In conclusion, thresholds are not harmful for training in the long term but hit good results faster at the beginning stage.

\clearpage

\subsection*{Training using other optimization methods}\indent
\addcontentsline{toc}{subsection}{\protect\numberline{}Training using other optimization methods}\indent

As introduced in the paper, the training for RL was not continuous and was completed by the evolution of training loops: the later training loop loaded the saved state of previous training loops and still had limited explorations.
Owing to the continuous optimization process of these methods (similar to a single training loop), the first 1024 steps were compared.

\begin{figure}[htbp]
    \centering
    \begin{subfigure}[b]{0.32\textwidth}
        \includegraphics[width=0.96\textwidth]{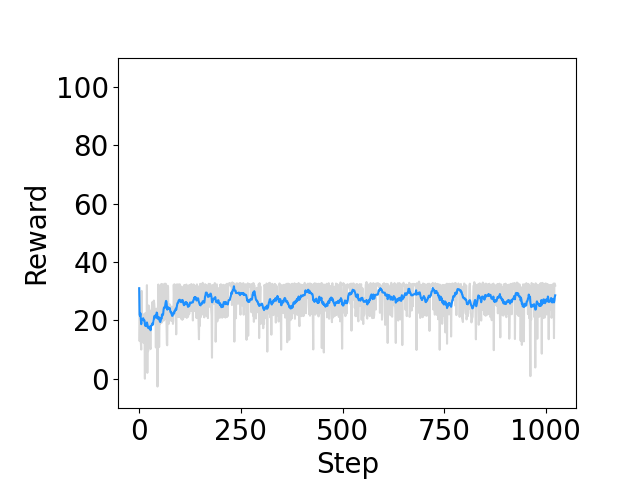}
        \subcaption{}
    \end{subfigure}
    \begin{subfigure}[b]{0.32\textwidth}
        \includegraphics[width=0.96\textwidth]{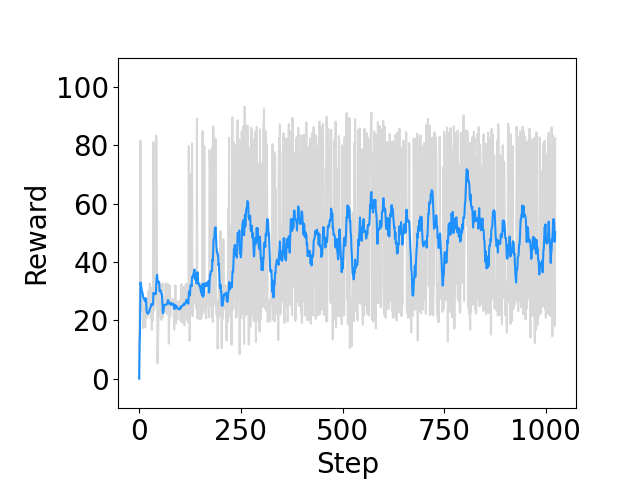}
        \subcaption{}
    \end{subfigure}
    \\
    \begin{subfigure}[b]{0.32\textwidth}
        \includegraphics[width=0.96\textwidth]{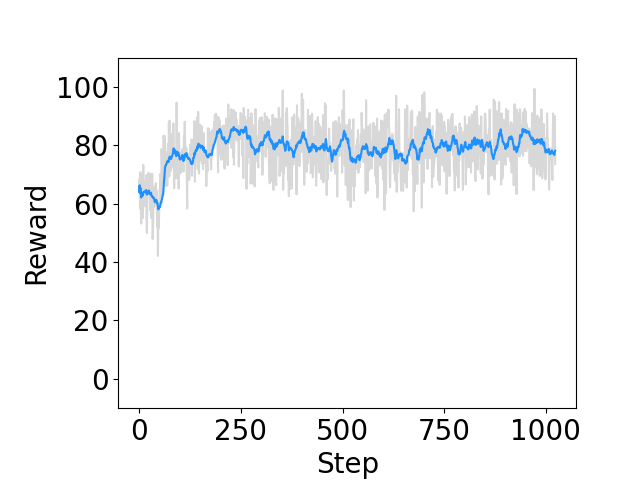}
        \subcaption{}
    \end{subfigure}
    \begin{subfigure}[b]{0.32\textwidth}
        \includegraphics[width=0.96\textwidth]{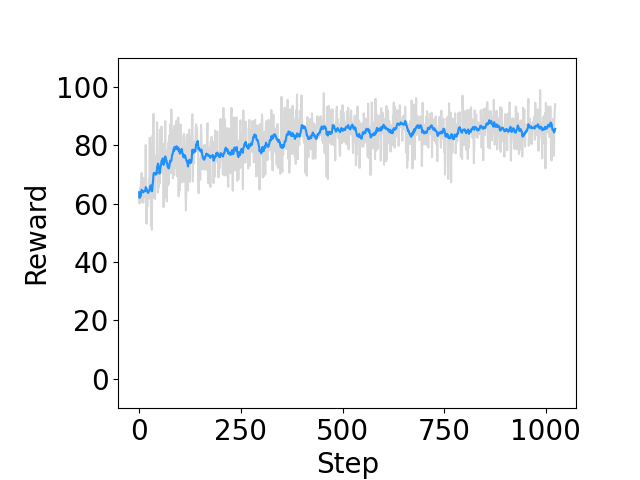}
        \subcaption{}
    \end{subfigure}
    \\
    \begin{subfigure}[b]{0.32\textwidth}
        \includegraphics[width=0.96\textwidth]{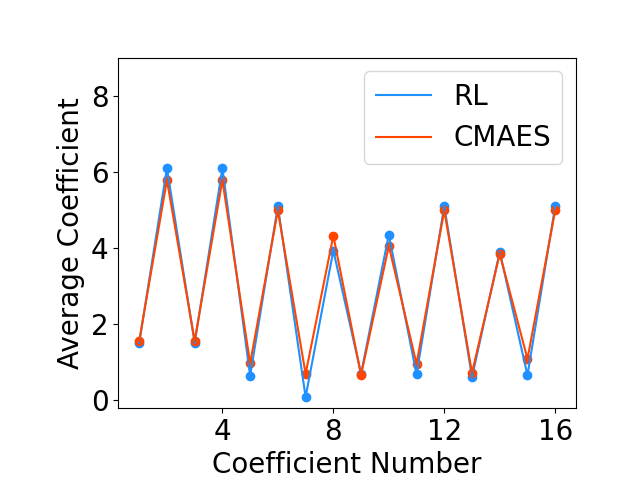}
        \subcaption{}
    \end{subfigure}
    \begin{subfigure}[b]{0.32\textwidth}
        \includegraphics[width=0.96\textwidth]{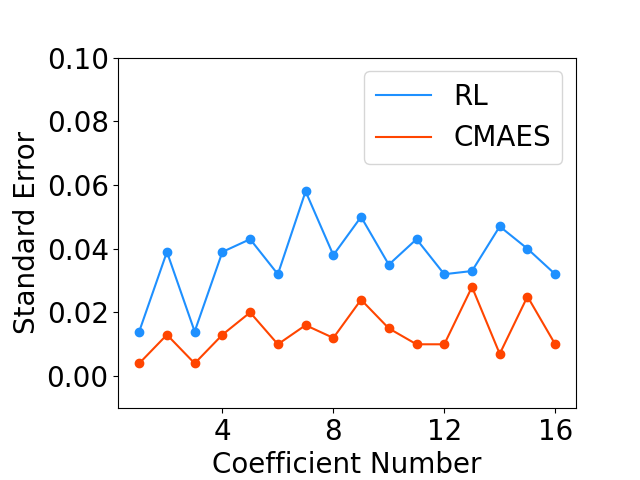}
        \subcaption{}
    \end{subfigure}
    \captionsetup{font=scriptsize}
    \caption{
    Training convergence and statistics from TPE and CMAES.
    (a) Training reward-step curve for TPE and (b) CMAES with thresholds.
    TPE did not find coefficients with high reward values when the thresholds were enabled.
    (c) Training reward-step curve for TPE and (d) CMAES without thresholds.
    (e) Average and (f) standard error of the best 20 coefficient sets (17 coefficients) from RL and CMAES with thresholds.
    TPE and CMAES were chosen to further validate the effectiveness of RL for potential parameterization.
    Owing to their continuous training and the phased RL training loops, only the first 1024 steps were compared.
    When using thresholds, TPE cannot achieve high reward values, which stressed the capability of RL to deal with strong nonlinear problems.
    The critical comparison was the best coefficients from CMAES and RL with thresholds.
    The deviations of the coefficients from RL were significantly higher than those from CMAES.
    As emphasized in the main paper, the transverse strength and toughness were only sampled once during training which required manual post-training performance checks.
    Thus higher deviations from RL were critical to compensate for the insufficient sampling and to find excellent coefficients.
    }
    \label{fig:reward_tpe_cmaes}
\end{figure}

As shown in the Figure \ref{fig:reward_tpe_cmaes}, TPE cannot find a coefficient set with a high reward value after the 1024 steps training.
The real advantage of RL was further demonstrated by the divergence of coefficient training with thresholds by RL and CMAES respectively.
The average and standard errors of the 20 best coefficient sets are compared in Figure~\ref{fig:reward_tpe_cmaes}(e)(f).
For the coefficients, no significant differences were observed between RL and CMAES;
for the standard error, all the coefficients of RL deviated more than those of CMAES.
As emphasized in the main paper, the key feature is the balance of reward and stochasticity, and manual tests must be performed to compensate for the insufficient sampling during training.
Thus, limited stochasticity is necessary for local minima prevention and the coefficient exploration when the sampling is insufficient.

\clearpage

\subsection*{Secondary summits of the bonded geometry parameter distributions}\indent
\addcontentsline{toc}{subsection}{\protect\numberline{}Secondary summits of the bonded geometry parameter distributions}\indent

\begin{figure}[htbp]
    \centering
    \scriptsize
    \begin{subfigure}[b]{0.32\textwidth}
        \centering
        \includegraphics[width=0.96\textwidth]{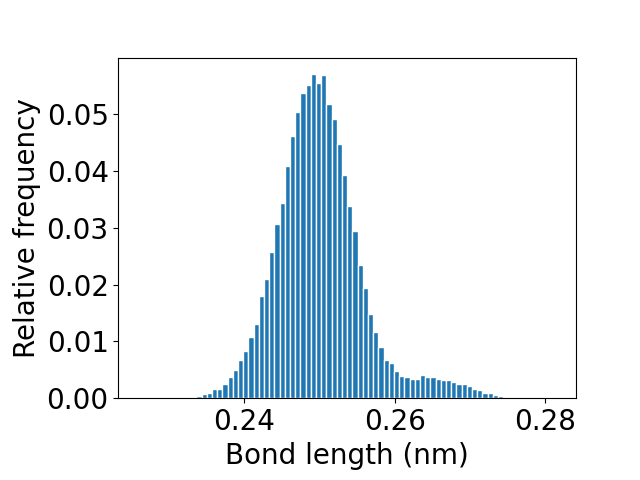}
        \subcaption{}
    \end{subfigure}
    \begin{subfigure}[b]{0.32\textwidth}
        \centering
        \includegraphics[width=0.96\textwidth]{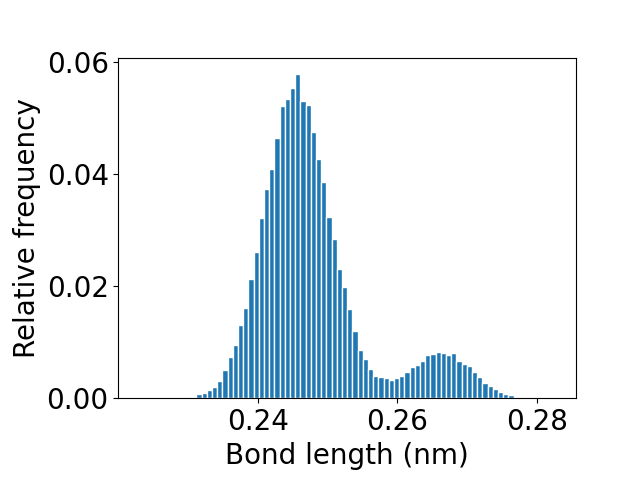}
        \subcaption{}
    \end{subfigure}
    \begin{subfigure}[b]{0.32\textwidth}
        \centering
        \includegraphics[width=0.96\textwidth]{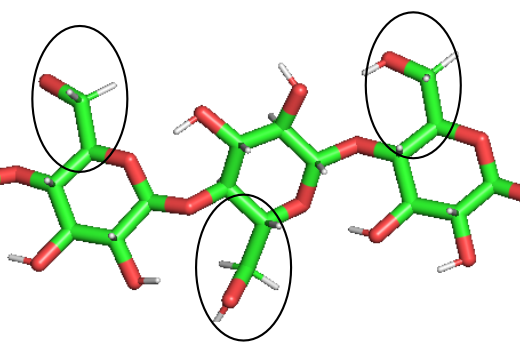}
        \subcaption{}
    \end{subfigure}
    \captionsetup{font=scriptsize}
    \caption{
    Secondary summits of the BD geometry parameter distributions.
    (a) Bond length distributions of CL1-CL2 from mapped crystal trajectories and (b) single-chain trajectories.
    The single-chain trajectories were simulated in a much larger simulation cell with vacuum spaces and governed by the NVT ensemble.
    (c) The secondary plateaus of the BD geometry parameters were caused by the rotational degrees of freedom of the branched hydroxyls (packed into CL2 beads), particularly in the single-chain simulations.
    This implies that the CG model cannot represent a specific rotational movement, that may be critical in certain situations.
    }
    \label{fig:secondary_summits_of_the_boned_geometry_parameter_distributions}
\end{figure}

The harmonic potentials were appropriate for symmetric BD geometry parameter distributions.
However, secondary plateaus or summits were observed in the BD geometry parameter distributions of CL2 beads (Figure~\ref{fig:secondary_summits_of_the_boned_geometry_parameter_distributions}).
This phenomenon was caused by the rotational degrees of freedom of the hydroxyls and was more evident in the single-chain simulations, in which the branch beads were not restrained by HBonds.
The single-chain simulation was under the NVT ensemble, and the only periodic chain is placed in the center of a much larger cell.
Ignoring the hydroxyl rotation is simple, but may cause problems in other cases.

